\documentclass[12pt]{article}
\usepackage{eqsection,subeqnarray,indent,amsfonts,amssymb,amsmath}
\usepackage{bm}    %% For bold math symbols
\usepackage{cite}  %% For writing intervals of citations as xx--yy
\usepackage{pstricks,pst-node,pst-text,pst-3d,pst-plot}  %% Plots
\usepackage{epic,eepic}
\usepackage{rotating}  %% Package \fancybox screws up \tableofcontents 
\usepackage{graphicx}
\usepackage{url} 

\begin{document}

\bibliographystyle{plain}

\date{February 19, 2010\\[1.5mm] revised June 6, 2011}

\title{\vspace*{-1cm}
       Transfer Matrices and Partition-Function Zeros \\
       for Antiferromagnetic Potts Models \\[5mm]
   \hspace*{-10mm}  %% Any less than 10mm and the line folds into two
   \large\bf VI.~Square Lattice with Extra-Vertex Boundary Conditions}

\author{
  {\small Jes\'us Salas}                       \\[-0.2cm]
  {\small\it Instituto Gregorio Mill\'an}      \\[-0.2cm]
  {\small\it and}                              \\[-0.2cm]
  {\small\it Grupo de Modelizaci\'on, Simulaci\'on Num\'erica y 
             Matem\'atica Industrial}          \\[-0.2cm]
  {\small\it Universidad Carlos III de Madrid} \\[-0.2cm]
  {\small\it Avda.\  de la Universidad, 30}    \\[-0.2cm]
  {\small\it 28911 Legan\'es, SPAIN}           \\[-0.2cm]
  {\small\tt JSALAS@MATH.UC3M.ES}              \\[4mm]
  {\small Alan D.~Sokal\thanks{Also at Department of Mathematics,
           University College London, London WC1E 6BT, England.}}  \\[-0.2cm]
  {\small\it Department of Physics}            \\[-0.2cm]
  {\small\it New York University}              \\[-0.2cm]
  {\small\it 4 Washington Place}               \\[-0.2cm]
  {\small\it New York, NY 10003 USA}           \\[-0.2cm]
  {\small\tt SOKAL@NYU.EDU}                    \\[-0.2cm]
  {\protect\makebox[5in]{\quad}}  % To force authors' names to be written
                                  %   vertically, one above another.
                                  % (\author seems to put them side-by-side
                                  %   if there is room.)
  \\
}

\maketitle
\thispagestyle{empty}   % Suppress page number on front page.

\begin{abstract}
We study, using transfer-matrix methods, the partition-function zeros of
the square-lattice $q$-state Potts antiferromagnet at zero temperature 
(= square-lattice chromatic polynomial)
for the boundary conditions
that are obtained from an $m \times n$ grid with free boundary conditions
by adjoining one new vertex adjacent to all the sites in the leftmost column
and a second new vertex adjacent to all the sites in the rightmost column.
We provide numerical evidence that the partition-function zeros
are becoming dense everywhere in the complex $q$-plane
outside the limiting curve $\mathcal{B}_\infty({\rm sq})$
for this model with ordinary (e.g.\ free or cylindrical) boundary conditions.
Despite this, the infinite-volume free energy is perfectly analytic
in this region.
\end{abstract}


\clearpage

\newcommand{\be}{\begin{equation}}
\newcommand{\ee}{\end{equation}}
\newcommand{\<}{\langle}
\renewcommand{\>}{\rangle}
\newcommand{\widebar}{\overline}
\def\reff#1{(\protect\ref{#1})}
\def\spose#1{\hbox to 0pt{#1\hss}}
\def\ltapprox{\mathrel{\spose{\lower 3pt\hbox{$\mathchar"218$}}
 \raise 2.0pt\hbox{$\mathchar"13C$}}}
\def\gtapprox{\mathrel{\spose{\lower 3pt\hbox{$\mathchar"218$}}
 \raise 2.0pt\hbox{$\mathchar"13E$}}}
\def\textprime{${}^\prime$}
\def\proof{\par\medskip\noindent{\sc Proof.\ }}
\def\qed{ $\square$ \bigskip}
\def\proofof#1{\bigskip\noindent{\sc Proof of #1.\ }}
\def\half{ {1 \over 2} }
\def\third{ {1 \over 3} }
\def\twothird{ {2 \over 3} }
\def\smfrac#1#2{{\textstyle{#1\over #2}}}
\def\smhalf{ \smfrac{1}{2} }
\newcommand{\real}{\mathop{\rm Re}\nolimits}
\renewcommand{\Re}{\mathop{\rm Re}\nolimits}
\newcommand{\imag}{\mathop{\rm Im}\nolimits}
\renewcommand{\Im}{\mathop{\rm Im}\nolimits}
\newcommand{\sgn}{\mathop{\rm sgn}\nolimits}
\newcommand{\tr}{\mathop{\rm tr}\nolimits}
\newcommand{\diag}{\mathop{\rm diag}\nolimits}
\newcommand{\Gal}{\mathop{\rm Gal}\nolimits}
\newcommand{\mycup}{\mathop{\cup}}
\newcommand{\Arg}{\mathop{\rm Arg}\nolimits}
\def\hboxscript#1{ {\hbox{\scriptsize\em #1}} }
\def\hboxrm#1{ {\hbox{\scriptsize\rm #1}} }
\def\zhat{ {\widehat{Z}} }
\def\phat{ {\widehat{P}} }
\def\qtilde{ {\widetilde{q}} }
\renewcommand{\mod}{\mathop{\rm mod}\nolimits}
\renewcommand{\emptyset}{\varnothing}
\newcommand{\toto}{\leftrightarrow}

% Script font
\def\scra{\mathcal{A}}
\def\scrb{\mathcal{B}}
\def\scrc{\mathcal{C}}
\def\scrd{\mathcal{D}}
\def\scrf{\mathcal{F}}
\def\scrg{\mathcal{G}}
\def\scrl{\mathcal{L}}
\def\scro{\mathcal{O}}
\def\scrp{\mathcal{P}}
\def\scrq{\mathcal{Q}}
\def\scrr{\mathcal{R}}
\def\scrs{\mathcal{S}}
\def\scrt{\mathcal{T}}
\def\scrv{\mathcal{V}}
\def\scrz{\mathcal{Z}}

% Blackboard font
\def\Z{{\mathbb Z}}
\def\R{{\mathbb R}}
\def\C{{\mathbb C}}
\def\Q{{\mathbb Q}}
\def\N{{\mathbb N}}

% Transfer matrices: use sans-serif font
\def\T{{\mathsf T}}
\def\H{{\mathsf H}}
\def\V{{\mathsf V}}
\def\D{{\mathsf D}}
\def\J{{\mathsf J}}
\def\P{{\mathsf P}}
\def\QQ{{\mathsf Q}}
\def\RR{{\mathsf R}}

% Boldface vectors
\def\bone{{\mathbf 1}}
\def\bv{{\bf v}}
\def\basise{{\bf e}}   % Previously called {\bf v}
\def\basisf{{\bf f}}   % Previously called {\bf w}
\def\startv{{\boldsymbol{\alpha}}}   % Previously called {\bf v}
\def\endv{{\boldsymbol{\omega}}}     % Previously called {\bf u}
\def\bsigma{{\boldsymbol{\sigma}}}

\newtheorem{theorem}{Theorem}[section]
\newtheorem{definition}[theorem]{Definition}
\newtheorem{proposition}[theorem]{Proposition}
\newtheorem{lemma}[theorem]{Lemma}
\newtheorem{corollary}[theorem]{Corollary}
\newtheorem{conjecture}[theorem]{Conjecture}
\newtheorem{question}[theorem]{Question}

% Array for subscripts

\newenvironment{sarray}{
          \textfont0=\scriptfont0
          \scriptfont0=\scriptscriptfont0
          \textfont1=\scriptfont1
          \scriptfont1=\scriptscriptfont1
          \textfont2=\scriptfont2
          \scriptfont2=\scriptscriptfont2
          \textfont3=\scriptfont3
          \scriptfont3=\scriptscriptfont3
        \renewcommand{\arraystretch}{0.7}
        \begin{array}{l}}{\end{array}}

\newenvironment{scarray}{
          \textfont0=\scriptfont0
          \scriptfont0=\scriptscriptfont0
          \textfont1=\scriptfont1
          \scriptfont1=\scriptscriptfont1
          \textfont2=\scriptfont2
          \scriptfont2=\scriptscriptfont2
          \textfont3=\scriptfont3
          \scriptfont3=\scriptscriptfont3
        \renewcommand{\arraystretch}{0.7}
        \begin{array}{c}}{\end{array}}

%
% Variants of \binom  (defined using the AMS "genfrac" command)
%
\newcommand{\stirlingsubset}[2]{\genfrac{\{}{\}}{0pt}{}{#1}{#2}}
\newcommand{\stirlingcycle}[2]{\genfrac{[}{]}{0pt}{}{#1}{#2}}
\newcommand{\assocstirlingsubset}[3]{%
{\genfrac{\{}{\}}{0pt}{}{#1}{#2}}_{\! \ge #3}}
\newcommand{\assocstirlingcycle}[3]{{\genfrac{[}{]}{0pt}{}{#1}{#2}}_{\ge #3}}
\newcommand{\euler}[2]{\genfrac{\langle}{\rangle}{0pt}{}{#1}{#2}}
\newcommand{\eulergen}[3]{{\genfrac{\langle}{\rangle}{0pt}{}{#1}{#2}}_{\! #3}}
\newcommand{\eulersecond}[2]{\left\langle\!\! \euler{#1}{#2} \!\!\right\rangle}
\newcommand{\eulersecondgen}[3]{%
{\left\langle\!\! \euler{#1}{#2} \!\!\right\rangle}_{\! #3}}
\newcommand{\binomvert}[2]{\genfrac{\vert}{\vert}{0pt}{}{#1}{#2}}

%
% \pentagon AND \hexagon FOR FIGURE CAPTIONS
%
\newcommand{\pentagon}{
\psset{xunit=0.15cm}
\psset{yunit=0.15cm}
\pspicture(-0.7,-0.8)(1.5,1.5)
\psline*(-0.951,0.309)(-0.588,-0.809)(0.588,-0.809)(0.951,0.309)(0,1)%
(-0.951,0.309)
\endpspicture
}

\newcommand{\hexagon}{
\psset{xunit=0.15cm}
\psset{yunit=0.15cm}
\pspicture(-0.7,-0.8)(1.5,1.5)
\psline*(-0.866,0.5)(-0.866,-0.5)(0,-1)(0.866,-0.5)(0.866,0.5) (0,1)(-0.866,0.5)
\endpspicture
}

\tableofcontents
\clearpage
%%%%%%%%%%%%%%%%%%%%%%%%%%%%%%%%%%%%%%%%%%%%%%%%%%%%%%%%%%%
%
% INTRODUCTION
%
%%%%%%%%%%%%%%%%%%%%%%%%%%%%%%%%%%%%%%%%%%%%%%%%%%%%%%%%%%%
\section{Introduction} \label{sec.intro}

It is well known that phase transitions do not occur
in statistical-mechanical systems in finite volume,
but occur only in the infinite-volume limit.
One approach to studying phase transitions was introduced
by Yang and Lee \cite{Yang-Lee_52} in 1952,
and involves investigating the zeros of the finite-volume partition function
when one or more physical parameters (e.g.\ temperature or magnetic field)
are allowed to take {\em complex}\/ values.
Lee and Yang showed, under mild conditions,
that the accumulation points of these zeros in the infinite-volume limit
constitute the only possible loci of phase transitions;
away from such accumulation points of zeros,
the infinite-volume free energy is analytic.

We would like to stress the word ``possible'' in the preceding sentence.
The finite-volume free energy is of course singular
at any zero of the finite-volume partition function
(since $\log 0 = -\infty$);
but it can happen that such a singularity disappears
in the infinite-volume limit.
To take a trivial example, suppose that the partition function
in volume~$n$ is $Z_n(x) = x$, where $x$ is a parameter;
then the finite-volume free energy
$f_n(x) = n^{-1} \log Z_n(x) = n^{-1} \log x$ is singular at $x=0$,
but the infinite-volume free energy
$f(x) = \lim\limits_{n \to\infty} f_n(x) = 0$ is analytic at $x=0$.\footnote{
   One could object that in this case we actually have
   $$ f(x)  \;=\;  \begin{cases}
                        0       & \text{for $x \neq 0$}  \\
                        -\infty & \text{for $x = 0$} 
                   \end{cases}
   $$
   which is not analytic at $x=0$
   (though the singularity at $x=0$ is removable).
   Here is a variant that avoids this objection: $Z_n(x) = x - 1/n$.
   Then the finite-volume free energy $f_n(x) = n^{-1} \log Z_n(x)$
   is singular at $x=1/n$,
   but it is easy to see that $f(x) = \lim\limits_{n \to\infty} f_n(x) = 0$
   for all $x$ (including $x=0$).
}

In this paper we would like to give an extreme example of this
phenomenon, in which the zeros of the finite-volume partition functions
appear to be {\em dense}\/ in a large region of the complex plane,
while the infinite-volume free energy is perfectly analytic there.\footnote{
   See also Section~\ref{subsec.density} for a brief discussion
   of the general question:
   When does an accumulation of partition-function zeros
   signal a nonanalyticity of the infinite-volume free energy?
}
Our example consists of the square-lattice Potts antiferromagnet
at zero temperature ---
a model that we have studied in a series of previous papers
\cite{transfer1,transfer2,transfer3,transfer4,Jacobsen-Salas_toroidal,transfer5}
--- but with some unusual boundary conditions.
We begin by reviewing the needed background on Potts models
and chromatic polynomials.

The Potts model \cite{Potts_52,Wu_82,Wu_84}
on a regular lattice ${\cal L}$ is characterized by two parameters:
the number $q$ of Potts spin states,
and the nearest-neighbor coupling $v = e^{\beta J}-1$.\footnote{
   Here we are considering only the {\em isotropic}\/ model,
   in which each nearest-neighbor edge is assigned the same coupling $v$.
   In a more refined analysis, one could put (for example)
   different couplings $v_1,v_2$ on the horizontal and vertical edges
   of the square lattice, different couplings $v_1,v_2,v_3$ on the
   three orientations of edges of the triangular or hexagonal lattice, etc.
}
Initially $q$ is a positive integer
and $v$ is a real number in the interval $[-1,+\infty)$,
but the Fortuin--Kasteleyn representation
(reviewed in Section~\ref{sec.setup.FK} below)
shows that the partition function $Z_G(q,v)$ of the $q$-state Potts model
on any finite graph $G$ is in fact a {\em polynomial}\/ in $q$ and $v$.
This allows us to interpret $q$ and $v$ as taking arbitrary
real or even complex values,
and to study the phase diagram of the Potts model in
the real $(q,v)$-plane or in complex $(q,v)$-space.
In particular, by studying the zeros of $Z_G(q,v)$ in complex $(q,v)$-space
for larger and larger pieces of the lattice ${\cal L}$,
we can locate the possible loci of phase transitions
in the real $(q,v)$-plane and more generally in complex $(q,v)$-space.

The partition function for $m \times n$ lattices
can be efficiently computed using {\em transfer matrices}\/.
Though the dimension of the transfer matrix
(and thus the computational complexity)
grows exponentially in the width $m$ ---
thereby restricting us in practice to widths $m \ltapprox 10$--30 ---
it is straightforward, by iterating the transfer matrix,
to handle quite large lengths $n$.
Indeed, by implementing the transfer-matrix method {\em symbolically}\/
(i.e., as polynomials in $q$ and/or $v$)
and using the Beraha--Kahane--Weiss theorem
(reviewed in Section~\ref{sec.setup.BKW}),
we can handle directly the limit $n \to\infty$
and compute the limiting curves $\scrb_m$ of partition-function zeros.
At a second stage we attempt to extrapolate these curves to $m=\infty$.

Since the problem of computing the phase diagram in
complex $(q,v)$-space is difficult, it has proven convenient
to study first certain ``slices'' through $(q,v)$-space,
in which one parameter is fixed (usually at a real value)
while the remaining parameter is allowed to vary in the complex plane.
One very interesting special case is the chromatic polynomial ($v=-1$),
which corresponds to the zero-temperature limit
of the Potts antiferromagnet ($\beta J=-\infty$).
In previous papers
\cite{transfer1,transfer2,transfer3,transfer4,Jacobsen-Salas_toroidal,transfer5}
we have used symbolic transfer-matrix methods
to study the square-lattice and triangular-lattice chromatic polynomials
for free, cylindrical, cyclic and toroidal boundary conditions.\footnote{
   See also the bibliographies of
   \cite{transfer1,transfer2,transfer3,transfer4,Jacobsen-Salas_toroidal,transfer5}
   for reference to the important related works of
   Shrock and collaborators.
}${}^,$\footnote{
   We adopt Shrock's \cite{Shrock_BCC99} terminology
   for boundary conditions:
   free ($m_{\rm F} \times n_{\rm F}$),
   cylindrical ($m_{\rm P} \times n_{\rm F}$),
   cyclic ($m_{\rm F} \times n_{\rm P}$),
   toroidal ($m_{\rm P} \times n_{\rm P}$),
   M\"obius ($m_{\rm F} \times n_{\rm TP}$) and
   Klein bottle ($m_{\rm P} \times n_{\rm TP}$).
   Here the first dimension ($m$) corresponds to the transverse (``short'')
   direction, while the second dimension ($n$) corresponds to the
   longitudinal (``long'') direction.
   The subscripts F, P and TP denote free, periodic and twisted-periodic
   boundary conditions, respectively.
}
Here we shall study the square-lattice model for the boundary conditions
that are obtained from an $m_{\rm F} \times n_{\rm F}$ square grid
($m$~columns, $n$~rows, free boundary conditions in both directions)
by adjoining one new vertex adjacent to all the sites in the leftmost column
and a second new vertex adjacent to all the sites in the rightmost column
(see Figure~\ref{fig_Smn}a).

\begin{figure}
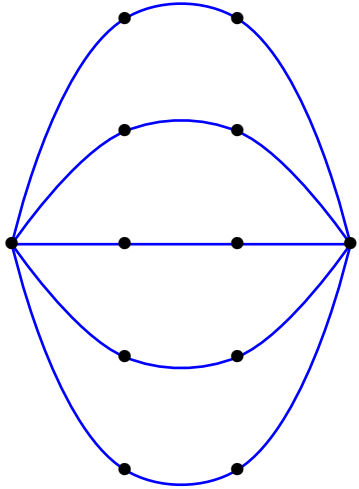

\centering
     \psset{xunit=1.5cm}
     \psset{yunit=1.5cm}
     \pspicture(0,0)(3,4)
     \pscurve[linecolor=blue,linewidth=1.0pt](0,2)(1,0)(2,0)(3,2)
     \pscurve[linecolor=blue,linewidth=1.0pt](0,2)(1,1)(2,1)(3,2)
     \pscurve[linecolor=blue,linewidth=1.0pt](0,2)(1,3)(2,3)(3,2)
     \pscurve[linecolor=blue,linewidth=1.0pt](0,2)(1,4)(2,4)(3,2)
     \psline[linecolor=blue,linewidth=1.0pt] (0,2)(1,2)(2,2)(3,2)
     \multirput{0}(1,0)(0,1){5}{$\bullet$}
     \multirput{0}(2,0)(0,1){5}{$\bullet$}
     \multirput{0}(0,2)(3,0){2}{$\bullet$}
     \endpspicture

     \vspace*{1cm}
\caption{\label{fig_thetasp}
Generalized theta graph $\Theta^{(3,5)}$ formed by $p=5$ chains in parallel,
each consisting in $s=3$ edges in series. 
}
\end{figure}

The motivation for studying this family of graphs
comes in part from work of one of the authors \cite{Sokal_chromatic_roots}
on the {\em generalized theta graphs}\/ \cite{gen_theta}:
namely, $\Theta^{(s,p)}$ is defined to be
the graph consisting of $p$~chains in parallel between a pair of endvertices,
each chain consisting of $s$~edges in series (Figure~\ref{fig_thetasp}).
For this family, the following result holds:

\begin{theorem}
  {\bf \protect\cite[Theorems~1.1, 1.2 and 1.4]{Sokal_chromatic_roots}}
   \label{thm1.1}
The set of chromatic roots of all the graphs $\Theta^{(s,p)}$
is dense in the entire complex $q$-plane
with the possible exception of the disc $|q-1| < 1$.
\end{theorem}

The proof of Theorem~\ref{thm1.1} proceeds in two steps:
First, one fixes $s$, and shows that as $p \to\infty$
the chromatic roots of the graphs $\Theta^{(s,p)}$
accumulate densely on an explicitly computable real algebraic curve $\scrc_s$.
This argument is a straightforward application of the
Beraha--Kahane--Weiss theorem.
Second, one shows that as $s \to\infty$
the curves $\scrc_s$ become dense in $\C \setminus \{|q-1| < 1\}$.
This requires an {\em ad hoc}\/ (but not terribly difficult) argument.
See Section~\ref{sec.setup.theta} for a slightly more detailed summary.

The computations arising in the proof of Theorem~\ref{thm1.1}
are based on the series and parallel reduction laws for the
Potts-model partition function
\cite{Sokal_chromatic_roots,Sokal_bcc2005}.
This approach works because the graphs $\Theta^{(s,p)}$
are {\em series-parallel}\/:
we can compute the partition function $Z_{\Theta^{(s,p)}}(q,v)$
by computing the ``effective coupling'' $v_{\rm eff}$
arising from putting $s$ edges (each of weight $v$) in series
and then placing $p$ such chains in parallel.

No such simple approach can work for the graphs $S_{m,n}$,
because they are not series-parallel.
Still, the graphs $S_{m,n}$ and the generalized theta graphs
are closely related:
indeed, the generalized theta graph $\Theta^{(m+1,n)}$
is obtained from $S_{m,n}$ simply by deleting all the vertical edges.
So it would not be surprising if the density-of-roots phenomenon
observed for the generalized theta graphs occurred also for the
family $S_{m,n}$.  In this paper we would like to present some
convincing evidence --- which, however, falls short at present
of a rigorous mathematical proof --- that this is indeed the case
(see in particular Conjecture~\ref{conj.dense} below).

As in the analysis \cite{Sokal_chromatic_roots}
of the generalized theta graphs, we proceed in two steps.
First we fix $m$, and show that as $n \to\infty$
the chromatic roots of the graphs $S_{m,n}$
accumulate densely on a real algebraic curve $\scrb_m$
(as well as on certain isolated points).
This argument is once again a straightforward application of the
Beraha--Kahane--Weiss theorem.
The more difficult problem is to compute the curves $\scrb_m$
and to study their behavior as $m \to\infty$.
In this paper we shall use the transfer-matrix formalism
developed in \cite{Blote_82,transfer1,transfer2,transfer3,transfer5}
to compute the curves $\scrb_m$ for $m \le 6$
(see Figures~\ref{figure_sq_1F}--\ref{figure_sq_6F}
 in Sections~\ref{sec.numerical.bm} and \ref{sec.thermo.asym} below).
We will see that the curve $\scrb_m$ contains $2m$ outward branches
running to infinity in the complex $q$-plane
and equally spaced in asymptotic angle.
Moreover, these branches appear to become dense
as $m \to\infty$ everywhere in the complex $q$-plane
outside the limiting curve $\scrb_\infty({\rm sq})$
for this model with ordinary (e.g.\ free or cylindrical)
boundary conditions --- a curve that we have estimated numerically
in previous papers \cite{transfer1,transfer2}
and which is entirely contained inside the disc $|q-1| \le 3$
(to give a crude bound).
If this is indeed the case, it follows that the zeros
of the chromatic polynomials of the graphs $S_{m,n}$
are likewise becoming dense everywhere outside $\scrb_\infty({\rm sq})$
and in particular everywhere in the region $|q-1| > 3$.

As will be explained in Section~\ref{sec.thermo.asym} below
(drawing on results from Section~\ref{sec.numerical}),
these outward branches arise from the fact that the
transfer matrix $\T''(m)$ for this model has two eigenvalues
--- coming from two sectors in the block-diagonalization of $\T''(m)$ ---
that are almost equal at large $q$,
with a relative difference of order $q^{-m}$,
and which compete for dominance:
the outward branches of the curves $\scrb_m$ are nothing other than
the loci where these two eigenvalues are equal in modulus.
On the other hand, the difference between these two eigenvalues disappears
as $m \to\infty$ when $|q|$ is large enough,
with the result that the infinite-volume free energy
is perfectly analytic at large $|q|$.

A behavior of this type was observed recently
by Jacobsen, Richard and Salas \cite{Jacobsen_RSOS}
for the Potts model with {\em cyclic}\/ boundary conditions
(i.e.\ {\em periodic}\/ boundary conditions in the longitudinal direction)
when $q$ is fixed at a Beraha number
and the temperature variable $v$ is taken complex.
On the square (resp.\ triangular) lattice,
there are two eigenvalues
--- coming once again from two sectors in the block-diagonalization of
 the transfer matrix ---
that are almost equal at large $v$,
with a relative difference of order $v^{-m}$ (resp.\ $v^{-(2m-1)}$).
This leads to $2m$ (resp.\ $4m-2$) outward branches in the curve $\scrb_m$
that appear to become dense as $m \to\infty$
everywhere outside a bounded region in the complex $v$-plane.
These same outward branches were observed earlier by Shrock and Chang
for $m=1,2$ \cite{Shrock_00_TEMP} and $m=3$ \cite{Chang-Shrock_01_TEMP}
on the square lattice
and for $m=2$ on the triangular lattice \cite{Chang-Shrock_00_TRI_TEMP};
they exist for {\em all}\/ $q \neq 0,1$, not only Beraha numbers.

The principle underlying all these examples is a general one:
Suppose that the transfer matrix at some parameter value $x_0$
has two dominant eigenvalues of equal modulus,
call them $\lambda_1(x_0)$ and $\lambda_2(x_0) = e^{i\theta} \lambda_1(x_0)$,
and that for parameters $x \approx x_0$ we have
$\lambda_2(x) - e^{i\theta} \lambda_1(x) = A (x-x_0)^k + o((x-x_0)^k)$
with $A \neq 0$ and some integer $k \ge 1$
[with the obvious modifications in case $x_0 = \infty$].
Then the curve $\scrb_m$ contains $2k$ branches passing through $x_0$,
with an asymptotic angle $\pi/k$ between neighboring branches.
If $k=1$ this is simply the generic situation for equimodular curves;
if $k>1$ it corresponds to a higher-order crossing.
In particular, if $k$ tends to infinity as $m \to\infty$,
these branches are likely to become dense in a neighborhood of $x_0$.

The plan of this paper is as follows:
In Section~\ref{sec.setup} we review some needed background
on chromatic and Tutte polynomials,
transfer matrices, and the Beraha--Kahane--Weiss theorem.
In Section~\ref{sec.transfer} we present the transfer-matrix theory
for the family $S_{m,n}$ and we prove some properties of the
dominant diagonal elements and the dominant eigenvalues.
In Section~\ref{sec.numerical} we present the results of
our transfer-matrix computation of the large-$q$ expansion
of the leading eigenvalues for $1 \le m \le 11$
and note some surprising properties.
In Section~\ref{sec.thermo} we study the thermodynamic limit
$m\to\infty$ of the corresponding free energies.
In Section~\ref{sec.numerical.bm} we present the numerically computed
limiting curves $\scrb_m$ for widths $1 \le m \le 6$.
In Section~\ref{sec.thermo.asym} we present some conjectures
on the behavior of these limiting curves as $m \to \infty$
and we relate these conjectures to the properties of the eigenvalues
that were empirically observed in Section~\ref{sec.numerical}.
Finally, in Section~\ref{sec.discussion} we conclude with some
brief discussion and suggestions for future research.
In Appendix~\ref{appendix.polymermodel}
we prove Proposition~\ref{prop.TMblockdiag.3}
concerning the dominant diagonal entries in the transfer matrix.
In Appendix~\ref{sec.dim} we compute a general formula
for the dimensions of our transfer matrices as a function of the width $m$.
In Appendix~\ref{sec.transfer.bis} we present the transfer-matrix theory
for the family $\widehat{S}_{m,n}$ obtained by a $90^\circ$ rotation
of the grid; this plays an important role in checking the correctness
of our transfer-matrix computations.

%%%%%%%%%%%%%%%%%%%%%%%%%%%%%%%%%%%%%%%%%%%%%%%%%%%%%%%%%%%
%
% SET UP
%
%%%%%%%%%%%%%%%%%%%%%%%%%%%%%%%%%%%%%%%%%%%%%%%%%%%%%%%%%%%
\section{Preliminaries} \label{sec.setup}

In this section we review briefly some needed background
on chromatic and Tutte polynomials (Section~\ref{sec.setup.FK}),
transfer matrices (Section~\ref{sec.setup.transfer}),
and the Beraha--Kahane--Weiss theorem (Section~\ref{sec.setup.BKW}).
We also recall the easy analysis of the chromatic roots of
bi-fans and bipyramids (Section~\ref{sec.setup.bipyr})
and briefly summarize the argument from \cite{Sokal_chromatic_roots}
concerning the chromatic roots of
generalized theta graphs (Section~\ref{sec.setup.theta});
both of these will serve as models for our analysis here
of the graphs $S_{m,n}$.

\subsection{Chromatic polynomials, Potts models, and all that}
    \label{sec.setup.FK}

Let $G=(V,E)$ be a finite undirected graph, and let $q$ be a positive integer.
Then the {\em $q$-state Potts-model partition function}\/
for the graph $G$ is defined by the Hamiltonian
\begin{equation}
   H_{\rm Potts}(\bsigma)  \;=\;
   - \sum_{e=ij \in E} J_e \, \delta(\sigma_i, \sigma_j)
   \;,
 \label{def.HPotts}
\end{equation}
where the spins $\bsigma = \{\sigma_i\}_{i \in V}$
take values in $\{ 1,2,\ldots,q \}$,
the $J_e$ are coupling constants, and the $\delta$ is the Kronecker delta
\begin{equation}
   \delta(a,b)   \;=\;
   \begin{cases} 1  & \quad \mbox{\rm if $a=b$} \\ 
                 0  & \quad \mbox{\rm if $a \neq b$} 
   \end{cases} 
\end{equation}
The partition function can then be written as
\begin{equation}
   Z_G^{\rm Potts}(q, {\bf v})   \;=\;
   \sum_{ \sigma \colon\, V \to \{ 1,2,\ldots,q \} }
   \; \prod_{e=ij \in E}  \,
      \bigl[ 1 + v_e \delta(\sigma_i, \sigma_j) \bigr]
   \;,
 \label{def.ZPotts}
\end{equation}
where $v_e = e^{\beta J_e} - 1$.
Please note, in particular, that if we set $v_e = -1$ for all edges $e$,
then $Z_G^{\rm Potts}$ gives weight 1 to each proper coloring
and weight 0 to each improper coloring,
and so counts the proper colorings.
Proper $q$-colorings ($v_e=-1$) thus correspond to
the zero-temperature ($\beta \to +\infty$) limit
of the antiferromagnetic ($J_e < 0$) Potts model.

It is far from obvious that $Z_G^{\rm Potts}(q, {\bf v})$,
which is defined separately for each positive integer $q$,
is in fact the restriction to $q \in \Z_+$
of a {\em polynomial}\/ in $q$.
But this is in fact the case, and indeed we have:

\begin{theorem}[Fortuin--Kasteleyn \protect\cite{Kasteleyn_69,Fortuin_72}
                representation of the Potts model]
   \label{thm.FK}
\hfill\break
\vspace*{-4mm}
\par\noindent
For integer $q \ge 1$,
\begin{equation}
   Z_G^{\rm Potts}(q, {\bf v}) \;=\;  
   \sum_{ A \subseteq E }  q^{k(A)}  \prod_{e \in A}  v_e
   \;,
 \label{eq.FK.identity}
\end{equation}
where $k(A)$ denotes the number of connected components
in the subgraph $(V,A)$.
\end{theorem}

\proof
In \reff{def.ZPotts}, expand out the product over $e \in E$,
and let $A \subseteq E$ be the set of edges for which the term
$v_e \delta(\sigma_i, \sigma_j)$ is taken.
Now perform the sum over maps $\sigma \colon\, V \to \{ 1,2,\ldots,q \}$:
in each component of the subgraph $(V,A)$
the color $\sigma_i$ must be constant,
and there are no other constraints.
We immediately obtain \reff{eq.FK.identity}.
\qed

{\bf Historical Remark.}
The subgraph expansion \reff{eq.FK.identity} was discovered by
Birkhoff \cite{Birkhoff_12} and Whitney \cite{Whitney_32a}
for the special case $v_e = -1$ (see also Tutte \cite{Tutte_47,Tutte_54});
in its general form it is due to
Fortuin and Kasteleyn \cite{Kasteleyn_69,Fortuin_72}
(see also \cite{Edwards-Sokal}).

\bigskip

The foregoing considerations motivate defining the
{\em multivariate Tutte polynomial}\/ of the graph $G$:
\begin{equation}
   Z_G(q, {\bf v})   \;=\;
   \sum_{A \subseteq E}  q^{k(A)}  \prod_{e \in A}  v_e
   \;,
 \label{def.ZG}
\end{equation}
where $q$ and ${\bf v} = \{v_e\}_{e \in E}$ are commuting indeterminates.
If we set all the edge weights $v_e$ equal to the same value $v$,
we obtain a two-variable polynomial that is equivalent to the
standard Tutte polynomial $T_G(x,y)$ after a simple change of variables
(see \cite{Sokal_bcc2005}).
If we set all the edge weights $v_e$ equal to $-1$,
we obtain the {\em chromatic polynomial} $P_G(q) = Z_G(q,-1)$.

Further information on the multivariate Tutte polynomial
$Z_G(q,{\bf v})$ can be found in a recent survey article
\cite{Sokal_bcc2005}.

\subsection{Transfer matrices}   \label{sec.setup.transfer}

For any family of graphs $G_n = (V_n,E_n)$
consisting of $n$ identical ``layers''
with identical connections between adjacent layers,
the multivariate Tutte polynomials of the $G_n$
(with edge weights likewise repeated from layer to layer)
can be written in terms of a transfer matrix \cite{Blote_82,transfer1}.
Here we briefly summarize the needed formalism \cite{transfer1}
specialized to the case of an $m \times n$ square lattice
with free boundary conditions in both directions.
The modifications needed to handle the extra sites at left and right
in the graphs $S_{m,n}$ will be discussed in Section~\ref{sec.transfer}.

Consider the $m \times n$ square grid
with edge weights $v_{i,i+1}$ on the horizontal edges ($1 \le i \le m-1$)
and $v_i$ on the vertical edges ($1 \le i \le m$).
We fix the ``width'' $m$ and consider the family of graphs $G_n$
obtained by varying the ``length'' $n$;
our goal is to calculate the multivariate Tutte polynomials
$Z_{G_n}(q, {\bf v})$ for this family by building up the graph $G_n$
layer by layer.
What makes this a bit tricky is the nonlocality of the factor
$q^{k(A)}$ in \reff{def.ZG}.
At the end we will need to know the number of connected components
in the subgraph $(V_n,A)$;  in order to be able to compute this,
we shall keep track, as we go along, of which sites in the
current ``top'' layer are connected to which other sites in that layer
by a path of occupied edges (i.e.\ edges of $A$) in lower layers.
Thus, we shall work in the basis of connectivities of the top layer,
whose basis elements $\basise_{\scrp}$ are indexed by
partitions $\scrp$ of the single-layer vertex set $\{1,\ldots,m\}$.
The elementary operators we shall need are:
\begin{itemize}
   \item The {\em join operators}\/
\be
   \J_{ij} \basise_{\scrp} \;=\; \basise_{\scrp \bullet ij}  \;,
\ee
where $\scrp \bullet ij$ is the partition obtained from $\scrp$
by amalgamating the blocks containing $i$ and $j$
(if they were not already in the same block).
Note that all these operators commute.
   \item The {\em detach operators}\/
\be
   \D_i \basise_{\scrp} \;=\;
        \begin{cases} 
        \basise_{\scrp \setminus i} & \text{if $\{i\} \notin \scrp$} \\[2mm] 
        q \basise_{\scrp}           & \text{if $\{i\} \in \scrp$} 
        \end{cases} 
\label{def_detach}
\ee
where $\scrp \setminus i$ is the partition obtained from $\scrp$
by detaching $i$ from its block (and thus making it a singleton).
Note that these operators commute as well.
\end{itemize}
Note, finally, that $\D_k$ commutes with $\J_{ij}$ whenever
$k \notin \{i,j\}$.

The horizontal transfer matrix, which adds a row of horizontal edges, is
\be
   \H   \;=\;
     \prod_{i=1}^{m-1} (1 + v_{i,i+1} \J_{i,i+1})
   \label{def_H}
\ee
(note that all the operators in the product commute).
The vertical transfer matrix, which adds a new row of sites along
with the corresponding vertical edges,
is
\be
   \V  \;=\;  \prod_{i=1}^{m} (v_i I + \D_i)
\label{def_V}
\ee
(note once again that all the operators commute).
The multivariate Tutte polynomial for $G_n$ is then given \cite{transfer1}
by the formula
\be
   Z_{G_n}(q, {\bf v})   \;=\;
     \endv^{\rm T} \H (\V \H)^{n-1} \basise_{\rm id}
   \;,
   \label{Z_free_FK}
\ee
where ``id'' denotes the partition in which each site $i \in \{1,\ldots,m\}$
is a singleton, and the ``end vector'' $\endv^{\rm T}$ is defined by
\be
   \endv^{\rm T} \basise_{\scrp}   \;=\;  q^{|\scrp|}
   \;.
 \label{def_uT}
\ee
The transfer matrix is thus
\be
   \T  \;=\;  \V \H   \;.
\label{def_T}
\ee

In what follows we shall use a convenient shorthand notation for the
basis vectors $\basise_\scrp$:
namely, we denote by $\bone$ the basis element $\basise_{\rm id}$
corresponding to the partition in which each site
is a singleton, and we denote the action on $\basise_{\rm id}$
of a join operator $\J_{ij}$ by the Kronecker deltas $\delta_{ij}$.
Thus, for instance, for $m=3$ we have the five basis vectors
\begin{subeqnarray}
   \basise_{\rm id} \,\equiv\,
   \basise_{ \{\, \{1\}, \{2\}, \{3\} \,\} }   & = &  \bone  \\[1mm]
   \basise_{ \{\, \{1,2\}, \{3\} \,\} }        & = &  \delta_{12}  \\[1mm]
   \basise_{ \{\, \{1,3\}, \{2\} \,\} }        & = &  \delta_{13}  \\[1mm]
   \basise_{ \{\, \{1\}, \{2,3\} \,\} }        & = &  \delta_{23}  \\[1mm]
   \basise_{ \{\, \{1,2,3\} \,\} }             & = &  \delta_{123}
       \,\equiv\, \delta_{12} \delta_{13} \delta_{23}
       \,=\, \delta_{12} \delta_{13}
       \,=\, \delta_{12} \delta_{23}
       \,=\, \delta_{13} \delta_{23}
 \label{def.delta.shorthand}
\end{subeqnarray}
For $m \ge 4$ we also have basis vectors that are products of deltas,
such as $\delta_{13} \delta_{24}$ and so forth.
This notation allows describing partitions and their
corresponding basis vectors in a fairly compact way.

In principle we are working here in the space spanned by the basis vectors
$\basise_{\scrp}$ for all partitions $\scrp$ of $\{1,\ldots,m\}$;
the dimension of this space is given by the Bell number $B_m$
\cite{Stanley_86,Stanley_99,deBruijn_61,Sloane_on-line}.
However, it is easy to see,
on topological grounds (thanks to the planarity of the $G_n$),
that only {\em non-crossing}\/ partitions can arise.
(A partition is said to be {\em non-crossing}\/ if
 $a < b < c < d$ with $a,c$ in the same block and $b,d$ in the same block
 imply that $a,b,c,d$ are all in the same block.)
The number of non-crossing partitions of $\{1,\ldots,m\}$
is given by the Catalan number \cite{Stanley_99,Sloane_on-line}
\be
   C_m  \;=\; {(2m)! \over m! \, (m+1)!}  \;=\; {1 \over m+1} \, {2m \choose m}
   \;.
\label{def_Cm}
\ee

When the horizontal couplings $v_{i,i+1}$ are all equal to $-1$
(which is the case for the chromatic polynomial),
then the horizontal operator $\H$ is a {\em projection}\/,
and we can work in its image subspace by using the modified transfer matrix
$\T' = \H \V \H$ in place of $\T = \V \H$,
and using the basis vectors
\be
   \basisf_{\scrp}   \;=\;   \H \basise_{\scrp}
 \label{def_wP}
\ee
in place of $\basise_{\scrp}$.
Please note that $\basisf_{\scrp} = 0$ if $\scrp$ has
any pair of nearest neighbors in the same block.
We thus work in the space spanned by the basis vectors $\basisf_{\scrp}$
where $\scrp$ is a non-crossing non-nearest-neighbor partition of
$\{1,\ldots,m\}$.
The dimension of this space is given
\cite[Proposition 3.6]{Simion_91} \cite{Klazar_98}
by the Motzkin number $M_{m-1}$,
where \cite{Riordan_75,Donaghey_77,Gouyou_88,Bernhart_99,%
Stanley_99,Sloane_on-line}\footnote{
   {\em Warning:}\/  Several references use the notation $m_n$
   to denote what we call $M_n$;
   and one reference \cite{Donaghey_77} writes $M_n$
   to denote a {\em different}\/ sequence.
}
\be
   M_n  \;=\;  \sum\limits_{j=0}^{\lfloor n/2 \rfloor}
                  {n \choose 2j} \, C_j
   \;.
 \label{def_Motzkin}
\ee

Finally, spatial symmetries further restrict the subspace
whenever the couplings $v_{i,i+1}$ and $v_i$
are invariant under the symmetry.
Here the relevant symmetry is reflection in the center line of the strip.
For reflection-invariant couplings,
we can work in the space of equivalence classes of
non-crossing non-nearest-neighbor partitions modulo reflection.
We denote the corresponding symmetry-reduced transfer matrix by $\T''$,
and its dimension $\mbox{\rm SqFree}(m)$
is then given by the number of these equivalence classes.
The exact expression for $\mbox{\rm SqFree}(m)$ was obtained in 
\cite[Theorem 2]{Tutte_sq_02}:
\be
\mbox{\rm SqFree}(m) \;=\; {1\over 2} M_{m-1} + {(m'-1)! \over 2} 
 \sum\limits_{j=0}^{\lfloor m'/2 \rfloor} {m'-j \over (j!)^2 (m'-2j)!}
\label{def_SqFree}
\ee
where 
\be
m' \;=\; \left\lfloor {m+1 \over 2} \right\rfloor
\label{def_mprime}
\ee
and $\lfloor p \rfloor$ stands for the largest integer $\le p$. 
(We give a new proof of this formula in Appendix~\ref{sec.dim}.)
The asymptotic behavior of
$\mbox{\rm SqFree}(m)$ is given by \cite[Corollary 1]{Tutte_sq_02}
\be
\mbox{\rm SqFree}(m)\;\sim\; {\sqrt{3} \over 4\sqrt{\pi} } \, 3^m \, m^{-3/2} 
\left[ 1 + O\left( {1\over m} \right) \right] \qquad \mbox{\rm as $m\to\infty$}
\label{def_SqFree_asymptotics}
\ee
The values of all these dimensions for $m\leq 16$ are displayed in 
Table~\ref{table_dimensions}.

Some further structural properties of the transfer matrices
can be found in Ref.~\cite{transfer5}.

\subsection{Beraha--Kahane--Weiss theorem}   \label{sec.setup.BKW}

A central role in our work is played by a theorem
on analytic functions due to
Beraha, Kahane and Weiss \cite{BKW_75,BKW_78,Beraha_79,Beraha_80}
and generalized slightly by one of us \cite{Sokal_chromatic_roots}.
The situation is as follows:
Let $D$ be a domain (connected open set) in the complex plane,
and let $\alpha_1,\ldots,\alpha_M,\lambda_1,\ldots,\lambda_M$ ($M \ge 2$)
be analytic functions on $D$, none of which is identically zero.
For each integer $n \ge 0$, define
\be
   f_n(z)   \;=\;   \sum\limits_{k=1}^M \alpha_k(z) \, \lambda_k(z)^n
   \;.
   \label{def_fn}
\ee
We are interested in the zero sets
\be
   \scrz(f_n)   \;=\;   \{z \in D \colon\;  f_n(z) = 0 \}
\ee
and in particular in their limit sets as $n\to\infty$:
\begin{eqnarray}
   \liminf \scrz(f_n)   & = &  \{z \in D \colon\;
   \hbox{every neighborhood $U \ni z$ has a nonempty intersection} \nonumber\\
      & & \qquad \hbox{with all but finitely many of the sets } \scrz(f_n) \}
   \\[4mm]
   \limsup \scrz(f_n)   & = &  \{z \in D \colon\;
   \hbox{every neighborhood $U \ni z$ has a nonempty intersection} \nonumber\\
      & & \qquad \hbox{with infinitely many of the sets } \scrz(f_n) \}
\end{eqnarray}
Let us call an index $k$ {\em dominant at $z$}\/ if
$|\lambda_k(z)| \ge |\lambda_l(z)|$ for all $l$ ($1 \le l \le M$);
and let us write
\be
   D_k  \;=\;  \{ z \in D \colon\;  k \hbox{ is dominant at } z  \}
   \;.
\ee
Then the limiting zero sets can be completely characterized as follows:

\begin{theorem}
  {\bf \protect\cite{BKW_75,BKW_78,Beraha_79,Beraha_80,Sokal_chromatic_roots}}
   \label{BKW_thm}
Let $D$ be a domain in $\C$,
and let $\alpha_1,\ldots,\alpha_M$, $\lambda_1,\ldots,\lambda_M$ ($M \ge 2$)
be analytic functions on $D$, none of which is identically zero.
Let us further assume a ``no-degenerate-dominance'' condition:
there do not exist indices $k \neq k'$
such that $\lambda_k \equiv \omega \lambda_{k'}$ for some constant $\omega$
with $|\omega| = 1$ and such that $D_k$ ($= D_{k'}$)
has nonempty interior.
For each integer $n \ge 0$, define $f_n$ by
$$
   f_n(z)   \;=\;   \sum\limits_{k=1}^M \alpha_k(z) \, \lambda_k(z)^n
   \;.
$$
Then $\liminf \scrz(f_n) = \limsup \scrz(f_n)$,
and a point $z$ lies in this set if and only if either
\begin{itemize}
   \item[(a)]  There is a unique dominant index $k$ at $z$,
       and $\alpha_k(z) =0$;  or
   \item[(b)]  There are two or more dominant indices at $z$.
\end{itemize}
\end{theorem}
Note that case (a) consists of isolated points in $D$;
we refer to these as {\em isolated limiting points}\/.
Case (b) consists of curves
(plus possibly isolated points where
all the $\lambda_k$ vanish simultaneously);
we call these {\em limiting curves}\/ or {\em dominant equimodular curves}\/.
Henceforth we shall denote by $\scrb$ the locus of points
satisfying condition (b).

We shall often refer to the functions $\lambda_k$ as ``eigenvalues'',
because that is how they arise in the transfer-matrix formalism.

\subsection{Chromatic roots of bi-fans and bipyramids}  \label{sec.setup.bipyr}

The simplest case of the graphs $S_{m,n}$ is width $m=1$:
the graphs $S_{1,n}$ are called {\em bi-fans}\/ (Figure~\ref{fig_bipyr}a),
and their chromatic polynomials can be easily computed.
Here we would like to recall this computation
and to discuss the behavior of the chromatic roots as $n \to\infty$.
Similar methods apply to a closely related class of graphs
called {\em bipyramids}\/ (Figure~\ref{fig_bipyr}b).

We recall that the {\em join}\/ $G+H$ of two graphs is
the graph obtained from the disjoint union of $G$ and $H$
by adding one edge connecting each pair of vertices
$i \in V(G)$, $j \in V(H)$.

Let $\bar{K}_2$ be the graph with two vertices and no edges,
and let $G$ be any graph.
Then the chromatic polynomial of the join $G + \bar{K}_2$
is given in terms of the chromatic polynomial of $G$
by the easy formula
\be
   P_{G + \bar{K}_2}(q)  \;=\;
   q P_G(q-1) \,+\, q(q-1) P_G(q-2)
   \;.
 \label{eq.PGjoin}
\ee
Indeed, the first (resp.\ second) term on the right-hand side
of \reff{eq.PGjoin}
counts the proper $q$-colorings of $G + \bar{K}_2$
in which the two vertices of $\bar{K}_2$ are colored
the same (resp.\ different).

Using the well-known chromatic polynomials of the $n$-vertex path $P_n$
and the $n$-vertex cycle $C_n$,
\begin{eqnarray}
   P_{P_n}(q)  & = &   q(q-1)^{n-1}   \\[2mm]
   P_{C_n}(q)  & = &   (q-1)^n \,+\, (q-1) (-1)^n 
\end{eqnarray}
we obtain the chromatic polynomials of the bi-fan $P_n + \bar{K}_2$
and the bipyramid $C_n + \bar{K}_2$:
\begin{eqnarray}
   \label{exact_P1xn}
   P_{P_n + \bar{K}_2}(q)  & = &
       q(q-1)(q-2)^{n-1} \,+\, q(q-1)(q-2)(q-3)^{n-1}   \\[2mm]
   P_{C_n + \bar{K}_2}(q)  & = &
       q(q-2)^n \,+\, q(q-1)(q-3)^n  \,+\, q(q^2 - 3q + 1) (-1)^n
\end{eqnarray}

Using the Beraha--Kahane--Weiss theorem,
we see immediately that the chromatic roots of the bi-fans
accumulate as $n \to\infty$ on the curve $|q-2| = |q-3|$,
i.e.\ the vertical line $\real q = 5/2$ \cite{Shrock_97c}.
(There are also isolated limiting points at $q=0,1,2$.)
The analysis for the bipyramids is slightly more complicated,
because there are three ``eigenvalues'' ($q-2$, $q-3$ and $-1$)
that need to be compared.
Elementary computations show \cite{Read_Royle,Shrock_97a}
that the limiting curve $\scrb$ consists of the pair of semi-infinite
vertical line segments $\real q = 5/2$, $|\imag q| \ge \sqrt{3}/2$
together with the pair of circular arcs
$q = 2 + e^{i \theta}$ and $q = 3 - e^{i \theta}$
for $-\pi/3 \le \theta \le \pi/3$.\footnote{
   Read and Royle's formula for the chromatic polynomial
   of the bipyramids \cite[p.~1017]{Read_Royle}
   has a typographical error: $z^2 - 3z + 3$ should be $z^2 - 3z + 1$.
}

\begin{figure}%%%[hbtp]
\centering
\includegraphics[width=350pt]{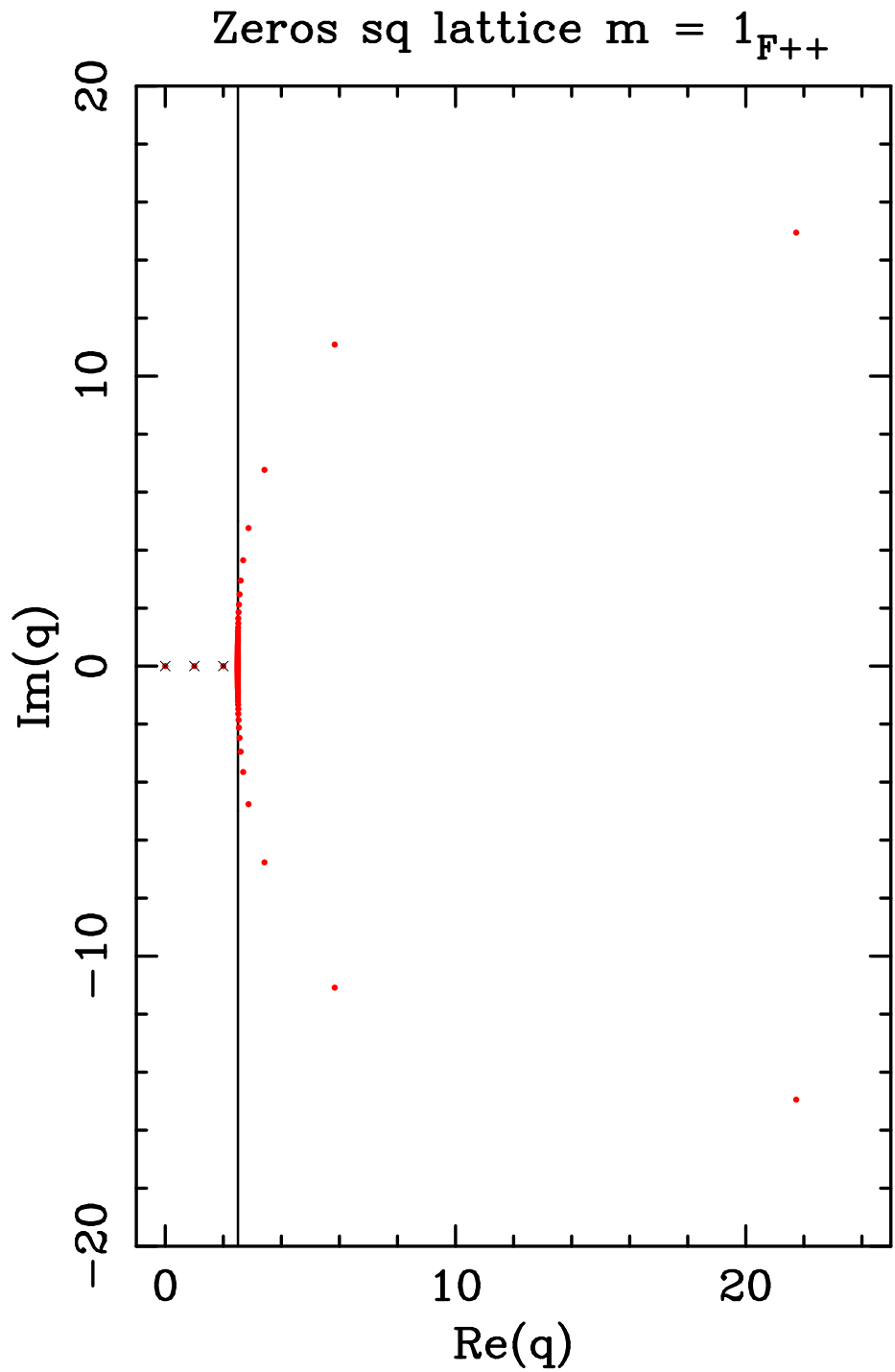}
\caption{\label{figure_sq_1Fx100F}
Chromatic roots for the bi-fan $P_{100} + \bar{K}_2 = S_{1,100}$ 
(red $\bullet$). 
We also show the limiting curve $\mathcal{B}_1$, and the isolated
limiting points ($\times$).
}
\end{figure}

It is worth remarking that if one plots the chromatic roots
of the bi-fans or bipyramids, one finds a large number of
roots accumulating on the parts of $\scrb$ fairly close to the real axis,
together with a few roots at large imaginary part lying
{\em far to the right}\/ of the line $\Re q = 5/2$.\footnote{
   For the bipyramids this was noticed already two decades ago
   by Read and Royle \cite[pp.~1019--1020]{Read_Royle}.
   It was further observed by Shrock and Tsai
   \cite[Sections~III.A and III.B]{Shrock_97c}
   for some classes of graphs that generalize the bi-fans and bipyramids.
}
(In Figure~\ref{figure_sq_1Fx100F} we depict the chromatic roots
of the $n=100$ bi-fan. The rightmost zeros are located at 
$q\approx 21.8463758923\pm 14.9443992222\, i$.) 
This behavior is at first surprising (at least it was to us),
but after some thought one realizes that it is perfectly consistent
with the Beraha--Kahane--Weiss theorem:
the limiting curve $\Re q = 5/2$ contains the point at infinity,
and roots can tend to infinity in the topology of the Riemann sphere
(which turns out to be the relevant sense)
in many different ways;
in particular, their real parts need not tend to 5/2.
In fact, it turns out that the rightmost root has
\be
   q  \;=\;  {n \over \log n \,-\, \log\log n}
             \left[ 1 \,\pm\, {\pi i \over \log n}
                      \,+\, O\!\left( {\log\log n \over \log^2 n} \right)
             \right]
   \;,
\ee
so that its real and imaginary parts {\em both}\/ tend to infinity
as $n \to\infty$.
This asymptotic formula can be proven by a minor modification
of the analysis given in \cite[Section 5]{gen_theta}
for the chromatic roots of the complete bipartite graphs
$K_{2,n} \simeq \bar{K}_n + \bar{K}_2$.

\subsection{Chromatic roots of generalized theta graphs}
  \label{sec.setup.theta}

Let us now summarize briefly the argument from \cite{Sokal_chromatic_roots}
concerning the chromatic roots of the generalized theta graphs.

The chromatic polynomial of $\Theta^{(s,p)}$ is
\cite{Shrock_98b,gen_theta,Sokal_chromatic_roots}
\be
   P_{\Theta^{(s,p)}}(q)  \;=\;
   q \lambda_{=,s}(q)^p  \,+\,  q(q-1) \lambda_{\neq,s}(q)^p
 \label{PG.theta}
\ee
where
\begin{eqnarray}
   \lambda_{=,s}(q)    & = &  {(q-1)^s + (q-1) (-1)^s  \over q}   \\[2mm]
   \lambda_{\neq,s}(q) & = &  {(q-1)^s - (-1)^s  \over q}
\end{eqnarray}
Indeed, $\lambda_{=,s}(q)$ [resp.\ $\lambda_{\neq,s}(q)$]
is the number of ways of coloring the internal vertices
of one $s$-edge path when the two endvertices are colored
the same (resp.\ different).
Please note that the leading large-$q$ term in both
$\lambda_{=,s}(q)$ and $\lambda_{\neq,s}(q)$ is $q^{s-1} \equiv q^m$,
and that they start to differ at order $q^0$:  indeed, we have
the curious formula
\be
   \lambda_{=,s}(q) \,-\, \lambda_{\neq,s}(q) \;=\; (-1)^s
   \;.
 \label{eq.curious.1}
\ee

Let us begin by fixing $s$ and considering the limit $p \to \infty$.
Using the Beraha--Kahane--Weiss theorem,
we see immediately that the chromatic roots of the graphs $\Theta^{(s,p)}$
accumulate as $p \to\infty$ on the real algebraic curve curve $\scrc_s$
defined by $|\lambda_{=,s}(q)| = |\lambda_{\neq,s}(q)|$, or equivalently
\be
   \scrc_s \colon\quad
   \left| 1 \,+\, (q-1) (1-q)^{-s} \right|
   \;=\;
   \left| 1 \,-\, (1-q)^{-s} \right|
   \;.
\label{def_Cs}
\ee

To handle the limit $s \to\infty$, we use the following lemma:
\begin{lemma}  {\bf \protect\cite[Lemma~1.6]{Sokal_chromatic_roots}}
Let $F_1,F_2,G$ be analytic functions on a disc $|z| < R$
satisfying $|G(0)| \le 1$ and $G \not\equiv \hbox{constant}$.
Then, for each $\epsilon > 0$, there exists $s_0 < \infty$
such that for all integers $s \ge s_0$ the equation
\be
   |1 + F_1(z) G(z)^s|   \;=\;  |1 + F_2(z) G(z)^s|
\ee
has a solution in the disc $|z| < \epsilon$.
\end{lemma}
This lemma immediately implies that the curves $\scrc_s$
accumulate as $s \to\infty$ at every point in the region $|q-1| \ge 1$.
A strengthened version of this lemma is proven in
\cite[Theorem~4.1]{Sokal_chromatic_roots}.

\begin{figure}%%%[hbtp]
\centering
\begin{tabular}{cc}
\includegraphics[width=230pt]{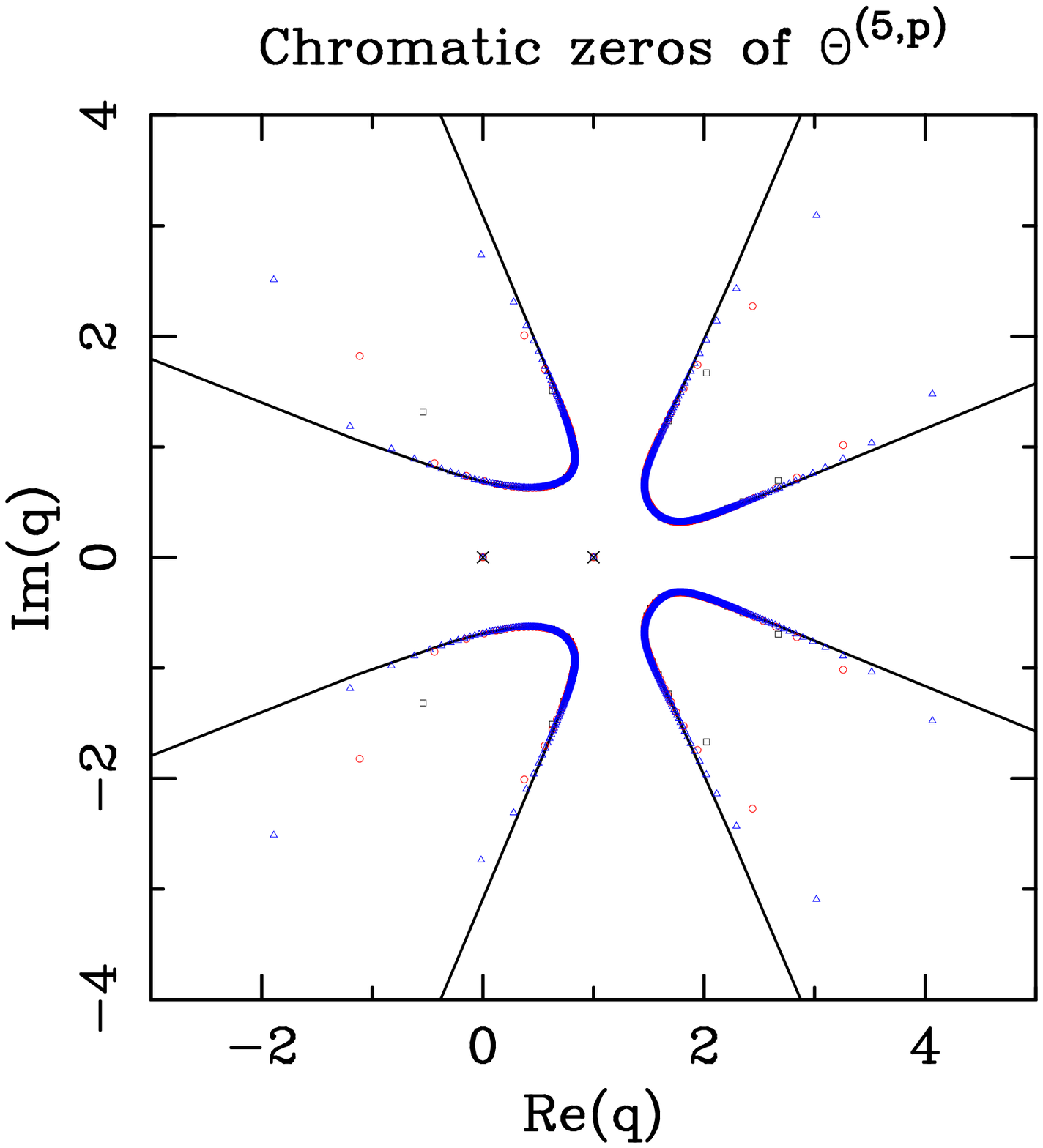} & 
\includegraphics[width=230pt]{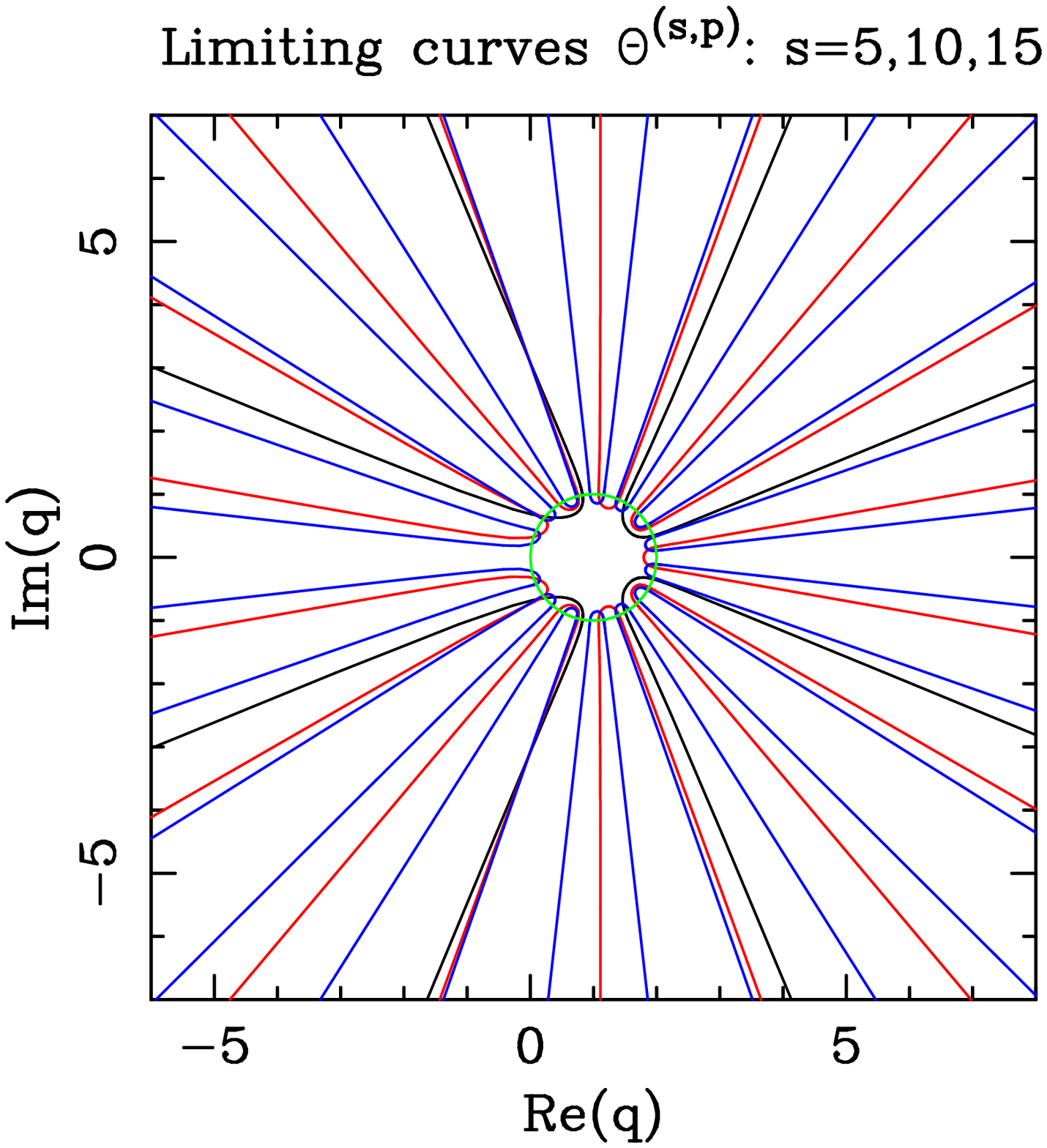} \\[1mm]
   \phantom{(((a)}(a)    & \phantom{(((a)}(b) \\[5mm]
\end{tabular}
\caption{\label{fig_theta}
Chromatic roots for the generalized theta graphs $\Theta^{(s,p)}$.
(a) Chromatic roots for the graphs $\Theta^{(5,p)}$ with $p=25$ 
(black $\square$), $p=100$ (red $\bullet$) and $p=400$ (blue $\triangle$), 
and the limiting curve $\mathcal{C}_5$ [cf.\ \protect\reff{def_Cs}].
We also show the isolated limiting points for this family (black $\times$).
(b) Limiting curves $\mathcal{C}_s$ for $s=5$ (black), 
$s=10$ (red) and $s=15$ (blue). The circle (depicted in green)
is $|q-1| = 1$.  
}
\end{figure}

In Figure~\ref{fig_theta}a we show the curve $\scrc_s$
together with the chromatic roots of $\Theta^{(s,p)}$
for $s=5$ and $p=25,100,400$.
Just as for the bi-fans, we see clearly the accumulation of roots
on the inner parts of the curve $\scrc_s$, together with a few points
at larger $|q|$ lying far off this curve.
All this can presumably be explained by a generalization of the analysis
given in \cite[Section 5]{gen_theta} for the special case $s=2$.
In particular, we expect that the root of largest modulus
grows like $(p/\log p)^{1/(s-1)}$ as $p \to\infty$.\footnote{
   Brown, Hickman, Sokal and Wagner \cite[Theorem~1.3]{gen_theta}
   proved that the roots are bounded in modulus by
   ${\rm const} \times p/\log p$, uniformly in $s$ and $p$.
   But this bound is presumably far from sharp when $s > 2$.
}
This explains the extremely slow convergence of the roots
to the outer parts of the curve $\scrc_s$ in Figure~\ref{fig_theta}a.
Note in particular that a naive computational approach,
without the insight provided by the Beraha--Kahane--Weiss theorem,
would give little indication that the roots are in fact
unbounded as $p \to\infty$.

In Figure~\ref{fig_theta}b we show the curves $\scrc_s$
for $s=5,10,15$.  We see clearly how these curves are becoming dense
in the entire complex plane with the exception of the disc $|q-1| < 1$.
The curves $\scrc_s$ do enter this disc, but as $s \to\infty$
they retreat to its boundary, so that no point of the open disc
is a limit point of the curves $\scrc_s$.

Finally, let us note, for future reference, the formula for
the multivariate Tutte polynomial of $\Theta^{(s,p)}$
when the {\em same}\/ weights $v_1,\ldots,v_s$
(but not necessarily $v_1 = \ldots = v_s = -1$)
are assigned to each path \cite{Sokal_chromatic_roots}:
\be
   Z_{\Theta^{(s,p)}}(q, \bv)  \;=\;
   q \lambda_{=,s}(q,\bv)^p  \,+\,  q(q-1) \lambda_{\neq,s}(q,\bv)^p
 \label{ZG.theta}
\ee
where
\begin{eqnarray}
   \lambda_{=,s}(q,\bv)    & = &
        {1 \over q}  \left[ \prod\limits_{i=1}^s (q+v_i)  \,+\,
                            (q-1) \prod\limits_{i=1}^s  v_i  \right] \\[2mm]
   \lambda_{\neq,s}(q,\bv) & = &
        {1 \over q}  \left[ \prod\limits_{i=1}^s (q+v_i)  \,-\,
                            \prod\limits_{i=1}^s  v_i  \right]
\end{eqnarray}
We have the curious formula
\be
   \lambda_{=,s}(q,\bv) \,-\, \lambda_{\neq,s}(q,\bv) \;=\;
   \prod\limits_{i=1}^s  v_i
   \;,
 \label{eq.curious.2}
\ee
which generalizes \reff{eq.curious.1}.

\section{Transfer-matrix theory for the family $\bm{S_{m,n}}$}
   \label{sec.transfer}

In Section~\ref{sec.setup.transfer} we explained the
transfer-matrix formalism for computing the
multivariate Tutte polynomial of an $m \times n$ square lattice
with free boundary conditions in both directions.
In this section we begin (Section~\ref{sec.transfer.basic})
by discussing the modifications needed to handle the extra sites
at left and right in the graphs $S_{m,n}$.
We then (Section~\ref{sec.block1}) go on to prove
some structural properties of the transfer matrix
for the family $S_{m,n}$,
by analogy with what was done in Ref.~\cite{transfer5}
for free and cylindrical boundary conditions;
in particular we show that the transfer matrix $\T$
can be decomposed into two independent blocks $\T_{=}$ and $\T_{\neq}$.
Next (Section~\ref{sec.block2}) we introduce
the families of graphs $S^{=}_{m,n}$ and $S^{\neq}_{m,n}$,
which shed further light on the block-diagonalization of the
transfer matrix for the family $S_{m,n}$.
Then (Section~\ref{sec.eigen}) we prove that the large-$|q|$
expansion for the dominant eigenvalue of the transfer matrix
$\T_{=}$ or $\T_{\neq}$ holds uniformly in the strip width $m$
(throughout the antiferromagnetic regime),
by invoking results from \cite{Sokal_chromatic_bounds}.
Finally (Section~\ref{sec.comp}), we prove a key identity
concerning the difference between the dominant eigenvalues
$\lambda_{\star,=}(m,q,{\bf v})$ and $\lambda_{\star,\neq}(m,q,{\bf v})$
corresponding to $S^{=}_{m,n}$ and $S^{\neq}_{m,n}$.

\subsection{Basic method}
   \label{sec.transfer.basic}

\begin{figure}
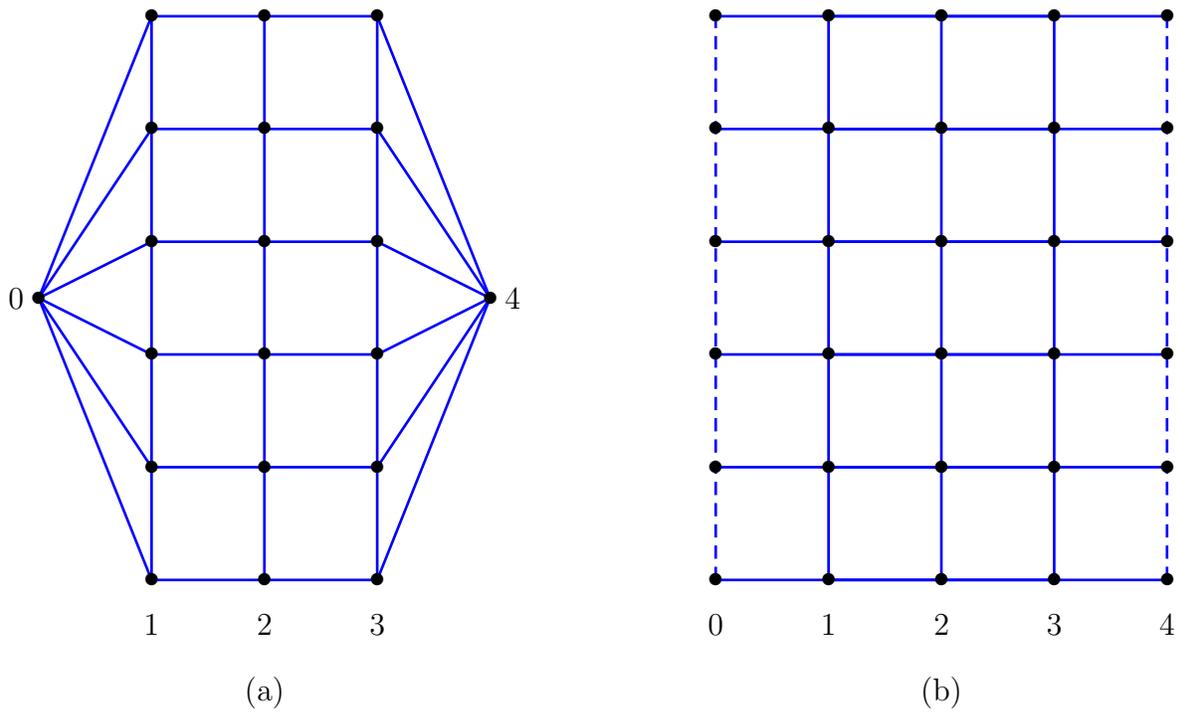

\centering
     \psset{xunit=1.5cm}
     \psset{yunit=1.5cm}
     \pspicture(0,0)(10,5)
     \psline[linecolor=blue,linewidth=1.0pt,linestyle=dashed](6,0)(6,5)
     \psline[linecolor=blue,linewidth=1.0pt](7,0)(7,5)
     \psline[linecolor=blue,linewidth=1.0pt](8,0)(8,5)
     \psline[linecolor=blue,linewidth=1.0pt](9,0)(9,5)
     \psline[linecolor=blue,linewidth=1.0pt,linestyle=dashed](10,0)(10,5)

     \psline[linecolor=blue,linewidth=1.0pt](1,0)(1,5)
     \psline[linecolor=blue,linewidth=1.0pt](2,0)(2,5)
     \psline[linecolor=blue,linewidth=1.0pt](3,0)(3,5)

     \psline[linecolor=blue,linewidth=1.0pt](6,0)(10,0)
     \psline[linecolor=blue,linewidth=1.0pt](6,1)(10,1)
     \psline[linecolor=blue,linewidth=1.0pt](6,2)(10,2)
     \psline[linecolor=blue,linewidth=1.0pt](6,3)(10,3)
     \psline[linecolor=blue,linewidth=1.0pt](6,4)(10,4)
     \psline[linecolor=blue,linewidth=1.0pt](6,5)(10,5)

     \psline[linecolor=blue,linewidth=1.0pt](7,0)(9,0)
     \psline[linecolor=blue,linewidth=1.0pt](7,1)(9,1)
     \psline[linecolor=blue,linewidth=1.0pt](7,2)(9,2)
     \psline[linecolor=blue,linewidth=1.0pt](7,3)(9,3)
     \psline[linecolor=blue,linewidth=1.0pt](7,4)(9,4)
     \psline[linecolor=blue,linewidth=1.0pt](7,5)(9,5)

     \psline[linecolor=blue,linewidth=1.0pt](1,0)(3,0)
     \psline[linecolor=blue,linewidth=1.0pt](1,1)(3,1)
     \psline[linecolor=blue,linewidth=1.0pt](1,2)(3,2)
     \psline[linecolor=blue,linewidth=1.0pt](1,3)(3,3)
     \psline[linecolor=blue,linewidth=1.0pt](1,4)(3,4)
     \psline[linecolor=blue,linewidth=1.0pt](1,5)(3,5)

     \psline[linecolor=blue,linewidth=1.0pt](0,2.5)(1,0)
     \psline[linecolor=blue,linewidth=1.0pt](0,2.5)(1,1)
     \psline[linecolor=blue,linewidth=1.0pt](0,2.5)(1,2)
     \psline[linecolor=blue,linewidth=1.0pt](0,2.5)(1,3)
     \psline[linecolor=blue,linewidth=1.0pt](0,2.5)(1,4)
     \psline[linecolor=blue,linewidth=1.0pt](0,2.5)(1,5)

     \psline[linecolor=blue,linewidth=1.0pt](4,2.5)(3,0)
     \psline[linecolor=blue,linewidth=1.0pt](4,2.5)(3,1)
     \psline[linecolor=blue,linewidth=1.0pt](4,2.5)(3,2)
     \psline[linecolor=blue,linewidth=1.0pt](4,2.5)(3,3)
     \psline[linecolor=blue,linewidth=1.0pt](4,2.5)(3,4)
     \psline[linecolor=blue,linewidth=1.0pt](4,2.5)(3,5)

     \multirput{0}(6,0) (0,1){6}{$\bullet$}
     \multirput{0}(7,0) (0,1){6}{$\bullet$}
     \multirput{0}(8,0) (0,1){6}{$\bullet$}
     \multirput{0}(9,0) (0,1){6}{$\bullet$}
     \multirput{0}(10,0)(0,1){6}{$\bullet$}

     \multirput{0}(1,0)(0,1){6}{$\bullet$}
     \multirput{0}(2,0)(0,1){6}{$\bullet$}
     \multirput{0}(3,0)(0,1){6}{$\bullet$}

     \multirput{0}(0,2.5)(4,0){2}{$\bullet$}

     \rput{0}(6, -0.4){$0$}
     \rput{0}(7, -0.4){$1$}
     \rput{0}(8, -0.4){$2$}
     \rput{0}(9, -0.4){$3$}
     \rput{0}(10,-0.4){$4$}

     \rput{0}(1,-0.4){$1$}
     \rput{0}(2,-0.4){$2$}
     \rput{0}(3,-0.4){$3$}

     \rput{0}( 8  ,-1.0){(b)}
     \rput{0}( 2  ,-1.0){(a)}
     \rput{0}( 4.2, 2.5){$4$}
     \rput{0}(-0.2, 2.5){$0$}
     \endpspicture
     \vspace*{2cm}
\caption{\label{fig_Smn}
   (a)  Square-lattice strip $S_{3,6}$ of width $m=3$ and length $n=6$
      with two extra sites.
   (b)  Alternate representation of $S_{3,6}$ in terms of a
      square-lattice strip of width $m+2=5$, length $n=6$
      and free boundary conditions, with edge weights $v \to +\infty$
      on the dashed edges.
}
\end{figure}

We want to compute the chromatic polynomial $P_{S_{m,n}}(q)$
--- or more generally, the multivariate Tutte polynomial
$Z_{S_{m,n}}(q, {\bf v})$ with periodically repeated edge weights ---
for the graph $S_{m,n}$ that is obtained from
a square-lattice strip of width $m$ and length $n$
(with free boundary conditions in both directions)
by connecting every site in the leftmost (resp.\ rightmost) column
to an extra site $s$ (resp.\  $s'$).
Thus, the number of sites of this strip is $mn+2$
(see Figure~\ref{fig_Smn}a).

We shall also denote the strip $S_{m,n}$ as $m_{\rm F++}\times n_{\rm F}$:
here the subscript F denotes free boundary conditions,
while the subscript F++ denotes free boundary conditions
with the two extra sites attached at left and right.
A square-lattice strip with standard free boundary conditions
in both directions will be denoted as $m_{\rm F}\times n_{\rm F}$.

In the strip $S_{m,n}$
we assign weight $v_{0,1}$ to all the edges connecting the
leftmost site (which we can think of as ``column 0'')
to the sites of column 1,
weights $v_{i,i+1}$ to the horizontal edges
linking column $i$ to column $i+1$ ($1 \le i \le m-1$),
weight $v_{m,m+1}$ to all the edges connecting the sites of column $m$
to the rightmost site (which we can think of as ``column $m+1$''),
and weight $v_i$ to the vertical edges in column $i$ ($1 \le i \le m$).
Of course, for the chromatic polynomial all these weights
will be taken to be $-1$.

The computation of the multivariate Tutte polynomial
for this strip family can be achieved by using a minor modification of
the transfer-matrix method explained in Section~\ref{sec.setup.transfer}.
As before, we fix the ``width'' $m$ and consider the family of graphs
$S_{m,n}$ obtained by varying the ``length'' $n$;
our goal is to calculate the multivariate Tutte polynomials
$Z_{S_{m,n}}(q, {\bf v})$ for this family by building up the graphs
$S_{m,n}$ layer by layer.

One way of constructing the transfer matrix $\T(m)$ for the family $S_{m,n}$
is to treat this family as a special case of
a square-lattice strip $(m+2)_{\rm F}\times n_{\rm F}$:
see Figure~\ref{fig_Smn}b. 
Thus, we choose to work in the space of connectivities
$\{\basise_\mathcal{P}\}$ associated to the non-crossing partitions 
$\mathcal{P}$
of the set $\{0,1,2,\ldots,m,m+1\}$. The dimension of this space is 
the Catalan number $C_{m+2}$. 

On the edges of the $(m+2)_{\rm F}\times n_{\rm F}$ strip
that correspond to the edges of the starting graph $S_{m,n}$,
we simply use the desired weights $v_{i,i+1}$ ($0 \le i \le m$)
and $v_i$ ($1 \le i \le m$).
On the vertical edges in the first and last columns
(depicted with dashed lines in Figure~\ref{fig_Smn}b),
we associate a weight $v_e\to +\infty$,
which corresponds to contracting all the sites
in each of these columns into a single site.
In other words, we use two different vertical-bond operators
depending on which column we are in:
\begin{equation}
\P_i  \;=\; \begin{cases} 
            v_i I + \D_i   & \text{for $1 \le i \le m$}  \\ 
            I              & \text{for $i=0,m+1$} 
            \end{cases} 
\end{equation}
Therefore, the horizontal transfer matrix is given by
\be
   \H  \;=\;
     \prod_{i=0}^{m} (1 + v_{i,i+1} \J_{i,i+1})
   \label{def_Hbis}
\ee 
[compare \reff{def_H}]
and the vertical transfer matrix is given by 
\be
   \V  \;=\;  \prod_{i=1}^{m} (v_i I + \D_i)
   \;.
\label{def_Vbis}
\ee 
Note that the formula for $\V$ is {\em identical}\/ to
the formula \reff{def_V} for an ordinary $m_{\rm F}\times n_{\rm F}$ grid;
the only difference is that here the operators act on the space of
partitions of $\{0,1,2,\ldots,m,m+1\}$ rather than $\{1,2,\ldots,m\}$.

The full transfer matrix is then $\T = \V \H$,
and the multivariate Tutte polynomial $Z_{S_{m,n}}(q, {\bf v})$ is given by  
\be
   Z_{S_{m,n}}(q, {\bf v})   \;=\;
     \endv^{\rm T} \H \T^{n-1} \basise_{\rm id}
   \;,
\ee
where $\basise_{\rm id}$ denotes the basis vector associated to the partition
\be
  {\rm id} \;=\; \left\{ \{0\},\{1\},\ldots,\{m\},\{m+1\}\right\}
\ee
and the left vector $\endv$ acts on a connectivity state as in \reff{def_uT}.

When the horizontal couplings $v_{i,i+1}$ are all equal to $-1$
(as is the case for the chromatic polynomial),
the horizontal operator $\H$ is a {\em projection}\/,
so that (as explained in Section~\ref{sec.setup.transfer})
we can work in its image subspace by using the modified transfer matrix
$\T' = \H \V \H$
and the basis vectors $\basisf_\mathcal{P}=\H\basise_\mathcal{P}$,
where $\scrp$ runs over the non-crossing non-nearest-neighbor partitions
of $\{0,\ldots,m+1\}$.
The dimension of this space is the Motzkin number $M_{m+1}$
[cf.\ \reff{def_Motzkin}].
We then write the chromatic polynomial for $S_{m,n}$ as
\be
  P_{S_{m,n}}(q) \;=\; \endv^{\rm T} (\T')^{n-1} \basisf_{\rm id}
  \;.
\label{P_Smn}
\ee

Finally, when the couplings are also reflection-symmetric
(e.g., when they are all $-1$),
we can work in the space of equivalence classes of
non-crossing non-nearest-neighbor partitions modulo reflection.
The dimension of this space is given by $\mbox{\rm SqFree}(m+2)$
[cf.\ \reff{def_SqFree}]. Hereafter, we will denote by $\T''(m)$ this 
symmetry-reduced transfer matrix.

\subsection{Block-diagonalization of the transfer matrix (I)}
   \label{sec.block1}

One important property of the transfer matrix $\T$ (or $\T'$ or $\T''$)
for the family $S_{m,n}$ is that it can be decomposed into
two independent transfer matrices of a lower dimension.
To see this, it suffices to note that the projection
$\J_{0,m+1}$ commutes with both $\H$ and $\V$:
it commutes with $\H$ because all join operators commute with
each other, and it commutes with $\V$ because \reff{def_Vbis}
does not contain the detach operators $\D_0$ or $\D_{m+1}$.
Therefore, $\T$ (or $\T'$ or $\T''$) block-diagonalizes
if we choose a basis in which the first group of basis elements
corresponds to the range of $\J_{0,m+1}$,
and the second group corresponds to the range of $I - \J_{0,m+1}$.
We call these two blocks $\T_{=}$ and $\T_{\neq}$, respectively.
Indeed, this block-diagonalization holds for any transfer matrix
built out of operators of the general form
\begin{subeqnarray}
   \H  & = &   \sum\limits_{A \subseteq \{0,\ldots,m\}}  c_A \,
                    \prod\limits_{i \in A} \J_{i,i+1}      \\[2mm]
   \V  & = &   \sum\limits_{B \subseteq \{1,\ldots,m\}}  d_B \,
                    \prod\limits_{i \in B} \D_i
 \slabel{eq.HV.general.b}
 \label{eq.HV.general}
\end{subeqnarray}
irrespective of the coefficients $\{c_A\}$ and $\{d_B\}$.
The point, once again, is simply that $\V$
does not contain the detach operators $\D_0$ or $\D_{m+1}$.

Let us now recall that ${\rm id}$ denotes the partition of
$\{0,\ldots,m+1\}$ in which each element is a singleton,
and that ${\rm id} \bullet (0,m+1)$ denotes the partition
in which $\{0,m+1\}$ is a block and all the other elements are singletons.
Let us call a partition $\scrp$ of $\{0,\ldots,m+1\}$ {\em non-trivial}\/
if it is neither ${\rm id}$ nor ${\rm id} \bullet (0,m+1)$.

\begin{proposition}
   \label{prop.TMblockdiag.1}
For any operators $\H$ and $\V$ of the form
\reff{eq.HV.general}, we can write
\begin{subeqnarray}
   \V \H \, \J_{0,m+1} \basise_{\rm id}  & = &
       t_{=} \, \J_{0,m+1} \basise_{\rm id}  \;\;+
       \sum\limits_{\scrp\, \hboxrm{non-trivial}}
              a^{=}_\scrp \, \J_{0,m+1} \basise_\scrp     \\[2mm]
   \V \H \, (I - \J_{0,m+1}) \basise_{\rm id}  & = &
       t_{\neq} \, (I - \J_{0,m+1}) \basise_{\rm id}  \;\;+
       \sum\limits_{\scrp\, \hboxrm{non-trivial}}
              a^{\neq}_\scrp \, (I - \J_{0,m+1}) \basise_\scrp
\label{eq.VHid}
\end{subeqnarray}
for some coefficients $t_{=}$, $t_{\neq}$, $\{a^{=}_\scrp\}$
and $\{a^{\neq}_\scrp\}$
that are polynomials in $q$, $\{c_A\}$ and $\{d_B\}$.
Furthermore, these coefficients satisfy the identity
\be
   t_{=} - t_{\neq}  \;=\;  c_{\{0,\ldots,m\}} d_{\{1,\ldots,m\}}
\label{eq.curious.3}
\ee
reminiscent of \reff{eq.curious.1}/\reff{eq.curious.2}.
\end{proposition}

\proof
Let us begin by considering the quantity $\H \basise_{\rm id}$:
\be
   \H \basise_{\rm id}
   \;=\;
   \sum\limits_{A \subseteq \{0,\ldots,m\}}  c_A
       \left( \prod\limits_{i \in A} \J_{i,i+1} \right)
       \basise_{\rm id}
   \;.
 \label{eq.He}
\ee
Let us now apply the operator $\V$ to \reff{eq.He}. 
It is clear that we obtain three types of terms: a) terms that contain
non-trivial partitions $\mathcal{P}$, b) terms that contain the trivial 
partition id, and c) a single term that contains the trivial partition
$\rm{id}\bullet(0,m+1)$.
Indeed, this latter term can arise in only one way:
first we must join all the sites $\{0,\ldots,m+1\}$ into a single block
(i.e., take $A = \{0,\ldots,m\}$ in $\H$),
and then we must detach all the inner sites
(i.e., take $B = \{1,\ldots,m\}$ in $\V$).
It follows that
\be
\V\H \basise_{\rm id} \;=\; 
    c_{\{0,\ldots,m\}}  d_{\{1,\ldots,m\}} \J_{0,m+1} \basise_{\rm id}
    \,+\,  a_{\rm id} \basise_{\rm id} 
    \,+\, \sum\limits_{\scrp\, \hboxrm{non-trivial}} a_\scrp \, \basise_\scrp
 \label{eq.VHe}
\ee
for some coefficients $a_{\rm id}$ and $\{a_\scrp\}$
that will clearly be polynomials in $q$, $\{c_A\}$ and $\{d_B\}$.
{}From this equation is easy to derive \reff{eq.VHid}, as the join operator
$\J_{0,m+1}$ commutes with all the join and detach operators
arising in \reff{eq.VHe}.
Indeed, using $\J_{0,m+1}^2=\J_{0,m+1}$ we find
\begin{subeqnarray}
t_{=}    &=& a_{\rm id} + c_{\{0,\ldots,m\}} d_{\{1,\ldots,m\}} \\
t_{\neq} &=& a_{\rm id}
\end{subeqnarray}
{}From this relation, \reff{eq.curious.3} follows immediately.
\qed

When $\H$ and $\V$ are the transfer matrices \reff{def_Hbis}/\reff{def_Vbis}
for the multivariate Tutte polynomial, we have
$c_{\{0,\ldots,m\}}= \prod_{i=0}^m v_{i,i+1}$
and $d_{\{1,\ldots,m\}}=1$, so that \reff{eq.curious.3} becomes
\be 
   t_{=} - t_{\neq}  \;=\; \prod_{i=0}^m v_{i,i+1}  \;.
\label{eq.curious.4}
\ee
Note that this formula holds irrespective of the values of the
vertical couplings $v_i$.
In particular, it holds when $v_i = 0$;
this explains the ``curious formula'' \reff{eq.curious.2}
for the generalized theta graphs.\footnote{
  There is a small missing step here, because
  \protect\reff{eq.curious.4} refers to the dominant diagonal 
  entries of the transfer matrix, while \protect\reff{eq.curious.2} refers
  to the transfer-matrix eigenvalues.
  But in the case of the generalized theta graphs $\Theta^{(s,p)}$
  these two quantities coincide, as can easily be seen
  by expressing $Z_{\Theta^{(s,p)}}(q,\bv)$ using a transfer-matrix formalism.
  Labelling the vertices along each of the $p$ chains from $0$ to $s$,
  it is not hard to
  see that the partition function can be written in the basis   
  $\{ \bone ,\delta_{0,s}\}$ as follows:
$$
Z_{\Theta^{(s,p)}}(q,\bv) \;=\; \left(\! \begin{array}{c} 
                                     q^2 \\ q \end{array}\!\right)^{\!\rm T} 
    \cdot 
    \left(\! \begin{array}{cc}
        \lambda_{\neq,s}(q,\bv)  & 0 \\
        \prod\limits_{i=0}^{s-1} v_{i,i+1} & \lambda_{=,s}(q,\bv) 
        \end{array}\!\right)^{\! p} \cdot \left(\! \begin{array}{c}
                                        1 \\ 0 \end{array}\!\right) \,.
$$
  Since the transfer matrix is lower-triangular,
  the leading diagonal entries of the transfer matrix
  coincide with its eigenvalues.

  Let us remark that if we choose the basis 
  $\{ \delta_{0s},\bone-\delta_{0s}\} = 
  \{\J_{0s},(I-\J_{0s})\}\basise_{\rm id}$, then the above expression 
  simplifies to
 $$
Z_{\Theta^{(s,p)}}(q,\bv) \;=\; q \, \left(\! \begin{array}{c} 
                                     1 \\ q-1 \end{array}\!\right)^{\!\rm T} 
    \cdot 
    \left(\! \begin{array}{cc}
        \lambda_{=,s}(q,\bv)  &   0 \\
        0                     & \lambda_{\neq,s}(q,\bv) 
        \end{array}\!\right)^{\! p} \cdot \left(\! \begin{array}{c}
                                        1 \\ 1 \end{array}\!\right) \,, 
$$
  and we obtain \protect\reff{ZG.theta}. 
}
When all the horizontal couplings $v_{i,i+1}$ equal $-1$, we have
\be 
   t_{=} - t_{\neq}  \;=\; (-1)^{m+1}  \;,
\label{eq.curious.5}
\ee
which includes \reff{eq.curious.1} as a special case. 

The next result is similar to Proposition~3.1 in \cite{transfer5}.

\begin{proposition}
   \label{prop.TMblockdiag.2}
Consider operators $\H$ and $\V$ of the form \reff{eq.HV.general}
where the coefficients $\{c_A\}$ and $\{d_B\}$ are \emph{numbers}
(i.e., independent of $q$).
Then, using the notation of Proposition~\ref{prop.TMblockdiag.1},
\begin{subeqnarray}
   t_{=}     & = &  c_{\emptyset} d_{\{1,\ldots,m\}} q^m
        \,+\, \hbox{terms of order at most $q^{m-1}$}       \\[2mm]
   t_{\neq}  & = &  c_{\emptyset} d_{\{1,\ldots,m\}} q^m
        \,+\, \hbox{terms of order at most $q^{m-1}$}
 \label{eq.prop.TMblockdiag.2}
\end{subeqnarray}
Furthermore, all the coefficients $\{a^{=}_\scrp\}$ and $\{a^{\neq}_\scrp\}$
defined in Proposition~\ref{prop.TMblockdiag.1} 
are polynomials in $q$ of degree at most $m-1$.
\end{proposition}

\proof
First of all, it is obvious that each entry in the transfer matrix
$\T = \V \H$ is a polynomial in $q$. Indeed, from 
\reff{def_detach} it is clear that we get a factor of $q$
every time we apply the operator $\D_i$ to a partition in which $i$ is
a singleton. Indeed, we can {\em maximize}\/ the number of factors of $q$ 
by applying the vertical transfer matrix $\V$ to a partition
in which every site $1,\ldots,m$ is a singleton,
i.e.\ either $\basise_{\rm id}$ or $\J_{0,m+1} \basise_{\rm id}$.
In particular, from \reff{eq.HV.general.b} we have
\begin{subeqnarray}
   \V \basise_{\rm id} &=&
      \sum_{B\subseteq \{1,\ldots,m\}} d_B q^{|B|} \, \basise_{\rm id} \\
   &=&  \bigl( d_{\{1,\ldots,m\}} q^m +
           \hbox{terms of order at most $q^{m-1}$} \bigr) \, \basise_{\rm id}
        \qquad
  \label{eq.tdom.1}
\end{subeqnarray}
and analogously for $\J_{0,m+1} \basise_{\rm id}$.
If we apply the vertical transfer matrix to any non-trivial partition,
we get a polynomial in $q$ of degree at most $m-1$.
Now let $M$ denote any linear combination of $\J_{0,m+1}$ and $I - \J_{0,m+1}$,
and consider
\begin{subeqnarray}
   \H M \basise_{\rm id} &=&
      \sum_{A\subseteq \{0,\ldots,m\}} c_A
             \left( \prod\limits_{i \in A} \J_{i,i+1} \right)  \basise_{\rm id} \\
   &=&  c_\emptyset M \basise_{\rm id} \,+\,
       \sum\limits_{\scrp\, \hboxrm{non-trivial}} b_\scrp M \basise_\scrp
  \label{eq.tdom.2}
\end{subeqnarray}
for some quantities $b_\scrp$ that are polynomials in $\{c_A\}$
(and of course independent of $q$).
Using \reff{eq.tdom.1}/\reff{eq.tdom.2} it is obvious that
\begin{equation}
   \V \H M \basise_{\rm id}  \;=\;
      c_\emptyset \sum_{B\subseteq \{1,\ldots,m\}} d_B q^{|B|} \, 
      M \basise_{\rm id}
         \,+\,  \sum\limits_{\scrp} b'_\scrp(q) \, \basise_\scrp
\end{equation}
where the coefficients $b'_\scrp(q)$ are polynomials in $q$
of degree at most $m-1$.
\qed

In view of Proposition~\ref{prop.TMblockdiag.2},
we shall henceforth refer to $t_{=}$ and $t_{\neq}$
as the ``dominant diagonal terms'' in the transfer matrix,
as they are indeed dominant at large $|q|$.
Furthermore, we can deduce from 
Propositions~\ref{prop.TMblockdiag.1} and \ref{prop.TMblockdiag.2}
the leading large-$|q|$ behavior of the eigenvalues.
We begin with a simple perturbation lemma proved in 
Ref.~\cite{transfer5}:

\begin{lemma}  {\bf \protect\cite[Lemma~3.2]{transfer5}}
   \label{lemma.perturbation}
Consider an $N \times N$ matrix $M(\xi) = (M_{ij}(\xi))_{i,j=1}^N$
whose entries are analytic functions of $\xi$ in some disc $|\xi| < R$.
Suppose that $M_{11} = 1$ and that $M_{ij} = O(\xi)$ for $(i,j) \neq (1,1)$.
Then, in some disc $|\xi| < R'$, $M(\xi)$ has a simple leading eigenvalue
$\mu_\star(\xi)$ that is given by a convergent expansion
\begin{equation}
   \mu_\star(\xi)  \;=\;  1 \,+\, \sum_{k=2}^\infty \alpha_k \xi^k
 \label{eq.lemma.perturbation.1}
\end{equation}
[note that $\alpha_1 = 0$] with associated eigenvector
\begin{equation}
   \bv_\star(\xi)  \;=\;  \basise_1 \,+\, \sum_{k=1}^\infty \bv_k \xi^k
   \;,
 \label{eq.lemma.perturbation.2}
\end{equation}
while all other eigenvalues are $O(\xi)$.
\end{lemma}

\noindent
{\bf Remarks.} 1. The key fact here is that the eigenvalue shift
in \reff{eq.lemma.perturbation.1} begins at order $\xi^2$, not order $\xi$.

2. The ``small'' eigenvalues need not be analytic in $\xi$.
For instance,
\begin{equation}
   M(\xi)  \;=\;   \left( \! \begin{array}{ccc}
                                  1 & 0 & 0  \\
                                  0 & 0 & \xi \\
                                  0 & \xi^2 & 0
                             \end{array}
                   \!\right)
\end{equation}
has eigenvalues $\mu=1$ and $\mu = \pm \xi^{3/2}$.
\qed

\medskip

To apply Lemma~\ref{lemma.perturbation} to our transfer matrix $\T_{=}$ 
(resp.\  $\T_{\neq}$), we set $\xi = q^{-1}$ and $M = \T_{=}/t_{=}$ 
(resp.\  $\T_{\neq}/t_{\neq}$). We then have:\footnote{
 This Corollary is similar to Corollary~3.3 in \protect\cite{transfer5}.
}

\begin{corollary}
   \label{cor.TMblockdiag.1and2}
Consider operators $\H$ and $\V$ of the form \reff{eq.HV.general}
where the coefficients $\{c_A\}$ and $\{d_B\}$ are \emph{numbers}
(i.e., independent of $q$)
and $c_\emptyset d_{\{1,\ldots,m\}} \neq 0$.
Then $\T_{=}$ and $\T_{\neq}$ each have a single eigenvalue
that is analytic for large $|q|$ and behaves there like 
${\rm const} \times q^m$:
more precisely, these eigenvalues have convergent expansions
\begin{subeqnarray}
   {\lambda_{\star,=} \over t_{=}}  & = &
         1 \,+\, \sum_{k=2}^\infty \alpha^{=}_k q^{-k}    \\[2mm]
   {\lambda_{\star,\neq} \over t_{\neq}}  & = &
         1 \,+\, \sum_{k=2}^\infty \alpha^{\neq}_k q^{-k}
\end{subeqnarray}
so that in particular
\begin{subeqnarray}
   \lambda_{\star,=}     -  t_{=}     & = &    O(|q|^{m-2})   \\
   \lambda_{\star,\neq}  -  t_{\neq}  & = &    O(|q|^{m-2})
\end{subeqnarray}
All other eigenvalues are $O(|q|^{m-1})$.
\end{corollary}

Let us now return to the case of main interest,
in which $\H$ and $\V$ are the transfer matrices
\reff{def_Hbis}/\reff{def_Vbis} for the chromatic polynomial (i.e., 
$v_{i,i+1}=-1$ for $0\le i \le m$, and $v_i=-1$ for $1\le i \le m$).
In this case we can sharpen \reff{eq.prop.TMblockdiag.2}
by providing explicit expressions for the lower-order terms:\footnote{
 This Proposition is similar to Propositions~3.4 and~3.6 
 in \protect\cite{transfer5}.
}

\begin{proposition}
   \label{prop.TMblockdiag.3}
Let $\H$ and $\V$ be the transfer matrices
\reff{def_Hbis}/\reff{def_Vbis} for the chromatic polynomial 
$v_{i,i+1}=v_i =-1$. Then for $m\geq 1$, we have that 
\begin{subeqnarray}
t_{\neq}(m)    &=&  \sum\limits_{k=0}^m  (-1)^k a_{k}(m) \, q^{m-k}
    \slabel{conj.tneq.EARLY} \\[2mm]
t_{=}(m) &=&  \sum\limits_{k=0}^m  (-1)^k a_{k}(m) \, q^{m-k}
                     \,+\, (-1)^{m+1} 
    \slabel{conj.teq.EARLY} 
    \label{conj.t.EARLY}
\end{subeqnarray} 
where each $a_k(m)$ is, for fixed $k\geq 0$, the restriction to integers 
$m\ge \max(1,k)$ of a polynomial in $m$ of degree $k$ given by
\begin{equation}
   a_k(m)  \;=\; 
   \sum_{p=0}^{k}  (-1)^p {m-p\choose p} \sum_{q=0}^{k-p} 3^q
       {m-2p \choose q} {p \choose k-p-q}
   \;. 
\label{eq.akm_final_neq}
\end{equation}
\end{proposition}

The first few coefficients $a_k(m)$ are 
\begin{subeqnarray}
  a_0(m) &=& 1 \\[1mm]
  a_1(m) &=& 2m + 1 \\[1mm]
  a_2(m) &=& 2m^2 + m - 2 \\[1mm]
  a_3(m) &=& \smfrac{4}{3} m^3 - \smfrac{16}{3} m + 1  
\label{def_akm_neq}
\end{subeqnarray}

The proof of Proposition~\ref{prop.TMblockdiag.3} is given in
Appendix~\ref{appendix.polymermodel};
it is based on computing the partition function for a special
one-dimensional polymer model. 

Since we have proven that $a_k(m)$ is a polynomial in $m$
of degree $k$, it is also of interest to obtain explicit expressions for the
coefficients of this polynomial, which we write as
\begin{equation}
a_k(m) \;=\; \sum_{\ell=0}^k
             \frac{ (-1)^\ell 2^{k-2\ell+1}}{(k-\ell)! (\ell+2)!} \;
             a_{k,\ell} \; m^{k-\ell}  \;;
\label{def_akl}
\end{equation}
here the prefactors have been chosen to make many (though not all)
of the coefficients $a_{k,\ell}$ integers
(in fact, they are all integers for $\ell\leq 5$ but not for $\ell=6$, see below).
Now we use the well-known expansion of the falling powers in terms of
Stirling cycle numbers \cite{Graham_94},
\begin{equation}
x^{\underline{r}} \;=\; \sum_{c\geq 0} \stirlingcycle{r}{c}  (-1)^{r-c} x^c \;,
\label{def_xfalling}
\end{equation}
and expand all the binomials in \reff{eq.akm_final_neq} involving $m$.
We arrive after some algebra at the following expression:
\begin{subeqnarray}
a_{k,\ell} & \equiv &
   \frac{ (k-\ell)!(\ell+2)!}{(-1)^\ell 2^{k-2\ell+1}} \,
               [m^{k-\ell}] a_k(m) \\[2mm]
 &=&
\frac{(k-\ell)!(\ell+2)! (-1)^k}{2^{k-2\ell+1}} \,  
           \sum_{p=0}^k \sum_{q=0}^{k-p}
           {p \choose k-p-q} \frac{(-3)^q }{p! q!} \nonumber \\
       & & \qquad\quad \times \sum_{a=0}^p \sum_{c=0}^q
            \stirlingcycle{p}{a}  \stirlingcycle{q}{c}
            \sum_{d=0}^{k-\ell} {a\choose k-\ell-d}
             {c \choose d} 2^{c-d} p^{a+c-k+\ell} 
\label{def_akl_bis_neq}
\slabel{def_akl_bis_neq.b}
\end{subeqnarray}
By computing \reff{def_akl_bis_neq.b} for integers $k \ge \ell \ge 0$,
we find {\em empirically}\/ that $a_{k,\ell}$ is in fact,
for each fixed $\ell$, (the restriction of)
a {\em polynomial}\/ in $k$ of degree $\ell$.
The first few of these polynomials are:
\begin{subeqnarray}
a_{k,0} &=& 1  \slabel{eq.ak0.neq} \\[2mm]
a_{k,1} &=& 3k-9 \\[2mm]
a_{k,2} &=& 6k^2-62k+4 \\[2mm]
a_{k,3} &=& 10k^3-220k^2+410k + 0 \\[2mm]
a_{k,4} &=& 15k^4 - 570k^3 +3245 k^2 + 1662 k +2872 \\[2mm]
a_{k,5} &=& 21k^5 -1225 k^4+ 14245k^3 - 6111 k^2 + 9982k +70560\\[2mm]
a_{k,6} &=& 28k^6 - 2324k^5 + 45780 k^4 - \smfrac{1166396}{9}k^3
  - \smfrac{653968}{3}k^2 \nonumber \\[2mm]
                & & \qquad - \smfrac{3313360 }{9}k + \smfrac{1935488}{3}
\end{subeqnarray}

\medskip

{\bf Full disclosure.}
We did not begin by proving
Propositions~\ref{prop.TMblockdiag.1}--\ref{prop.TMblockdiag.3}.
Rather, we computed the transfer matrices for small widths $m$
and noticed some patterns.
Only then did we try to provide proofs.

\subsection{Block-diagonalization of the transfer matrix (II)}
    \label{sec.block2}

This block-diagonalization can be explained in another way,
without reference to transfer matrices.
We begin by observing that the addition-contraction relation
(which is a variant of the deletion-contraction relation)
for multivariate Tutte polynomials gives
\be
   Z_{S_{m,n}}(q, {\bf v})  \;=\;
      Z_{S^{=}_{m,n}}(q, {\bf v}) \,+\,  Z_{S^{\neq}_{m,n}}(q, {\bf v})
 \label{eq.add-contr}
\ee
where $S^{=}_{m,n}$ is the graph obtained from $S_{m,n}$ by contracting
the two extra vertices into one, while $S^{\neq}_{m,n}$ is the graph
obtained from $S_{m,n}$ by adding a $v=-1$ edge between the two extra
vertices.
Indeed, \reff{eq.add-contr} simply says, in coloring language,
that the two extra vertices in $S_{m,n}$
must receive either the same color or different colors.
The block-diagonalization of $\T$ into $\T_{=}$ and $\T_{\neq}$
simply corresponds to carrying out separate transfer-matrix calculations
for the two families $S^{=}_{m,n}$ and $S^{\neq}_{m,n}$
and observing that
\begin{subeqnarray}
\slabel{def_Zeq}
Z_{S_{m,n}^{=}}(q, {\bf v})     &=& \endv^{\rm T} \, \T_{=}(m)^{n-1} \, 
                           \basise_{\rm id}\\ 
Z_{S_{m,n}^{ \neq}}(q, {\bf v}) &=& \endv^{\rm T} \, \T_{\neq}(m)^{n-1} \, 
                           \basise_{\rm id}
\slabel{def_Zneq}
\label{def_Zblocks}
\end{subeqnarray}

\begin{figure}
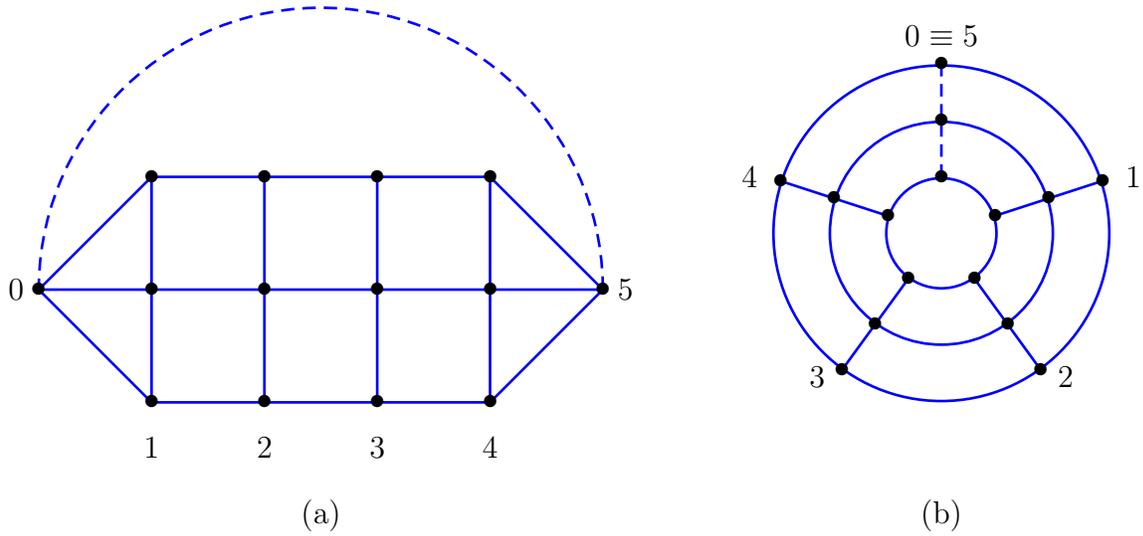

\centering
     \psset{unit=1.5cm}
     \pspicture(0,0)(10,4)
     \psline[linecolor=blue,linewidth=1.0pt](1,0)(1,2)
     \psline[linecolor=blue,linewidth=1.0pt](2,0)(2,2)
     \psline[linecolor=blue,linewidth=1.0pt](3,0)(3,2)
     \psline[linecolor=blue,linewidth=1.0pt](4,0)(4,2)

     \psline[linecolor=blue,linewidth=1.0pt](1,0)(4,0)
     \psline[linecolor=blue,linewidth=1.0pt](1,1)(4,1)
     \psline[linecolor=blue,linewidth=1.0pt](1,2)(4,2)

     \psline[linecolor=blue,linewidth=1.0pt](0,1)(1,0)
     \psline[linecolor=blue,linewidth=1.0pt](0,1)(1,1)
     \psline[linecolor=blue,linewidth=1.0pt](0,1)(1,2)

     \psline[linecolor=blue,linewidth=1.0pt](5,1)(4,0)
     \psline[linecolor=blue,linewidth=1.0pt](5,1)(4,1)
     \psline[linecolor=blue,linewidth=1.0pt](5,1)(4,2)

     \psarc[linecolor=blue,linewidth=1.0pt,linestyle=dashed](2.5,1){2.5}%
          {0}{180}
     \pscircle[linecolor=blue,linewidth=1.0pt](8,1.5){0.5}
     \pscircle[linecolor=blue,linewidth=1.0pt](8,1.5){1}
     \pscircle[linecolor=blue,linewidth=1.0pt](8,1.5){1.5}

     \psline[linecolor=blue,linewidth=1.0pt,linestyle=dashed](8,2)(8,3)
     \psline[linecolor=blue,linewidth=1.0pt](8.47553,1.65451)(9.42658,1.96353)
     \psline[linecolor=blue,linewidth=1.0pt](7.52447,1.65451)(6.57342,1.96353)
     \psline[linecolor=blue,linewidth=1.0pt](7.70611,1.09549)(7.11832,0.286475)
     \psline[linecolor=blue,linewidth=1.0pt](8.29389,1.09549)(8.88168,0.286475)

     \multirput{0}(1,0)(0,1){3}{$\bullet$}
     \multirput{0}(2,0)(0,1){3}{$\bullet$}
     \multirput{0}(3,0)(0,1){3}{$\bullet$}
     \multirput{0}(4,0)(0,1){3}{$\bullet$}
     \multirput{0}(0,1)(5,0){2}{$\bullet$}

     \multirput[c]{0}(8,2)(0,0.5){3}{$\bullet$}
     \multirput[c]{0}(8.47553,1.65451)(0.475528,0.154508){3}{$\bullet$}
     \multirput[c]{0}(7.52447,1.65451)(-0.475528,0.154508){3}{$\bullet$}
     \multirput[c]{0}(7.70611,1.09549)(-0.293893,-0.404508){3}{$\bullet$}
     \multirput[c]{0}(8.29389,1.09549)(0.293893,-0.404508){3}{$\bullet$}

     \rput{0}( 1  ,-0.4){$1$}
     \rput{0}( 2  ,-0.4){$2$}
     \rput{0}( 3  ,-0.4){$3$}
     \rput{0}( 4  ,-0.4){$4$}
     \rput{0}( 5.2, 1){$5$}
     \rput{0}(-0.2, 1){$0$}

     \rput{0}(8.0,3.25){$0\equiv 5$}
     \rput{0}(9.7,2.00){$1$}
     \rput{0}(9.1,0.23){$2$}
     \rput{0}(6.9,0.23){$3$}
     \rput{0}(6.3,2.00){$4$}

     \rput{0}( 8  ,-1.0){(b)}
     \rput{0}( 2.5,-1.0){(a)}
     \endpspicture
     \vspace*{2cm}
\caption{
   (a) Square-lattice strip $S_{4,3}^{=}$ of width $m=4$ and length $n=3$
      with the two extra sites (0 and 5) contracted.
      This contraction is indicated by a dashed line.
   (b) Alternate representation of $S_{4,3}^{=}$ in terms of a
      square-lattice strip of width $m+1=5$, length $n=3$
      and cylindrical boundary conditions, with edge weights $v \to +\infty$
      on the dashed edges.
  \label{fig_SmnEq}
}
\end{figure}

Note also that $S^{=}_{m,n}$ is equivalent to a lattice of width $m+1$
with \emph{cylindrical} boundary conditions
in which $m$ of the columns have the ordinary vertical edges $v_i$
while one column (corresponding to the contracted extra sites)
has $v \to +\infty$;
the horizontal edges are given by the $v_{i,i+1}$ as usual
(see Figure~\ref{fig_SmnEq}).
Likewise, $S^{\neq}_{m,n}$ is equivalent to a lattice of width $m+2$
with cylindrical boundary conditions
in which $m$ of the columns have the ordinary vertical edges $v_i$
while two adjacent columns (corresponding to the two extra sites)
have $v \to +\infty$;
the horizontal edges are given by the $v_{i,i+1}$ as usual
except that the horizontal edges between the two adjacent special columns
are set to $-1$ (see Figure~\ref{fig_SmnNotEq}).

For the chromatic polynomials $P_{S_{m,n}^{=}}(q)$ and $P_{S_{m,n}^{\neq}}(q)$,
we can use the modified transfer matrix $\T'=\H\V\H$ because all the
horizontal weights are $v=-1$:
\begin{subeqnarray}
\slabel{def_Teq}
P_{S_{m,n}^{=}}(q)     &=& \endv^{\rm T} \, \T_{=}'(m)^{n-1} \, 
                           \basisf_{\rm id}\\ 
P_{S_{m,n}^{ \neq}}(q) &=& \endv^{\rm T} \, \T_{\neq}'(m)^{n-1} \, 
                           \basisf_{\rm id}
\slabel{def_Tneq}
\label{def_Tblocks}
\end{subeqnarray}
The transfer matrix $\T_{=}'(m)$ acts
on the space of non-crossing non-nearest-neighbor partitions of the
set $\{0,\ldots,m\}$ with periodic boundary conditions (i.e., $0$ and $m$ 
are also considered to be nearest neighbors). 
However, because of the reflection symmetry,
we can work in the space of equivalence classes modulo reflection
(with respect a diameter passing through the vertex $0$)
of non-crossing non-nearest-neighbor partitions
of the set $\{0,\ldots,m\}$ on a circle.
We write $N_{=}=\dim\T_{=}''(m)$ for the dimension of this space, where
$\T''(m)$ is the symmetry-reduced transfer matrix for the family $S_{m,n}^{=}$.
Likewise, the transfer matrix  $\T_{\neq}'(m)$ acts on the
space of non-crossing non-nearest-neighbor partitions of the
set $\{0,\ldots,m+1\}$ with periodic boundary conditions. 
Once again exploiting the reflection symmetry,
we can work in the space of equivalence classes modulo reflection
(with respect a diameter passing between the vertices $0$ and $m+1$)
of non-crossing non-nearest-neighbor
partitions of the set $\{0,\ldots,m+1\}$ on a circle.
We write $N_{\neq}=\dim\T_{\neq}''(m)$ for the dimension of this space.
In Table~\ref{table_dimensions} we quote the dimensionalities 
$N_{=}(m)$ and $N_{\neq}(m)$ for $1\leq m \leq 14$.

In Appendix~\ref{sec.dim} we compute the following general formulae
for $N_{=}(m)$ and $N_{\neq}(m)$:

\begin{theorem} \label{theorem_Neq}
The number $N_{=}(m)$ of equivalence classes,
modulo reflection with respect an axis going through vertex $0$,
of non-crossing non-nearest-neighbor partitions of 
the set $\{0,1,\ldots,m\}$ on a circle is given by
\be
N_{=}(m) \;=\; {1\over 2}R_{m+1} + 
{1\over 2} \sum_{k=0}^{\lfloor m'/2 \rfloor} {m' \choose k}
{m' -k \choose k + I[\mbox{\rm $m$ is even}]} 
\label{def_Neq}
\ee
where $m' = \lfloor (m+1)/2 \rfloor$
and
\be
   I[\mbox{\em condition}]
   \;\equiv\;
   \begin{cases} 1  & \text{if {\em condition}\/ is true} \\
                 0  & \text{if {\em condition}\/ is false} 
         \end{cases}
\ee
The generating function for this sequence is
\be
G_{=}(z) \;=\; {1\over z}\left[-1 - {1\over 4z} \sqrt{1-3z \over 1+z} +
               {1+2z-z^2 \over 4z \sqrt{(1+z^2)(1-3z^2)}} \right] 
   \;.
\label{def_Geq}
\ee
\end{theorem} 
In the above theorem, $R_m$ denotes the $m$th Riordan number 
(see Appendix~\ref{sec.dim.1} for the main properties of the Riordan numbers
\cite{Bernhart_99}). 
We remark that the sequence $N_{=}(m)$ coincides with sequence {\tt A005218}
in \cite{Sloane_on-line} shifted by two units.

\begin{theorem} \label{theorem_Nnoteq}
The number $N_{\neq}(m)$ of equivalence classes,
modulo reflection with respect an axis
going between the vertices $0$ and $m+1$,
of non-crossing non-nearest-neighbor partitions of
the set $\{0,1,\ldots,m,m+1\}$ on a circle is given by
\be
N_{\neq}(m) \;=\; {1\over 2}R_{m+2} + 
{1\over 2} \sum_{k=0}^{\lfloor m'/2 \rfloor} {m' \choose k}
{m' -k \choose k + I[\mbox{\rm $m$ is odd}]} 
\label{def_Nnoteq}
\ee
where $m' = \lfloor (m+1)/2 \rfloor$.
The generating function for this sequence is
\be
G_{\neq}(z) \;=\; {1\over 4z^3} \left[ 
    {2z^3-z^2+1 \over \sqrt{(1+z^2)(1-3z^2)}} 
  - \sqrt{1-3z \over 1+z} - 2z(1+2z^2) \right] 
   \;.
\label{def_Gnoteq}
\ee
\end{theorem} 

\medskip

It follows that the dimension of the full symmetry-reduced 
transfer matrix $\T''(m)$ is
\be
\dim\T''(m) \;=\; \mbox{\rm SqFree}(m+2) \;=\; N_{\neq}(m) + N_{=}(m)
  \;.
\label{def_dimT0}
\ee
In particular, by using Theorems~\ref{theorem_Neq} and~\ref{theorem_Nnoteq}
we have rederived formula \reff{def_SqFree}
and have found a few equivalent expressions:

\begin{corollary} \label{corollary_dimT}
The number $\dim\T''(m)={\rm SqFree}(m+2)$ of equivalence classes,
modulo reflection in the center of the strip,
of non-crossing non-nearest-neighbor partitions of 
the set $\{0,1,\ldots,m,m+1\}$ is given by
\begin{subeqnarray}
{\rm SqFree}(m+2) &=&
 {1\over 2}M_{m+1} + 
{1\over 2} \sum_{k=0}^{\lfloor m'/2 \rfloor} {m' \choose k}
{m' -k +1 \choose k + 1}
   \slabel{def_dimT_a} \\[2mm]
           &=&
 {1\over 2}M_{m+1} + 
{1\over 2} \sum_{k=0}^{\lfloor m'/2 \rfloor} {m' \choose k}
{m' -k +1 \choose k}
   \slabel{def_dimT_b} \\[2mm]
           &=&
  {1\over 2}M_{m+1} +
{1\over 4} \sum_{k=0}^{\lfloor m'/2 \rfloor} {m' \choose k}
{m' -k +2 \choose k + 1} 
   \slabel{def_dimT_c}
\label{def_dimT}
\end{subeqnarray}
where $m'=\lfloor (m+1)/2\rfloor$. The generating function for this
sequence is
\be
G(z) \;=\; {1\over 2z^2} \left[ 
   {1+z\over z}\, \sqrt{1+z^2\over 1-3z^2} 
   -{\sqrt{(1+z)(1-3z)} \over z} - (1+2z+2z^2) \right] 
\label{def_G}
\ee
\end{corollary} 

We defer to Appendix~\ref{sec.dim} the detailed proof of this Corollary  
and the fact that (\ref{def_dimT}a,b,c) and \reff{def_SqFree}
do in fact coincide.

\subsection{Relating the partition function to the transfer-matrix eigenvalues}
    \label{sec.eigen}

For each boundary condition $\sharp$ ($=$ or $\neq$),
the partition function $Z_{S_{m,n}^\sharp}(q,{\bf v})$
can be expanded in the form
\be
Z_{S_{m,n}^\sharp}(q,{\bf v}) \;=\; \sum\limits_k 
\alpha_{k,\sharp}(m,q,{\bf v}) \, \lambda_{k,\sharp}(m,q,{\bf v})^n \;,
\label{def.Z_eigen}
\ee
where $\lambda_{k,\sharp}(m,q,{\bf v})$ are the eigenvalues (labelled 
arbitrarily) of the transfer matrix $\T_\sharp(m,q,{\bf v})$,
and $\alpha_{k,\sharp}(m,q,{\bf v})$ are the corresponding amplitudes.
Please note that the $\alpha_{k,\sharp}$ and $\lambda_{k,\sharp}$ are 
algebraic functions of $q$ and ${\bf v}$,
but we have to be careful about labels as we go around branch points.

For large $|q|$ everything becomes simple. 
By Corollary~\ref{cor.TMblockdiag.1and2} there exists,
for each $m$ and ${\bf v}$,
a value $Q(m,{\bf v})$ such that (for both boundary conditions)
the dominant eigenvalue $\lambda_{\star,\sharp}(m,q,{\bf v})$ is analytic
in the region $|q|> Q(m,{\bf v})$ and has there a convergent expansion
in powers of $q^{-1}$,
\begin{equation}
    \lambda_{\star,\sharp}(m,q,{\bf v})  \;=\;
    \sum\limits_{k=0}^\infty  (-1)^k b^{\sharp}_{k}(m,{\bf v}) \, q^{m-k}
   \;,
     \label{def_bkm}
\end{equation}
with all other eigenvalues being strictly smaller in modulus.
We have $b_0^\sharp(m,{\bf v})=1$ by Proposition~\ref{prop.TMblockdiag.2} 
specialized to the particular operators $\H$ and $\V$ given 
by \reff{def_Hbis}/\reff{def_Vbis}.
By increasing $Q(m,{\bf v})$ if necessary,
we may also suppose that the amplitude $\alpha_{\star,\sharp}(m,q,{\bf v})$
corresponding to the dominant eigenvalue is nonvanishing
for $|q| > Q(m,{\bf v})$.
Inserting these facts into \reff{def.Z_eigen},
we conclude that for $|q| > Q(m,{\bf v})$ we have 
\begin{subeqnarray}
\label{def.ratio_bk1}
\lim_{n\to\infty} \left( \frac{ Z_{ S_{m,n}^\sharp}(q,{\bf v}) }
                              {q^{|V(S_{m,n}^\sharp)|}} \right)^{1/n}  
  &=&  q^{-m} \lambda_{\star,\sharp}(m,q,{\bm v}) \\
  &=&  1 \,+\, \sum\limits_{k=1}^\infty (-1)^k b_k^\sharp(m,{\bf v}) \, q^{-k} 
\slabel{def.ratio_bk}
\label{def.ratio_bk2}
\end{subeqnarray} 
uniformly on compact subsets of $|q| > Q(m,{\bf v})$.
[Here we used $\lim_{n\to \infty} |V(S_{m,n}^\sharp)|/n = m$.]
It follows from this that \reff{def.ratio_bk2} also holds
order-by-order as a power series in $1/q$.  

Unfortunately, the foregoing argument does not control
the dependence of $Q(m,{\bf v})$ on $m$ and ${\bf v}$.
However, the uniformity in $m$ (and ${\bf v}$) can be proven,
at least for the antiferromagnetic case $-1 \le v_e \le 0$,
by invoking the following result:\footnote{
  Actually, \cite[Theorem~6.3]{Sokal_chromatic_bounds}
  also gives a more general result applicable to {\em complex}\/ edge weights
  in the ``complex antiferromagnetic regime'' $|1+v_e| \le 1$,
  albeit with slightly weaker bounds.
}

\begin{theorem}
{\bf \protect \cite[Theorem~6.3]{Sokal_chromatic_bounds}}
   \label{thm.sokal3}
Let $G=(V,E)$ be a loopless finite undirected graph in which all vertices
have degree $\le \Delta'$ except perhaps for an $N$-clique $y_1,\ldots,y_N$.
Let $G$ be equipped with edge weights $\{ v_e \}_{e \in E}$
satisfying $-1 \le v_e \le 0$ for all $e$
and $v_{\<y_i y_j\>} = -1$ for all $i \neq j$.
Then all the zeros of $Z_G(q, \bv)$
lie in the disc $|q| < 7.963907 \Delta' + N$.
\end{theorem}

For the family $S^{=}_{m,n}$ we have $\Delta'=4$ and $N=1$,
where $y_1$ is the extra site;
therefore, all zeros of $P_{S^{=}_{m,n}}(q)$ lie within the disc
$|q| < 7.963907 \Delta' + 1 = 32.855628$.
For the family $S^{\neq}_{m,n}$ we have $\Delta'=4$ and $N=2$,
where the 2-clique is formed by the two extra sites;
therefore, all zeros of
$P_{S^{\neq}_{m,n}}(q)$ lie within the slightly larger disc
$|q| < 7.963907 \Delta' + 2 = 33.855628$.

Using the Beraha--Kahane--Weiss theorem (Theorem~\ref{BKW_thm})
we can conclude that, outside such a disc,
the transfer matrix
$\T_\sharp(m,q,{\bf v})$
for each width $m$ and each antiferromagnetic coupling ${\bf v}$
must have one and only one eigenvalue of largest modulus 
$\lambda_{\star,\sharp}(m,q,{\bf v})$.
Since this dominant eigenvalue cannot collide with any other 
eigenvalue of the same transfer matrix,
it must be an analytic function of $q$ (and nonvanishing)
outside the given disc;
and the series \reff{def_bkm} must be convergent outside the given disc.
That is, we can take $Q(m,{\bf v}) = 33.855628$ for all $m$
and all antiferromagnetic ${\bf v}$.

In summary, the transfer matrix for a square-lattice strip of width $m$
and with $=$ or $\neq$ boundary conditions has a single dominant
eigenvalue $\lambda_{\star,=}(m,q,{\bf v})$ or 
$\lambda_{\star,\neq}(m,q,{\bf v})$
that is a nonvanishing analytic function of $q$
(in fact, $q^m$ times an analytic function of $q^{-1}$)
whenever $|q| > 33.855628$, uniformly in $m$ and in ${\bf v}$ belonging to the
antiferromagnetic regime.

\subsection[Comparing $S_{m,n}^{\neq}$ with $S_{m,n}^{=}$]
           {Comparing $\bm{S_{m,n}^{\neq}}$ with $\bm{S_{m,n}^{=}}$}
    \label{sec.comp}

A {\em 2-terminal graph}\/ $(G,s,t)$ is a (loopless and connected) graph
$G=(V,E)$ with two distinguished vertices $s$ and $t$ ($s\neq t$),
called the {\em terminals}\/.
We define the partial partition functions of $(G,s,t)$
as follows \cite[Section~2.1]{Sokal_chromatic_roots}:
\begin{subeqnarray}
Z_G^{(s\not\leftrightarrow t)}(q,{\bf v})  &=&
\sum_{\begin{scarray}
A \subseteq E \\
A \, {\rm does\,not\,connect} \, s \, {\rm to} \, t
\end{scarray}
}
q^{k(A)} \;  \prod_{e \in A} v_e \slabel{def.zgsnt} \\[4mm]
Z_G^{(s\leftrightarrow t)}(q,{\bf v})   &=& 
\sum_{\begin{scarray}
A \subseteq E \\
A \, {\rm connects} \, s \, {\rm to} \, t
\end{scarray}
}
q^{k(A)} \;  \prod_{e \in A}  v_e
\label{def.zgst}
\end{subeqnarray} 
{}From \reff{eq.FK.identity} it follows trivially that
\begin{equation}
   Z_G(q, {\bf v}) \;=\;  
   Z_G^{(s\leftrightarrow t)}(q,{\bf v})  \,+\, 
   Z_G^{(s\not\leftrightarrow t)}(q,{\bf v}) \,. 
 \label{eq.Z.identity1}
\end{equation}
We define $G \bullet st$ to be the graph in which $s$ and $t$ are 
contracted into a single vertex. (Note that if $G$ contains one or more
edges $st$, then these edges are {\em not}\/ deleted, but become loops
in $G \bullet st$.) Then from \reff{eq.FK.identity} it also follows 
\cite{Sokal_chromatic_roots} that
\begin{equation}
   Z_{G\bullet st}(q, {\bf v}) \;=\;  
             Z_G^{(s\leftrightarrow t)}(q,{\bf v})  \,+\, 
   q^{-1} \, Z_G^{(s\not\leftrightarrow t)}(q,{\bf v}) \,. 
 \label{eq.Z.identity2}
\end{equation}
Finally, let $G+st$ define the graph $G$ with an extra edge $st$. Then it
also follows from \reff{eq.FK.identity} that
\begin{equation}
   Z_{G + st}(q, {\bf v}) \;=\;  
   (1+ v_{st}) Z_G^{(s\leftrightarrow t)}(q,{\bf v})  \,+\, 
   \left( 1 + \frac{v_{st}}{q}\right) \, 
               Z_G^{(s\not\leftrightarrow t)}(q,{\bf v}) \,, 
 \label{eq.Z.identity3}
\end{equation}
which equals $Z_G(q, {\bf v}) + v_{st}Z_{G\bullet st}(q, {\bf v})$
in agreement with the deletion-contraction formula.  

Both $Z_G^{(s\not\leftrightarrow t)}(q,{\bf v})$ and 
$Z_G^{(s\leftrightarrow t)}(q,{\bf v})$
are polynomials in $q$ and ${\bf v}$.
At large $|q|$, the leading behavior of $Z_G^{(s\leftrightarrow t)}$ and
$Z_G^{(s\not\leftrightarrow t)}$ is
\begin{subeqnarray}
Z_G^{(s\not\leftrightarrow t)}(q,{\bf v}) &=& q^{|V(G)|} \:+\: \ldots \\[2mm]
Z_G^{(s\leftrightarrow t)}(q,{\bf v})     &=& 
   q^{|V(G)|-d_G(s,t)} \!\!\!\! \sum_{\begin{scarray}
{\rm shortest\,paths} \\
s\to t 
\end{scarray}} 
\prod_{e \in {\rm path}} v_e  \:+\: \ldots 
   \label{eq.shortest1}
\end{subeqnarray}
where $d_G(s,t)$ is the length of the shortest path from $s$ to $t$ using
edges having $v_e\ne 0$. 

Let us now specialize these general formulae to the case $G=S_{m,n}$,
with the terminals $s,t$ taken to be the two special vertices. Then
\begin{subeqnarray}
G \bullet st &=& S_{m,n}^{=} \\[1mm]
G +       st &=& S_{m,n}^{\neq}  
\end{subeqnarray} 
so that
\begin{subeqnarray}
   Z_{S_{m,n}^{=}}(q,{\bf v})
   & = & 
     Z_{S_{m,n}}^{(s\leftrightarrow t)}(q,{\bf v})  \:+\:
     q^{-1} \,  Z_{S_{m,n}}^{(s\not\leftrightarrow t)}(q,{\bf v})
     \slabel{eq.ZSmn.eq}   \\[2mm]
   Z_{S_{m,n}^{\neq}}(q,{\bf v})
   & = & 
     (1 - q^{-1}) \,  Z_{S_{m,n}}^{(s\not\leftrightarrow t)}(q,{\bf v})
     \slabel{eq.ZSmn.neq}
     \label{eq.ZSmn}
\end{subeqnarray}
We have $|V(G)|=mn+2$, $|V(G \bullet st)|=mn+1$ and $|V(G + st)|=mn+2$.
Furthermore, we have $d_G(s,t)=m+1$,
and there are $n$ shortest paths from $s$ to $t$:
each of them consists of the $m-1$ horizontal edges in a row of $S_{m,n}$
together with the two additional edges joining this row to the vertices $s,t$.
Therefore
\be
\sum_{\begin{scarray}
{\rm shortest\,paths} \\
s\to t
\end{scarray}}
\prod_{e \in {\rm path}} v_e  \;=\; n \prod_{i=0}^m v_{i,i+1} \;.
   \label{eq.shortest2}
\ee
Finally, in the graph $G+st = S_{m,n}^{\neq}$ we have $v_{st}=-1$. 

For $|q| > Q(m,{\bf v})$, using \reff{def.ratio_bk1} applied to $S_{m,n}^{\neq}$
together with \reff{eq.ZSmn.neq}, we see that
\begin{subeqnarray}
\slabel{eq.lambdaNeq1}
\lambda_{\star,\neq}(m,q,{\bf v}) &=& \lim_{n\to\infty} \left( 
       Z_{S_{m,n}}^{(s\not\leftrightarrow t)}(q,{\bf v}) \right)^{1/n} \\
       & =& q^m \left[1 + \sum\limits_{k=1}^\infty 
                (-1)^k b_k^{\neq} (m,{\bf v})\, q^{-k} \right] 
   \;.
\slabel{eq.lambdaNeq2}
\label{eq.lambdaNeqFull}
\end{subeqnarray}
Likewise, using \reff{def.ratio_bk1} applied to $S_{m,n}^{=}$
together with \reff{eq.ZSmn.eq}, \reff{eq.shortest1} and \reff{eq.shortest2},
we see that
\begin{subeqnarray}
\slabel{eq.lambdaEq1}
\lambda_{\star,=}(m,q,{\bf v}) &=& \lim_{n\to\infty} \left(
       Z_{S_{m,n}}^{(s\leftrightarrow t)}(q,{\bf v}) +  
       q^{-1} \,  
       Z_{S_{m,n}}^{(s\not\leftrightarrow t)}(q,{\bf v}) \right)^{1/n} \\
       & =& \lambda_{\star,\neq}(m,q,{\bf v}) \, \lim_{n\to\infty}  
       \left[ 1 + q \frac{ Z_{S_{m,n}}^{(s\leftrightarrow t)}(q,{\bf v})}
                         {Z_{S_{m,n}}^{(s\not\leftrightarrow t)}(q,{\bf v})}
       \right]^{1/n} \\
       &=& \lambda_{\star,\neq}(m,q,{\bf v}) \, \lim_{n\to\infty} 
       \left[ 1 + q^{-m} n \prod_{i=0}^{m} v_{i,i+1} + O(q^{-m-1}) 
       \right]^{1/n} \\
       &=& \lambda_{\star,\neq}(m,q,{\bf v}) \, 
       \left[ 1 + q^{-m} \prod_{i=0}^m v_{i,i+1} + O(q^{-m-1})
       \right] \\
       &=& \lambda_{\star,\neq}(m,q,{\bf v}) \:+\: \prod_{i=0}^m v_{i,i+1} \:+\:
       O(q^{-1}) 
     \;.
\slabel{eq.lambdaEq2}
\label{eq.lambdaEqFull}
\end{subeqnarray}
Please note the logic in the step from
(\ref{eq.lambdaEqFull}c) to (\ref{eq.lambdaEqFull}d):
we know from \reff{def.ratio_bk1} that the limit exists
order-by-order as a power series in $q^{-1}$;
so the term of order $q^{-m}$ is indeed as claimed,
and everything else is of order $q^{-m-1}$ or smaller.
We do not need to have any explicit bounds on the $n$-dependence
of the $O(q^{-m-1})$ terms in (\ref{eq.lambdaEqFull}c).
We have therefore proven:

\begin{proposition} \label{prop.diff.eigen.0}
The difference at large $|q|$ between the leading eigenvalues
$\lambda_{\star,=}(m,q,{\bf v})$ and $\lambda_{\star,\neq}(m,q,{\bf v})$
of the transfer matrices $\T''_{=}(m,q,{\bf v})$ and $\T''_{\neq}(m,q,{\bf v})$
for $m\ge 1$ satisfies
\be
\lambda_{\star,=}(m,q,{\bf v}) -  \lambda_{\star,\neq}(m,q,{\bf v}) \;=\; 
\prod_{i=0}^m v_{i,i+1} + O(q^{-1}) \;.
\label{eq.diff.lambda}
\ee
In particular, for the chromatic-polynomial case $v_e=-1$ we have
\begin{equation}
   \lambda_{\star,=}(m,q) - \lambda_{\star,\neq}(m,q)  \;=\;
   (-1)^{m+1} \,+\, O(q^{-1})
   \;.
\end{equation}
\end{proposition}
The formula \reff{eq.diff.lambda} generalizes the result \reff{eq.curious.2}
for theta graphs [which corresponds to setting the vertical couplings $v_i=0$].
It has the same form as the result \reff{eq.curious.4}
for the dominant diagonal entries
--- the only difference being that \reff{eq.curious.4} is exact,
while \reff{eq.diff.lambda} is for the leading term only.

%
% Numerical results I
%
\section{Numerical results I: Large-$\bm{q}$ expansion of the
   leading eigenvalue}   \label{sec.numerical}

In this section we compute the large-$q$ expansion of
the leading eigenvalues $\lambda_{\star,=}(m)$ and $\lambda_{\star,\neq}(m)$
and determine empirically some of their remarkable properties.
In Section~\ref{sec.thermo} we shall provide theoretical explanations
of some (but not all!) of these empirical observations. 

\subsection{Overview of method and results}

In Proposition~\ref{prop.TMblockdiag.3} (proven in 
Appendix~\ref{appendix.polymermodel})
we computed in closed form the dominant diagonal entry in the transfer matrix,
$t_{=}$ or $t_{\neq}$,
for a strip of width $m\geq 1$ with either $=$ or $\neq$ boundary conditions.
We found that this entry is in each case a polynomial in $q$ of degree $m$:
\begin{subeqnarray}
   t_{\neq}(m) & = &  \sum\limits_{k=0}^m  (-1)^k a_{k}(m) \, q^{m-k}
  \slabel{tneq_atilde_c}
        \\[2mm]
   t_{=}(m) & = & \sum\limits_{k=0}^m (-1)^k a_{k}(m) \, q^{m-k} \,+\, 
   (-1)^{m+1} \slabel{teq_atilde_c}
\end{subeqnarray}
We furthermore computed in closed form the coefficients $a_{k}(m)$,
which are in fact, for each fixed $k\geq 0$, the restriction to integers
$m\geq \max(1,k)$ of polynomials in $m$ of degree $k$
[cf.\ \reff{eq.akm_final_neq}].

In this section we want to carry out an analogous
(though ultimately less explicit) computation
for the leading {\em eigenvalue}\/ of the transfer matrix,
which we call $\lambda_{\star,=}$ or $\lambda_{\star,\neq}$.
{}From Section~\ref{sec.eigen} we recall that 
each $\lambda_\star(m)$ has, for large $|q|$,
a convergent expansion in powers of $q^{-1}$,
\begin{equation}
    \lambda_{\star,\sharp}(m)  \;=\;
        \sum\limits_{k=0}^\infty  (-1)^k b^{\sharp}_{k}(m) \, q^{m-k}
    \;,
\end{equation}
where $\sharp$ denotes $=$ or $\neq$.
It is also illuminating to pass from
the eigenvalue $\lambda_\star(m)$ to its logarithm,
which is a free energy, and define
\begin{equation}
   \log {\lambda_{\star,\sharp}(m)  \over q^m}
   \;=\;
   \sum\limits_{k=1}^\infty  c^{\sharp}_k(m) \, q^{-k}
   \;.
 \label{def_ckm}
\end{equation}

Corollary~\ref{cor.TMblockdiag.1and2} guarantees that
the first two terms in the expansion of the leading eigenvalue
coincide with those in the dominant diagonal entry:
\begin{equation}
   b^{=}_k(m)  \;=\;  b^{\neq}_k(m)  \;=\;  a_k(m)
   \qquad\hbox{for $k=0,1$}
 \label{eq.bkm.akm.k01}
\end{equation}
(with $m \ge 2$ required for $b_1^{=}$
 in order to avoid the $(-1)^{m+1}$ term),
or more specifically
\begin{subeqnarray}
   b^{=}_0(m)  \;=\;  b^{\neq}_0(m)  & = &   1   
       \slabel{eq.bkm.akm.k0} \\[1mm]
   b^{=}_1(m)  \;=\;  b^{\neq}_1(m)  & = &   2m+1   
       \qquad \hbox{for $m \ge 2$}
\end{subeqnarray}
Here we shall go further and compute the coefficients $b^{=}_k(m)$
and $b^{\neq}_k(m)$ for $1 \le m \le 12$ and $1 \le m \le 11$,
respectively, with $0 \le k \le 40$.\footnote{
   It would not be difficult to extend this computation to
   much larger values of $k$, if we really cared.
   Extension to larger values of $m$ is, however, an extremely demanding
   computational task.
   The dimensions of the transfer matrices $\T''_{=}(13)$ and 
   $\T''_{\neq}(12)$ are $15465$ and $15339$, respectively,
   which are beyond the capabilities of our current computer facilities.    
}
We shall find empirically that
the agreement between the eigenvalue and the dominant diagonal entry
extends one term beyond that
guaranteed by Corollary~\ref{cor.TMblockdiag.1and2}:
namely, we have\footnote{
   In \cite{transfer5} we found, for free and cylindrical boundary
   conditions, an agreement extending {\em two}\/ terms
   beyond that guaranteed by Corollary~\ref{cor.TMblockdiag.1and2}.
}
\begin{subeqnarray}
   b^{\neq}_2(m)  & = &  a_2(m)   \qquad\hbox{for $m\geq 2$}  \\[1mm]
   b^{=}_2(m)     & = &  a_2(m)   \qquad\hbox{for $m \ge 3$}
\end{subeqnarray}
Beyond this, the eigenvalues deviate from the dominant diagonal terms.
However, from Proposition~\ref{prop.diff.eigen.0} we know ---
and our computations of course confirm ---
that the deviations of the eigenvalues from the dominant diagonal terms
are the {\em same}\/ for $=$ and $\neq$ boundary conditions,
all the way down to order 1, so that the {\em difference}\/ between
the two eigenvalues has the same leading behavior as the
difference between the two dominant diagonal terms:
\begin{equation}
   \lambda_{\star,=}(m) - \lambda_{\star,\neq}(m)  \;=\;
   (-1)^{m+1} \,+\, O(q^{-1})
 \label{eq.curious.lambda}
\end{equation}
as compared with
\be 
   t_{=}(m) - t_{\neq}(m)  \;=\; (-1)^{m+1}  \;.
\ee
In terms of the coefficients, this says that
\begin{equation}
   b^{=}_k(m) - b^{\neq}_k(m)  \;=\;
      \begin{cases}
          0   & \text{for $0 \le k \le m-1$}  \\
          -1  & \text{for $k=m$}
      \end{cases}
\end{equation}
and 
\begin{equation}
   c^{=}_k(m) - c^{\neq}_k(m)  \;=\;
      \begin{cases}
          0   & \text{for $1 \le k \le m-1$}  \\
          (-1)^{m+1}  & \text{for $k=m$}
      \end{cases}
\end{equation}

This amazing behavior suggests that we look more closely at the differences
between $=$ and $\neq$ boundary conditions for orders $k \ge m$.
We therefore define coefficients $d_\ell(m)$ by
\begin{equation}
   (-1)^{m+1} [\lambda_{\star,=}(m) - \lambda_{\star,\neq}(m)]
   \;=\;
   1 + \sum_{\ell=1}^\infty (-1)^\ell d_\ell(m) \, q^{-\ell}
 \label{def_series_dl}
\end{equation}
or equivalently
\begin{equation}
   d_\ell(m)  \;=\;  - \, [b^{=}_{m+\ell}(m) - b^{\neq}_{m+\ell}(m)]
   \;.
\end{equation}
It furthermore turns out to be interesting to look at the
{\em logarithm}\/ of this difference of eigenvalues:
\begin{equation}
   \log\bigl\{
   (-1)^{m+1} [\lambda_{\star,=}(m) - \lambda_{\star,\neq}(m)]
   \bigr\}
   \;=\;
   \sum_{\ell=1}^\infty e_\ell(m) \, q^{-\ell}
   \;.
 \label{def_series_el}
\end{equation}
The following variant is also of interest:
\begin{equation}
   - \log\left[
   (-1)^{m+1} q^m \log\Bigl( {\lambda_{\star,=}(m) \over
                              \lambda_{\star,\neq}(m)
                             }
                      \Bigr)
   \right]
   \;=\;
   \sum_{\ell=1}^\infty f_\ell(m) \, q^{-\ell}
   \;.
 \label{def_series_fl}
\end{equation}
We shall obtain empirical formulae for the coefficients
$d_\ell(m)$, $e_\ell(m)$ and $f_\ell(m)$ for $1 \le \ell \le 13$.

In this section we shall proceed as follows
(imitating what we did in \cite[Section~4]{transfer5}
 for free and cylindrical boundary conditions):
First we shall compute the transfer matrices for strips of width
$m \le 11$ (resp.\ $m \le 12$) for $\neq$ (resp.\ $=$) boundary conditions,
using the methods of \cite{transfer1,transfer2}
as modified for the family $S_{m,n}$
(see Section~\ref{sec.transfer} above).
{}From these matrices we then extract the dominant eigenvalue as a 
power series in $q^{-1}$, i.e.\ for each available $m$ we compute
as many coefficients $b^{\sharp}_k(m)$ and $c^{\sharp}_k(m)$
as we please.\footnote{
   To compute the dominant eigenvalue as a power series in $q^{-1}$,
   we have used the power method \cite[Section~7.3.1]{Golub_book}
   {\em in symbolic form}\/.  Each iteration gives one additional term
   in the expansion of the dominant eigenvalue in powers of $q^{-1}$.
   We can therefore compute the {\em exact}\/ expansion up to
   any desired order in a {\em finite}\/ number of steps.
}
We then observe {\em empirically}\/ that, for each $k \ge 0$,
the coefficients $b^{=}_k(m)$ and $b^{\neq}_k(m)$
[resp.\ $c^{=}_k(m)$ and $c^{\neq}_k(m)$]
are {\em equal}\/ and are represented by
a polynomial $B_k(m)$ [resp.\ $C_k(m)$] of degree $k$ [resp.\ degree 1] in $m$,
{\em provided that we restrict to integers}\/
$m \ge m_{\rm min}(k)$, where
\begin{subeqnarray}
   m_{\rm min}^{\neq}(k) & = &  \max(k,1)  \\[1mm]
   m_{\rm min}^{=}(k)    & = &  k+1
\end{subeqnarray}
for the two boundary conditions.\footnote{
   By contrast, for the dominant diagonal entry we have {\em proven}\/
   in Proposition~\ref{prop.TMblockdiag.3}
   that $a_k(m)$ are polynomials in $m$ of degree $k$;
   and in this case the polynomial form holds for {\em all}\/ allowable
   integers $m$, i.e.\ $m \ge \max(k,1)$.
}
{\em Assuming}\/ that this empirical observation is accurate
(i.e., that the polynomial behavior observed for $m \le 11,12$
 persists to all larger $m$),
we can infer the exact expressions for the polynomials $B_k$ and $C_k$
for $k \le 17$ for $\neq$ and $=$ boundary conditions.
The fact that the polynomials $C_k$ are of degree 1 in $m$
--- hence consist of a bulk term $\alpha_k m$ and a surface term $\beta_k$ ---
can be explained (non-rigorously) by finite-size-scaling theory:
see Section~\ref{sec.thermo} below
and especially the discussion around
eqns.~\reff{def_FSS_Ansatz2}/\reff{def_FSS_Ansatz2_bis}
and at the beginning of Section~\ref{sec.thermo.neqeq}.

Analogously --- and rather more surprisingly ---
we find that the coefficients $d_\ell(m)$, $e_\ell(m)$ and $f_\ell(m)$ 
are also represented by polynomials $D_\ell(m)$, $E_\ell(m)$ and $F_\ell(m)$, 
of degrees $\ell$, 1 and 1, respectively,
provided that we restrict to integers $m \ge \ell+1$.
This behavior --- in which the difference of two eigenvalues
(which are essentially free energies)
behaves {\em like a partition function}\/
in the sense that its {\em logarithm}\/ grows asymptotically linearly
in the strip length --- is wholly unexpected, at least to us.

\subsection[Results for $S_{m,n}^{\neq}$]{Results for $\bm{S_{m,n}^{\neq}}$}
   \label{sec4.2}

Using the methods just described, we have obtained
the leading eigenvalue $\lambda_{\star,\neq}(m)$
for $1 \leq m \leq 11$ as a power series in $q^{-1}$
[cf.\ \reff{def_bkm}] through order $k=40$.
The resulting coefficients $b^{\neq}_k(m)$ are displayed
for $0 \le k \le 15$ in Table~\ref{table_coef_b_neq},
and the corresponding coefficients $c^{\neq}_k(m)$ [cf.\ \reff{def_ckm}]
are displayed in Table~\ref{table_coef_c_neq}.
(The complete data set for $0 \le k \le 40$
 is contained in the {\sc Mathematica} file {\tt data\_Neq.m}
 that is included in the on-line version of this paper at arXiv.org.)
It is interesting to note that for all pairs $(k,m)$ that we have computed
(namely, $1\leq m \leq 11$ and $0\leq k \leq 40$), the coefficients
$b^{\neq}_k(m)$ and $k c^{\neq}_k(m)$ are integers.
Furthermore, we observe {\em empirically}\/ that, for each fixed $k$,
the coefficients $b^{\neq}_k(m)$ [resp.\ $c^{\neq}_k(m)$]
are the restriction to integers $m$
of a polynomial $B_k(m)$ [resp.\ $C_k(m)$]
of degree $k$ (resp.\ degree 1) in $m$,
{\em provided that we restrict attention to
$m\geq m_{\rm min}^{\neq}(k)$}\/  with
\begin{equation}
m_{\rm min}^{\neq}(k) \;=\; \max(k,1)  \;.
\label{def_m_min_neq}
\end{equation}
Below this threshold $m^{\neq}_{\rm min}(k)$,
the coefficients deviate from polynomial behavior.
With our available data together with a few tricks described below,
we are able to determine these polynomials for $0\leq k \leq 17$.

First we start by trying to fit the coefficients $b^{\neq}_k(m)$ with
$m \ge m^{\neq}_{\rm min}(k)$ to a polynomial $B_k(m)$
of degree $k$. As we need $k+1$
coefficients for such a polynomial, we are able to obtain these
polynomials when $m^{\neq}_{\rm min}(k) + k \le 11$, i.e.\ $k \le 5$.
Please observe that in all cases we have at least one data point
more than the number of unknowns, so that every fit can be tested
on at least one extra data point.\footnote{
   In our view this test is important, because the formula \reff{def_m_min_neq}
   for $m_{\rm min}^{\neq}(k)$ is only empirical;  and if it were to be
   mistaken by one unit (i.e.\ the real threshold for polynomial behavior
   were one unit higher than we thought), then a fit with no extra data points
   would give a definite result (one can {\em always}\/ fit $k+1$ points to a
   polynomial of degree $k$) {\em but this result would be total nonsense}\/.
}
Our results are:
\begin{subeqnarray}
   B_0(m)  & = &   1       \\[2mm]
   B_1(m)  & = &   2m+1       \\[2mm]
   B_2(m)  & = &   2m^2 + m - 2       \\[2mm]
   B_3(m)  & = &   \smfrac{4}{3} m^3 - \smfrac{16}{3} m + 3   \\[2mm]
   B_4(m)  & = &   \smfrac{2}{3} m^4 - \smfrac{2}{3} m^3  
                  -\smfrac{37}{6} m^2 + \smfrac{61}{6} m -3  \\[2mm]
   B_5(m)  & = &   \smfrac{4}{15} m^5 - \smfrac{2}{3} m^4 - \smfrac{13}{3}m^3
                  +\smfrac{91}{6}m^2 - \smfrac{343}{30}m - 2 
\label{def_bk_poly_neq}
\end{subeqnarray}
We see that the three highest-order coefficients in these polynomials
agree with those of the corresponding polynomial $a_k(m)$
[cf.\ \reff{eq.akm_final_neq}/\reff{def_akm_neq}], i.e.
\begin{equation}
B_k(m) \;=\;
     \begin{cases}
        a_k(m)                   & \text{for $0 \le k \le 2$} \\[2mm]
        a_k(m) + O(m^{k-3})      & \text{for $k\geq 3$} 
     \end{cases}
\end{equation}

There is, however, a better way of extracting the desired information from
our numerical data: instead of using the coefficients $b^{\neq}_k(m)$ as
our basic quantities, we can use the related coefficients $c^{\neq}_k(m)$
[cf.\ \reff{def_ckm}].
The latter coefficients are empirically found to be,
for each fixed $k$,
the restriction to integer $m$ of a polynomial $C_k(m)$
{\em of degree 1}\/, i.e.\
\begin{equation}
   C_k(m)  \;=\;  \alpha_k m + \beta_k  \;,
 \label{def.Ck}
\end{equation}
provided again that we restrict attention to
$m\ge$ the same $m^{\neq}_{\rm min}(k)$ defined in \reff{def_m_min_neq}.
As we now need only the {\em two}\/ coefficients $\alpha_k$ and $\beta_k$
to specify such a polynomial, we are able to obtain these polynomials
up to $k=9$ (if we want at least one extra data point to test the fit)
or $k=10$ (if we don't).
The first polynomials $C_k$ are given by
\begin{subeqnarray}
   C_1(m)    & = &              -2m - 1              \\[2mm]
   C_2(m)    & = &              - m - \smfrac{5}{2}  \\[2mm]
   C_3(m)    & = & \smfrac{1}{3}  m - \smfrac{16}{3}  \\[2mm]
   C_4(m)    & = & \smfrac{5}{2}  m - \smfrac{41}{4} \\[2mm]
   C_5(m)    & = & \smfrac{28}{5} m - \smfrac{81}{5} %%%\\[2mm]
%%%%%%%%%%%%%%%%%%%%%%%%%%%%%%%%%%%%%%%%%%%%%%%%%%%%%%%%%%%%%%%
%   C_6(m)    & = & \smfrac{55}{6} m - \smfrac{49}{3} \\[2mm]
%   C_7(m)    & = & \smfrac{89}{7} m + \smfrac{55}{7} \\[2mm]
%   C_8(m)    & = & \smfrac{81}{4} m + \smfrac{719}{8} \\[2mm]
%   C_9(m)    & = & \smfrac{505}{9}m + \smfrac{2459}{9}   \\[2mm]
%   C_{10}(m) & = & \smfrac{1029}{5}m + \smfrac{1239}{2}
%%%%%%%%%%%%%%%%%%%%%%%%%%%%%%%%%%%%%%%%%%%%%%%%%%%%%%%%%%%%%%%
\label{def_ck_poly_neq}
\end{subeqnarray}
and the results for $6 \le k \le 10$ are shown in Table~\ref{table_new_Ck}.
The polynomials $B_k$ for $k\le 10$ can then be
determined from the $C_k$ using \reff{def_bkm}/\reff{def_ckm}.

Actually, we can do better than this.
We believe (or at least conjecture!)\ that the coefficients
$c_k^{\neq}(m)$ are, for each fixed $k \ge 0$,
the restriction to integers $m \ge m^{\neq}_{\rm min}(k)$
of a polynomial $C_k(m)$ of degree 1.
If we now compute the difference
\begin{equation}
\Delta^{\neq}_k(m) \;=\; c^{\neq}_k(m) - C_k(m)
\label{def_Delta}  
\end{equation}
between the numerical coefficients $c^{\neq}_k(m)$
and the corresponding polynomials $C_k(m)$,
we find, not surprisingly,
that these differences are nonzero whenever $m< m^{\neq}_{\rm min}(k)$:
see Table~\ref{table_diff_coef_c_neq},
where we initially know $\Delta^{\neq}_k(m)$ only for $k \le 10$
(or $k \le 9$ if we wish to be conservative).
If we could somehow guess an analytic form for at least some of
these coefficients $\Delta^{\neq}_k(m)$,
we could then define ``improved'' coefficients $\widehat{c}^{\neq}_k(m)$ by
\begin{equation}
  \widehat{c}^{\neq}_k(m) \;=\; c^{\neq}_k(m) - \Delta^{\neq}_k(m) \;,
  \label{def_chat}
\end{equation}
so that these coefficients $\widehat{c}_k^{\neq}(m)$ would be,
for each fixed $k$, the restriction to integers
$m \ge \widehat{m}^{\neq}_{\rm min}(k)$
of the same polynomial $C_k(m)$,
but with a {\em smaller}\/ threshold
$\widehat{m}^{\neq}_{\rm min}(k) < m^{\neq}_{\rm min}(k)$.
The important point here is that a smaller threshold
$\widehat{m}^{\neq}_{\rm min}(k)$ implies that we can obtain more
polynomials $C_k$ with the same raw data.

By inspecting Table~\ref{table_diff_coef_c_neq}, it is not difficult
to realize that there are some patterns in $\Delta^{\neq}_k(m)$
just below the threshold $m^{\neq}_{\rm min}(k)$,
and more specifically along diagonals $m=k-\ell$
with fixed integer $\ell \ge 1$.
For instance, for $\ell=1$ we see that $\Delta^{\neq}_{k}(k-1)=(-1)^{k+1}$
for $2 \le k \le 10$; and it is reasonable to {\em conjecture}\/
that this holds also for larger $k$, i.e.
\be
   \Delta^{\neq}_{k}(k-1) \;=\;
      (-1)^{k+1}  \quad \hbox{\rm for all $k\geq 2$} \;.
  \label{eq.Deltaneq.1}
\ee
Likewise, for $\ell=2,3$ we find for $k \le 10$ ---
and conjecture for all larger $k$ --- that $\Delta^{\neq}_k(k-\ell)$
can be written as $(-1)^{k+1}$ times a polynomial in $k$ of degree $\ell-1$,
provided that $k \ge 2\ell-1$:
\begin{subeqnarray}
\Delta^{\neq}_{k}(k-2) &=& (-1)^{k+1} \left(k-7\right)
      \quad \hbox{\rm for $k\geq 3$}\\[2mm]
\Delta^{\neq}_{k}(k-3) &=& (-1)^{k+1} \left( \smfrac{1}{2}k^2 
               - \smfrac{27}{2}k + 57\right)
      \quad \hbox{\rm for $k\geq 5$}
  \label{eq.Deltaneq.23}
\end{subeqnarray}
Each of these fits can be tested on at least three additional data points.
[We can also obtain a fit for $\Delta^{\neq}_k(k-4)$ for $k \ge 7$,
 but at this stage we would have no extra data points to test it.]

Then, assuming the correctness of \reff{eq.Deltaneq.1}/\reff{eq.Deltaneq.23},
we can infer the polynomials $C_k$ for $k=11,12$ 
(if we demand at least one extra data point to test the fit)
and $k=13$ (if we don't).
In turn, knowing the polynomials $C_{11}$ and $C_{12}$,
we can obtain a fit for $\Delta^{\neq}_k(k-4)$ that can be tested
on two extra data points:
\be
\Delta^{\neq}_{k}(k-4) \;=\; (-1)^{k+1} \left(\smfrac{1}{6}k^3 -10 k^2 
               + \smfrac{821}{6}k - 487 \right)
      \quad \hbox{\rm for $k\geq 7$}
   \;.
 \label{eq.Deltaneq.4}
\ee

Next, assuming the correctness also of \reff{eq.Deltaneq.4},
we get one additional data point to test the fit to $C_{13}$
(not surprisingly, it checks)
and we can infer $C_{14}$ if we don't demand any extra data points to test it.
In turn, knowing the polynomial $C_{13}$,
we can guess one further correction term $\Delta^{\neq}_k(k-\ell)$:
\begin{equation}
\Delta^{\neq}_{k}(k-5) \;=\; (-1)^{k+1}\left( \smfrac{1}{24}k^4 
  - \smfrac{53}{12}k^3 + \smfrac{3107}{24}k^2 - \smfrac{15817}{12}k 
  + 4253\right)
   \quad \hbox{\rm for $k\geq 9$}  \;,
 \label{eq.Deltaneq.5}
\end{equation} 
where here there are {\em no}\/ additional data points to test the fit
if we use only $C_{13}$, but there is one additional data point
if we trust also the inferred $C_{14}$.
Finally, assuming the correctness of \reff{eq.Deltaneq.5},
we can infer $C_{15}$ (but without any extra data points to test the fit).
All the polynomials $C_k$ for $k \leq 15$
are shown in Table~\ref{table_new_Ck}.

We can extend slightly our results by noticing the following two 
empirical results from the formulae for $\Delta^{\neq}_{k}(k-\ell)$ in
\reff{eq.Deltaneq.1}--\reff{eq.Deltaneq.5}:
(a) the leading coefficient in  $\Delta^{\neq}_{k}(k-\ell)$ is $1/(\ell-1)!$,
and (b) the 
next-to-leading coefficient is $-1/(\ell-1)! \times [0,7,27,60,106]$, 
which is fitted by $\smfrac{1}{2} (\ell-1)(13\ell-12)$. 
If these patterns persist, then we need only 4 coefficients
(rather than 6) to get $\Delta^{\neq}_{k}(k-6)$, i.e.\ $k=11,12,13,14$,
so we can then get $C_{15}$ with one extra data point,
or $C_{16}$ without. The result of the fit is
\begin{equation}
\Delta^{\neq}_{k}(k-6) \;=\; (-1)^{k+1}\left( \smfrac{1}{120}k^5 
  - \smfrac{11}{8}k^4 + \smfrac{1693}{24}k^3 - \smfrac{11565}{8}k^2 
  + \smfrac{248889}{20}k - 378651\right)
   \quad \hbox{\rm for $k\geq 11$}  \;,
 \label{eq.Deltaneq.6}
\end{equation} 
where here there is one additional data point if we trust also the inferred 
$C_{14}$ and $C_{15}$.

{\bf Remark.}  In Section~\ref{sec.thermo.neqeq} we will infer also
$C_{17}$ by exploiting the known large-$q$ series
for the bulk free energy \cite{Bakaev_94,transfer5},
which gives us the coefficients $\alpha_k$;
we then need only {\em one}\/ additional data point to infer $\beta_k$
and hence $C_k$.
All the polynomials $C_k$ for $k \leq 17$
are shown in Table~\ref{table_new_Ck}.

\subsection[Results for $S_{m,n}^{=}$]{Results for $\bm{S_{m,n}^{=}}$}
   \label{sec4.3}

We carried out a completely analogous analysis for
the case of $=$ boundary conditions;
we shall therefore be brief in describing it,
stressing only the differences from the preceding analysis
and the final results.

We were able to compute $\lambda_{\star,=}(m)$ for
for $1\leq m \leq 12$ (one term more than for $\neq$ boundary conditions).
The resulting coefficients $b^{=}_k(m)$ and $c^{=}_k(m)$
for $0 \le k \le 15$ are displayed in Tables~\ref{table_coef_b_eq}
and \ref{table_coef_c_eq}, respectively.\footnote{
  The complete data set for $0 \le k \le 40$
  is contained in the {\sc Mathematica} file {\tt data\_Eq.m}
  that is included in the on-line version of this paper at arXiv.org.
}
We observe {\em empirically}\/ that, for each fixed $k$,
the coefficients $b^{=}_k(m)$ [resp.\ $c^{=}_k(m)$]
are the restriction to integers $m$
of a polynomial $B_k(m)$ [resp.\ $C_k(m)$]
of degree $k$ (resp.\ degree 1) in $m$,
{\em provided that we restrict attention to
$m\geq m_{\rm min}^{=}(k)$}\/  with
\begin{equation}
m_{\rm min}^{=}(k) \;=\; k+1  \;.
\label{def_m_min_eq}
\end{equation}
Furthermore, the polynomials $B_k(m)$ and $C_k(m)$
are empirically found to be the {\em same}\/ as those for
$\neq$ boundary conditions.
Using our available data together with the tricks shown in the
preceding subsection,
we are able to confirm this equality for $k \le 17$.
Let us sketch quickly the logic of this computation
and the relevant intermediate results.

As before, we can extract the desired information
most efficiently by fitting the coefficients $c^{=}_k(m)$
to a polynomial $C_k(m)$ of degree 1,
for $m\ge m_{\rm min}^{=}(k) = k+1$.
As we need only two coefficients for such a polynomial,
we can obtain these polynomials
up to $k=9$ (if we want at least one extra data point to test the fit)
or $k=10$ (if we don't). The results are equal to those of $\neq$ 
boundary conditions.

Next we try to improve these results, as before, by defining the difference
$\Delta^{=}_k(m) = c^{=}_k(m) - C_k(m)$
and attempting to guess an analytic form for some of the coefficients
$\Delta^{=}_k(m)$:
see Table~\ref{table_diff_coef_c_eq}, where we initially know only $k \le 10$.
As in the case of $\neq$ boundary conditions, we find empirically
that the coefficients $\Delta^{=}_k(m)$ closest to the boundary
$m^{=}_{\rm min}(k)$
are the restriction to integers $m$ of certain polynomials.
Because $m^{=}_{\rm min}(k) = k+1$,
we are concerned this time with $\Delta^{=}_k(k-\ell)$
for $\ell \ge 0$ (rather than $\ell \ge 1$ as before).
We obtain:
\begin{subeqnarray}
\Delta^{=}_{k}(k) &=& (-1)^{k+1}   \quad \hbox{\rm for $k\geq 1$}\\[2mm]
\Delta^{=}_{k}(k-1) &=& (-1)^{k+1} \left(k-5\right)
      \quad \hbox{\rm for $k\geq 3$}\\[2mm]
\Delta^{=}_{k}(k-2) &=& (-1)^{k+1} \left( \smfrac{1}{2}k^2 
               - \smfrac{23}{2}k + 37\right)
      \quad \hbox{\rm for $k\geq 5$}\\[2mm]
\Delta^{=}_{k}(k-3) &=& (-1)^{k+1} \left(\smfrac{1}{6}k^3 -9 k^2 
               + \smfrac{623}{6}k - 303 \right)
      \quad \hbox{\rm for $k\geq 7$}
\slabel{eq.Deltaeq.1}
\end{subeqnarray}
We are able to test each fit on at least one additional data point.\footnote{
   For $\ell=3$ we do not {\em immediately}\/ have such an extra data point,
   but we can obtain it after using using $\ell=0,1,2$
   to infer $C_{11}$ and $C_{12}$.  The reasoning is completely analogous
   to what was explained in detail for the case of $\neq$ boundary conditions.
}
The pattern is slightly different from that for $\neq$ boundary conditions:
here $\Delta^{=}_k(k-\ell)$ is $(-1)^{k+1}$ times a polynomial of degree $\ell$
in $k$, valid for $k \ge 2\ell+1$.
By this method, we can obtain the polynomials $C_k$
up to $k=13$ (if we demand at least one extra data point to test the fit)
or $k=14$ (if we don't).

We can then guess one further correction term $\Delta^{=}_k(k-\ell)$:
\begin{equation}
\Delta^{=}_{k}(k-4) \;=\; (-1)^{k+1}\left( \smfrac{1}{24}k^4 
  - \smfrac{49}{12}k^3 + \smfrac{2555}{24}k^2 - \smfrac{11225}{12}k 
  + 2573\right)
   \quad \hbox{\rm for $k\geq 9$} \,.  
 \label{eq.Deltaeq.2}
\end{equation}
where there are no additional data points to test the fit
if we use only $C_{13}$, but there is one additional data point
if we trust also the inferred $C_{14}$.
Finally, assuming the correctness of \reff{eq.Deltaeq.2},
we can infer $C_{15}$ (but without any extra data points to test the fit).

As for $\neq$ boundary conditions, we can extend slightly our results 
by looking at the patterns for the first two terms in  
$\Delta^{=}_{k}(k-\ell)$ \reff{eq.Deltaeq.1}/\reff{eq.Deltaeq.2}: 
The leading coefficient in $\Delta^{=}_{k}(k-\ell)$ is $1/\ell!$, and 
the next-to-leading coefficient is $-1/\ell! \times [0,5,23,54,98]$, 
which is fitted by $\smfrac{1}{2}\ell(13\ell-3)$. 
If these patterns persist, then we need only 4 coefficients
(rather than 6) to get $\Delta^{=}_{k}(k-5)$,
so we can then get $C_{15}$ with one extra data point,
or $C_{16}$ without. The result of the fit is
\begin{equation}
\Delta^{=}_{k}(k-5) \;=\; (-1)^{k-4}\left( -\smfrac{1}{120} k^5 + 
\smfrac{31}{24}k^4
  - \smfrac{1457}{24}k^3 + \smfrac{26705}{24}k^2 
  - \smfrac{506477}{60}k + 22330\right)
   \quad \hbox{\rm for $k\geq 11$}  \;,
 \label{eq.Deltaeq.3}
\end{equation} 
where here there is one additional data point if we trust also the inferred 
$C_{14}$ and $C_{15}$.
We find, in all cases, the {\em same}\/ polynomials $C_k$
as for $\neq$ boundary conditions.

{\bf Remark.}  In Section~\ref{sec.thermo.neqeq} we will infer also
$C_{17}$ for $=$ boundary conditions
by exploiting the known large-$q$ series
for the bulk free energy \cite{Bakaev_94,transfer5}.

\subsection[Comparing $S_{m,n}^{\neq}$ with $S_{m,n}^{=}$]%
{Comparing $\bm{S_{m,n}^{\neq}}$ with $\bm{S_{m,n}^{=}}$}

We were able to compute the difference 
$(-1)^{m+1}[\lambda_{\star,=}(m)-\lambda_{\star,\neq}(m)]$ for
for $1\leq m \leq 11$, and we have extracted the coefficients 
$d_\ell(m)$ [cf.\ \reff{def_series_dl}]
for $1\le m \le 11$ and $1\le \ell\le 20$. 
The resulting coefficients are displayed in Table~\ref{table_diff_coef_d}
for $1 \le \ell\le 13$. (The complete data set for $1\le \ell\le 20$
is included in the {\sc Mathematica} file {\tt data\_Diff.m} that is
included in the on-line version of this paper at arXiv.org.) 
We observe {\em empirically}\/ that, for each fixed $\ell$,
the coefficients $d_\ell(m)$ are the restriction to integers $m$
of a polynomial $D_\ell(m)$ of degree $\ell$ in $m$,
{\em provided that we restrict attention to
$m\geq m_{\rm min}(\ell) = \ell+1$} [cf.\ \reff{def_m_min_eq}].
We start by trying to perform such fit: As we need $\ell+1$ coefficients 
for the polynomial $D_\ell(m)$, we are able to obtain these
polynomials for $\ell \le 4$ (if we allow for at least one extra data point 
to test the fit) or $\ell\le 5$ (if we don't).
Our results are:
\begin{subeqnarray}
   D_1(m)  & = &   3m-4       \\[2mm]
   D_2(m)  & = &   \smfrac{9}{2} m^2 - \smfrac{37}{2} m +14   \\[2mm]
   D_3(m)  & = &   \smfrac{9}{2} m^3 - \smfrac{75}{2} m^2 
                  + 84 m  -55   \\[2mm]
   D_4(m)  & = &   \smfrac{27}{8} m^4 - \smfrac{189}{4} m^3
                  +\smfrac{1681}{8} m^2 - \smfrac{1461}{4} m + 224  \\[2mm]
   D_5(m)  & = &   \smfrac{81}{40} m^5 - \smfrac{171}{4} m^4 
                  +\smfrac{2523}{8}m^3
                  -\smfrac{4147}{4}m^2 + \smfrac{15661}{10}m - 880 
\label{def_dk_poly}
\end{subeqnarray}

As in the preceding sections, we can extract more information by considering
the coefficients $e_\ell(m)$ arising in the {\em logarithm}\/
of the difference of eigenvalues [cf.\ \reff{def_series_el}].
These coefficients are displayed in
Table~\ref{table_diff_coef_e} for $1\le \ell\le 13$, and the complete set
is included in the file {\tt data\_Diff.m}. We again observe 
{\em empirically}\/ that, for each fixed $\ell$, 
the coefficients $e_\ell(m)$ are the restriction to integers $m$
of a polynomial $E_\ell(m)$ of degree $1$ in $m$,
{\em provided that we restrict attention to
$m\geq m_{\rm min}(\ell) = \ell+1$}. As we need two coefficients for
such a polynomial, we can obtain these polynomials up to $\ell=8$ 
(if we want at least one extra data point to test the fit) or $\ell=9$ 
(if we don't). The first polynomials are given by 
\begin{subeqnarray}
   E_1(m)  & = &   -3m+4       \\[2mm]
   E_2(m)  & = &   -\smfrac{13}{2} m + 6 \\[2mm] 
   E_3(m)  & = &   -16m + \smfrac{61}{3} \\[2mm] 
   E_4(m)  & = &   -\smfrac{181}{4}m + 66 \\[2mm] 
   E_5(m)  & = &   -\smfrac{658}{5} m + \smfrac{934}{5} 
\label{def_ek_poly}
\end{subeqnarray}
and the results for $6\le \ell\le 9$ are shown in Table~\ref{table_new_Ck}.
The polynomials $D_\ell$ for $\ell\le 9$ can then be determined from the
$E_\ell$ using \reff{def_series_dl}/\reff{def_series_el}.

Next we try to improve these results, as before, by defining the difference
\begin{equation}
\Delta^{(e)}_\ell(m) \;=\; e_\ell(m) - E_\ell(m)
\label{def_deltaE}
\end{equation}
and attempting to guess an analytic form for some of the coefficients
$\Delta^{(e)}_\ell(k)$: see Table~\ref{table_diff_coef_deltae}, 
where we initially know only $\ell \le 9$.
As in the previous cases, we find empirically
that the coefficients $\Delta^{(e)}_\ell(k)$ closest to the boundary
$k_{\rm min}(\ell)=\ell+1$
are the restriction to integers~$\ell$ of certain polynomials.
We obtain:
\begin{subeqnarray}
\Delta^{(e)}_{\ell}(\ell) &=& (-1)^{\ell}
      \quad \hbox{\rm for $\ell\geq 1$}\\[2mm]
\Delta^{(e)}_{\ell}(\ell-1) &=& (-1)^{\ell} \left(4\ell - 9\right)
      \quad \hbox{\rm for $\ell\geq 3$}\\[2mm]
\Delta^{(e)}_{\ell}(\ell-2) &=& (-1)^{\ell} \left(-\ell^2 -8\ell+39 \right)
      \quad \hbox{\rm for $\ell\geq 5$}\\[2mm]
\Delta^{(e)}_{\ell}(\ell-3) &=& (-1)^{\ell} \left(\smfrac{14}{3}\ell^3 
         -65 \ell^2 + \smfrac{985}{3}\ell - 613 \right)
      \quad \hbox{\rm for $\ell\geq 7$}
\label{eq.DeltaE}
\end{subeqnarray}
We are able to test each fit on at least one additional data point. By this
method, we are able to extract the polynomials $E_\ell$ up to $\ell=12$ 
(if we demand at least one extra data point to test the fit) or $\ell=13$
(if we don't).   

The same game can be played with the coefficients $f_\ell(m)$ 
[cf.\ \reff{def_series_fl}]. These coefficients are displayed in
Table~\ref{table_diff_coef_f} for $1\le \ell\le 12$, and the complete set
is included in the file {\tt data\_Diff.m}. From this table we see that
the coefficients $\ell f_\ell(m)$ are not integers, in contrast to the
observed behavior for the coefficients $\ell e_\ell(m)$ (see 
Table~\ref{table_diff_coef_e}). We observe 
{\em empirically}\/ that, for each fixed $\ell$, 
the coefficients $f_\ell(m)$ are the restriction to integers $m$
of a polynomial $F_\ell(m)$ of degree $1$ in $m$,
{\em provided that we restrict attention to
$m\geq m_{\rm min}(\ell) = \ell+1$}\/. Again,  
we can obtain these polynomials up to $\ell=8$ 
(if we want at least one extra data point to test the fit) or $\ell=9$ 
(if we don't). The first polynomials are given by 
\begin{subeqnarray}
   F_1(m)  & = &   m-5       \\[2mm]
   F_2(m)  & = &   \smfrac{11}{2} m -\smfrac{17}{2} \\[2mm] 
   F_3(m)  & = &   \smfrac{49}{3} m - \smfrac{77}{3} \\[2mm] 
   F_4(m)  & = &   \smfrac{191}{4}m - \smfrac{305}{4} \\[2mm] 
   F_5(m)  & = &   \smfrac{686}{5}m - 203  
\label{def_fk_poly}
\end{subeqnarray}
and the ones for $6\le \ell\le 9$ are shown in Table~\ref{table_new_Ck}. 
(In Section~\ref{sec.free_energy_series.exact} we {\em prove}\/
 that $f_1(m) = m-5$ for $m \ge 2$.)

We now try to improve these results by defining the difference
\begin{equation}
\Delta^{(f)}_\ell(m) \;=\; f_\ell(m) - F_\ell(m)
\label{def_deltaF}
\end{equation}
and attempting to guess an analytic form for some of the coefficients
$\Delta^{(f)}_\ell(m)$: see Table~\ref{table_diff_coef_deltaf}, 
where we initially know only $\ell \le 9$.
As in the previous cases, we find empirically
that the coefficients $\Delta^{(f)}_\ell(m)$ closest to the boundary
$m_{\rm min}(\ell)=\ell+1$
are the restriction to integers $\ell$ of certain polynomials.
We obtain:
\begin{subeqnarray}
\Delta^{(f)}_{\ell}(\ell) &=& (-1)^{\ell+1} \, \smfrac{3}{2} 
      \quad \hbox{\rm for $\ell\geq 1$}\\[2mm]
\Delta^{(f)}_{\ell}(\ell-1) &=& (-1)^{\ell+1} \left(\smfrac{9}{2}\ell 
              - 11\right)
      \quad \hbox{\rm for $\ell\geq 3$}\\[2mm]
\Delta^{(f)}_{\ell}(\ell-2) &=& (-1)^{\ell} \left(\smfrac{3}{2}\ell^2 
          +\smfrac{53}{4}\ell -54 \right)
      \quad \hbox{\rm for $\ell\geq 5$}\\[2mm]
\Delta^{(f)}_{\ell}(\ell-3) &=& (-1)^{\ell} \left(\smfrac{19}{4}\ell^3 
         -\smfrac{277}{4} \ell^2 + \smfrac{747}{2}\ell - 736 \right)
      \quad \hbox{\rm for $\ell\geq 7$}
\label{eq.DeltaF}
\end{subeqnarray}
We are able to test each fit on at least one additional data point. By this
method, we are able to extract the polynomials $F_\ell$ up to $\ell=12$ 
(if we demand at least one extra data point to test the fit) or $\ell=13$
(if we don't).   

If we look carefully at Table~\ref{table_new_Ck}, we realize that the
polynomials $C_\ell$, $E_\ell$ and $F_\ell$ are not independent, but they
satisfy (at least for $1\le \ell\le 13$) the relation
\begin{equation}
C_\ell(m) \;=\; E_\ell(m) + F_\ell(m) \,. 
\label{eq.poly_f_e_c}
\end{equation}
The same relation of course holds for the corresponding coefficients 
$c_\ell^{\neq}(m)$, $e_\ell(m)$ and $f_\ell(m)$ for $m\ge \ell +1$:
\be
   c_\ell^{\neq}(m) \;=\;  e_\ell(m) + f_\ell(m)
    \quad \hbox{for $m\ge \ell+1$} 
   \;.
   \label{eq.exact_f_e_c}
\ee
We can understand this latter relation by using
\reff{def_ckm}/\reff{def_series_el}/\reff{def_series_fl} and defining
\be
   G(m)
   \;=\;
   \sum\limits_{\ell=1}^\infty
     \biggl[ e_\ell(m) + f_\ell(m) - c_\ell^{\neq}(m) \biggr] q^{-\ell}
   \;=\;
   \log\left( \frac{\displaystyle 
           \frac{ \lambda_{\star,=}(m)} { \lambda_{\star,\neq}(m) } - 1 }
  {\displaystyle \log \frac{\lambda_{\star,=}(m)}
                           {\lambda_{\star,\neq}(m)}}  \right)
   \;.
\ee
Since $\lambda_{\star,=}(m)/\lambda_{\star,\neq}(m) = 1 + O(q^{-m})$
from \reff{def_bkm}/\reff{eq.bkm.akm.k0}/\reff{eq.curious.lambda},
it follows that $G(m) = O(q^{-m})$,
i.e.\ $[q^{-\ell}] G(m) = 0$ for $\ell < m$,
which is precisely \reff{eq.exact_f_e_c}.

\subsection{Summary of conjectured behavior}

Let us conclude this section by summarizing our empirical findings
concerning the eigenvalues
$\lambda_{\star,=}(m)$ and $\lambda_{\star,\neq}(m)$
and their difference. 
First recall that in Proposition~\ref{prop.diff.eigen.0} we {\em prove}\/
that 
\begin{equation}
   \lambda_{\star,=}(m) - \lambda_{\star,\neq}(m)  \;=\;
   (-1)^{m+1} \,+\, O(q^{-1})
   \;.
\end{equation}
We have found {\em empirically}\/ that stronger results hold
Our first finding concerns the behavior of the free energies
for $\neq$ and $=$ boundary conditions:

\begin{conjecture} \label{conj.diff.eigen.Ck}
There exist polynomials $C_1,C_2,\ldots$ with rational coefficients,
of degree~1, such that
\begin{subeqnarray}
   \log {\lambda_{\star,\neq}(m)  \over q^m}
   & = &
   \sum\limits_{k=1}^m  C_k(m) \, q^{-k}  \:+\: O(q^{-(m+1)})
         \\[2mm]
   \log {\lambda_{\star,=}(m)  \over q^m}
   & = &
   \sum\limits_{k=1}^m  C_k(m) \, q^{-k}  \:+\: (-1)^{m+1} q^{-m}
                                          \:+\:  O(q^{-(m+1)})
\end{subeqnarray}
\end{conjecture}
We have verified this conjecture for $m \le 10$,
using the polynomials $C_1,\ldots,C_{10}$
shown in Table~\ref{table_new_Ck}.
A special case of Conjecture~\ref{conj.diff.eigen.Ck} is 
Proposition~\ref{prop.diff.eigen.0} concerning the {\em difference}\/ of 
eigenvalues.

Secondly, the empirical observation that the coefficients $d_\ell(m)$
defined in \reff{def_series_dl} are the restriction 
to integers $m \ge \ell+1$ of a polynomial of degree~$\ell$
in the variable~$m$ can formalized as a conjecture as follows: 

\begin{conjecture} \label{conj.diff.eigen}
There exist polynomials $D_1,D_2,\ldots$ with rational coefficients,
with $\deg D_\ell = \ell$, such that
\begin{equation}
\lambda_{\star,=}(m) - \lambda_{\star,\neq}(m) \;=\;
 (-1)^{m+1}\left[ 1  + \sum_{\ell=1}^{m-1} (-1)^\ell D_\ell(m) q^{-\ell}
           \right]+  O(q^{-m}) \;.
\label{eq.conj.diff}
\end{equation}
\end{conjecture}
We have verified this conjecture for $m \le 6$,
using the polynomials $D_1,\ldots,D_5$ given in \reff{def_dk_poly}.
Moreover, the same polynomials $D_1,\ldots,D_5$ give correctly
the expansion through order $q^{-5}$ for all $m \le 11$.

Finally, the empirical observation that the coefficients
$e_\ell(m)$ and $f_\ell(m)$ defined in
\reff{def_series_el}/\reff{def_series_fl}
are the restriction to integers $m \ge \ell+1$
of a polynomial {\em of degree~1}\/ in the variable~$m$
can formalized as a conjecture as follows: 

\begin{conjecture} \label{conj.diff.eigen2}
There exist polynomials $E_1,E_2,\ldots$ and $F_1,F_2,\ldots$
with rational coefficients, of degree~1, such that
\begin{eqnarray}
\log\left[ (-1)^{m+1} \left( \lambda_{\star,=}(m) - 
    \lambda_{\star,\neq}(m)\right)\right] &=& 
  \sum_{\ell=1}^{m-1} E_\ell(m) q^{-\ell} + O(q^{-m}) \\[1mm]
-\log \left[ (-1)^{m+1} q^m \log\left( 
  \frac{ \lambda_{\star,=}(m)}{\lambda_{\star,\neq}(m)} \right)\right] 
  &=& \sum_{\ell=1}^{m-1} F_\ell(m) q^{-\ell}
     + O(q^{-m})
\label{eq.conj.diff2}
\end{eqnarray}
\end{conjecture}
We have verified this conjecture for $m \le 10$,
using the polynomials $E_1,\ldots,E_9$ and $F_1,\ldots,F_9$
shown in Table~\ref{table_new_Ck}.

%
% THERMODYNAMIC LIMIT 
%
\section{Thermodynamic limit $\bm{m\to \infty}$ of the free energies}
   \label{sec.thermo}

In the previous section we studied the large-$q$ expansion
of the leading eigenvalues
$\lambda_{\star,=}(m)$ and $\lambda_{\star,\neq}(m)$
--- or what is essentially equivalent,
the strip free energies $f_m^{=}$ and $f_m^{\neq}$
to be defined in \reff{def_Fm}/\reff{def_Fm_bis} below ---
for strips of fixed width $m$ with $=$ or $\neq$ boundary conditions.
In this section we will study the thermodynamic limit $m\to\infty$
of these free energies.

The plan of this section is as follows:
In Section~\ref{sec.thermo.gen} we introduce some preliminary definitions
and discuss the expected behavior of the strip free energies per
unit length, $f_m^{=}(q)$ and $f_m^{\neq}(q)$,
as a function of the strip width $m$.
We then give, in Section~\ref{sec.thermo.overview},
an overview of the computations we have performed
and the results we have obtained.
In Section~\ref{sec.free_energy_series}
we present a hand computation of the first four terms
in the large-$q$ expansion of the bulk, surface and corner free energies
for $=$ and $\neq$ boundary conditions.
In Section~\ref{sec.free_energy_series.exact}
we prove some exact results, valid to all orders in $1/q$,
for the difference between the free energies
with $=$ and $\neq$ boundary conditions.
Finally, in Section~\ref{sec.thermo.neqeq}
we explain one of the empirical observations from
Sections~\ref{sec4.2} and \ref{sec4.3},
and we extract additional information on the vertical-surface free energies
from our transfer matrices.

\subsection{Generalities and finite-size-scaling theory} 
    \label{sec.thermo.gen}

Corollary~\ref{cor.TMblockdiag.1and2} shows that,
for each width $m$ and each boundary condition ($=$ or $\neq$),
the transfer matrix has, for sufficiently large $|q|$,
a {\em single}\/ dominant eigenvalue $\lambda_\star(q)$
that moreover is an analytic function of $q$
(in fact, it is $q^m$ times an analytic function of $q^{-1}$).
Theorem~\ref{thm.sokal3} actually proves that $\lambda_{\star,\sharp}(q)$
is $q^m$ times an analytic function of $q^{-1}$ whenever $|q| > 33.855628$,
uniformly in $m$. 

Let us now introduce the free energy per site\footnote{
   Actually {\em minus}\/ the free energy per site
   in the usual thermodynamic convention
   --- but we prefer not to encumber our formulae
   with unnecessary minus signs, or to encumber our text
   with constant repetition of the word ``minus''.
}
for a finite strip with $=$ or $\neq$ boundary conditions,
\begin{equation}
 f_{m,n}^{\sharp}(q)  \;=\;
     {1 \over m n} \log P_{S_{m,n}^{\sharp}}(q)
\label{def_Fmn}
\end{equation}
where $\sharp$ denotes $=$ or $\neq$,
and its limiting value for a semi-infinite strip,
\begin{equation}
 f_{m}^{\sharp}(q)  \;=\;
   \lim_{n\to\infty} {1 \over m n} \log P_{S_{m,n}^{\sharp}}(q) \;.
\label{def_Fm}
\end{equation}
Finally, let us introduce the free energy per site for the infinite lattice,
\begin{equation}
 f^{\sharp}(q)  \;=\;
  \lim_{m,n\to\infty} {1 \over m n} \log P_{S_{m,n}^{\sharp}}(q) \;.
\label{def_F}
\end{equation}
Here we are assuming that the indicated limits exist and that in
\reff{def_F} the limit is independent of the way that
$m$ and $n$ tend to infinity.
Furthermore, it is natural to expect that in \reff{def_F}
the limiting free energy is independent of boundary conditions,
in which case we can omit the superscripts $=$ or $\neq$ and 
write simply $f(q)$.

In fact, some of these assumptions can be proven.
Indeed, the above discussion guarantees that, at least for $|q| > 33.855628$,
the limiting strip free energy $f_m(q)$ exists for all $m$ and is given by
\begin{equation}
f_m^{\sharp}(q) \;=\; \frac{1}{m} \log\lambda_{\star,\sharp}(q) \;,
\label{def_Fm_bis}
\end{equation}
which is moreover an analytic function of $q$ in the indicated domain.
Moreover, Procacci {\em et al.}\/ \cite[Theorem 2]{Procacci_03}
have proven that, when $|q|$ is large enough
(namely, $|q|> 8e^3 \approx 160.684295$),
the infinite-volume limiting free energy $f(q)$ exists and is analytic in $1/q$
and is the same for all sequences of graphs $G_{m\times n}$
with free boundary conditions;
in particular, it is independent of the way
that $m$ and $n$ tend to infinity.\footnote{
   In fact, Procacci {\em et al.}\/ prove that this limit is the same for
   all F\o{}lner--van Hove sequences of finite subvolumes
   of the infinite lattice, not just rectangles;
   and they prove this result for a wide variety of lattices
   (namely, locally-finite connected quasi-transitive amenable
    infinite graphs), not just the square lattice.
   On the other hand, we mistakenly asserted in
   \cite[Section~5.1]{transfer5}
   that Procacci {\em et al.}\/ had proven this result also for
   cylindrical, cyclic and toroidal boundary conditions.
   We suspect that such a result can indeed be obtained by a
   suitable modification of their proof,
   but we were wrong to assert that it is contained in their paper.
}
We {\em expect}\/ that the same result holds
for $=$ or $\neq$ boundary conditions
and gives rise to the {\em same}\/
infinite-volume limiting free energy $f(q)$ as for free boundary conditions
(indeed, we expect that this can be proven by a modification of the
 Procacci {\em et al.}\/ argument).
In this paper we will take $n \to\infty$ first
and then take $m\to\infty$, so that
\begin{equation}
 f(q)  \;=\;  \lim_{m\to\infty} f_m^{\sharp}(q)  \;.
\label{def_F_bis}
\end{equation}

Finite-size-scaling theory \cite[Section~2.5]{Privman_90}
gives a rather precise prediction for the form
of the free energy \reff{def_Fmn}/\reff{def_Fm}
for a finite or semi-infinite system
away from a critical point (and in the absence of soft modes).
In particular, for an $m\times n$ strip with $=$ or $\neq$  
boundary conditions and bulk correlation length $\xi_{\rm bulk} \ll m,n$,
the predicted behavior is
\begin{equation}
f_{m,n}^{\sharp} \;=\; f_{\rm bulk}       \;+\; 
        \frac{n f_{\rm surf,vert}^{\sharp} + 
              m f_{\rm surf,horiz}^{\sharp}}{mn}  \;+\; 
\frac{1}{mn} \, f_{\rm corner}^{\sharp} \;+\;
            O( e^{-\min(m,n)/\xi_{\rm bulk}} )
   \;,
\label{def_FSS_Ansatz}
\end{equation}
where $\sharp$ denotes $=$ or $\neq$,
and $f_{\rm bulk} = f$, $f_{\rm surf,vert}^{\sharp}$,
$f_{\rm surf,horiz}^{\sharp}$ and $f_{\rm corner}^{\sharp}$ are,
respectively, the free energies for the bulk,
the two vertical surfaces (at left and right),
the two horizontal surfaces (at top and bottom),
and the four corners plus the extra sites.
For a semi-infinite strip $m\times\infty$ with $=$ or $\neq$ boundary 
conditions, we have
\begin{equation}
f_{m}^{\sharp} \;=\; f_{\rm bulk}       \;+\; 
        \frac{1}{m}  f_{\rm surf,vert}^{\sharp}  \;+\; 
          O( e^{-m/\xi_{\rm bulk}} )
   \;.
\label{def_FSS_Ansatz2}
\end{equation}
Since the families $S^{\sharp}_{m,n}$ are not invariant
under $90^\circ$ rotation,
we expect the vertical and horizontal free energies to be different.
Finally, we cannot take for granted that the surface and corner free
energies are the same for $=$ and $\neq$ boundary conditions.

In this section we {\em assume}\/ that the
behaviors \reff{def_FSS_Ansatz}/\reff{def_FSS_Ansatz2} hold
and that expansion in $1/m$ and $1/n$ can be commuted freely
with expansion in $1/q$.
(We expect that this may be provable by an extension
 of the Procacci {\em et al.}\/ argument.)
This assumption justifies the manipulations
to be made in the following subsections.

The relation \reff{def_FSS_Ansatz2} of course holds for the
chromatic polynomials at {\em fixed}\/ large $q$.
But we can also argue heuristically what it should imply
for the series expansion in powers of $1/q$.
It is not difficult to see that,
for large $q$, we have
\begin{equation}
   e^{-1/\xi_{\rm bulk}(q)}  \;=\;
   {1 \over q} \,+\, O\Bigl( {1 \over q^2} \Bigr)
 \label{eq.xibulk}
\end{equation}
(just as for a {\em one-dimensional}\/
 Potts antiferromagnet at zero temperature).
As explained in detail in \cite{transfer5}, 
we can therefore interpret $O( e^{-m/\xi_{\rm bulk}(q)} )$
as meaning $O(q^{-m})$.  Therefore, we expect that
\begin{equation}
f_{m}^{\sharp}(q) \;=\; 
  f_{\rm bulk}(q) \,+\, \frac{1}{m} f_{\rm surf,vert}^{\sharp}(q) \,+\, O(q^{-m})
  \;,
   \label{def_FSS_Ansatz2_bis}
\end{equation}
or in other words we predict $m_{\rm min}^{\sharp}(k) = k+1$.
Of course, we should not take too seriously the ``$+1$'' here,
since the {\em amplitude}\/ of the correction term in \reff{def_FSS_Ansatz2}
could be proportional to a positive or negative power of $q$.
But we do predict that $m_{\rm min}^{\sharp}(k) \approx k$
in the sense that $\lim_{k \to\infty} m_{\rm min}^{\sharp}(k)/k = 1$.
This is indeed what we found for $=$ and $\neq$ boundary conditions.

\subsection{Overview of computations and results}
    \label{sec.thermo.overview}

Before beginning the detailed computations, it is useful
to give an overview of the methods to be used and the results obtained,
taking into account both the computations reported below for
$=$ and $\neq$ boundary conditions and those reported in a previous paper
\cite[Sections~5.2--5.4]{transfer5} for free (F) and cylindrical (P)
boundary conditions.

The computations we are able to carry out are as follows:
\begin{itemize}
   \item For the {\em bulk}\/ free energy (which is of course the same
      for all four boundary conditions), we obtained in \cite{transfer5}
      long series (through order $q^{-47}$) using the finite-lattice
      method for free boundary conditions.
      We are also able to check these series up to moderate order
      (namely, order $q^{-33}$, $q^{-16}$, $q^{-15}$, $q^{-15}$ respectively
       for F, P, $=$ and $\neq$ boundary conditions)
      by transfer-matrix calculations of the leading eigenvalue.
   \item For the {\em vertical surface}\/ free energy
      with $=$ or $\neq$ boundary conditions,
      we have a hand calculation through order $q^{-4}$,
      which we can extend to order $q^{-16}$ or $q^{-17}$
      by transfer-matrix calculations of the leading eigenvalue.
   \item For the {\em horizontal surface}\/ free energy
      with $=$ or $\neq$ boundary conditions,
      we have only a hand calculation through order $q^{-4}$.
      [The horizontal surface free energy does not appear in the
       strip eigenvalue \reff{def_Fm_bis}/\reff{def_FSS_Ansatz2}.]
   \item For the {\em corner}\/ free energy
      with $=$ or $\neq$ boundary conditions,
      we have likewise only a hand calculation through order $q^{-4}$.
\end{itemize}

Our conclusions (from the available orders) are as follows:
\begin{itemize}
   \item The {\em bulk}\/ free energy
      agrees for all four boundary conditions (no surprise).
   \item The {\em horizontal surface}\/ free energy
      agrees for $=$ and $\neq$ boundary conditions.
      This is again no surprise, as the horizontal surfaces are the same
      for the two boundary conditions.
   \item The {\em vertical surface}\/ free energy
      agrees for $=$ and $\neq$ boundary conditions.
      {\em A priori}\/ this might be somewhat of a surprise,
      as it is far from obvious whether the effect of the edge
      connecting the two extra sites
      (which is $v=\infty$ or $-1$ according as the boundary conditions
       are $=$ or $\neq$)
      is of order 1 (it is, after all, only a single edge)
      or is of order $n$ (its endpoints are directly connected
        to the whole vertical sides).
      It turns out that the former is the case.
      But on closer examination this is no surprise at all:
      the equality $f_{\rm surf,vert}^{=} = f_{\rm surf,vert}^{\neq}$
      follows immediately from \reff{def_Fm_bis}/\reff{def_FSS_Ansatz2_bis}
      together with Proposition~\ref{prop.diff.eigen.0},
      since
      $\log[\lambda_{\star,=}(m)/\lambda_{\star,\neq}(m)] =
       (-1)^{m+1} q^{-m} + O(q^{-m-1})$
      vanishes to all orders in $q^{-1}$ as $m \to\infty$.
   \item The horizontal and vertical surface free energies are different
      from each other, and different from the surface free energy
      for free (F) boundary conditions.
   \item The {\em corner}\/ free energies are different for
      $=$, $\neq$ and free boundary conditions.  However,
      the {\em difference}\/ between the corner free energies
      for $=$ and $\neq$ boundary conditions is simple:
      $f_{\rm corner}^{=}(q) - f_{\rm corner}^{\neq}(q) = - \log(1 - q^{-1})$.
\end{itemize}

Let us conclude this subsection by recalling the large-$q$ series
for the bulk free energy that was obtained in 
\cite{transfer5} through order $q^{-47}$ (extending previous series
by Bakaev and Kabanovich \cite{Bakaev_94})
by using the finite-lattice method
\cite{Neef-Enting_77,Enting_78,Kim-Enting_79,Enting_96,Enting_05,Enting_06}
for free boundary conditions.\footnote{
  After the completion of this work, we learned that Jacobsen 
  \protect\cite{Jacobsen_10} had extended the series expansions for the bulk, 
  surface and corner free energies to orders $O(q^{-79})$, $O(q^{-79})$
  and $O(q^{-78})$, respectively.
} 
As $|q|\to\infty$, the 
exponential of the free energy per site for an infinite
square lattice is given by the series expansion
\begin{eqnarray}
e^{f(q)} &=& \frac{(q-1)^2}{q} \left[ 1 + z^3 + z^7 + 3z^8 + 4z^9 +
 3z^{10} + 3z^{11} + 11z^{12} + 24z^{13} + 8z^{14} \right. 
\nonumber \\
         & & \quad -91z^{15} 
 -261z^{16} -290z^{17} + \ldots 
\left.  -598931311074z^{47} + O(z^{48}) \right] \,,
\label{series_1}
\end{eqnarray}
where $z$ is defined as
\begin{equation}
z \;=\; \frac{1}{q-1}  \;.
\label{def_z}
\end{equation}
We have here copied the first few terms and the last one; the remaining
terms are reported in \cite[Table~9]{transfer5}.
In terms of the variable $1/q$, we obtain
\begin{eqnarray}
e^{f(q)} &=&  q \left[ 1 - 2q^{-1} + q^{-2} + q^{-3} + q^{-4} + q^{-5}
  + q^{-6} + 2q^{-7} + 9q^{-8} + 38q^{-9} + 130 q^{-10}  \right.
\nonumber \\
         & & \quad + 378q^{-11} + 987q^{-12} + \ldots 
+1311159363081366872 q^{-47} + \left. O(q^{-48})\right] \,. \quad
\label{series_1bis}
\end{eqnarray}
Finally, for the bulk free energy $f(q)$ itself
(rather than its exponential)
in terms of the variable $1/q$, we obtain
\begin{eqnarray}
f(q) &=& \log q - \smfrac{2}{q} - \smfrac{1}{q^2} +
   \smfrac{1}{3q^3} + \smfrac{5}{2q^4} + \smfrac{28}{5q^5}+ \smfrac{55}{6q^6}+
   \smfrac{89}{7q^7} + \smfrac{81}{4q^8} + \smfrac{505}{9q^9} 
   +\smfrac{1029}{5q^{10}} + \smfrac{7742}{11q^{11}} \nonumber \\
    & & \qquad
   + \smfrac{25291}{12q^{12}}
   + \smfrac{73552}{13q^{13}} 
   + \smfrac{197755}{14q^{14}} + \ldots 
   + \smfrac{190018276619486037135}{47q^{47}}
+ O(q^{-48})
   \;. \quad
\label{series_fbulk}
\end{eqnarray}
Clearly, $e^{f(q)}$ has a much simpler expansion than $f(q)$;
in particular, its coefficients are integers
(at least through the order calculated thus far).\footnote{
   The coefficients of $f(q)$ are not integers,
   but $k \, [q^{-k}] f(q)$ is an integer
   (at least through the order calculated thus far).
   Indeed, it is not hard to show that if $F(z)$ is a power series
   with integer coefficients and constant term 1,
   then $k \, [z^{k}] \log F(z)$ is always an integer.
}
A further simplification is obtained by using the variable $z = 1/(q-1)$
in place of $1/q$:  the integer coefficients become much smaller.
Finally, a slight extra simplification arises from
extracting the prefactor $(q-1)^2/q$ in \reff{series_1}.

Unfortunately, the finite-lattice method ---
which is an extraordinarily efficient method for calculating series
expansions --- does not appear to be applicable to $=$ and $\neq$
boundary conditions.
We therefore resorted to a hand calculation,
to be reported in the next subsection, which we carried through order $q^{-4}$.

%%%%%%%%%%%%%%%%%%%%%%%%%%%%%%%%%%%%%%%%%%%%%%%%%%%%%%%%%%%
%
% SUBSECTION: LARGE-Q EXPANSIONS 
%
%%%%%%%%%%%%%%%%%%%%%%%%%%%%%%%%%%%%%%%%%%%%%%%%%%%%%%%%%%%
\subsection[Hand calculation of large-$q$ expansion for the bulk, surface and
 corner free energies]
    {Hand calculation of large-$\bm{q}$ expansion for the bulk, surface and
 corner free energies}
\label{sec.free_energy_series}

First of all, as we are interested in the large-$q$ limit, it is
convenient to explicitly remove the leading term $\log q$ in the
free energy by considering the modified chromatic polynomial
$\widetilde{P}_{G}$ for a loopless graph $G=(V,E)$:
\begin{equation}
\widetilde{P}_{G}(q) \;=\; q^{-|V|} P_{G}(q) \;.
\label{def.Ptilde}
\end{equation}
Using the Fortuin--Kasteleyn representation \reff{eq.FK.identity} we get
\begin{subeqnarray}
\widetilde{P}_G(q) &=& \sum\limits_{A\subseteq E} (-1)^{|A|} \, q^{k(A)-|V|}
   \\[1mm]
         &=& \sum\limits_{A\subseteq E} (-1)^{|A|} \, (1/q)^{|A|-c(A)}  \;,
\slabel{eq.FK.identity.Bis.b}
\label{eq.FK.identity.Bis}
\end{subeqnarray}
where $c(A) = |A| - |V| + k(A)$ is the cyclomatic number
of the subgraph $(V,A)$.

It is instructive to begin by computing ``by hand''
the first few terms of the large-$q$ expansion for
the bulk, surface and corner free energies.
To do this, let us first consider an $m\times n$ square lattice
with $\neq$ boundary conditions:
it has $|V|=mn+2$ sites, $|E|=2mn-m+n+1$ edges, $|F_\text{sq}|=(m-1)(n-1)$ 
square faces, and $|F_\text{tri}|=2(n-1)$ triangular faces.
There are also two extra faces, each bounded by a cycle of length $m+2$,
cf.\ the inner and outer faces in Figure~\ref{fig_SmnNotEq}b;
but these faces will play no role in the calculation
if $m$ is large enough (see below).
We can compute the first few first terms in the large-$q$ expansion
for the modified chromatic polynomial
$\widetilde{P}_{S^{\neq}_{m,n}}\!(q)$
by using \reff{eq.FK.identity.Bis.b} and explicitly identifying the
subsets $A$ having a given small value of $|A| - c(A)$:
\begin{quote}
\begin{itemize}
   \item[] $|A| - c(A) = 0$: Only $A = \emptyset$.
   \item[] $|A| - c(A) = 1$: $A = $ any single edge.
   \item[] $|A| - c(A) = 2$: $A = $ two distinct edges {\em or}\/
                three edges forming a triangular face. 
   \item[] $|A| - c(A) = 3$: $A = $ three distinct edges not forming a
              triangular face, {\em or}\/
              four edges forming a cycle of length four
              (i.e.\ a square face or the boundary cycle of 
              two adjacent triangular faces), {\em or}\/
              five edges consisting of all the edges of
              two adjacent triangular faces. 
\end{itemize}
\end{quote}
We have also identified the contributions of the subsets with
$|A| - c(A) = 4$ and will include them in our calculation,
but we refrain from writing them down as their description is rather lengthy.
For small $m$ we also have to worry about terms $A$
that wind horizontally around the lattice using the two extra sites;
the smallest of these are cycles of length $m+2$.
But since all such terms have $|A| - c(A) \ge m+1$,
we can avoid them simply by assuming that $m$
is large enough, i.e.\ $m \ge k$ if we want an expansion
valid through order $q^{-k}$.
In particular, we can obtain the expansion through order $q^{-4}$
by assuming that $m \ge 4$ and ignoring ``winding'' subsets $A$.
We therefore have
\begin{eqnarray}
\widetilde{P}_{S^{\neq}_{m,n}}\!(q) &=&  
  1 - \frac{|E|}{q} + 
      \left[ \frac{|E|(|E|-1)}{2} - F_\text{tri}\right]\frac{1}{q^2} 
   \nonumber\\ 
  & & \qquad - \left[ \frac{|E|(|E|-1)(|E|-2)}{6} - |F_\text{sq}| - 
           |F_\text{tri}|(|E|-2)\right] \frac{1}{q^3}   
  \nonumber \\
  & & \qquad + \left[ 
   \frac{|E|(|E|-1)(|E|-2)(|E|-3)}{24} - 
  (|F_\text{tri}|+|F_\text{sq}|)(|E|-3) \right. \nonumber \\
  & & \qquad \qquad \left.  
   - |F_\text{tri}|\frac{(|E|-3)(|E|-4)}{2} + 
     \frac{|F_\text{tri}|(|F_\text{tri}|-1)}{2} \right] \frac{1}{q^4} 
  \nonumber \\
               & & \qquad + O(q^{-5})
 \label{eq.exp_Pneq_mn}
\end{eqnarray}
and hence
\begin{eqnarray}
   \log \widetilde{P}_{S^{\neq}_{m,n}}\!(q) &=&  
   - \frac{|E|}{q} 
   \,-\, \frac{1}{2q^2} \Bigl( |E| - 2 |F_\text{tri}| \Bigr)
   \,-\, \frac{1}{3q^3} \Bigl( |E| - 3 |F_\text{sq}| + 6 |F_\text{tri}| \Bigr)
      \nonumber \\
  & & \qquad\;\:
   \,-\, \frac{1}{4q^4} \Bigl( |E| - 12 |F_\text{sq}| + 14 |F_\text{tri}| \Bigr)
   \,+\, O(q^{-5})  \;.
 \label{eq.exp_logPneq_mn}
\end{eqnarray}
Note that in the expression for $\log \widetilde{P}_{S^{\neq}_{m,n}}\!(q)$,
all geometrical quantities occur {\em linearly}\/ (as they should).
Inserting the values of $|E|$, $|F_\text{sq}|$ and $|F_\text{tri}|$,
dividing by $mn$, expanding in $1/n$ and $1/m$, 
and putting back the leading term $\log q$, one finds the large-$q$ expansion 
for the free energy
\begin{eqnarray}
f_{m,n}^{\neq}(q) &=& \log q -\frac{2}{q} - \frac{1}{q^2} + \frac{1}{3q^3} 
     + \frac{5}{2q^4} 
     + \left[ -\frac{1}{q} - \frac{5}{2q^2} - \frac{16}{3q^3} 
             -\frac{41}{4q^4} \right] \frac{1}{m} 
  \nonumber \\
     & & \quad + \left[ \frac{1}{q} + \frac{1}{2q^2} - \frac{2}{3q^3} 
             -\frac{11}{4q^4} \right] \frac{1}{n} 
     + \left[-\frac{1}{q} + \frac{3}{2q^2} + \frac{14}{3q^3}+
             \frac{39}{4q^4} \right] \frac{1}{mn} \nonumber \\
    & & \quad  + O(q^{-5})\,.
\label{series.sq.neq}
\end{eqnarray}
Comparing to the finite-size-scaling Ansatz \reff{def_FSS_Ansatz}, we obtain
\begin{subeqnarray}
f_{\rm bulk}(q) &=& \log q -\frac{2}{q} - \frac{1}{q^2} + \frac{1}{3q^3}
     + \frac{5}{2q^4} + O(q^{-5})   \\[1mm]
f_{\rm surf,vert}^{\neq}(q) &=&  -\frac{1}{q} - \frac{5}{2q^2} - \frac{16}{3q^3}
             -\frac{41}{4q^4} + O(q^{-5})   \slabel{fsurf_vert_neq} \\[1mm]
f_{\rm surf,horiz}^{\neq}(q) &=&  \phantom{-}\frac{1}{q} 
             + \frac{1}{2q^2} - \frac{2}{3q^3}
             -\frac{11}{4q^4} + O(q^{-5})   \\[1mm]
f_{\rm corner}^{\neq}(q) &=&  -\frac{1}{q} 
             + \frac{3}{2q^2} + \frac{14}{3q^3}
             +\frac{39}{4q^4} + O(q^{-5})
\label{series.sq.neq.components}
\end{subeqnarray}

We can do the same computation for the family $S^{=}_{m,n}$: 
it has $|V|=mn+1$ sites, $|E|=2mn-m+n$ edges, $|F_\text{sq}|=(m-1)(n-1)$ 
square faces, and $|F_\text{tri}|=2(n-1)$ triangular faces.
There are also two extra faces, each bounded by a cycle of length $m+1$,
cf.\ the inner and outer faces in Figure~\ref{fig_SmnEq}b;
but these faces will play no role in the computation if $m$ is large enough.
The subsets $A$ contributing through order $O(q^{-4})$
are exactly the same as for the family $S^{\neq}_{m,n}$. 
The only difference is that the ``winding'' terms now start with
cycles of length $m+1$ rather than $m+2$,
and they have $|A| - c(A) \ge m$ instead of $m+1$.
Therefore, we must take $m \ge k+1$ if we want an expansion
valid through order $q^{-k}$. In particular, we can obtain the expansion 
through order $q^{-4}$ by assuming that $m \ge 5$. The large-$q$ 
expansion for $\log\widetilde{P}_{S^{=}_{m,n}}\!(q)$ is given by the
{\em same}\/ formula \reff{eq.exp_logPneq_mn}
but with the new value of $|E|$ inserted
(note that $|F_\text{sq}|$ and $|F_\text{tri}|$ remain the same).
Dividing by $mn$, expanding in $1/n$ and $1/m$,
and putting back the leading term $\log q$, one finds the large-$q$ expansion
for the free energy
\begin{eqnarray}
f_{m,n}^{=}(q) &=& \log q -\frac{2}{q} - \frac{1}{q^2} + \frac{1}{3q^3} 
     + \frac{5}{2q^4} 
     + \left[ -\frac{1}{q} - \frac{5}{2q^2} - \frac{16}{3q^3} 
             -\frac{41}{4q^4} \right] \frac{1}{m} 
  \nonumber \\
     & & \quad + \left[ \frac{1}{q} + \frac{1}{2q^2} - \frac{2}{3q^3} 
             -\frac{11}{4q^4} \right] \frac{1}{n} 
     + \left[\frac{2}{q^2} + \frac{5}{q^3}+
             \frac{10}{q^4} \right] \frac{1}{mn} \nonumber \\
    & & \quad  + O(q^{-5})\,.
\label{series.sq.eq}
\end{eqnarray}
Comparing to the finite-size-scaling Ansatz \reff{def_FSS_Ansatz}, we obtain
\begin{subeqnarray}
f_{\rm bulk}(q) &=& \log q -\frac{2}{q} - \frac{1}{q^2} + \frac{1}{3q^3}
     + \frac{5}{2q^4} + O(q^{-5})   \\[1mm]
f_{\rm surf,vert}^{=}(q) &=&  -\frac{1}{q} - \frac{5}{2q^2} - \frac{16}{3q^3}
             -\frac{41}{4q^4} + O(q^{-5})  \slabel{fsurf_vert_=}  \\[1mm]
f_{\rm surf,horiz}^{=}(q) &=&  \phantom{-}\frac{1}{q}
             + \frac{1}{2q^2} - \frac{2}{3q^3}
             -\frac{11}{4q^4} + O(q^{-5})   \\[1mm]
f_{\rm corner}^{=}(q) &=&  \phantom{-}\frac{2}{q^2} + \frac{5}{q^3}
             +\frac{10}{q^4} + O(q^{-5})
\label{series.sq.eq.components}
\end{subeqnarray}

If we compare the series \reff{series.sq.neq.components}
with \reff{series.sq.eq.components},
we conclude that:

1) The bulk contributions are the same for $\neq$ and $=$
boundary conditions, and they agree also with the result for
free and cylindrical boundary conditions \cite[Eq.~(5.14a)]{transfer5}.  

2) The surface contributions are the same for both boundary conditions,
and they depend on the orientation of the ``surface''
(i.e.\ vertical or horizontal).
Moreover, $f_{\rm surf,horiz}^{=}=f_{\rm surf,horiz}^{\neq}$ agrees
with the surface free energy for free boundary conditions
\cite[Eq.~(5.14b)]{transfer5}.
This is not unexpected, as the top and bottom rows are identical
for all these boundary conditions.

3) The corner free energies do depend on the boundary conditions
(i.e.\ $f_{\rm corner}^{=}\neq f_{\rm corner}^{\neq}$) and they also differ
from the corner free energy for free boundary conditions
\cite[Eq.~(5.14c)]{transfer5}.
We observe the curious fact
\be
   f_{\rm corner}^{=}(q) - f_{\rm corner}^{\neq}(q)
   \;=\;
   \frac{1}{q} + \frac{1}{2q^2} + \frac{1}{3q^3} + \frac{1}{4q^4} + O(q^{-5})
   \;.
 \label{eq.fcorner.diff}
\ee
In the next subsection we will establish this equality
to all orders in $1/q$.

\subsection{Exact results for the bulk, surface and corner free energies}
\label{sec.free_energy_series.exact}

We can obtain some exact results concerning the {\em difference}\/
between the free energies $f_{m,n}^{=}(q)$ and $f_{m,n}^{\neq}(q)$
--- and hence between the bulk, surface and corner free energies
for $=$ and $\neq$ boundary conditions ---
by using the results of Section~\ref{sec.comp}.
Let us write the modified chromatic polynomials
$\widetilde{P}_{S_{m,n}^\sharp}(q)$ [cf.\ \reff{def.Ptilde}]
in terms of the partial chromatic polynomials of the
two--terminal graphs $(S_{m,n},s,t)$ 
introduced in Section~\ref{sec.comp};  we obtain
\begin{subeqnarray}
\widetilde{P}_{S_{m,n}^=}(q) &=& q^{-mn-1} 
\left[ P_{S_{m,n}}^{(s \leftrightarrow t)}(q) + q^{-1}\, 
       P_{S_{m,n}}^{(s \not\leftrightarrow t)}(q) \right] \\[2mm]
\widetilde{P}_{S_{m,n}^{\neq}}(q) &=& q^{-mn-2}
\, (1 - q^{-1}) \, P_{S_{m,n}}^{(s \not\leftrightarrow t)}(q) 
\end{subeqnarray} 
Therefore, the difference between the finite-size free energies
for the two boundary conditions can be written as
\begin{subeqnarray}
f_{m,n}^=(q) - f_{m,n}^{\neq}(q) &=& \frac{1}{mn} \left[ 
   \log \widetilde{P}_{S_{m,n}^=}(q) - \log \widetilde{P}_{S_{m,n}^{\neq}}(q)
   \right] \\[1mm]
   &=& - \frac{1}{mn} \log \left( 1 - \frac{1}{q} \right) 
       + \frac{1}{mn} \log \left[ 1 + q \, 
         \frac{\widetilde{P}_{S_{m,n}}^{(s \leftrightarrow t)}(q)}
              {\widetilde{P}_{S_{m,n}}^{(s \not\leftrightarrow t)}(q)}
              \right] 
\end{subeqnarray}
where $\widetilde{P}_{S_{m,n}}^{(s \leftrightarrow t)}$ and
$\widetilde{P}_{S_{m,n}}^{(s \not\leftrightarrow t)}$ are the 
corresponding modified chromatic polynomials. 
{}From \reff{eq.shortest1} and \reff{eq.shortest2} we see immediately that
\begin{subeqnarray}
   \widetilde{P}_{S_{m,n}}^{(s \not\leftrightarrow t)}(q)
      & = &  1 \,+\, O(q^{-1})  \\[2mm]
   \widetilde{P}_{S_{m,n}}^{(s \leftrightarrow t)}(q)
      & = &  n \, (-1)^{m+1} q^{-(m+1)} \, [1 \,+\, O(q^{-1})]
 \label{eq.Ptilde.st}
\end{subeqnarray}
and hence
\be
f_{m,n}^=(q) - f_{m,n}^{\neq}(q) \;=\; 
- \frac{1}{mn} \log \left( 1 - \frac{1}{q} \right)
 \:+\: \frac{(-1)^{m+1}}{m} \, q^{-m}  \:+\: O( q^{-(m+1)}) \,.
\ee
The term proportional to $q^{-m}$ disappears to all orders in $1/q$
as $m \to\infty$, hence we have
\begin{subeqnarray}
   f_{\rm surf, horiz}^{=}(q) &=&f_{\rm surf, horiz}^{\neq}(q)  \\[1mm]
   f_{\rm surf, vert}^{=}(q)  &=&f_{\rm surf, vert}^{\neq}(q)  \\[1mm]
   f_{\rm corner}^{=}(q)      &=&f_{\rm corner}^{\neq}(q) 
   - \log \left( 1 - \frac{1}{q} \right)
 \label{eq.f.diff.exact}
\end{subeqnarray}
to all orders in $1/q$.\footnote{
   We can interpret the correction term $O(q^{-m})$ as 
   $O(e^{-m/\xi_{\rm bulk}})$ [cf.\ \reff{eq.xibulk}]. 
}
In particular, this proves \reff{eq.fcorner.diff} to all orders in $1/q$.

We can actually push the expansions (\ref{eq.Ptilde.st}a,b)
to higher order in $1/q$ by methods similar to those used
in the preceding subsection.
Indeed, $\widetilde{P}_{S_{m,n}}^{(s \not\leftrightarrow t)}$
is exactly equal to $\widetilde{P}_{S_{m,n}}^{\neq}$
whenever $m$ is large enough (compared to the order in $1/q$ being considered)
so that winding subsets $A$ (i.e.\ sets connecting the two terminals $s,t$)
are absent.
In particular, $\widetilde{P}_{S_{m,n}}^{(s \not\leftrightarrow t)}$
is given through order $q^{-k}$ ($k \le 4$) by \reff{eq.exp_Pneq_mn}
whenever $m \ge k$.
On the other hand, a straightforward hand computation yields
\be
\widetilde{P}_{S_{m,n}}^{(s \leftrightarrow t)}(q) \;=\; 
\frac{(-1)^{m+1}}{q^{m+1}} \left[ n  \,-\,
   \frac{n|E|+ mn-5n -m +4}{q}  \,+\, O( q^{-2}) \right]
\ee 
whenever $m \ge 2$.
It follows that
\begin{subeqnarray}
\log\lambda_{\star,=}(m) - \log \lambda_{\star,\neq}(m) &=&
\lim_{n\to \infty} \frac{1}{n} \log \left( 
1 + q \, 
\frac{\widetilde{P}_{S_{m,n}}^{(s \leftrightarrow t)}(q)}
              {\widetilde{P}_{S_{m,n}}^{(s \not\leftrightarrow t)}(q)}
              \right) \\[1mm]
 &=& \frac{ (-1)^{m+1}}{q^m} \left[ 1 - \frac{m-5}{q} + 
              O( q^{-2}) \right]  \;.
\end{subeqnarray}
In terms of the quantities $f_\ell(m)$ defined in \reff{def_series_fl},
this says that $f_1(m)=m-5$ for $m \ge 2$,
which agrees with Table~\ref{table_diff_coef_f}.

%
% \neq bc's 
%
\subsection[Families $S_{m,n}^{\neq}$ and $S_{m,n}^{=}$ via strip free energies]%
{Families $\bm{S_{m,n}^{\neq}}$ and $\bm{S_{m,n}^{=}}$ via strip free energies}
  \label{sec.thermo.neqeq}

Using finite-size-scaling theory,
we can understand (non-rigorously)
a fact observed empirically in Sections~\ref{sec4.2} and \ref{sec4.3}:
namely, that the coefficients $c_k^\sharp(m)$
arising in the large-$|q|$ expansion of the free energy \reff{def_ckm}
are represented for $m \ge m_{\rm min}^\sharp(k)$
by a polynomial of degree 1 in the strip width $m$, i.e.
\begin{equation}
   c_k^{=}(m)  \;=\;  c_k^{\neq}(m)  \;=\;  C_k(m)  \;=\;  \alpha_k m + \beta_k
   \qquad\hbox{for $m \ge m_{\rm min}(k)$}
   \;.
 \label{eq.sec.thermo.neq.Ckm}
\end{equation}
To see this, it suffices to compare
the large-$|q|$ expansion of the limiting free energy
for a semi-infinite strip,
\begin{eqnarray}
f_{m}^\sharp(q)&=& \frac{1}{m} \log \lambda_{\star,\sharp}(q)  \nonumber \\
         &=& \log q + \frac{1}{m}\log\left[ \sum\limits_{k=0}^\infty
             (-1)^k b^{\sharp}_k(m) q^{-k} \right] \nonumber \\
&=& \log q + \frac{1}{m}  \sum\limits_{k=1}^\infty
             (-1)^k c^{\sharp}_k(m) q^{-k}
   \;,
\label{def_fm_sharp_largeq}
\end{eqnarray}
with the finite-size-scaling Ansatz
\reff{def_FSS_Ansatz2}/\reff{def_FSS_Ansatz2_bis}.
The behavior \reff{eq.sec.thermo.neq.Ckm} is an immediate consequence,
where $\alpha_k$ (resp.\ $\beta_k$) is the coefficient of $q^{-k}$
in the bulk (resp.\ vertical surface) free energy.
As discussed after \reff{def_FSS_Ansatz2_bis},
we expect $m_{\rm min}^\sharp(k) \approx k$.

We can also use our transfer matrices to
check in part the result \reff{series_fbulk} for the bulk free energy
and to notably extend the results \reff{fsurf_vert_neq}/\reff{fsurf_vert_=}
for the vertical surface free energy.
Indeed, the coefficients in both these expansions
can be immediately read off from the coefficients $\alpha_k$ and $\beta_k$ 
in Table~\ref{table_new_Ck}.
We thus obtain an independent check
of the first $15$ terms of the series \reff{series_fbulk}.
We also obtain the first 15 terms in the large-$q$ expansion of
the vertical surface free energy
$f^{\neq}_{\rm surf,vert} = f^{=}_{\rm surf,vert}$:
\begin{eqnarray}
 f_{\rm surf,vert}^{\sharp}(q) &=& 
 - \frac{1}{q} - \frac{5}{2q^2} -
       \frac{16}{3q^3} - \frac{41}{4q^4} - \frac{81}{5q^5} - \frac{49}{3q^6}
     +\frac{55}{7 q^7} + \frac{719}{8q^8} + \frac{2459}{9q^9} \nonumber\\
   & & \qquad  
     +\frac{1239}{2q^{10}} + \frac{15168}{11q^{11}} +\frac{23051}{6q^{12}} +
      \frac{171677}{13q^{13}} + \frac{647719}{14q^{14}} + 
      \frac{744743}{5q^{15}} \nonumber \\
   & & \qquad + \frac{6898415}{16q^{16}} + O(q^{-17})
\slabel{def_fsurf_neq}
\end{eqnarray}

We can slightly extend this latter series by using the series
\reff{series_fbulk} for $f_{\rm bulk}$ as an {\em input}\/:
in this case, each polynomial $C_k$ contains a single unknown
coefficient to be determined (rather than two unknown coefficients).
We then obtain the coefficient of the term $q^{-17}$ in 
$f^{\sharp}_{\rm surf,vert}$,
\begin{equation}
\left[q^{-17} \right] f_{\rm surf,vert}^{\sharp} \;=\;\smfrac{19118828}{17}
   \;,
\label{surf_term_17}
\end{equation}
and thus the polynomial $C_{17}$ shown in Table~\ref{table_new_Ck}.

Finally, the simplest expression for the surface free energy is given
by its exponential in terms of the variable $z = 1/(q-1)$:
\begin{eqnarray}
e^{f_{\rm surf,vert}^{\sharp}(q)} &=& 1 -z -z^2 +z^4 +3z^5 +2z^6-5z^7 -11z^8 
   +3z^9 +43z^{10} + 57z^{11}  -34z^{12} \nonumber \\
  & & \quad-178z^{13} -122z^{14} +220z^{15}
            +200 z^{16} - 1170 z^{17} + O(z^{18}) \,. 
\label{def_fsurf_neq_z}
\end{eqnarray}

%
% Numerical results II
%
\section{Numerical results II: Limiting curves $\bm{\scrb_m}$}
   \label{sec.numerical.bm}

For each $1\leq m \leq 10$, we have computed the {\em symbolic}\/ 
transfer matrix $\T''(m)$, as well as the vectors $\basisf_{\rm id}$ and
$\endv^{\rm T}$.
In this computation we exploited the fact that $\T''(m)$ is block-diagonal 
[cf.\ \reff{eq.add-contr}/\reff{def_Tblocks}], which allows us to obtain
{\em separately}\/ the two diagonal blocks $\T''_{=}(m)$ and $\T''_{\neq}(m)$
and, for each block,
the corresponding left and right vectors
$\endv^{\rm T}$ and $\basisf_{\rm id}$.
 
The symbolic computations were done
by means of a {\sc Mathematica} program for $1 \le m \le 5$
and by a {\tt perl} script for $1 \le m \le 10$.
The {\tt perl} script runs faster and uses a smaller 
amount of memory than the {\sc Mathematica} program for the same width $m$.
We have checked that both programs give the same answer for $1 \le m\leq 5$. 
We also performed several checks on our symbolic results using the identity
\begin{equation}
P_{S_{m,n}}(q) \;=\; P_{\widehat{S}_{n,m}}(q)  \;,
\label{check_poly}
\end{equation}
where $\widehat{S}_{m,n}$ is the graph that is just like $S_{m,n}$
except that the two extra sites lie at top and bottom rather than
left and right (see Figure~\ref{fig_hatSmn}).
Of course $S_{m,n}$ and $\widehat{S}_{n,m}$ are isomorphic,
so that the identity \reff{check_poly} holds trivially,
but the transfer-matrix approaches to calculating their chromatic
polynomials are rather different, because of a reversal in which
direction is considered ``longitudinal'' and which ``transverse''
(see Appendix~\ref{sec.transfer.bis}).
The identity \reff{check_poly} thus constitutes a nontrivial check
on the correctness of our computations.
We checked it for $1 \le m \le 8$ and $1 \le n \le 6$.

For each $1\leq m \leq 6$, we computed the zeros of the 
chromatic polynomials $P_{S_{m,\rho m}}(q)$ for strips of aspect 
ratio $\rho=10,20$, and we also computed the accumulation set 
$\mathcal{B}_m$ of chromatic roots in the limit $\rho\to\infty$.  
For $m\leq 4$, we used the resultant method \cite[Section 4.1.1]{transfer1}
to compute the limiting curve $\mathcal{B}_m$,
together with the direct-search method \cite[Section 4.1.2]{transfer1}
to refine some details.
For $m=5$, we used the resultant method to compute the endpoints of
the limiting curve, while the rest of the curve
was obtained via the direct-search method.
For $m=6$, the full limiting curve was obtained using the direct-search method.
We took advantage of a few slight improvements to these methods
as described in detail in Ref.~\cite[Section~2]{transfer3}.  
In principle we could have determined the limiting curve $\mathcal{B}_m$
also for $7 \le m \le 10$;
but it is extremely laborious (in both human and CPU time)
to determine accurately the small details in $\mathcal{B}_m$
using the direct-search method, so we decided to stop at $m=6$.

Since $\H$ is a projection, the basis elements
in our transfer-matrix calculation are given by
$\basisf_{\mathcal P}=\H \basise_{\mathcal P}$, and {\em not}\/ by 
$\basise_{\mathcal P}$, as explained in Section~\ref{sec.setup.transfer}. 
However, to lighten the notation, we will represent the basis for a given 
strip as a collection of vectors $\basise_{\mathcal P}$;
but it should be understood that the basis vectors are actually 
$\basisf_{\mathcal P}$. 
We shall use the delta-function shorthand
\reff{def.delta.shorthand} to write the vectors $\basise_{\mathcal P}$.

%
% m= 1
%
\subsection[$m=1$]{$\bm{m=1}$}

The transfer matrix $\T''_{=}(1)$ is one-dimensional,
and a basis is given by $\bm{B}_{=}=\{ \bone \}$.
This matrix and the corresponding
left and right vectors are, in this basis,
\begin{subeqnarray}
\T''_{=}(1)               &=& q-2   \\
\endv^{\rm T}            &=& q(q-1)\\
\basisf_{\rm id}^{\rm T} &=& 1
\label{res_sq_eq_1F}
\end{subeqnarray}

The transfer matrix $\T''_{\neq}(1)$ is also one-dimensional,
and a basis is given by $\bm{B}_{\neq} =\{ \bone \}$.
In this basis, the transfer matrix and the corresponding vectors are
given by
\begin{subeqnarray}
\T''_{\neq}(1)            &=& q-3        \\
\endv^{\rm T}            &=& q(q-1)(q-2)\\
\basisf_{\rm id}^{\rm T} &=& 1
\label{res_sq_neq_1F}
\end{subeqnarray}

The two eigenvalues are evidently
$\lambda_{=} = q-2$ and $\lambda_{\neq} = q-3$,
which become equimodular along the line
\begin{equation}
\Re q \;=\; \frac{5}{2}   \;.
\end{equation}
Thus, this is the limiting curve for this strip.
The amplitudes can be
easily read off from \reff{res_sq_eq_1F}/\reff{res_sq_neq_1F}: i.e.,
$\alpha_{=}=q(q-1)$ and $\alpha_{\neq}=q(q-1)(q-2)$,
in agreement with the exact result \reff{exact_P1xn}.

The dominant eigenvalue in the half-plane $\Re q < 5/2$
is $\lambda_{\neq}=q-3$,
while the dominant eigenvalue in the half-plane $\Re q > 5/2$
is $\lambda_{=}=q-2$.
The amplitude of the dominant eigenvalue vanishes at $q=0,1,2$
(and only there);
hence, these three points are the only isolated limiting points
for this strip.
The limiting curve $\mathcal{B}_1$ and the chromatic roots for $n=10,20$ are
depicted in Figure~\ref{figure_sq_1F}. 

%
% m= 2
%
\subsection[$m=2$]{$\bm{m=2}$}

The transfer matrix $\T''_{\neq}(2)$ is one-dimensional.
In the basis $\bm{B}_{=}=\{ \bone \}$,
this matrix and the corresponding vectors take the form
\begin{subeqnarray}
\T_{=}''(2)               &=& q^2-5q+7 \\
\endv^{\rm T}            &=& q(q-1)(q-2) \\
\basisf_{\rm id}^{\rm T} &=& 1
\label{res_sq_eq_2F}
\end{subeqnarray}

The transfer matrix $\T''_{\neq}(2)$ is two-dimensional.
Let us choose the basis as
$\bm{B}_{\neq}=\{ \delta_{0,2}+\delta_{1,3}, \bone \}$.
The transfer matrix and the left and right vectors for this strip are then
\begin{subeqnarray}
\T''_{\neq}(2)            &=& \left( \begin{array}{cc}
                               2-q    & -1       \\
                               2(q-2) & q^2-5q+8 \\
                                    \end{array} \right) \\
\endv^{\rm T}            &=& q(q-1) \, \left( 2(q-1),q^2-3q+3 \right) \\
\basisf_{\rm id}^{\rm T} &=& \left( 0,1\right)
\label{res_sq_neq_2F}
\end{subeqnarray}

The eigenvalues are given by
\begin{subeqnarray}
\lambda_{=}      &=& q^2-5q+7 \\
\lambda_{\neq,1} &=& \frac{1}{2}\left( 10 -6q+q^2 - \sqrt{Q_2} \right) \\
\lambda_{\neq,2} &=& \frac{1}{2}\left( 10 -6q+q^2 + \sqrt{Q_2} \right)
\end{subeqnarray}
where we have used the shorthand notation
\be
Q_2(q) \;=\; q^4-8q^3+28q^2-56q+52   \;.
\label{def_Q2}
\ee
The eigenvalues $\lambda_{\neq,1}$ and $\lambda_{\neq,2}$
are the solutions of the quadratic equation
\be
x^2 - x(q^2 -6q+10) -q^3+ 7q^2-16q+12 \;=\; 0   \;.
\ee
The amplitudes are given by
\begin{subeqnarray}
\alpha_{=}      &=& q(q-1)(q-2) \\
\alpha_{\neq,1} &=& \frac{q(q-1)}{2}\,
                 \frac{ (3-3q+q^2) \sqrt{Q_2}  -
                        (22-34q+21q^2-7q^3+q^4)}{\sqrt{Q_2}}\\
\alpha_{\neq,2} &=& \frac{q(q-1)}{2}\,
                 \frac{ (3-3q+q^2) \sqrt{Q_2}  +
                        (22-34q+21q^2-7q^3+q^4)}{\sqrt{Q_2}}
\end{subeqnarray}

We have computed the limiting curve $\scrb_2$ by using the resultant method.
This curve crosses the real $q$-axis at
$q=3$, which is a quadruple point.
There are two pairs of complex-conjugate T points, namely
$q\approx 2.6099757836 \pm 1.8423725343\,i$  and % #1
$q\approx 2.8900242164 \pm 0.5194968788\,i$.     % #2

In the two regions having nonempty intersection with the real $q$-axis,
the dominant eigenvalue comes from $\T''_{\neq}(2)$;  
in the other four regions, the dominant eigenvalue is $\lambda_=$.
On the complex-conjugate curves connecting the
T points $q\approx 2.6099757836 \pm 1.8423725343\,i$ and
$q\approx 2.8900242164 \pm 0.5194968788\,i$, $\lambda_{\neq,1}$ and 
$\lambda_{\neq,2}$ are equimodular.
Hence, at the four T points the three eigenvalues become
equimodular. Finally, at the quadruple point $\lambda_==\lambda_{\neq,2}=1$ and
$\lambda_{\neq,1}=0$.\footnote{
  The eigenvalues $\lambda_{\neq,1}$ and $\lambda_{\neq,2}$ are analytic
  functions of $q$ except at the branch cuts of the function $\sqrt{Q_2(q)}$.
  The polynomial $Q_2$ has zeros at
  $q = 2 \pm i +\sqrt{-1\mp 2i} \approx 2.78615 \mp 0.27202i$
  and at
  $q = 2 \mp i -\sqrt{-1\pm 2i} \approx 1.21305 \mp 2.27202i$. 
  These four zeros belong to regions where $\lambda_{=}$ is dominant: namely,
  the first pair of zeros belong to the two oval-like closed regions 
  protruding from the quadruple point $q=3$, while the later pair 
  belong to the other two regions not intersecting the real $q$-axis.
  We have chosen the branch cuts to be horizontal lines starting at each
  zero and going to $q\to-\infty$: i.e., for each zero $q_i$ with 
  $i=1,\ldots,4$, we define $q=q_i + r_i e^{i\theta_i}$ with $r_i\geq 0$ 
  and $-\pi < \theta_i \le \pi$. With this definition of branch cuts, 
  $\lambda_{\neq,2}$ is dominant in the two regions having nonempty 
  intersection with the real $q$-axis, except whenever 
  $2.27202 \gtapprox |\Im q| \gtapprox 0.27202$ and $\Re q$ lies to the 
  left of the lines joining two T points.   
}

There are four isolated limiting points at $q=0,1,2,B_5$. At these values
the dominant amplitude $\alpha_{\neq,2}$ vanishes.
The limiting curve $\mathcal{B}_2$ and the chromatic roots for $n=20,40$ are
depicted in Figure~\ref{figure_sq_2F}.

%
% m= 3
%
\subsection[$m=3$]{$\bm{m=3}$}

The transfer matrix $\T''_{=}(3)$ has dimension 3. The basis is chosen as
$\bm{B}_{=}=\{ \delta_{0,2},\delta_{1,3}, \bone\}$.
In this basis, the transfer matrix and the left and right vectors for
this strip are
\begin{subeqnarray}
\T''_{=}(3)               &=&   \left( \begin{array}{ccc}
                               -L_{4,4} &-1     &5 - 2q  \\
                                0       & q-2   & 1      \\
                                L_{4,4} &L_{5,7}& T_1
                                       \end{array} \right)\\
\endv^{\rm T}            &=& q(q-1) \left( q-1,q-1,L_{3,3} \right) \\
\basisf_{\rm id}^{\rm T} &=& \left( 0,0,1 \right)
\label{res_sq_eq_3F}
\end{subeqnarray}
where we have used the shorthand notation
\begin{subeqnarray}
T_1(q)     &=& q^3-7q^2+19q-20 \\
L_{m,n}(q) &=& q^2-mq+n
\label{short_notation}
\end{subeqnarray}

The transfer matrix $\T''_{\neq}(3)$ has dimension 4. The basis is chosen as
$\bm{B}_{\neq}=\{ \delta_{0,2}+\delta_{4,2},\delta_{0,3}+\delta_{1,4},
\delta_{1,3}, \bone \}$.
In this basis, the transfer matrix and the left and right vectors for
this strip are
\begin{subeqnarray}
\T''_{\neq}(3)            &=&   \left( \begin{array}{cccc}
                                -L_{5,6} & q-3      & -1      & 3-q\\
                                q-2      & -L_{5,7} & q-2     & 1 \\
                                0        & 0        & q-2     & 1 \\
                                2L_{5,6} & 2L_{5,7} & L_{6,9} & T_1-1
                                       \end{array} \right)\\
\endv^{\rm T}            &=& q(q-1)(q-2) \left(2(q-1),2(q-1),q-1,L_{2,2}\right)
  \\
\basisf_{\rm id}^{\rm T} &=& \left( 0,0,0,1 \right)
\label{res_sq_neq_3F}
\end{subeqnarray}
where we have used \reff{short_notation}.

The three eigenvalues of $\T''_{=}(3)$ come from the third-order polynomial
\begin{eqnarray}
x^3 &-& x^2(q^3-8q^2+24q-26) - x(q^5-12q^4+59q^3-149q^2+193q-101) \nonumber\\
    & & + q^6-13q^5 + 70q^4-200q^3+320q^2 -272q+96
\end{eqnarray}
while the four eigenvalues of $\T''_{\neq}(3)$ come from the fourth-order
polynomial
\begin{eqnarray}
x^4 &-& x^3(q^3-9q^2+30q-36) - x^2(2q^5-26q^4+140q^3-388q^2+551q-318)\nonumber\\
    & & - x(q^7-19q^6+153q^5-681q^4+1815q^3-2901q^2+2577q-981) \nonumber \\
    & & +
q^8-19q^7+ 157q^6 - 738q^5+ 2161q^4- 4039q^3 \nonumber \\
    & & + 4707q^2 - 3128q +908 \,. 
\end{eqnarray}

We have computed the limiting curve by using the resultant method, and
fine-tuned the results using the direct-search method.
The limiting curve $\mathcal{B}$ crosses the real $q$-axis at
$q\approx 2.8177131118$.
There is one pair of complex-conjugate endpoints at
$q \approx 2.7852013976 \pm 0.7798713630\,i$.
There are seven pairs of complex-conjugate T points at
$q\approx 1.0229779807 \pm 2.4615628248\,i$,     %#6
$q\approx 2.3154363287 \pm 1.9575822608\,i$,     %#5
$q\approx 2.4538824566 \pm 1.8334262878\,i$,     %#7
$q\approx 2.5862741033 \pm 1.5583461255\,i$,     %#4
$q\approx 2.8988477958 \pm 0.4358504561\,i$,     %#1
$q\approx 2.9168615472 \pm 0.6195940190\,i$, and %#2
$q\approx 3.0476880109 \pm 0.4560842031\,i$.     %#3
There are four isolated limiting points at $q=0,1,2,B_5$.

The dominant eigenvalue comes from $\T''_{\neq}(3)$ in the
region containing the interval $q \in (-\infty,2.8177131118)$,
and in the two large regions bounded asymptotically by
$|\arg q| \in (\pi/6,\pi/2)$ as $|q|\to\infty$. The dominant eigenvalue
comes from $\T''_{=}(3)$ in the other three large regions:
the one that contains the interval $q\in(2.8177131118,\infty)$ and the
other two asymptotically bounded by $|\arg q| \in (\pi/2,5\pi/6)$ as
$|q|\to\infty$.

There are four isolated limiting points at $q=0,1,2,B_5$. 
The limiting curve $\mathcal{B}_3$ and the chromatic roots for $n=30,60$ are
depicted in Figure~\ref{figure_sq_3F}.

%
% m= 4
%
\subsection[$m=4$]{$\bm{m=4}$}

The transfer matrix $\T''_{=}(4)$ has dimension 5, and
$\T''_{\neq}(4)$ has dimension 9.
They are too lengthy to be reported here,
but they can be found in the {\sc Mathematica} file
{\tt square\_extra\_sites.m} (1.9 MB)
that is available on request from the authors.

We have computed the limiting curve by using the resultant method, and 
fine-tuned the results using the direct-search method.
The limiting curve $\mathcal{B}$ crosses the real $q$-axis at
$q\approx 2.9060325277$ and $q\approx 3.9030119682$.
There are two pairs of complex-conjugate endpoints at
$q \approx 2.6169563471 \pm 1.4357301392\,i$, and
$q \approx 2.8849105593 \pm 0.7356893954\,i$.

There are ten pairs of complex-conjugate T points at
$q\approx 0.3411402631 \pm 2.3118521368\,i$,     %#1
$q\approx 1.1260056618 \pm 2.5364550333\,i$,     %#2
$q\approx 1.3027336503 \pm 2.5447465991\,i$,     %#3
$q\approx 2.2640047808 \pm 1.9520686076\,i$,     %#4
$q\approx 2.4920548627 \pm 1.7925248200\,i$.     %#10
$q\approx 2.7002452362 \pm 1.5923412229\,i$,     %#6
$q\approx 2.8652031850 \pm 1.1531618171\,i$,     %#5
$q\approx 2.9601783483 \pm 0.5518584736\,i$,     %#9
$q\approx 2.9623803372 \pm 0.5807622346\,i$, and %#8
$q\approx 3.0624715624 \pm 0.6434188421\,i$.     %#7
There is a pair of bulb-like regions protruding from T points
$q\approx 2.7002452362 \pm 1.5923412229\,i$.

\begin{figure}
\centering
\begin{tabular}{cc}
\includegraphics[width=200pt]{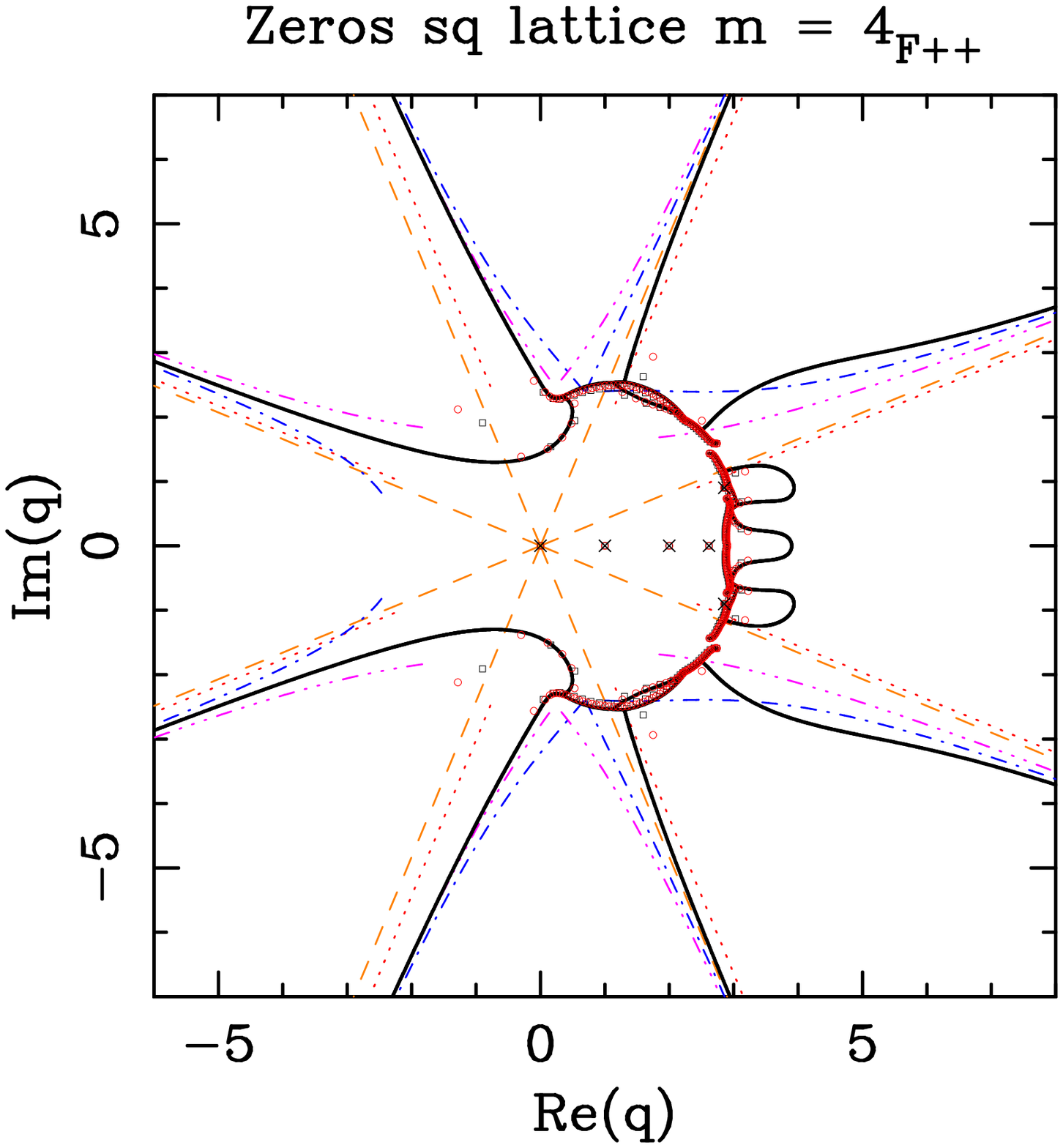} & 
\includegraphics[width=200pt]{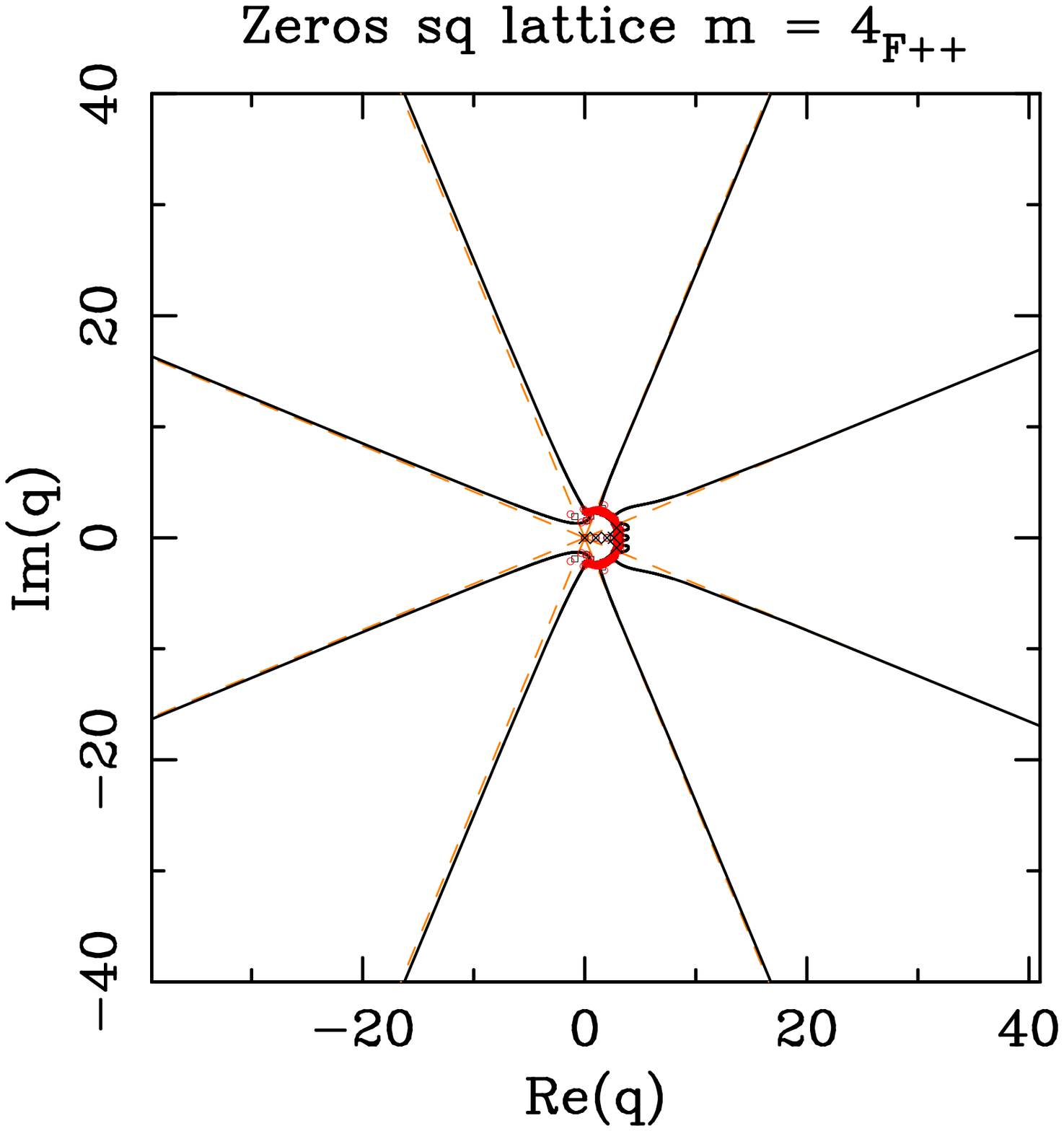} \\[1mm] 
\phantom{(((a)}(a) & \phantom{(((a)}(b) \\[5mm]
\end{tabular}
\caption{\label{figure_sq_4F}
Limiting curves for square-lattice strips of width $m=4$ with two extra sites. 
We also show the zeros for the strips $S_{4,40}$ 
(black $\square$) and $S_{4,80}$ (red $\circ$).
We depict the isolated limiting points with the symbol $\times$.
The zeroth-, first-, second- and third-order asymptotic expansions
for the outward branches [cf.\ \reff{eq.argq.asymp}]
are depicted as
dashed orange, dotted red, dot-dashed magenta, and dot-dot-dot-dashed blue
curves, respectively.
}
\end{figure}

The dominant eigenvalue comes from $\T''_{\neq}(4)$ in the
region containing the interval
$q \in (-\infty,2.9060325277)\cup(3.9030119682,\infty)$,
and in the two large regions bounded asymptotically by
$|\arg q| \in (3\pi/8,5\pi/8)$ as $|q|\to\infty$. The dominant eigenvalue
comes from $\T''_{=}(4)$ in the other four large regions: i.e.,
those asymptotically bounded by $|\arg q| \in (\pi/8,3\pi/8)$ and
$|\arg q| \in (5\pi/8,7\pi/8)$ as $|q|\to\infty$. The dominant eigenvalue
comes from this block also in the region containing the interval
$q\in(2.9060325277,3.9030119682)$, and the other two closed regions
pointing to the right in Figure~\ref{figure_sq_4F}.

There are four real isolated limiting points at $q=0,1,2,B_5$, and a pair
of complex-conjugate isolated limiting points at
$q\approx 2.8555521103  \pm 0.9018551071\,i$.
The limiting curve $\mathcal{B}_4$ and the chromatic roots for $n=40,80$ are
depicted in Figure~\ref{figure_sq_4F}.

%
% m= 5
%
\subsection[$m=5$]{$\bm{m=5}$}

The transfer matrix $\T''_{=}(5)$ has dimension 11, and the matrix
$\T''_{\neq}(5)$ has dimension 21.

This is the first case where we have computed the limiting curve using
the direct-search method, except for the endpoints that were computed using the
resultant method.
The limiting curve $\mathcal{B}$ crosses the real $q$-axis at
$q\approx 2.9268787368$.

There are 14 pairs of complex-conjugate T points at
$q\approx 0.0265421747 \pm 2.1428032766\,i$,     %#10
$q\approx 0.2211937394 \pm 2.2964381892\,i$,     %#6
$q\approx 0.5678532323 \pm 2.4981722822\,i$,     %#1
$q\approx 1.5459127222 \pm 2.4751057958\,i$,     %#2
$q\approx 1.5503506823 \pm 2.4742675615\,i$,     %#11
$q\approx 2.2165122888 \pm 2.0475480936\,i$,     %#8
$q\approx 2.3467238502 \pm 1.9855375152\,i$,     %#12
$q\approx 2.5384138871 \pm 1.7717517039\,i$,     %#7
$q\approx 2.8417052689 \pm 1.2459939540\,i$,     %#3
$q\approx 2.9278218101 \pm 0.9693176128\,i$,     %#5
$q\approx 2.9651600535 \pm 0.7574018277\,i$,     %#4
$q\approx 3.0055390406 \pm 0.9222249388\,i$,     %#9
$q\approx 3.1196012894 \pm 0.2666719871\,i$, and %#14
$q\approx 3.1587639645 \pm 0.5586982282\,i$.     %#13
There are two pairs of complex-conjugate endpoints at
$q\approx 2.0144392334 \pm 2.0277231065\,i$, and %#1
$q\approx 2.3246979483 \pm 1.9016376971\,i$.     %#2
Finally, there is a pair of complex-conjugate bulk-like regions protruding from
the T points $q\approx 2.3467238502 \pm 1.9855375152\,i$.

There are four real isolated limiting points at $q=0,1,2,B_5$, and two 
pairs of complex-conjugate isolated limiting points at
$q\approx 2.7190757419 \pm 1.4455587779\, i$, and 
$q\approx 2.8265910048 \pm 0.9420673312\, i$.
The limiting curve $\mathcal{B}_5$ and the chromatic roots for $n=50,100$ are
depicted in Figure~\ref{figure_sq_5F}.

%
% m= 6
%
\subsection[$m=6$]{$\bm{m=6}$}

The transfer matrix $\T''_{=}(6)$ has dimension 21, and the matrix
$\T''_{\neq}(6)$ has dimension 49.

The limiting curve has been completely determined by using the direct-search
method. This curve ${\mathcal B}$ crosses the real $q$-axis at
$q\approx 2.9477131589$, and $q\approx 4.2138764783$.

There are 16 pairs of complex-conjugate T points at
$q \approx 0.1195227877 \pm 2.2972440367\,i$,     %% #6
$q \approx 0.8945659676 \pm 2.5558662562\,i$,     %% #7
$q \approx 1.0187320850 \pm 2.5519732169\,i$,     %% #8
$q \approx 1.7361257994 \pm 2.4200684634\,i$,     %% #9
$q \approx 2.0496846729 \pm 2.2109675624\,i$,     %% #5
$q \approx 2.0659399006 \pm 2.1928587989\,i$,     %% #14
$q \approx 2.1816875305 \pm 2.1274251306\,i$,     %% #15
$q \approx 2.5037573375 \pm 1.8512451004\,i$,     %% #10
$q \approx 2.7553322546 \pm 1.4607519460\,i$,     %% #11
$q \approx 2.9072366988 \pm 1.0781535615\,i$,     %% #16
$q \approx 2.9302627055 \pm 1.0265712687\,i$,     %% #2
$q \approx 2.9319763271 \pm 1.0161098821\,i$,     %% #3
$q \approx 2.9600871035 \pm 0.8562446926\,i$,     %% #1
$q \approx 3.0829802406 \pm 0.7932348764\,i$,     %% #13
$q \approx 3.0859998597 \pm 0.4641385413\,i$, and %% #4
$q \approx 3.0934600966 \pm 0.4747067803\,i$.     %% #12
There are two pairs of complex-conjugate endpoints at
$q \approx 1.8355583156 \pm 2.1088360158\,i$, and
$q \approx 2.8978250409 \pm 1.0665199267\,i$.

In this case we have been able to find four isolated limiting points at
$q=0,1,2,B_5$; but we cannot rule out the existence of other isolated 
limiting points (especially complex ones).
The limiting curve $\mathcal{B}_6$ and the chromatic roots for $n=60,120$ are
depicted in Figure~\ref{figure_sq_6F}.

%
% Comparison of 1 \le m \le 6
%
\subsection[Comparison of $1 \le m \le 6$]{Comparison of $\bm{1 \le m \le 6}$}
  \label{subsec.comparison}

\begin{figure}%%%[hbtp]
\centering
\begin{tabular}{cc}
\includegraphics[width=200pt]{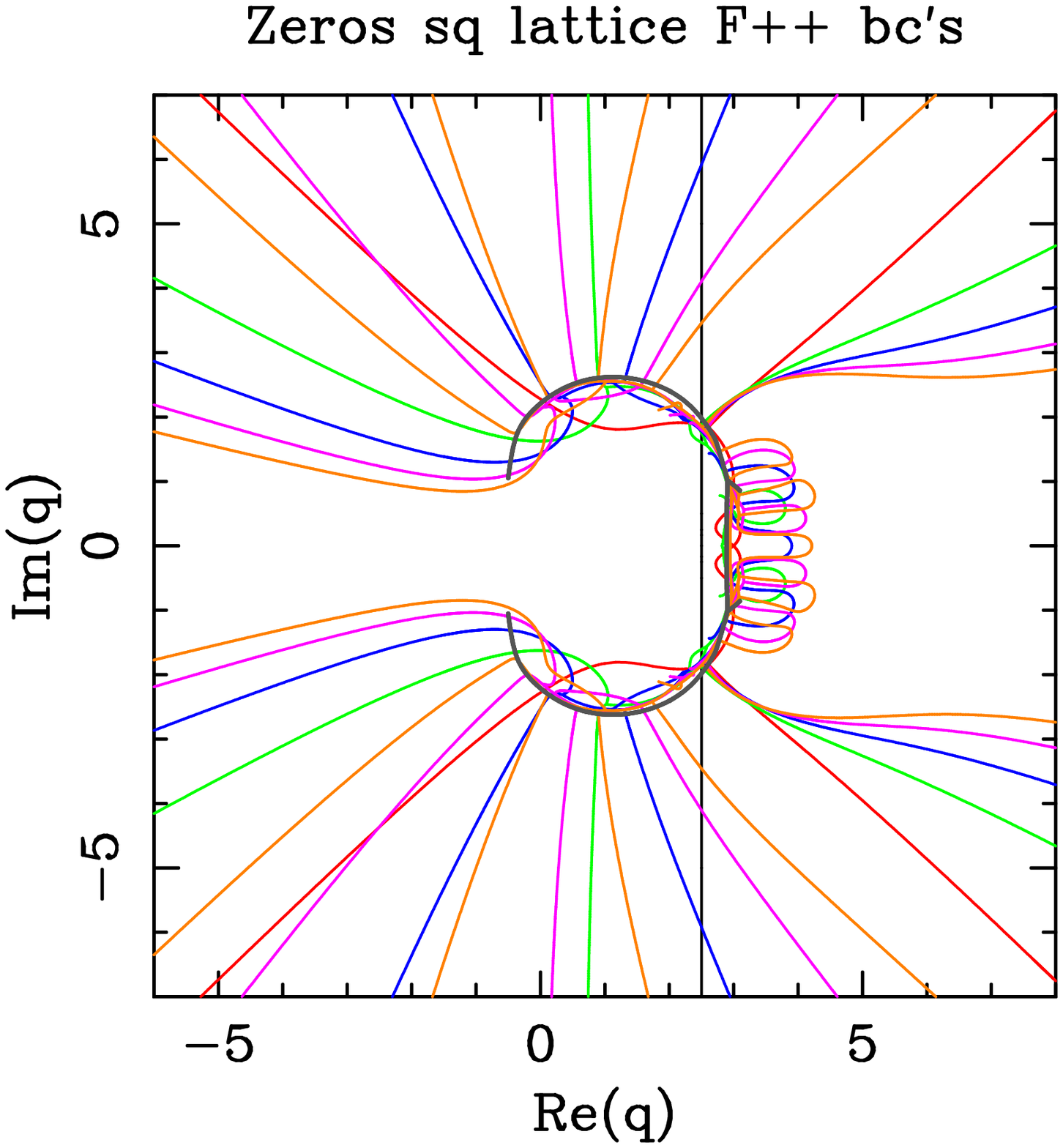} &
\includegraphics[width=200pt]{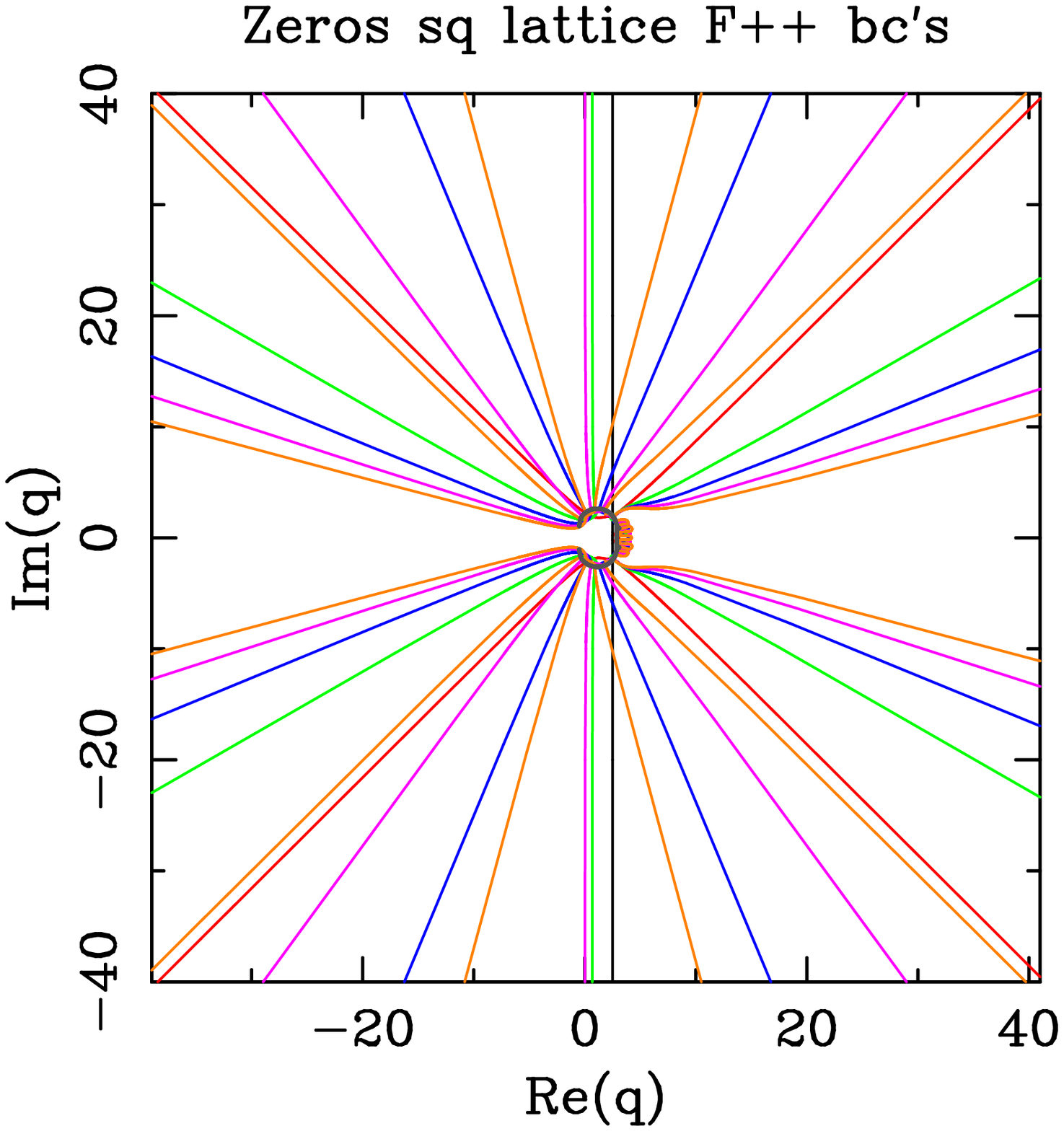} \\[1mm] 
\phantom{(((a)}(a) & \phantom{(((a)}(b) \\[5mm]
\end{tabular}
\caption{\label{figure_sq_allF}
Limiting curves $\mathcal{B}_m$ for square-lattice strips $S_{m,n}$ of widths
$m=1$ (black), $m=2$ (red), $m=3$ (green), $m=4$ (blue), $m=5$ (pink)
and $m=6$ (orange) with two extra sites. 
The solid dark gray curve corresponds to the limiting curve
$\scrb_{11}^{\rm cyl}$ for a strip of width $m=11$
with {\em cylindrical}\/ boundary conditions \protect\cite{transfer2}.
}
\end{figure}

In Figure~\ref{figure_sq_allF} we show the limiting curves $\scrb_m$
for $1 \le m \le 6$, plotted together.
We see clearly that:
\begin{itemize}
   \item[(a)]  There is an oval-shaped region at small $q$,
extending roughly from $q=0$ to $q=3$ in the real direction
and between $q \approx 1.1 \pm 2.6\,i$ in the imaginary direction,
where the limiting curves $\scrb_m$ do not enter,
or at least from which they retreat as $m\to\infty$.
   \item[(b)]  Outside of this region, the curves $\scrb_m$ appear to become
{\em dense}\/ as $m\to\infty$, except possibly near the real axis.
   \item[(c)]  The curve $\scrb_m$ also exhibits, for each $m \ge 3$,
$m-1$ small ``fingers'' extending outside the oval-shaped region
towards the right
(these ``fingers'' are more easily seen on
 Figures~\ref{figure_sq_3F}--\ref{figure_sq_6F}).
\end{itemize}

In Figure~\ref{figure_sq_allF} we have also superposed,
in dark gray, the limiting curve $\scrb_{11}^{\rm cyl}$
for a square-lattice strip of width $m=11$
and {\em cylindrical}\/ boundary conditions \cite{transfer2}.
This curve can be taken as a rough approximation of the expected
{\em infinite-volume}\/ limiting curve
$\scrb_\infty({\rm sq}) = \lim\limits_{m\to\infty} \scrb_m^{\rm cyl}$
for the square-lattice model with cylindrical boundary conditions,
with the exception that we expect the curves $\scrb_m^{\rm cyl}$
to move slightly outward as $m$ grows and to close up at $q=0$;
in particular we expect $\scrb_\infty({\rm sq})$
to be a {\em closed}\/ curve that crosses
the real $q$-axis at $q=0$ and $q_0=q_c=3$ \cite{transfer1,transfer2}.
We expect, furthermore, that the same limiting curve will be obtained
for free boundary conditions \cite{transfer1,transfer2}.

Figure~\ref{figure_sq_allF} clearly suggests that the
oval-shaped curve to which the curves $\scrb_m$ are retreating
as $m\to\infty$ is the {\em same}\/ as the curve $\scrb_\infty({\rm sq})$
obtained for cylindrical (or free) boundary conditions.
The region where the curves $\scrb_m$ are apparently becoming dense
is precisely the exterior of $\scrb_\infty({\rm sq})$.

%
% Limiting curves
%
\section{Conjectures on the limiting curves $\bm{\mathcal{B}_m}$
    as $\bm{m\to\infty}$} 
\label{sec.thermo.asym}

In this section we discuss in more detail the behavior at large $|q|$
of the limiting curves $\scrb_m$;
in particular, we formulate conjectures concerning
the behavior of these limiting curves as $m\to\infty$.

\subsection[Behavior at large $|q|$ for each $m$]%
{Behavior at large $\bm{|q|}$ for each $\bm{m}$}

\begin{figure}
\centering
\begin{tabular}{cc}
\includegraphics[width=200pt]{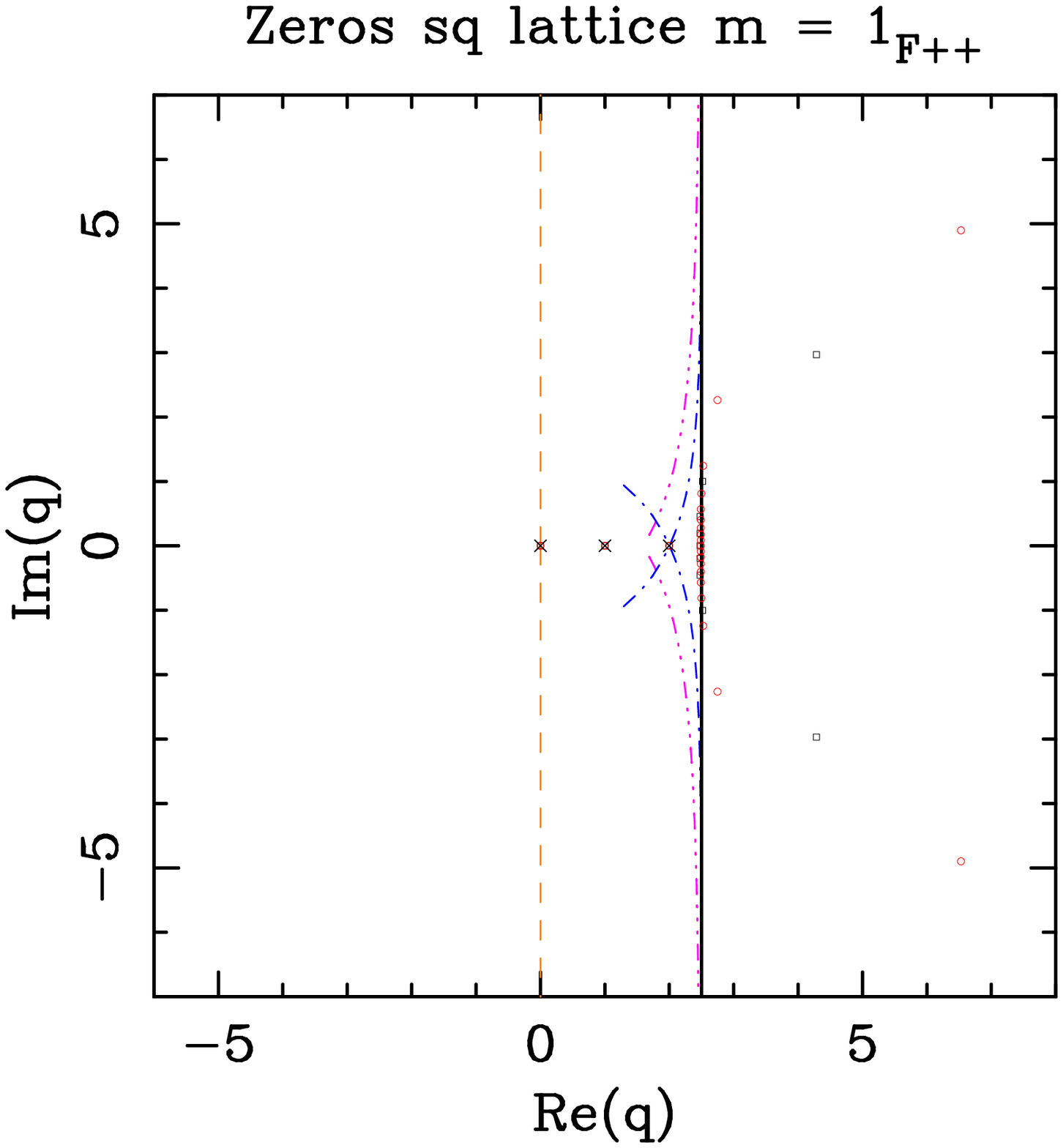} & 
\includegraphics[width=200pt]{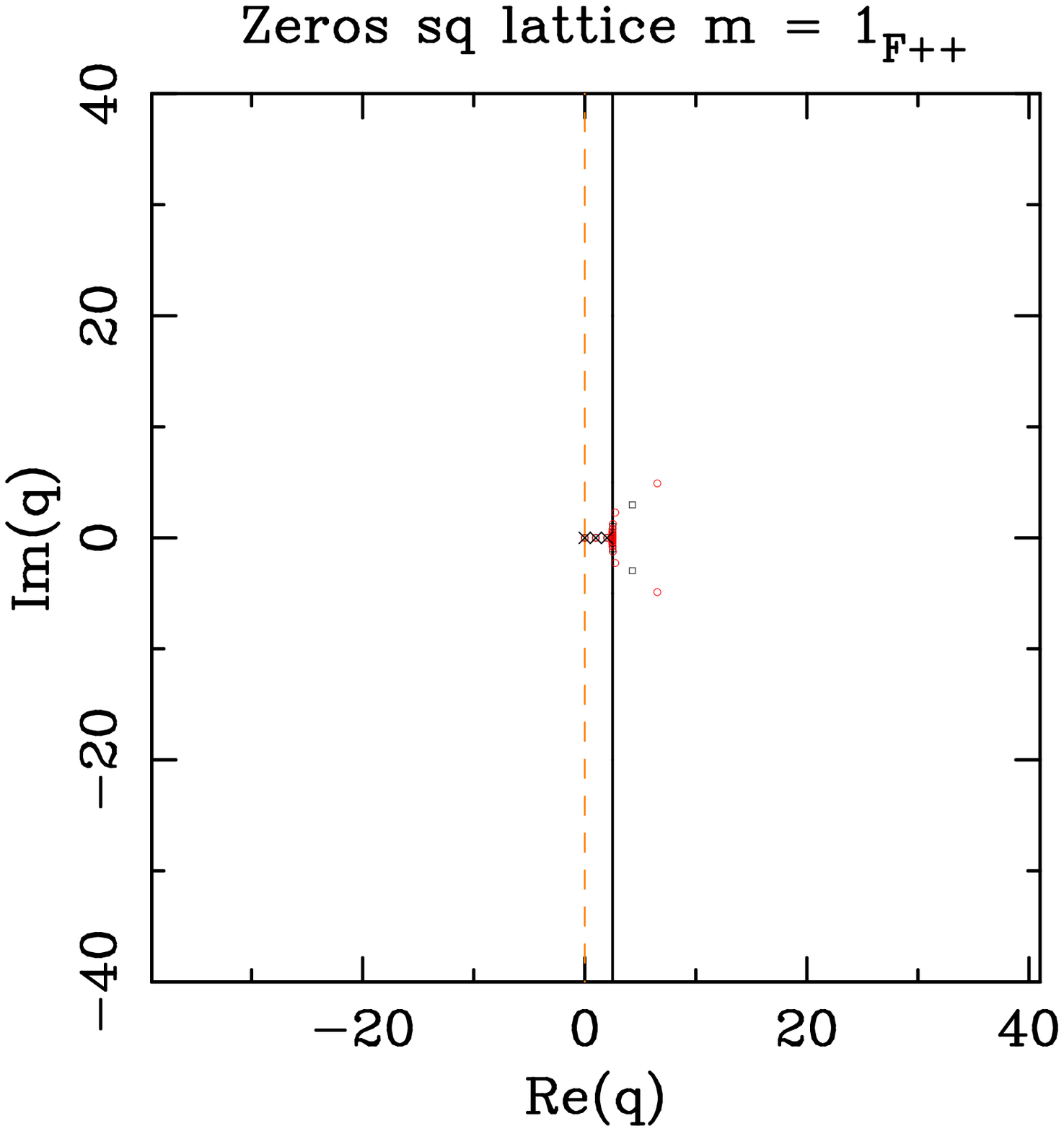} \\[1mm] 
\phantom{(((a)}(a) & \phantom{(((a)}(b) \\[5mm]
\end{tabular}
\caption{\label{figure_sq_1F}
Limiting curves for square-lattice strips of width $m=1$ with two extra sites. 
We also show the zeros for the strips $S_{1,10}$ 
(black $\square$) and $S_{1,20}$ (red $\circ$).
We depict the isolated limiting points with the symbol $\times$.
The zeroth-, first-, second- and third-order asymptotic expansions
for the outward branches [cf.\ \reff{eq.argq.asymp}]
are depicted as
dashed orange, dotted red, dot-dashed magenta, and dot-dot-dot-dashed blue
curves, respectively.
Please note that for $m=1$ the first- and second-order asymptotics coincide,
as the term of order $q^{-2}$ has the factor $\sin[(2k-1)\pi]=0$.
}
\end{figure}

\begin{figure}
\centering
\begin{tabular}{cc}
\includegraphics[width=200pt]{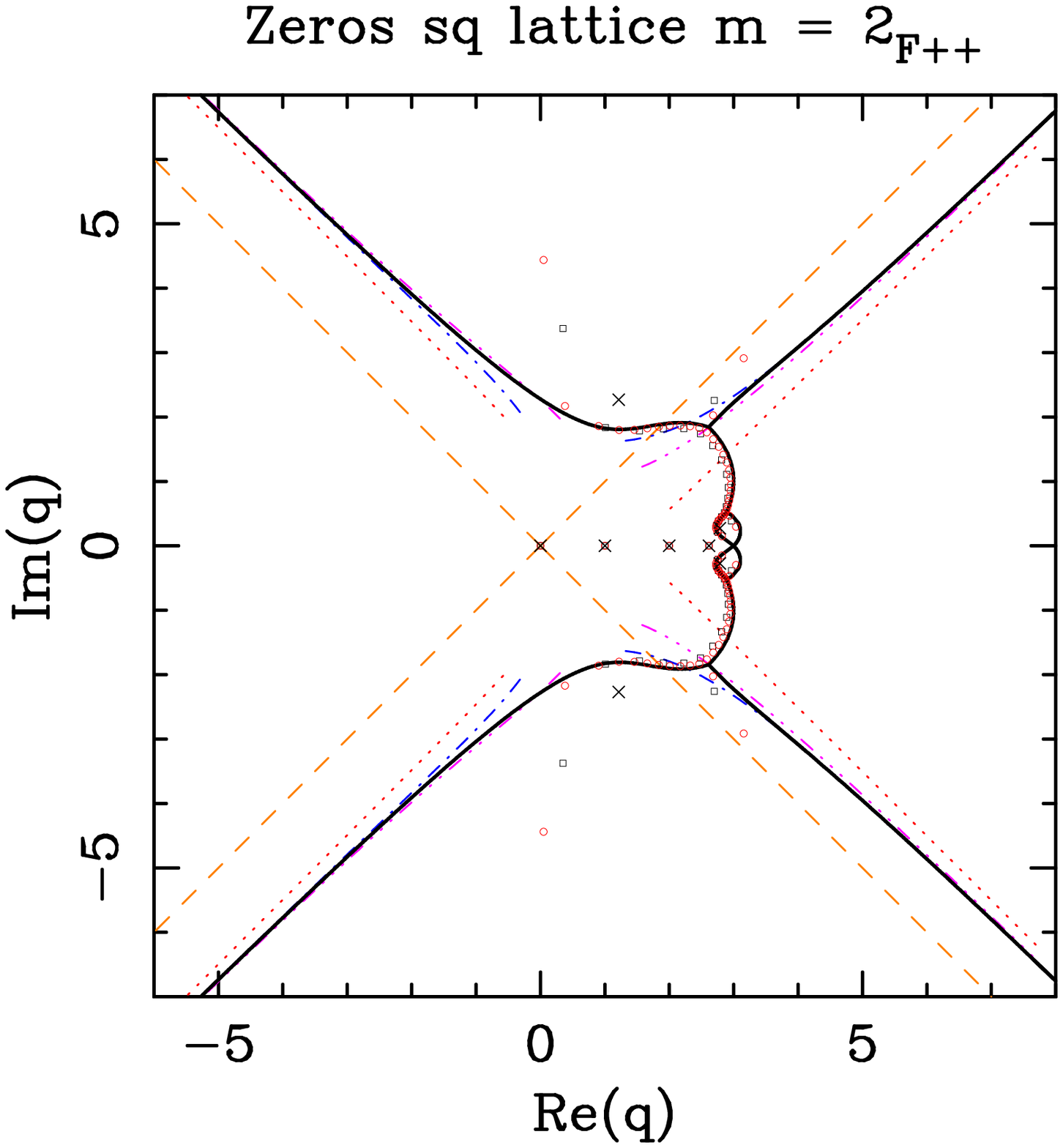} & 
\includegraphics[width=200pt]{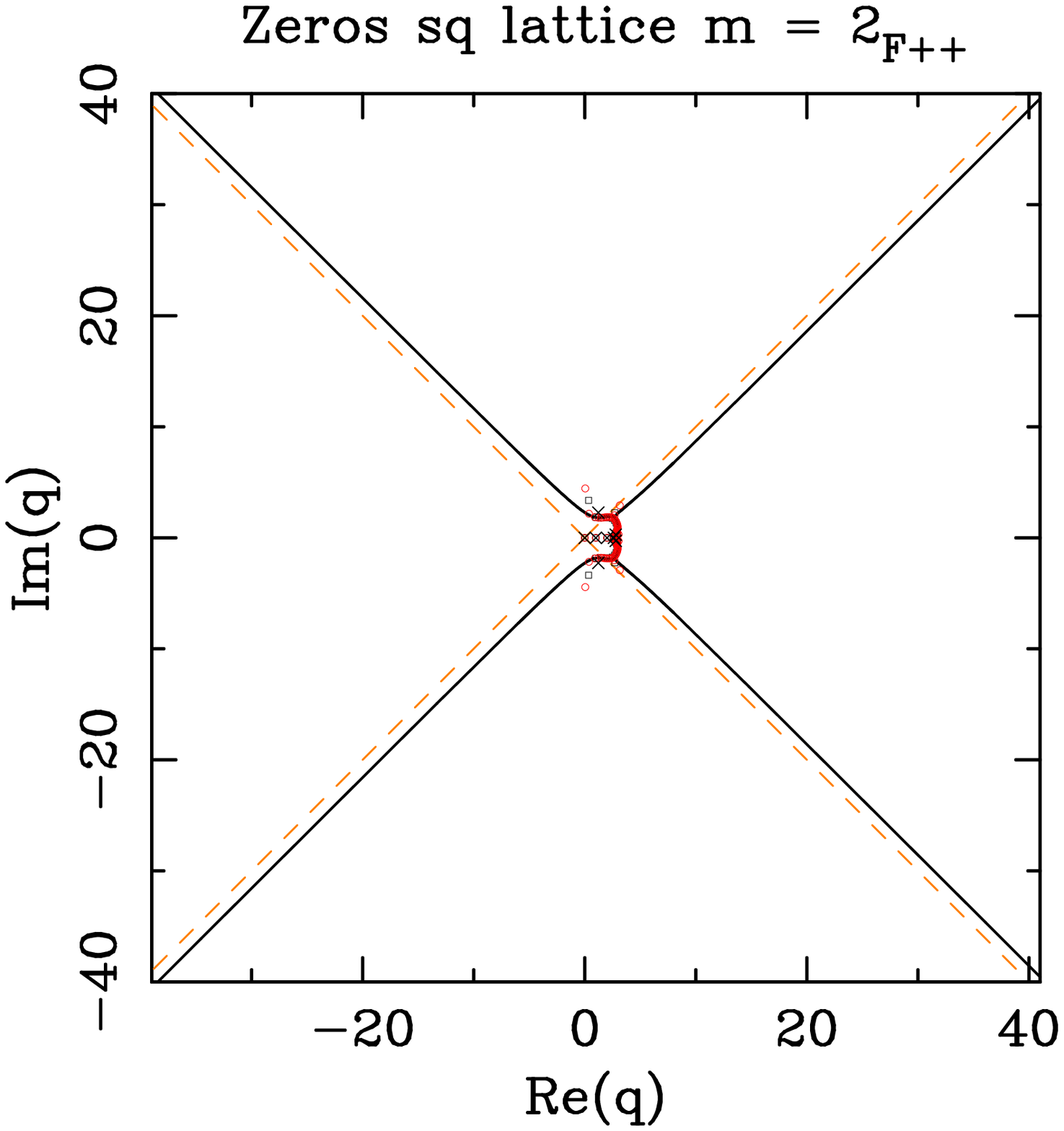} \\[1mm] 
\phantom{(((a)}(a) & \phantom{(((a)}(b) \\[5mm]
\end{tabular}
\caption{\label{figure_sq_2F}
Limiting curves for square-lattice strips of width $m=2$ with two extra sites.  We also show the zeros for the strips $S_{2,20}$ 
(black $\square$) and $S_{2,40}$ (red $\circ$).
We depict the isolated limiting points with the symbol $\times$.
The zeroth-, first-, second- and third-order asymptotic expansions
for the outward branches [cf.\ \reff{eq.argq.asymp}]
are depicted as
dashed orange, dotted red, dot-dashed magenta, and dot-dot-dot-dashed blue 
curves, respectively.
}
\end{figure}

\begin{figure}
\centering
\begin{tabular}{cc}
\includegraphics[width=200pt]{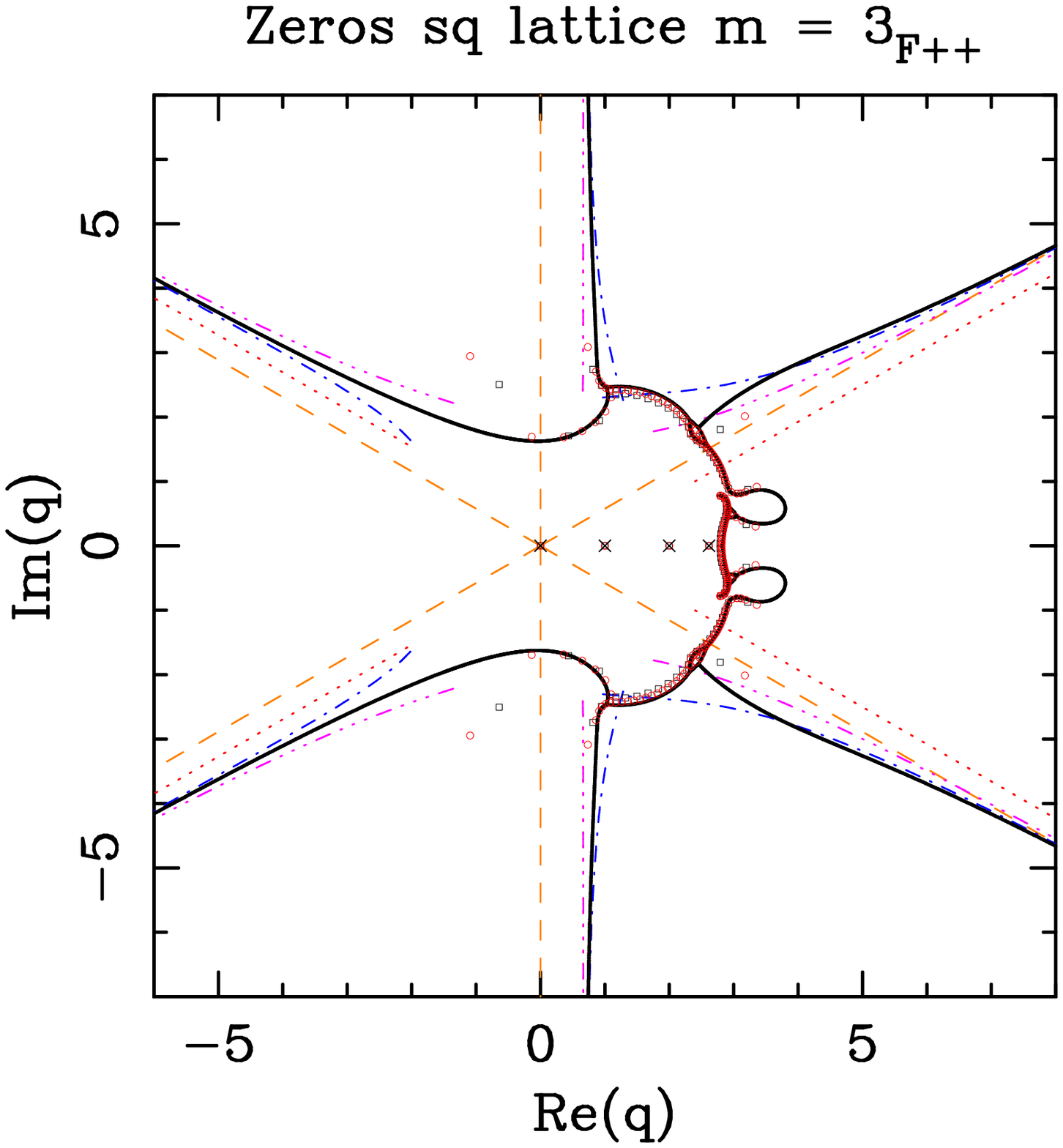} & 
\includegraphics[width=200pt]{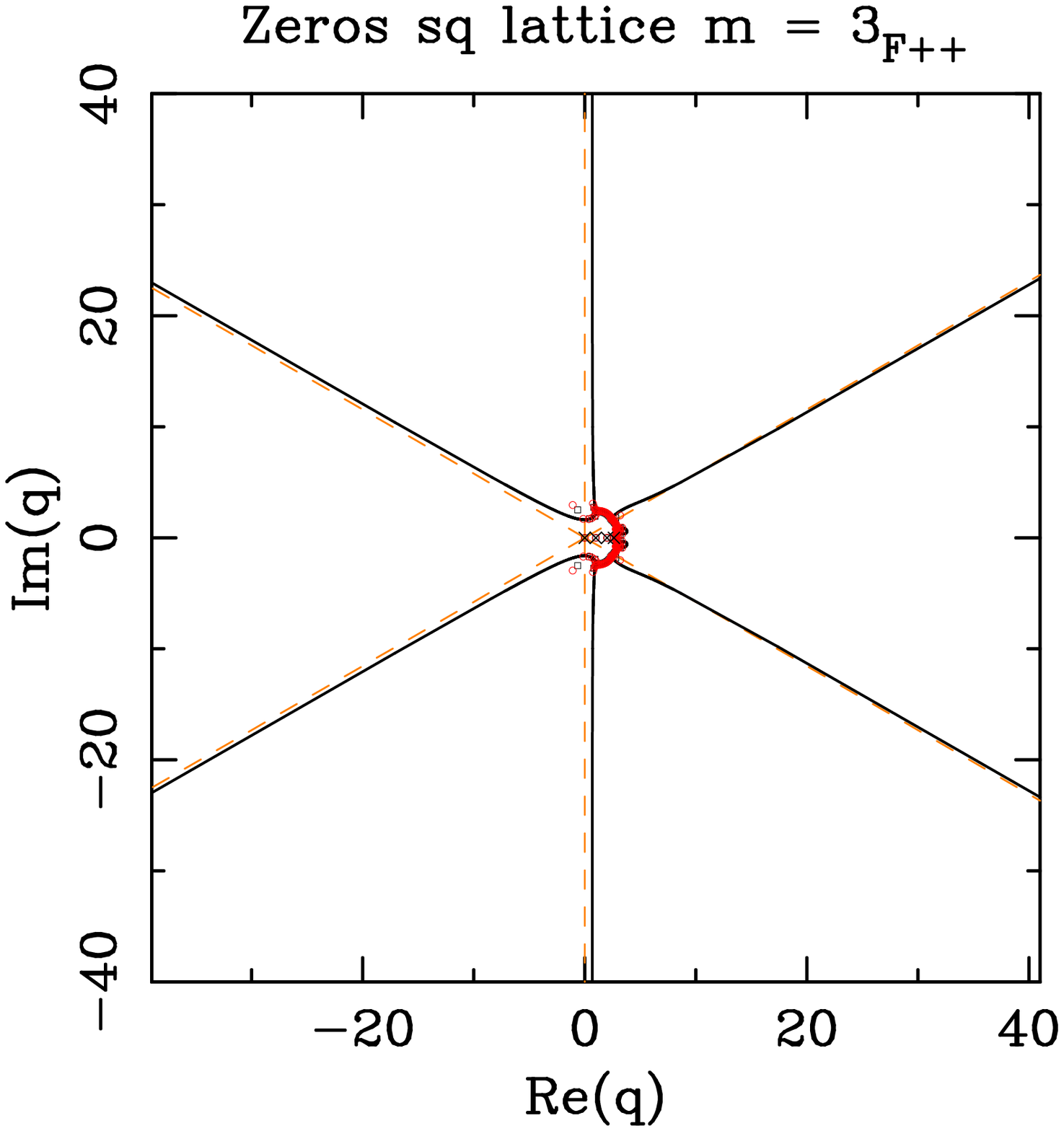} \\[1mm] 
\phantom{(((a)}(a) & \phantom{(((a)}(b) \\[5mm]
\end{tabular}
\caption{\label{figure_sq_3F}
Limiting curves for square-lattice strips of width $m=3$ with two extra sites. 
We also show the zeros for the strips $S_{3,30}$ 
(black $\square$) and $S_{3,60}$ (red $\circ$).
We depict the isolated limiting points with the symbol $\times$.
The zeroth-, first-, second- and third-order asymptotic expansions
for the outward branches [cf.\ \reff{eq.argq.asymp}]
are depicted as
dashed orange, dotted red, dot-dashed magenta, and dot-dot-dot-dashed blue 
curves, respectively.
Please note that for $m=3$ and $k=2,5$,
the first- and second-order asymptotics coincide,
as the term of order $q^{-2}$ has the factor $\sin[(2k-1)\pi/3]=0$,
which vanishes for $k=2,5$.
}
\end{figure}

\begin{figure}
\centering
\begin{tabular}{cc}
\includegraphics[width=200pt]{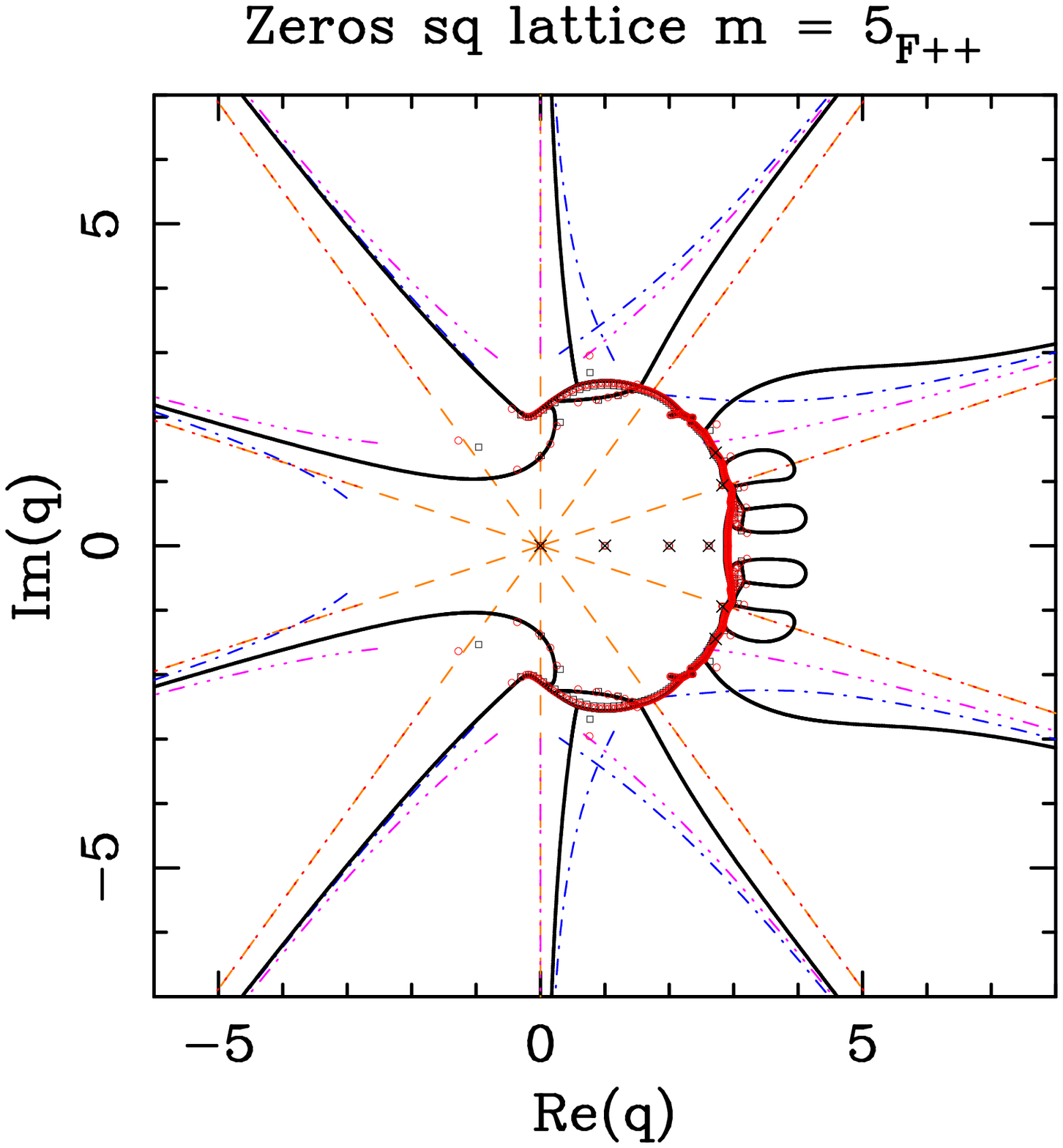} & 
\includegraphics[width=200pt]{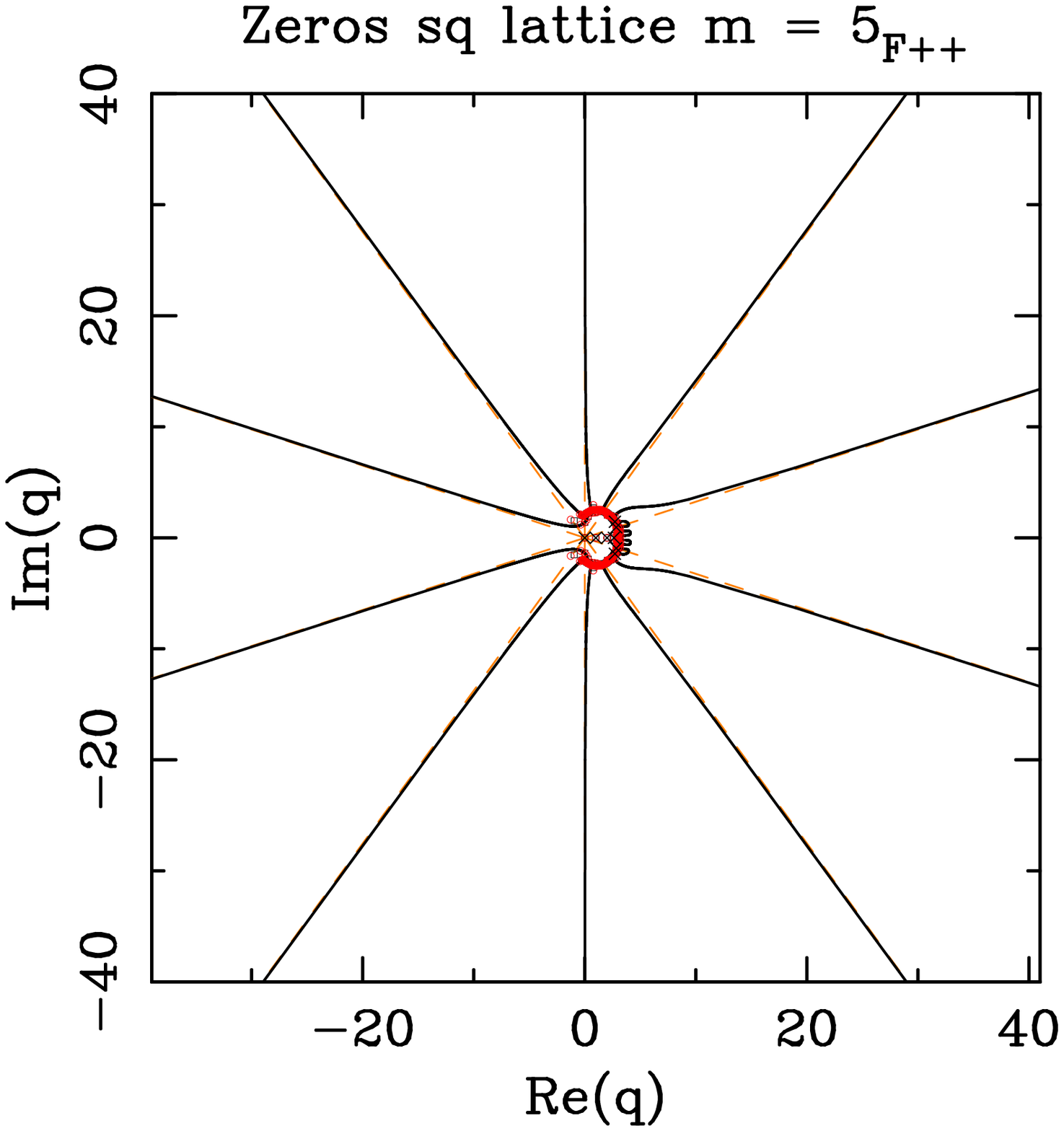} \\[1mm] 
\phantom{(((a)}(a) & \phantom{(((a)}(b) \\[5mm]
\end{tabular}
\caption{\label{figure_sq_5F}
Limiting curves for square-lattice strips of width $m=5$ with two extra sites. 
We also show the zeros for the strips $S_{5,50}$ 
(black $\square$) and $S_{5,100}$ (red $\circ$).
We depict the isolated limiting points with the symbol $\times$.
The zeroth-, first-, second- and third-order asymptotic expansions
for the outward branches [cf.\ \reff{eq.argq.asymp}]
are depicted as
dashed orange, dotted red, dot-dashed magenta, and dot-dot-dot-dashed blue
curves, respectively.
Please note that for $m=5$ the zeroth- and first-order asymptotics coincide,
as the term of order $q^{-1}$ has the factor $f_1(5)=0$. 
}
\end{figure}

\begin{figure}
\centering
\begin{tabular}{cc}
\includegraphics[width=200pt]{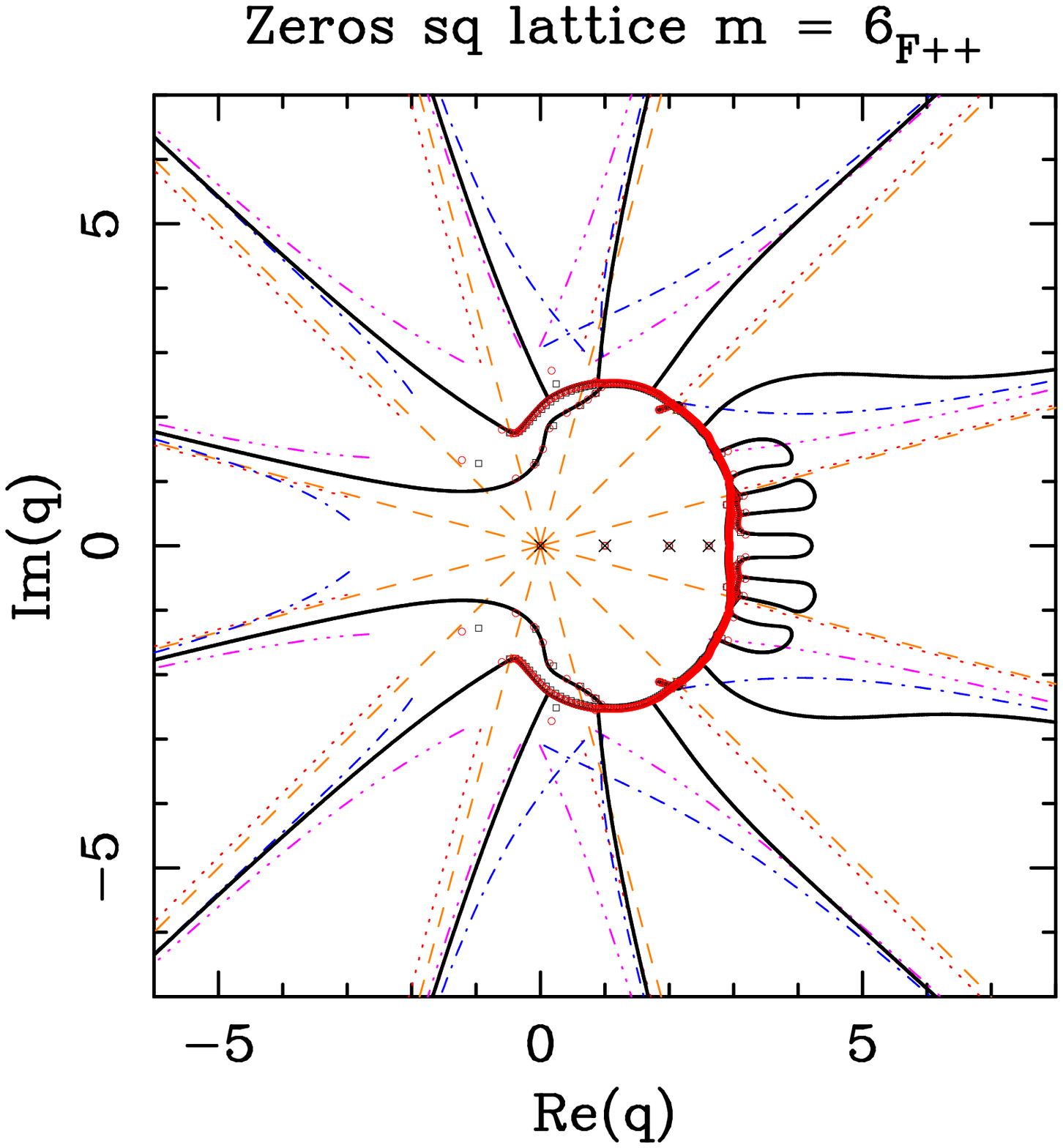} & 
\includegraphics[width=200pt]{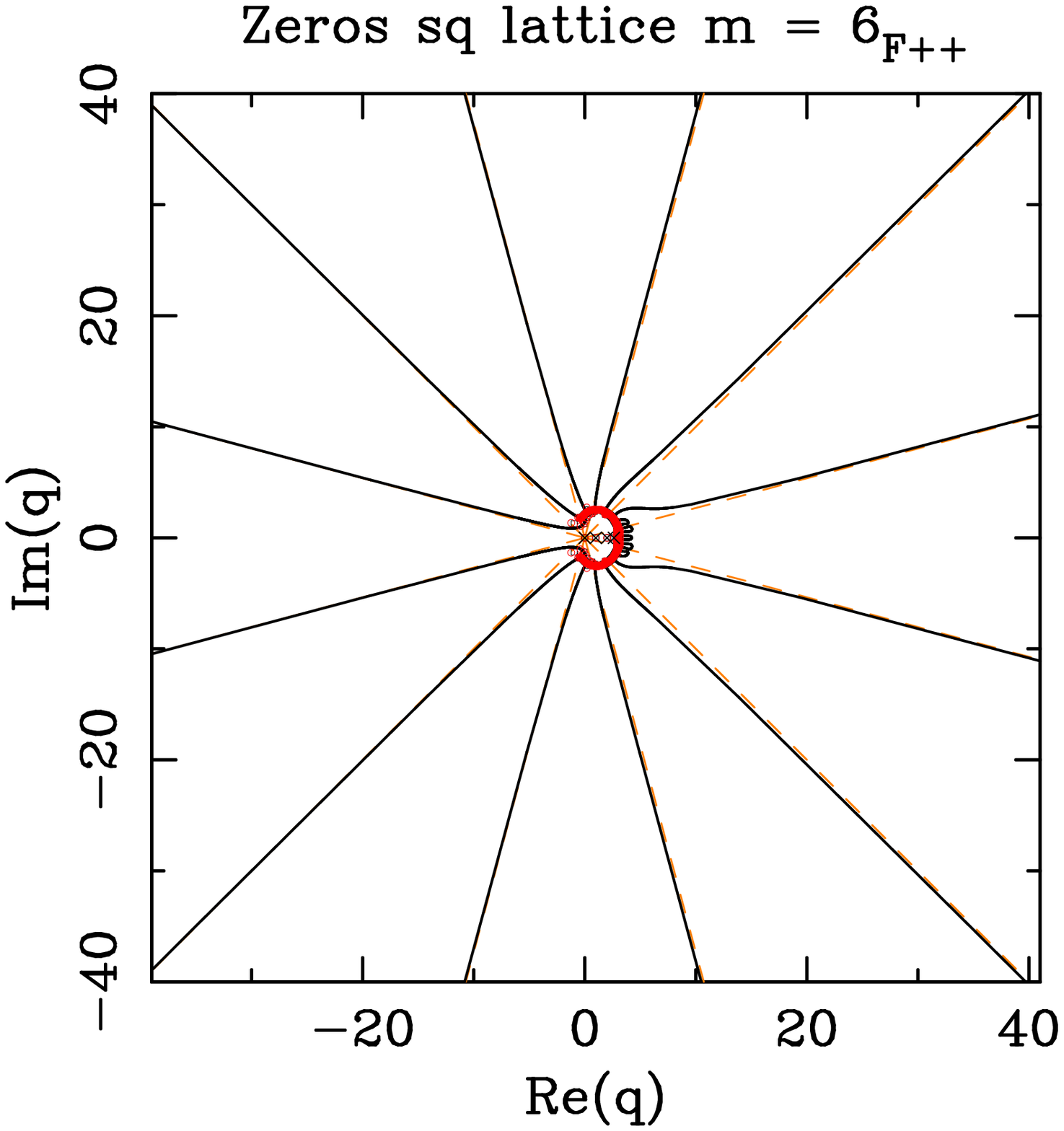} \\[1mm] 
\phantom{(((a)}(a) & \phantom{(((a)}(b) \\[5mm]
\end{tabular}
\caption{\label{figure_sq_6F}
Limiting curves for square-lattice strips of width $m=6$ with two extra sites. 
We also show the zeros for the strips $S_{6,60}$ 
(black $\square$) and $S_{6,120}$ (red $\circ$).
We depict the isolated limiting points with the symbol $\times$.
The zeroth-, first-, second- and third-order asymptotic expansions
for the outward branches [cf.\ \reff{eq.argq.asymp}]
are depicted as
dashed orange, dotted red, dot-dashed magenta, and dot-dot-dot-dashed blue
curves, respectively.
}
\end{figure}

Figures~\ref{figure_sq_1F}--\ref{figure_sq_6F} show the limiting curves
$\scrb_m$ for $1\leq m \leq 6$:
in each case we give both a ``close-up'' view 
that shows the details of the small-$|q|$ behavior
and a ``distance'' view that makes clear the large-$|q|$ asymptotics.
We notice, first of all, that the curve $\scrb_m$ has
$2m$ outward branches tending to $q=\infty$,
with asymptotic (as $|q| \to \infty$) angles
that are equally spaced around the circle.
These empirical findings for $1 \le m \le 6$ motivate the following 
result:

\medskip

\begin{proposition} \label{prop1}
Fix $m \ge 1$.
Then the limiting curve $\mathcal{B}_m$ for chromatic roots of the
square-lattice strips $S_{m,n}$ ($n \to\infty$)
has exactly $2m$ outward branches extending to infinity,
with asymptotic angles $\arg q = \theta_{k,m}$ where
\be
 \theta_{k,m} \;=\; \biggl( k - \frac{1}{2} \biggr) \frac{\pi}{m}
 \qquad\hbox{for } k=1,\ldots,2m  \;.
 \label{eq.conj1.1}
\ee
Moreover, the
dominant eigenvalue comes from $\T''_{=}(m)$ in the asymptotic regions
\be
    \theta_{k,m} < \arg q  < \theta_{k+1,m}
         \quad\hbox{for}\quad
         \left\{ \begin{array}{ll}
          k=1,3,\ldots,2m-1 & \quad \mbox{\rm if $m$ is even} \\
          k=2,4,\ldots,2m   & \quad \mbox{\rm if $m$ is odd}
         \end{array}\right.
\ee
while in the other asymptotic sectors the dominant eigenvalue comes from
$\T''_{\neq}(m)$.
\end{proposition}

\proof 
By the Beraha--Kahane--Weiss theorem (Theorem~\ref{BKW_thm}),
the limiting curve $\scrb_m$ consists of those $q$ for which
\be
  \left| {\lambda_{\star,=}(m) \over \lambda_{\star,\neq}(m) } \right|
  \;=\;
  1  \;,
\ee
or equivalently
\be
   \log\left( {\lambda_{\star,=}(m) \over
               \lambda_{\star,\neq}(m)
              }
       \right)
  \;=\;
  i\theta
  \qquad\hbox{with $\theta$ real}  \;,
\ee
or equivalently
\be
   \log\left[
   (-1)^{m+1} \log\Bigl( {\lambda_{\star,=}(m) \over
                              \lambda_{\star,\neq}(m)
                             }
                      \Bigr)
   \right]
   \;=\;
   -\, (2k-1) \frac{\pi}{2}
   \qquad\hbox{with $k$ integer}
 \label{eq.imloglog_k}
\ee
[the factor $(-1)^{m+1}$ is included here solely for future convenience].
If we now define
\be
   \scrf_m(q)
   \;=\;
   - \log\left[
   (-1)^{m+1} q^m \log\Bigl( {\lambda_{\star,=}(m) \over
                              \lambda_{\star,\neq}(m)
                             }
                      \Bigr)
   \right]
 \label{def.Fmq}
\ee
[compare \reff{def_series_fl}],
then \reff{eq.imloglog_k} can be rewritten as
\be
   \arg q \;=\; (2k-1) \frac{\pi}{2m} \,-\, \imag \frac{\scrf_m(q)}{m}
   \;.
 \label{eq.argq}
\ee
On the other hand, Proposition~\ref{prop.diff.eigen.0}
together with \reff{eq.bkm.akm.k0} tells us that
\be
   \scrf_m(q)  \;=\;  O(q^{-1})  \;,
\ee
so that the curve $\scrb_m$ is given by
\be
   \arg q \;=\; (2k-1) \frac{\pi}{2m} \,+\, O(|q|^{-1})
   \;.
 \label{eq.argq0}
\ee
This proves the first statement of Proposition~\ref{prop1}.

The second statement of Proposition~\ref{prop1} follows easily from
\be
  {\lambda_{\star,=}(m) \over \lambda_{\star,\neq}(m) }
  \;=\;
  1 \:+\: (-1)^{m+1} q^{-m}  \:+\: O(q^{-(m+1)})
  \;,
\ee
which implies
\be
  \left| {\lambda_{\star,=}(m) \over \lambda_{\star,\neq}(m) } \right|
  \;=\;
  1 \:+\: (-1)^{m+1} \real(q^{-m})  \:+\: O(q^{-(m+1)})
  \;.
\ee
\qed

{\bf Remark.}
This proof of Proposition~\ref{prop1} actually uses
a bit less than the full strength of Proposition~\ref{prop.diff.eigen.0}.
More specifically,
the first statement of Proposition~\ref{prop1} follows from
\be
   \lambda_{\star,=}(m) - \lambda_{\star,\neq}(m)
   \;=\;
   c_m \,+\, o(1)
   \qquad\hbox{with $c_m$ real $\neq 0$}   \;,
\ee
while the second statement uses $\sgn(c_m) = (-1)^{m+1}$.
In neither case do we need the exact value $c_m = (-1)^{m+1}$,
nor do we need to know that the error $o(1)$ is actually $O(q^{-1})$.
\qed

Using Proposition~\ref{prop.diff.eigen.0} we can actually go farther,
and compute the corrections to $\arg q = \theta_{k,m}$
as a series in inverse powers of $|q|$.
It suffices to use the fundamental relation \reff{eq.argq}
and insert the expansion
\be
   \scrf_m(q)
   \;=\;
   \sum_{\ell=1}^\infty f_\ell(m) \, q^{-\ell}
 \label{eq.Fmq.expansion}
\ee
[cf.\ \reff{def_series_fl}],
which is guaranteed by Corollary~\ref{cor.TMblockdiag.1and2}
together with Proposition~\ref{prop.diff.eigen.0}
to be convergent for all sufficiently large $|q|$
(how large may depend on $m$).
It is then straightforward to determine iteratively
$\arg q$ as a power series in $|q|^{-1}$.
For instance, the series through order $|q|^{-3}$ is given by
\begin{eqnarray}
   \arg q
   & = &
   \theta_{k,m}  \:+\:
   \left( \frac{f_1(m)}{m} \sin \theta_{k,m} \right) |q|^{-1}
   \:+\:
   \left[  \left( \frac{f_1(m)^2}{2m^2} +  \frac{f_2(m)}{m} \right)
           \sin 2\theta_{k,m}
   \right]  |q|^{-2}
       \nonumber \\
   & &
   \;+\:
   \left[  \left( \frac{f_1(m)^3}{8m^3} +  \frac{f_1(m) f_2(m)}{2m^2} \right)
           ( 3 \sin 3\theta_{k,m} \,-\,  \sin \theta_{k,m})
           \:+\:
            \frac{f_3(m)}{m} \sin 3\theta_{k,m}
   \right]  |q|^{-3}
       \nonumber \\[2mm]
   & &
   \;+\: O(|q|^{-4})  \;.
 \label{eq.argq.asymp}
\end{eqnarray}
In particular, for $m \le 11$ we know explicitly the coefficients $f_\ell(m)$.

If we assume Conjecture~\ref{conj.diff.eigen2},
then we can replace $f_\ell(m)$ by the polynomial $F_\ell(m)$
whenever $m \ge \ell + 1$,
and we can moreover use the specific forms of $F_\ell(m)$
shown in Table~\ref{table_new_Ck} at least for $\ell \le 9$.

In Figures~\ref{figure_sq_1F}--\ref{figure_sq_6F}
we compare the exact limiting curves $\scrb_m$
with the approximations \reff{eq.argq.asymp}
through orders $q^0$, $q^{-1}$, $q^{-2}$ and $q^{-3}$.
We have checked numerically the correctness of the asymptotic expansions
\reff{eq.argq.asymp},
by plotting the deviations of $\arg q$ from \reff{eq.argq.asymp}
truncated at order $|q|^{-\ell}$,
multiplied by $|q|^{\ell+1}$, versus $|q|$,
and verifying that they are tending to constants
(or at least bounded) as $|q| \to\infty$.

\subsection[Uniformity in $m$ and density of chromatic roots]%
{Uniformity in $\bm{m}$ and density of chromatic roots}

Since the asymptotic rays $\arg q = \theta_{k,m}$
arising in Proposition~\ref{prop1} become {\em dense}\/ as $m \to\infty$,
it is natural to expect that the chromatic roots of the graphs $S_{m,n}$
become dense as $m,n \to\infty$ in the whole complex plane
outside of a bounded set (analogously to what happens for the
generalized theta graphs \cite{Sokal_chromatic_roots}).
We can formalize this conjecture as follows:

\begin{conjecture} \label{conj.dense}
There exists a constant $Q < \infty$
such that the chromatic roots of the graphs $S_{m,n}$
become dense in the region $|q| > Q$ when $m,n \to\infty$.
\end{conjecture}

We stress that Conjecture~\ref{conj.dense}
does {\em not}\/ follow from Proposition~\ref{prop1} alone
--- or from the more fundamental Proposition~\ref{prop.diff.eigen.0}
and Conjectures~\ref{conj.diff.eigen}--\ref{conj.diff.eigen2} ---
because these results apply to each value of $m$ separately,
and there is no guarantee that the error bounds in these asymptotic
expansions are uniform in $m$;
in particular, there is no guarantee that they can be made to apply
to a region $|q| > Q$ where $Q$ is {\em independent of $m$}\/.
In this subsection we would like to discuss a refinement of
Proposition~\ref{prop.diff.eigen.0}
from which Conjecture~\ref{conj.dense} can indeed be deduced. 

We know rigorously that there exists $Q_0 < \infty$, independent of $m$,
such that $\lambda_{\star,=}(m)$ and $\lambda_{\star,\neq}(m)$
are analytic and nonvanishing in $|q| > Q_0$ for all $m \ge 1$
(see Theorem~\ref{thm.sokal3} and the discussion following it,
 where it is shown that $Q_0 = 33.855628$ suffices).
If we now use Proposition~\ref{prop.diff.eigen.0},
we can conclude that the functions $\scrf_m(q)$ defined in \reff{def.Fmq}
are likewise analytic in $|q| > Q_0$ for all $m \ge 1$
and moreover vanishing at $q=\infty$,
so that they are given for $|q| > Q_0$ by a convergent expansion
\reff{eq.Fmq.expansion}.
Let us now make the following hypothesis:

\begin{conjecture} \label{conj.uniform}
There exist $Q_1 \in (Q_0,\infty)$ and $C < \infty$ such that
\be
   \sup_{|q|=Q_1} |\scrf_m(q)|
   \;\le\; Cm  \qquad\hbox{for all $m$}  \;.
 \label{eq.conj.uniform}
\ee
Note that here $Q_1$ and $C$ are {\em independent of $m$}\/.
\end{conjecture}

This hypothesis is natural in view of Conjecture~\ref{conj.diff.eigen2},
which says that the coefficients $f_\ell(m)$ in \reff{eq.Fmq.expansion}
are given for $m \ge \ell+1$
by polynomials $F_\ell(m) = \epsilon_\ell m + \phi_\ell$ of degree 1 in $m$;
so it is natural to expect that $\scrf_m(q)$ will likewise
satisfy a bound that is linear in $m$.
The trouble is that the coefficients $f_\ell(m)$ for $\ell \ge m$
are completely uncontrolled, so we cannot deduce
Conjecture~\ref{conj.uniform} from Conjecture~\ref{conj.diff.eigen2}.
It is, nevertheless, a natural extension of Conjecture~\ref{conj.diff.eigen2}.

Let us now prove that Conjecture~\ref{conj.uniform}
implies Conjecture~\ref{conj.dense}:

\medskip

{\sc Proof of Conjecture~\ref{conj.dense},
   given Conjecture~\ref{conj.uniform}.}
Using the hypothesis \reff{eq.conj.uniform}
together with the Cauchy integral formula in the variable $z=1/q$,
we conclude that
\be
   |\scrf'_m(q)| \;\le\; 4Cm {Q_1 \over |q|^2}
   \qquad\hbox{whenever $|q| \ge 2Q_1$}
   \;.
 \label{eq.proof.conj.dense.star1}
\ee  
We now use the fundamental formula \reff{eq.argq}
determining the curve $\scrb_m$,
and we consider $\imag \scrf_m(q)/m$ as a function of $\arg q$
for each fixed $|q|$:
\be
   f_m(\arg q ; |q|)  \;=\; \imag { \scrf_m(q) \over m}  \;.
\ee
It follows from \reff{eq.proof.conj.dense.star1} that
\be
   \Bigl| f'_m(\arg q ; |q|) \Bigr|
   \;\le\; 4C {Q_1 \over |q|}
   \qquad\hbox{whenever $|q| \ge 2Q_1$}
\ee
and hence that
\be
   \Bigl| f'_m(\arg q ; |q|) \Bigr|
   \;\le\; \half
   \qquad\hbox{whenever $|q| \ge Q_2 \equiv \max(8CQ_1, 2Q_1)$}
   \;.
\ee
Now \reff{eq.argq} amounts to solving the equation
\be
   h_m(\arg q)  \;=\;  y
 \label{eq.proof.conj.dense.star2}
\ee
for a set of values of $y$ spaced by $\pi/m$, where
\be
   h_m(x)  \;=\;  x \,+\, f_m(x; |q|)
\ee
is a continuously differentiable (in fact infinitely differentiable)
function satisfying $h'_m(x) \ge 1/2$
(and also $h'_m(x) \le 3/2$, but we do not need this)
whenever $|q| \ge Q_2$.
It follows that the spacing between consecutive solutions
of \reff{eq.proof.conj.dense.star2} cannot exceed $2\pi/m$.
Taking $m \to\infty$, we conclude that the union of the curves $\scrb_m$
is dense in the region $|q| \ge Q_2$.
It easily follows from this and the Beraha--Kahane--Weiss theorem
that the union of the chromatic roots of $S_{m,n}$ as $m,n \to\infty$
is dense in the region $|q| \ge Q_2$ as well.
\qed

\subsection{Size of the discontinuity in the derivative of the free energy}

It is curious that the branches at large $|q|$
of the curves $\scrb_m$ are caused by {\em very small}\/
differences between the two eigenvalues ---
namely, $|\lambda_{\star,=} - \lambda_{\star,\neq}| \approx 1$
compared to
$|\lambda_{\star,=}| \approx |\lambda_{\star,\neq}| \approx |q|^m$
---
which moreover become {\em irrelevant}\/ in the limit $m \to \infty$.

\begin{figure}%%%[hbtp]
\centering
\begin{tabular}{cc}
\includegraphics[width=200pt]{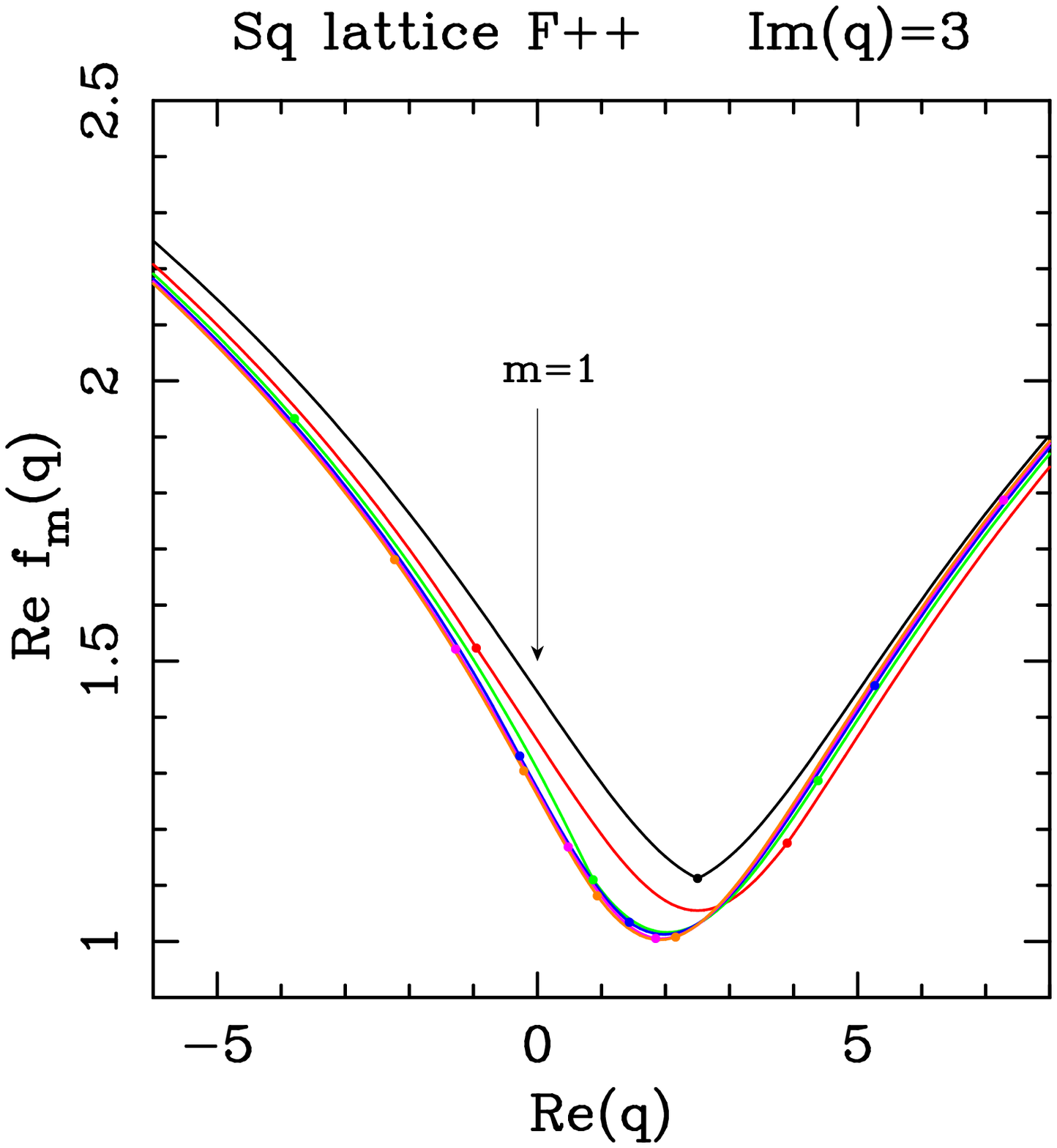} & 
\includegraphics[width=200pt]{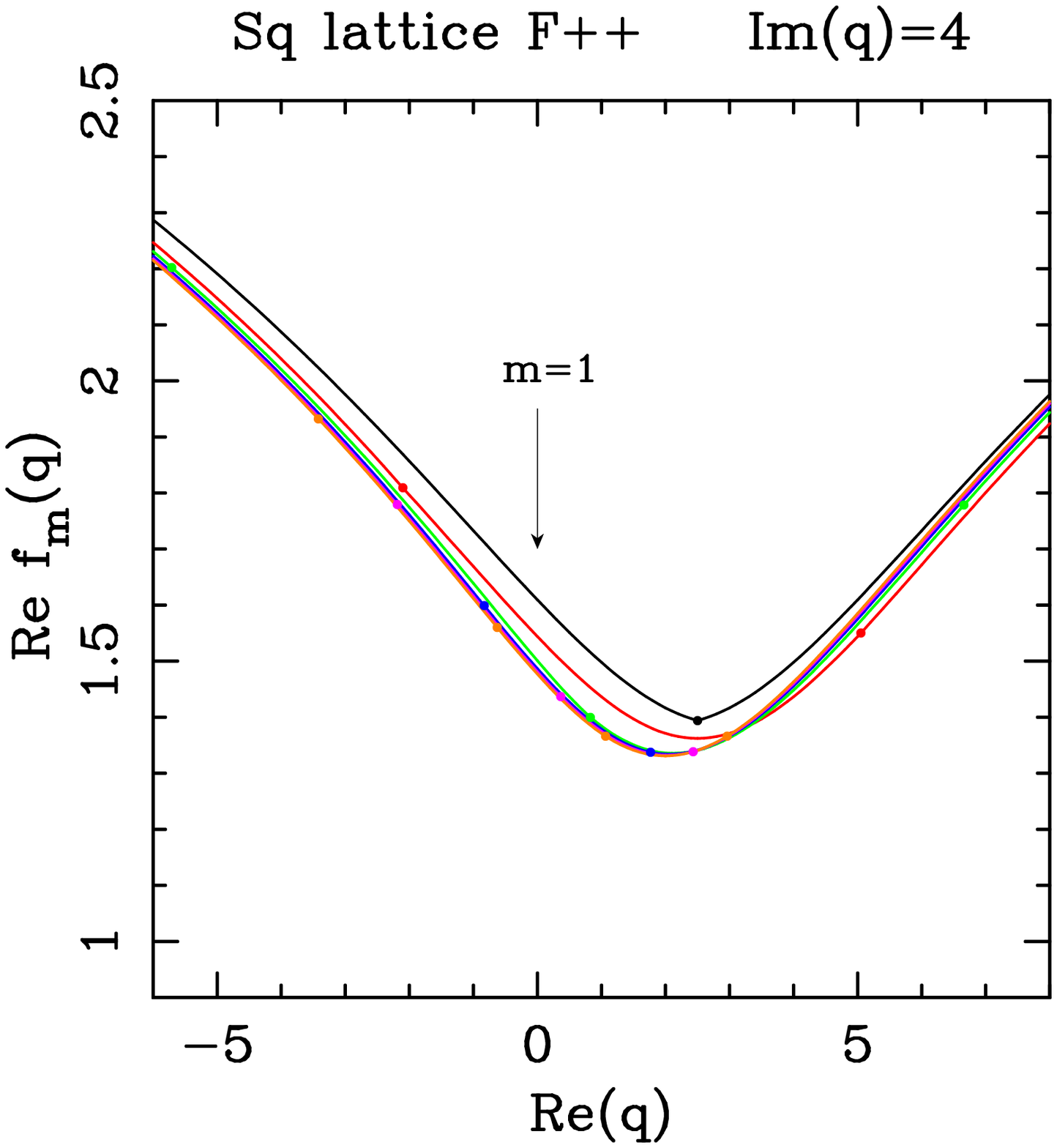} \\[1mm]
\phantom{(((a)}(a) & \phantom{(((a)}(b) \\[5mm]
\end{tabular}
\caption{\label{figure_Free_Imqlarge}
Real part of the free energy $\real f_m(q)$ [cf.\ \protect\reff{def_Fm}]
for square-lattice strips $S_{m,n}$ of width $m$ 
(and length $n\to\infty$) as a function of $\Re q$,
for (a) $\Im q = 3$ and (b) $\Im q = 4$.
Widths are
$m=1$ (black), $m=2$ (red), $m=3$ (green), $m=4$ (blue), $m=5$ (pink)
and $m=6$ (orange).
The solid dots show the points of
discontinuity in the derivative of the free energy.
}
\end{figure}

In Figure~\ref{figure_Free_Imqlarge} we plot the real part of the free energy
\begin{equation}
   f_m(q) \;=\; \max[f_m^{=}(q), f_m^{\neq}(q)]
   \;=\; \frac{1}{m} \max[ \log\lambda_{\star,=}(q),
                           \log\lambda_{\star,\neq}(q) ]
 \label{def_Fm_max}
\end{equation}
for a semi-infinite strip of width $m$ ($1 \le m \le 6$)
as a function of $\Re q$ for fixed $\Im q = 3$ or 4.
The solid dots show the points of
discontinuity in the derivative of the free energy,
which arise from the transition between the dominance of
$\lambda_{\star,=}$ and $\lambda_{\star,\neq}$.
The discontinuity is small already for $m=1$ and $\Im q = 3$,
and it decreases as $m$ and $\Im q$ grow;
indeed, it is essentially invisible for all $m \ge 2$.

\begin{figure}%%%[hbtp]
\centering
\begin{tabular}{cc}
\includegraphics[width=200pt]{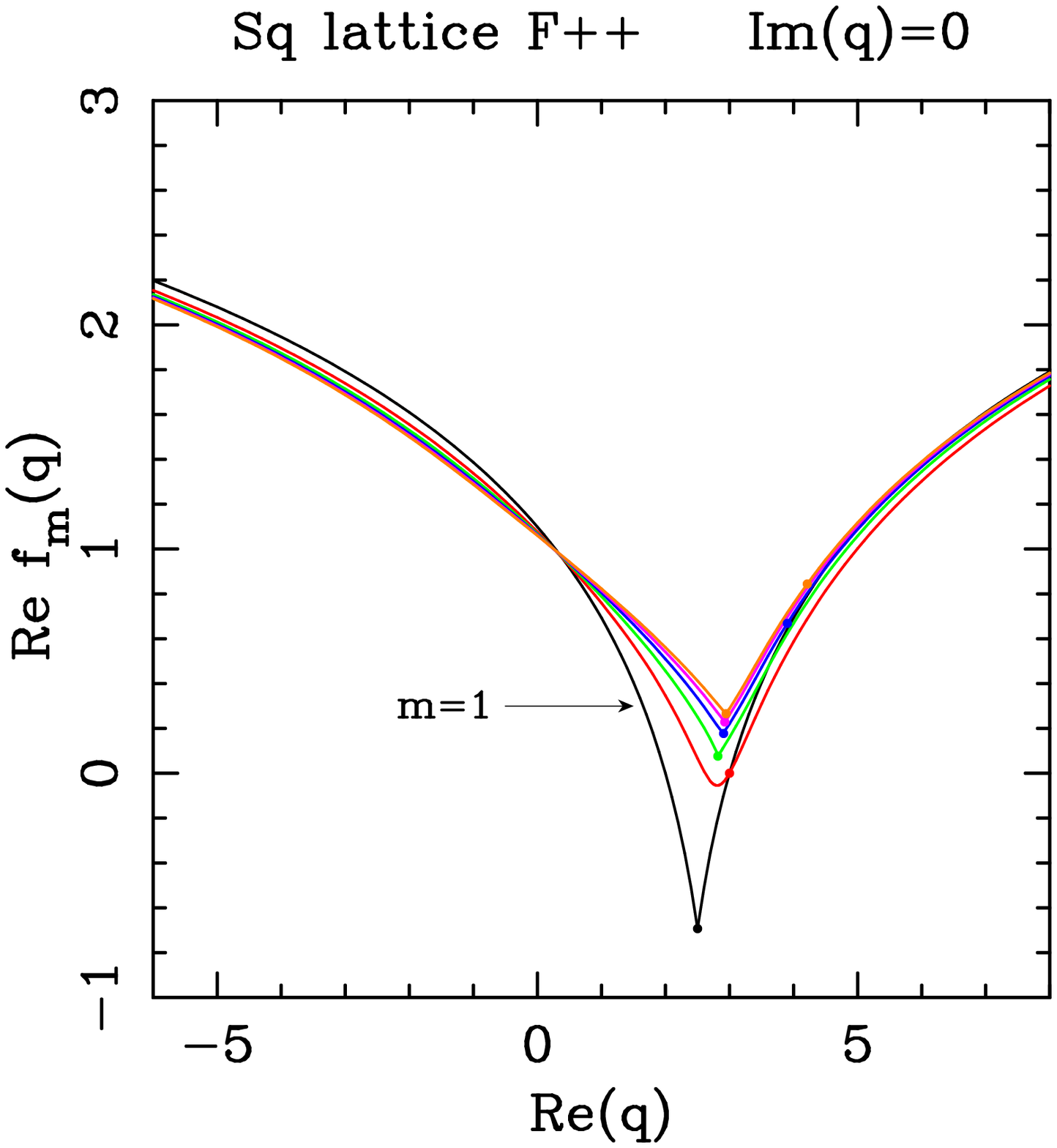} & 
\includegraphics[width=200pt]{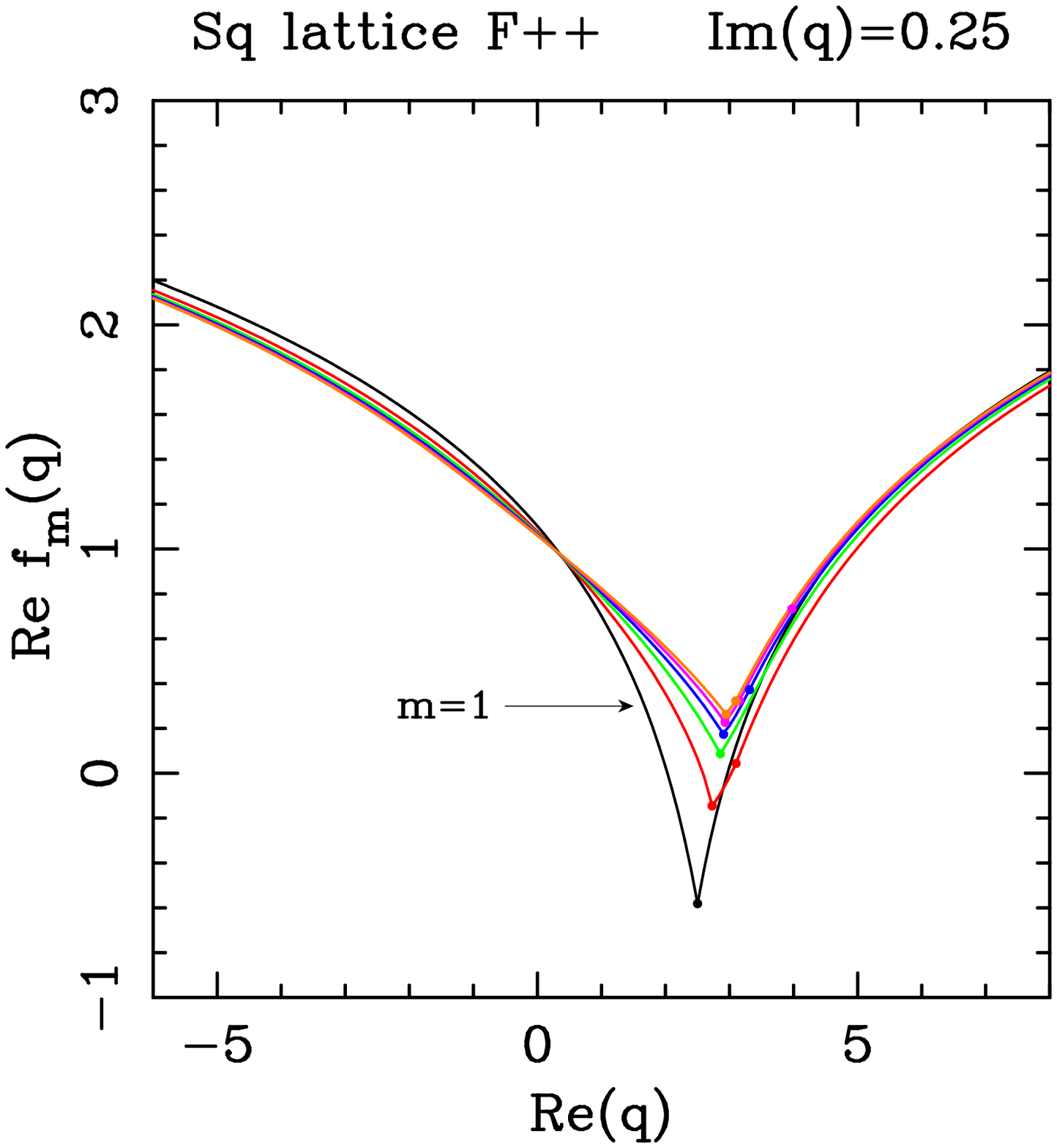} \\[1mm]
\phantom{(((a)}(a) & \phantom{(((a)}(b) \\[8mm]
\includegraphics[width=200pt]{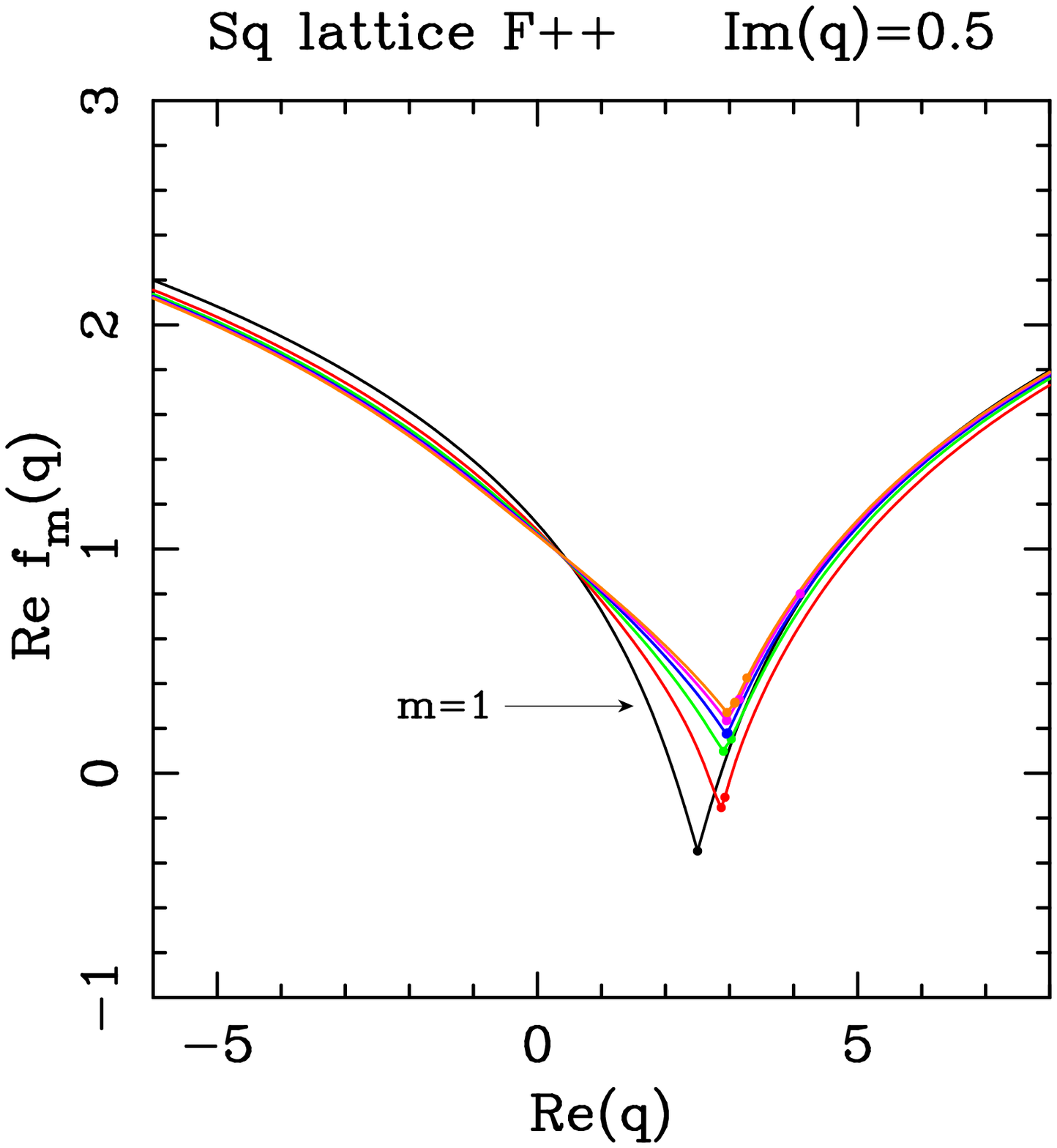} & 
\includegraphics[width=200pt]{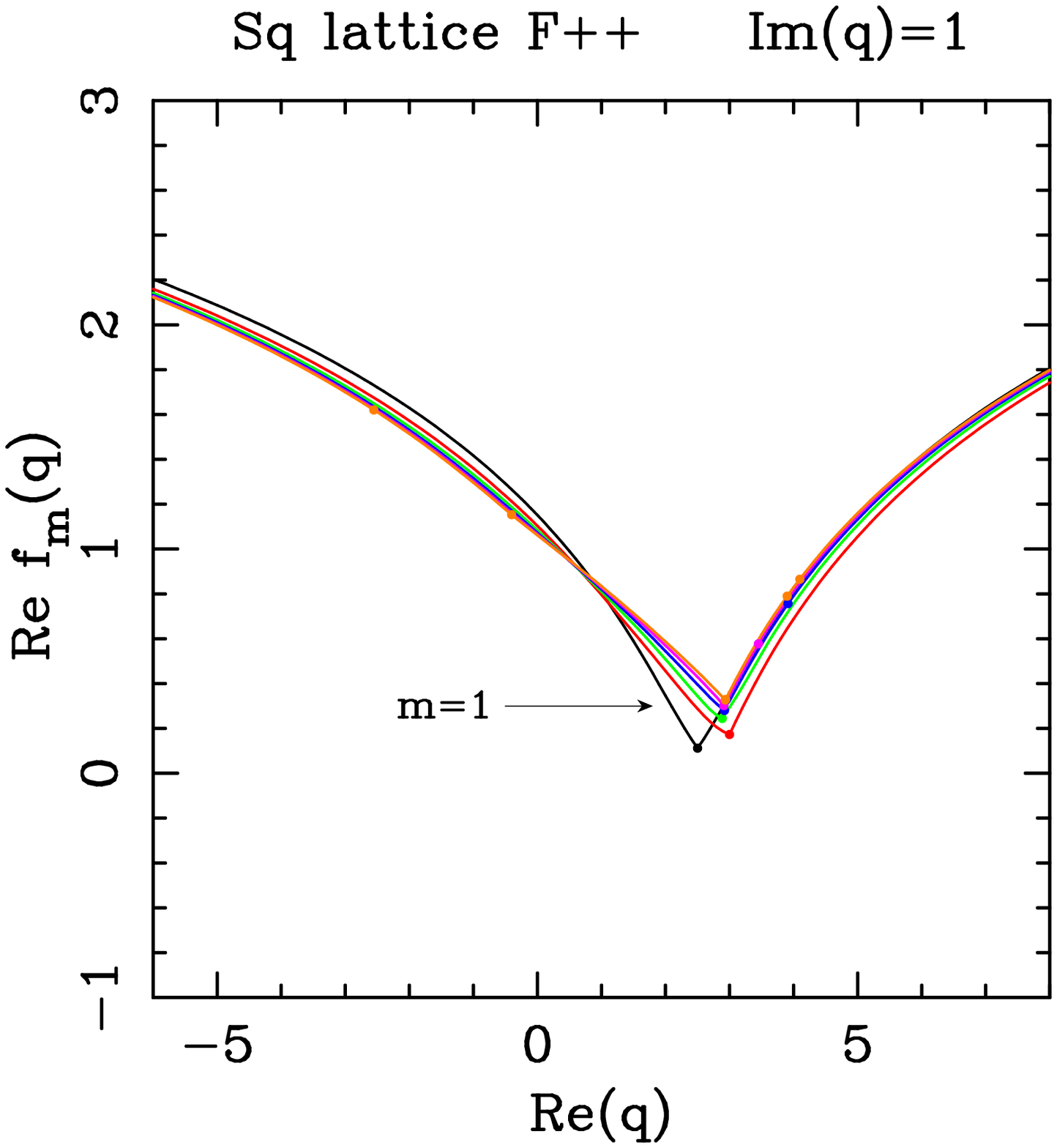} \\[1mm]
\phantom{(((a)}(c) & \phantom{(((a)}(d) \\[5mm]
\end{tabular}
\caption{\label{figure_Free_Imqsmall}
Real part of the free energy $\real f_m(q)$ [cf.\ \protect\reff{def_Fm}]
for square-lattice strips $S_{m,n}$ of width $m$ 
(and length $n\to\infty$) as a function of $\Re q$,
for (a) $\Im q = 0$, (b) $\Im q = 0.25$, (c) $\Im q = 0.5$ and (d) $\Im q = 1$.
Widths are
$m=1$ (black), $m=2$ (red), $m=3$ (green), $m=4$ (blue), $m=5$ (pink)
and $m=6$ (orange).
The solid dots show the points of
discontinuity in the derivative of the free energy.
}
\end{figure}

This should be contrasted with the behavior at the
phase transition between the ``disordered'' phase
(at large $|q|$) and the ``ordered phase'' (at small $|q|$),
which are separated by a curve lying close to
$\scrb_\infty({\rm sq}) = \lim\limits_{m\to\infty} \scrb_m^{\rm cyl}$
(see Section~\ref{subsec.comparison}).
At this transition ---
which occurs also for cylindrical and free boundary conditions,
and which is the fundamental phase transition for this model ---
the discontinuities in the derivative of the free energy
do {\em not}\/ disappear as $m \to\infty$.
This can be seen clearly in Figure~\ref{figure_Free_Imqsmall},
which is analogous to Figure~\ref{figure_Free_Imqlarge}
but for smaller values of $\Im q$ (namely, 0, 0.25, 0.5 and 1).

We conclude that the outward branches (tending to $q=\infty$)
of the curves $\scrb_m$ do {\em not}\/ correspond to
phase transitions of the infinite-volume free energy.
This is no surprise, because no such phase transitions
are observed for free or cylindrical boundary conditions,
and we believe that the infinite-volume free energy $f(q)$
is the {\em same}\/ for $=$, $\neq$, free and cylindrical boundary conditions
(at least in the ``disordered'' phase,
where we have checked it in the large-$q$ expansion
through order $q^{-15}$, cf.\ Section~\ref{sec.thermo.overview}).

\subsection{Largest magnitude of a chromatic root}

By Proposition~\ref{prop1} we know that, for each $m \ge 1$,
the chromatic roots of $S_{m,n}$ become dense as $n \to\infty$
on a curve $\scrb_m$ that contains $2m$ outward branches extending to infinity.
It is natural to ask:  At what rate does this curve get filled out
as $n \to\infty$?  In particular, what is the largest magnitude of a
chromatic root of $S_{m,n}$?

For the generalized theta graphs $\Theta^{(s,p)}$,
Brown, Hickman, Sokal and Wagner \cite[Theorem~1.3]{gen_theta}
proved that the chromatic roots are bounded in modulus by
$[1 + o(1)] p/\log p$, uniformly in $s$ and $p$,
where $o(1)$ denotes a constant $C(p)$ that tends to zero as $p \to\infty$.
They furthermore showed \cite[Corollary~5.2]{gen_theta}
that this bound is sharp when $s=2$.
However, when $s>2$ we expect that this bound is far from sharp;
in particular, we expect that the root of largest modulus
grows as $p \to\infty$ only like $(p/\log p)^{1/(s-1)}$.

For our graphs $S_{m,n}$, the only thing we know rigorously
is that the chromatic roots are bounded in modulus by $C \max(n,4)$,
where
\be
   C \;=\; {W(e/2) \over [1-W(e/2)]^2} \;\approx\; 6.907652
\ee
and $W$ is the Lambert $W$ function \cite{Corless_96},
i.e.\ the inverse function to $x \mapsto x e^x$;
this follows from a result of Fern\'andez and Procacci \cite{FP_paper2}
bounding the chromatic roots of arbitrary graphs in terms of the
graph's maximum degree (i.e., the maximum number of neighbors
of any vertex).\footnote{
   Earlier, Sokal \cite{Sokal_chromatic_bounds} had obtained
   the same result with the weaker constant 7.963907.
   The identification of the Fern\'andez--Procacci constant
   in terms of the Lambert $W$ function is due to
   Jackson, Procacci and Sokal \cite{JPS_08}.
}
However, by analogy with the generalized theta graphs
it is natural to expect that the chromatic roots of the graphs $S_{m,n}$
in fact obey an upper bound that is {\em sublinear}\/ in $n$,
e.g.\ $n/\log n$ or even $(n/\log n)^{1/m}$
[since $s$ corresponds to $m+1$].

It is an interesting open problem to verify (or refute) these conjectures,
first for the generalized theta graphs and then for the family $S_{m,n}$.

\section{Discussion}  \label{sec.discussion}

In this section we make some final comments
and some suggestions for future research.

\subsection{Comparison with the work of Jacobsen, Saleur and Dubail}

In a series of recent papers
\cite{Jacobsen_Saleur_08a,Jacobsen_Saleur_08b,Jacobsen_Saleur_08c,%
Dubail_Jacobsen_Saleur_09},
Jacobsen, Saleur and Dubail have studied models
of densely packed self-avoiding loops on the annulus
in which loops touching one or both rims of the annulus
receive different weights from ``bulk'' loops.
By a generalization of the Baxter--Kelland--Wu \cite{Baxter_76} mapping,
such models can be mapped onto Potts models in which
spins lying on the rims of the annulus lie in specified subsets
of $\{1,\ldots,q\}$,
or equivalently Fortuin--Kasteleyn models in which
clusters touching one or both rims of the annulus
receive different weights from ``bulk'' clusters.

These latter models can be defined in a very general way as follows
\cite{Dubail_Jacobsen_Saleur_09}:
Let $G=(V,E)$ be a finite graph,
and let $V_1, V_2 \subseteq V$ be {\em disjoint}\/ subsets of $V$.
Fix subsets $S_1, S_2 \subseteq \{1,\ldots,q\}$,
and let us define a Potts model in which
spins at a site in $V_i$ are required to lie in $S_i$ ($i=1,2$):
\begin{equation}
   Z_G^{\rm Potts}(q, S_1,S_2, {\bf v})   \;=\;
   \sum_{\begin{scarray}
            \sigma \colon\, V \to \{ 1,2,\ldots,q \} \\
            \sigma[V_1] \subseteq S_1 \\
            \sigma[V_2] \subseteq S_2
         \end{scarray}
        }
   \; \prod_{e=ij \in E}  \,
      \bigl[ 1 + v_e \delta(\sigma_i, \sigma_j) \bigr]
   \;.
 \label{def.ZPotts.boundary}
\end{equation}
Likewise, let us define a Fortuin--Kasteleyn model
in which clusters touching $V_1$ or $V_2$ or both
receive different weights from ``bulk'' clusters:
\begin{equation}
   Z_G(q,q_1,q_2,q_{12}, {\bf v}) \;=\;
   \sum_{ A \subseteq E }
      q^{k_0(A)} q_1^{k_1(A)} q_2^{k_2(A)} q_{12}^{k_{12}(A)}
      \prod_{e \in A}  v_e
   \;,
 \label{eq.FK.boundary}
\end{equation}
where $k_0(A), k_1(A), k_2(A), k_{12}(A)$ denote, respectively,
the number of connected components in the subgraph $(V,A)$
that intersect neither $V_1$ nor $V_2$, $V_1$ but not $V_2$,
$V_2$ but not $V_1$, or both $V_1$ and $V_2$.
It is then easy to see that
\be
   Z_G^{\rm Potts}(q, S_1,S_2, {\bf v})   \;=\;
   Z_G(q,q_1,q_2,q_{12}, {\bf v})
\ee
provided that we identify
\be
   q_1 \,=\, |S_1|, \qquad
   q_2 \,=\, |S_2|, \qquad
   q_{12} \,=\, |S_1 \cap S_2|
   \;.
\ee

Our partition functions
$Z_{S^{=}_{m,n}}(q, {\bf v})$ and $Z_{S^{\neq}_{m,n}}(q, {\bf v})$
are, up to trivial prefactors, special cases of this construction.
Indeed, let $G$ be the usual $m \times n$ square lattice
with free boundary conditions ($m \ge 2$),
and let $V_1$ (resp.\ $V_2$) consist of the sites in the
leftmost (resp.\ rightmost) column.
Then it is easy to see that
\begin{subeqnarray}
   Z_{S^{=}_{m,n}}(q, {\bf v})
      & = & q \, Z_G(q,q-1,q-1,q-1, {\bf v})  \\[1mm]
   Z_{S^{\neq}_{m,n}}(q, {\bf v})
      & = & q(q-1) \, Z_G(q,q-1,q-1,q-2, {\bf v})
\end{subeqnarray}

Dubail, Jacobsen and Saleur (DJS) have used the loop-model representation
associated to the Fortuin--Kasteleyn model \reff{eq.FK.boundary}
to obtain the conformal properties
(central charge, critical exponents, exact continuum-limit partition functions,
 etc.)\ 
both for the usual critical point $n = \sqrt{q}$
\cite{Dubail_Jacobsen_Saleur_09}
and for the Berker--Kadanoff phase $n = -\sqrt{q}$
\cite{Jacobsen_Saleur_08c},
where $n$ is the bulk loop weight.\footnote{
   The analysis of the Berker--Kadanoff phase in \cite{Jacobsen_Saleur_08c}
   is actually restricted to the one-boundary case
   $q_2 = q$, $q_{12} = q_1$,
   but it could presumably be generalized.
}
This analysis is relevant for values of $q$ lying {\em inside}\/
the limiting curve $\scrb_\infty({\rm sq})$.
By contrast, in the present paper we have focused on the
behavior at large~$|q|$,
i.e.\ on the disordered (and non-critical) phase
lying {\em outside}\/ the limiting curve $\scrb_\infty({\rm sq})$.

There are, however, some points of contact between our work
and that of DJS.
In particular, our transfer-matrix eigenvalues
should correspond to the sector with no non-contractible clusters
in the work of DJS (who used periodic longitudinal boundary conditions).

\subsection{Density of zeros}   \label{subsec.density}

Let us go back to the question raised at the beginning of this paper:
When does the accumulation of partition-function zeros at some parameter value
(or on some curve in the complex plane of some parameter)
signal a nonanalyticity of the infinite-volume free energy?
The answer is straightforward:
when the zeros have a {\em nonzero density per unit volume}\/
in the infinite-volume limit.
Thus, in the trivial example $Z_n(x) = x$ for a system in volume $n$,
the zero at $x=0$ has multiplicity 1,
so the density of zeros is $1/n$, which tends to zero as $n \to \infty$;
hence the infinite-volume free energy does {\em not}\/ exhibit
nonanalyticity at $x=0$.
Similarly for an isolated limiting point in the Beraha--Kahane--Weiss theorem:
it has fixed multiplicity, independent of the system length $n$,
so that the density of zeros again tends to zero as $n \to \infty$.

It would therefore be very interesting to try to understand analytically
the density of zeros in the various graph families considered in this paper:
starting with the bi-fans $P_n + \bar{K}_2$ and bipyramids $C_n + \bar{K}_2$
(Section~\ref{sec.setup.bipyr}),
continuing with the generalized theta graphs $\Theta^{(s,p)}$
(Section~\ref{sec.setup.theta}),
and finishing with the graphs $S_{m,n}$.
For the first three families one might hope to find exact expressions
for the density of zeros as $n$ or $p$ tends to infinity;
furthermore, for the generalized theta graphs
one might be able to study the subsequent limit $s \to\infty$.
For $S_{m,n}$ it seems unlikely that one could obtain exact expressions
(except for very small $m$);
but one might hope at least to obtain expansions in powers of $q^{-1}$.

\subsection{Some concluding remarks}  \label{subsec.remarks}

In this paper we have studied the graphs $S_{m,n}$
that are obtained from an $m \times n$ square grid
($m$ columns, $n$ rows, free boundary conditions in both directions)
by adjoining one new vertex adjacent to all the sites in the leftmost column
and a second new vertex adjacent to all the sites in the rightmost column.
By similar transfer-matrix methods one can study the graphs $T_{m,n}$
that are obtained analogously from a $m \times n$ grid of
the {\em triangular}\/ lattice.
(The triangular lattice is most easily viewed as a
square lattice with a SW--NE diagonal edge added to each face.)
All of these graphs are 3-connected and planar,
and we expect the square and triangular families to have a
similar qualitative behavior, though the details of the
limiting curves will differ \cite{transfer1,transfer2,transfer3}.
The triangular lattice has the advantage that there exists a precise
conjecture for at least the outer boundary of the limiting curve
$\scrb_\infty$ for ordinary (e.g., free or cylindrical)
boundary conditions:
see Baxter \cite{Baxter_86,Baxter_87} for this conjecture
and the remarkable Bethe-Ansatz computation supporting it,
and see \cite[especially Section~6]{transfer3} for further discussion.

More interestingly, one can also study the graphs
$S_{m,n}^{\rm per}$ (resp.\ $T_{m,n}^{\rm per}$)
that are obtained by this construction starting from an $m \times n$ grid
with {\em periodic}\/ boundary conditions in the longitudinal ($n$) direction.
These are 4-connected planar graphs (resp.\ 4-connected plane triangulations),
and for $m=1$ they reduce to the bipyramids $C_n + \bar{K}_2$.
Their chromatic polynomials can once again be analyzed
by transfer-matrix methods \cite{transfer4,Jacobsen-Salas_toroidal}.
The transfer matrices are larger
than in the free-longitudinal-boundary-condition case,
since they have to keep track of the connections
among the sites of the top {\em and bottom}\/ rows.
On the other hand, the transfer matrix is block-triangular:
firstly in terms of the number $\ell$ of disjoint paths
connecting the top and bottom rows,
and secondly in terms of the connectivity of the bottom row.
Moreover, the amplitudes $\alpha_k(q)$ are explicitly determinable
polynomials in $q$.
For this reason the computation is more involved
than for free longitudinal boundary conditions,
but not vastly computationally more demanding.
See \cite{transfer4,Jacobsen-Salas_toroidal,Jacobsen-Salas_flow}
for more details.

Let us conclude with a vague idea. 
It may be possible to prove Conjecture~\ref{conj.uniform} by using
the polymer expansion \cite{Procacci_03}
and exploiting the uniform (in $m$) convergence of
the polymer expansion for sufficiently large $|q|$.

\section*{Acknowledgments}

We wish to thank Bill Jackson, Jesper Jacobsen and Steve Noble
for helpful conversations.
We also wish to thank two referees for incisive comments
on the first version of this paper.

J.S.\ is grateful for the kind hospitality of
the Physics Department of New York University
and the Mathematics Department of University College London,
where part of this work was done.
Likewise, A.D.S.\ is grateful for the kind hospitality of
the M.S.M.I.\  group of the Universidad Carlos III de Madrid.
Both authors also thank the Isaac Newton Institute for Mathematical Sciences,
University of Cambridge, for hospitality during the programme on
Combinatorics and Statistical Mechanics (January--June 2008);
LPTHE-Jussieu and the \'Ecole Normale Sup\'erieure
for hospitality in June 2009;
and the \'Ecole Normale Sup\'erieure for hospitality in May--June 2011.
Finally, A.D.S.\ thanks the Institut Henri Poincar\'e -- Centre Emile Borel
for hospitality during the programmes on
Interacting Particle Systems, Statistical Mechanics and Probability Theory
(September--December 2008)
and Statistical Physics, Combinatorics and Probability
(September--December 2009).

The authors' research was supported in part
by U.S.\ National Science Foundation grant PHY--0424082 (A.D.S.)\ 
and by Spanish MEC grants FPA2009-08785 and MTM2008-03020 (J.S.).

\appendix

\section{Proof of Proposition~\ref{prop.TMblockdiag.3}}
\label{appendix.polymermodel}

In this appendix we prove Proposition~\ref{prop.TMblockdiag.3}
concerning the dominant diagonal entries in the transfer matrix.

In Ref.~\cite[Section 3.2]{transfer5}, we showed that the dominant diagonal 
entry $t_{\rm F}(m)$ of the transfer matrix for a square-lattice strip of 
width $m$ and {\em free}\/  boundary conditions is equal to the 
partition function for a certain one-dimensional $m$-site polymer gas 
(with free boundary conditions). In this polymer model, each polymer of length
$\ell \ge 1$ gets a fugacity $\mu_\ell = v^{\ell -1}(q+\ell v')$, 
where $v$ (resp.\ $v'$) is the weight of all horizontal (resp.\ vertical)
edges in the corresponding Potts model.
We solved this problem by using the generating function 
(``grand partition function'')
\begin{equation}
\Phi_{\rm F}(z) \;=\; \sum\limits_{m=1}^\infty z^m t_{\rm F}(m) 
                \;=\; \frac{\Psi(z)}{1-\Psi(z)} 
   \;,
\label{def_Phi_F0}
\end{equation}
where 
\begin{equation}
\Psi(z) \;=\; \sum\limits_{\ell=1}^\infty z^\ell v^{\ell-1} (q + \ell v') 
        \;=\; z\left[ \frac{q}{1-z v} + \frac{v'}{(1-z v)^2} \right]
\label{def_Psi_F}
\end{equation}
is the total weight for a single polymer of arbitrary size.
It follows from \reff{def_Phi_F0}/\reff{def_Psi_F} that
\begin{equation}
\Phi_{\rm F}(z) \;=\; \frac{(q-1)z + q z^2}{1 -(q+2v+v^\prime)z + v(q+v)z^2}\,.
\label{def_Phi_F}
\end{equation}
We then expanded \reff{def_Phi_F} in powers of $z$
to extract the coefficients $t_{\rm F}(m)$.
A similar analysis \cite[Section 3.3]{transfer5}
handled the case of cylindrical boundary conditions.

In this appendix we shall perform the analogous computations for the
transfer matrices associated to the graph families
$S_{m,n}^{=}$ and $S_{m,n}^{\neq}$,
culminating in a proof of Proposition~\ref{prop.TMblockdiag.3}.

Let us start with the family $S_{m,n}^{=}$. As explained in 
Section~\ref{sec.block2} above, the boundary conditions for this family can 
be handled by considering $m+1$ sites on a circle, where the site labelled
$0$ is special. Consider first the action of $\H$ on the start vector
$\basise_{\rm id}$. It generates $2^{m+1}$ terms, each of which corresponds
to a partition $\mathcal{P}$ in which all the blocks are sequential
sets of vertices on the $(m+1)$-cycle (we shall call these sets ``polymers'').  
Each polymer of size $\ell<m+1$ picks up a factor $v^{\ell-1}$, while a
polymer of size $m+1$ picks up a factor $v^{m+1} + (m+1)v^m$ (the
$v^{m+1}$ comes from the case in which all edges are occupied, while the
$(m+1)v^{m}$ comes from the $m+1$ cases in which all edges but one are
occupied).  

Consider next the action of $\V$ on a basis vector $\basise_\mathcal{P}$
corresponding to an arbitrary partition $\mathcal{P}=\{P_1,\ldots,P_k\}$.
If we are to end up with the partition $\basise_{\rm id}$, then for each block
$P_j$ we must either choose the detach operator $\D_i$ for all $i\in P_j$
(the last deletion gives a factor of $q$) or else choose the detach 
operator for all but one $i\in P_j$ and choose $v' I$ for the last site
(this can be done in $|P_j|$ ways). This means that the fugacity for any
polymer of size $1\le \ell\le m$ {\em not containing}\/  the special site $0$, 
the fugacity is given by 
\begin{equation}
\mu_\ell \;=\; v^{\ell -1} (q + \ell v') \,.  
\label{def_mul}
\end{equation}
However, the fugacity for a polymer of length $1\le\ell\le m+1$ containing the
special site $0$ is given by the formula\footnote{
   Let us remark that the analogous definition \cite[eq.~(3.51)]{transfer5}
   suffers from an unfortunate notational ambiguity:
   $\widehat{\mu}_\ell$ should have been written as $\widehat{\mu}_\ell^{(m)}$,
   because the weight in question depends on $m$ as well as $\ell$.
   The same superscript ${}^{(m)}$ should appear also in
   \cite[eqs.~(3.52) and (3.53)]{transfer5}.
   This notational clarity is important because,
   although at this stage we are considering one fixed value of $m$,
   we will soon be summing over $m$.
}
\begin{equation}
\nu_\ell^{(m)} \;=\; \begin{cases}
                v^{\ell -1}   & \text{for $1\leq \ell \leq m$}  \\
                v^m (v+m+1)   & \text{for $\ell = m+1$} 
               \end{cases} 
\label{def_nul}
\end{equation}
because the vertical operator $\V$ does not include $\D_0$ (therefore,
for any block $P_j$ containing the site $0$, we are forced to choose
the detach operator $\D_i$ for all $i \in P_j$
other than the special site $0\in P_j$).   

As we did for cylindrical boundary conditions in 
\cite[Section 3.3]{transfer5}, we can compute $t_{=}(m)$ by using a 
simple recursion relating our case to that of free boundary conditions:
\be
t_{=}(m) \;=\; \sum\limits_{k=1}^m k \nu_k^{(m)} \, t_{\rm F}(m+1-k) 
         \;+\; \nu_{m+1}^{(m)} \;.
\label{def_recursion_eq}
\ee
To prove \reff{def_recursion_eq},
let $k\ge 1$ the size of the polymer containing site $0$. If $k\le m$, there
are $k$ ways of placing this polymer such that the site $0$ belongs to it,
with fugacity $\nu_k^{(m)} = v^{k-1}$ for each placement;
and for the rest of ring, the total
weight of all admissible polymer configurations is $t_{\rm F}(m+1-k)$.
Finally, if $k=m+1$, there is only one way of placing the polymer,
and it receives a weight $\nu_{m+1}^{(m)}$. This proves \reff{def_recursion_eq}. 

In order to compute explicitly the $t_{=}(m)$, we introduce the generating 
function 
\begin{equation}
\Phi_{=}(z) \;=\; \sum\limits_{m=1}^\infty z^m \, t_{=}(m) 
   \;.
\label{def_Phi_eq}
\end{equation}
The upper limit on the sum in \reff{def_recursion_eq} can be replaced 
by $\infty$, provided that we define $t_{\rm F}(\ell)=0$ for $\ell\le 0$ 
[which is anyway implicit in the definition \reff{def_Phi_F}]. 
Multiplying both sides of \reff{def_recursion_eq} by $z^m$ and summing over
$m$, we arrive at the equation
\be
\Phi_{=}(z) \;=\; \frac{ \Phi_{\rm F}(z) }{(1-vz)^2}
                  \,+\, \frac{v^2 z}{1-vz}
                  \,+\, \frac{vz(2-vz)}{(1-vz)^2}
   \;.
\label{def_Phi_eq_1}
\ee
When $v=v'=-1$, we obtain the final formula
\be
\Phi_{=}(z) \;=\; -\frac{1}{1+z} + \frac{1}{1-(q-3)z -(q-1)z^2} 
   \;.
\label{def_Phi_eq_final}
\ee
By expanding this function in powers of $z$, we have checked that it agrees
with the dominant diagonal elements $t_{=}(m)$ for $m\leq 15$.

Let us next consider the family $S_{m,n}^{\neq}$.
Defining the generating function
\be
\Phi_{\neq}(z) \;=\; \sum\limits_{m=1}^\infty z^m t_{\neq}(m) \,,
\label{def_Phi_neq}
\ee
it follows easily from \reff{def_Phi_eq} and \reff{eq.curious.4} that
\be
\Phi_{\neq}(z) \;=\; \Phi_{=}(z)  \,-\, \frac{v^2 z}{1-vz}
   \;.
\label{def_Phi_neq_1}
\ee
When $v=v'=-1$, we obtain the final formula
\be 
\Phi_{\neq}(z) \;=\; -1 + \frac{1}{1-(q-3)z -(q-1)z^2} 
  \;.
\label{def_Phi_neq_final}
\ee
By expanding this function in powers of $z$, we have checked that it agrees
with the dominant diagonal elements $t_{\neq}(m)$ for $m\leq 15$.

It is now straightforward to extract from
\reff{def_Phi_eq}/\reff{def_Phi_eq_final}
and \reff{def_Phi_neq}/\reff{def_Phi_neq_final}
the coefficients $t_{=}(m)$ and $t_{\neq}(m)$.
In view of \reff{eq.curious.4}/\reff{eq.curious.5},
it suffices to study one of the two;
and it turns out that $t_{\neq}(m)$ takes a slightly simpler form
[compare \reff{conj.tneq.EARLY} with \reff{conj.teq.EARLY}].
Let us use the notation $[z^m] P(z)$ to denote the coefficient of $z^m$
in a polynomial or formal power series.
Using the identity \cite[eq.~(3.21a)]{transfer5}
\be
[z^m] \frac{1}{1-az-bz^2} \;=\; \sum\limits_{j=0}^{\lfloor m/2\rfloor} 
      {m-j \choose j} a^{m-2j} b^j \;,
\ee
we get for $m\ge 1$
\be
t_{\neq}(m) \;=\; [z^m]\Phi_{\neq}(z) \;=\;  
\sum\limits_{j=0}^{\lfloor m/2\rfloor}
      {m-j \choose j} (q-3)^{m-2j} (q-1)^j \,, 
\label{def_tneq}
\ee
which is manifestly a polynomial in $q$ of degree $m$,
as claimed in \reff{conj.t.EARLY}.

The next goal is to compute the coefficients $a_k(m)$ arising in
\reff{conj.t.EARLY}, where $m\geq 1$ and $0\le k \le m$. Expanding the 
binomials in \reff{def_tneq}, we have
\be
a_k(m) \;=\; \sum\limits_{j=0}^{\lfloor m/2\rfloor} \sum\limits_{\ell=0}^\infty
 \binom{m-j}{j}\binom{m-2j}{k+\ell-2j}\binom{j}{\ell} 3^{k+\ell-2j} (-1)^j
\,.
\label{def_akm_1}
\ee
We now want to substitute the $m$-dependent upper index in the sum
over $j$ by something independent of $m$, e.g.\ by $k$.

There are two nontrivial cases:
a) If $k<\lfloor m/2\rfloor$, then the second binomial vanishes whenever
$k+\ell-2j<0$, and the third binomial is non-vanishing only if $j\geq \ell$.
Therefore for $j>k$ and $j\geq \ell$ we have that
$k+\ell-2j < k +\ell -k-\ell=0$. So all these terms vanish.
b) If $k>\lfloor m/2\rfloor$, then the first binomial does not vanish only when
$0\leq j \leq \lfloor m/2\rfloor$ or when $j\geq m+1$. As we are adding
terms with $\lfloor m/2 \rfloor + 1 \leq j\leq k-1\leq m-1$, none of them
give rise to a non-vanishing contribution.
Therefore, we can re-write \reff{def_akm_1} as
\begin{equation}
a_k(m) \;=\; \sum\limits_{j=0}^{k} \sum\limits_{\ell=0}^\infty
 \binom{m-j}{j}\binom{m-2j}{k+\ell-2j}\binom{j}{\ell} 3^{k+\ell-2j} (-1)^j
\,.
\label{def_akm_2}
\end{equation}
where the independent variable $m$ does not appear in the summation limits.
After some straightforward but lengthy algebra we can rewrite the above
formula in a more compact form:
\begin{equation}
a_k(m) \;=\; \sum\limits_{p=0}^k \binom{m-p}{p} (-1)^p 
          \sum\limits_{q=0}^{k-p} 3^q \, \binom{m-2p}{q}\binom{p}{k-p-q} 
\,.
\label{def_akm_final}
\end{equation}
It is clear from \reff{def_akm_final} that $a_k(m)$
is for fixed $k\geq 0$, the restriction to integers $m\geq \max(1,k)$ 
of a polynomial in $m$ of degree at most $k$,
as $m$ appears only in the upper index of the binomials and 
\be
\binom{m}{j} \;=\; \frac{m^{\underline{j}}}{j!} \;=\; 
\frac{m(m-1)(m-2)\cdots(m-j+1)}{j!} 
\ee
is a polynomial in $m$ of degree $j$. [Here we use Knuth's \cite{Graham_94}
notation for falling powers: $x^{\underline{j}} = x(x-1)\cdots(x-j+1)$.]

To see that the degree of $a_k(m)$ is exactly $k$,
let us extract the term of order $m^k$:
\begin{equation}
[m^k] a_k(m) \;=\; \sum_{p=0}^k \frac{ (-1)^p 3^{k-p} }{p! (k-p)! } 
         \;=\; \frac{2^k}{k!} \;\neq\; 0 
   \;.
\end{equation}
This equation also implies that the coefficient $a_{k,0}=1$ 
[cf.\ \reff{eq.ak0.neq}] for all $k\ge 0$.

This completes the proof of Proposition~\ref{prop.TMblockdiag.3}.

Just to be safe, we have checked for $0 \le k \le m \le 30$
that \reff{def_akm_final} agrees with the expansion of
\reff{def_Phi_neq_final}.

\section{Dimension of the transfer matrices}
\label{sec.dim}

In this appendix we compute the dimensionalities
$N_=(m)$ and $N_{\neq}(m)$ of the transfer matrices
for $=$ and $\neq$ boundary conditions,
proving the formulae stated in
Theorems~\ref{theorem_Neq} and \ref{theorem_Nnoteq}.
As a corollary we obtain the dimensionality
$\dim \T''(m) = \mbox{\rm SqFree}(m+2) = N_{\neq}(m) + N_{=}(m)$
of the full symmetry-reduced transfer matrix,
confirming the formula \reff{def_SqFree}
and proving the alternative formulae (\ref{def_dimT}a,b,c).

\subsection{Catalan, Motzkin and Riordan numbers}   \label{sec.dim.1}

The number of non-crossing partitions of $\{1,\ldots,m\}$
is given by the Catalan number \cite{Stanley_99}
\be
   C_m  \;=\; {(2m)! \over m! \, (m+1)!}  \;=\; {1 \over m+1} \, {2m \choose m}
   \;.
\ee
The generating function for the Catalan numbers is
\be
   C(z) \;=\;  \sum\limits_{m=0}^\infty C_m z^m
   \;=\; {1 - \sqrt{1-4z} \over 2z}
   \;.
\ee

The number of non-crossing non-nearest-neighbor partitions of the set 
$\{1,2,\ldots,m\}$ is given by the Motkzin number $M_{m-1}$,
where \cite{Bernhart_99}\footnote{
   {\em Warning:}\/  Several references use the notation $m_n$
   to denote what we call $M_n$;
   and one reference \cite{Donaghey_77} writes $M_n$
   to denote a {\em different}\/ sequence.
}
\be
M_m \;=\;  \sum\limits_{k=0}^{\lfloor m/2 \rfloor} {m \choose 2k} \, C_k 
  \;.
\label{def_Mm}
\ee
If we consider the same vertex set {\em on a circle}\/,
then the number of non-crossing 
non-nearest-neighbor partitions is given by the numbers \cite{transfer1}
\be
d_m \;=\; \left\{ \begin{array}{ll}
                  1   & \quad \mbox{\rm for $m=1$} \\
                  R_m & \quad \mbox{\rm for $m\geq 2$}
                  \end{array}
          \right.
\label{def_dm}
\ee
where the $R_m$ are the Riordan numbers (or Motzkin alternating sums) 
defined by \cite{Bernhart_99}\footnote{
   Equation (3.59) of \cite{transfer1} erroneously runs the sum
   down to $k=0$.
}
\be
R_m \;=\; \left\{ \begin{array}{ll}
                  1   & \quad \mbox{\rm for $m=0$} \\[2mm]
                  \sum\limits_{k=1}^{m-1} (-1)^{m-k-1} M_k 
                      & \quad \mbox{\rm for $m\geq 1$}
                  \end{array}
          \right.
\label{def_Rm}
\ee
In the following we will need the linear recursion for the Riordan numbers
\be
R_m \;=\; -R_{m-1} + M_{m-1} + \delta_{m,0}
\label{recursion_Rm}
\ee
and the generating functions for $M_m$, $R_m$ and $d_m$:
\begin{eqnarray}
M(z) &=& \sum\limits_{m=0}^\infty M_m z^m \;=\;
   {1 - z - \sqrt{(1+z)(1-3z)} \over 2z^2 }
\label{def_GMotkzin} \\[1mm]
R(z) &=& \sum\limits_{m=0}^\infty R_m z^m \;=\; 
   {1 + z - \sqrt{(1+z)(1-3z)} \over 2z(1+z) } 
\label{def_GRiordan} \\[1mm]
D(z) &=& \sum\limits_{m=1}^\infty d_m z^m \;=\; 
   R(z) - 1 + z  \;=\;
   {1 - z + 2z^3 - \sqrt{(1+z)(1-3z)} \over 2z(1+z) } 
   \qquad
\label{def_Gdm}
\end{eqnarray}
The numbers $d_m$ for $1\leq m \leq 16$ are displayed in 
Table~\ref{table_dimensions2}.

\subsection{Partitions on a circle modulo reflection}

Our goal is to compute the number of equivalence classes modulo
reflection of non-crossing non-nearest-neighbor
(or ``ncnnn'' for short) partitions of the set 
$\{1,2,\ldots,m\}$ on a circle. However, on a circle there
are two distinct types of reflection:
\begin{itemize}
\item Reflection ${\mathcal R}_2$ with respect to an axis going 
      halfway between a pair of neighboring sites
     (which we shall fix to be $1$ and $m$)
\item Reflection ${\mathcal R}_3$ with respect to an axis going 
      through a site (which we shall fix to be the site $1$)
\end{itemize}
Therefore, we define $N_2(m)$ [resp.\ $N_3(m)$] to be the number of
equivalence classes modulo a reflection
of type ${\mathcal R}_2$ (resp.\ ${\mathcal R}_3$)
of ncnnn partitions of the set $\{1,2,\ldots,m\}$ on a circle. 
Of course, the two types of reflection coincide when $m$ is odd, so that
\be
N_2(2\ell+1) \;=\; N_3(2\ell+1) \qquad\hbox{for } \ell =0,1,2,\ldots \;.
\label{property1}
\ee
For $m$ even we have to solve the two problems separately.

The first step is to define
\begin{eqnarray}
X(m) & = & \# \hbox{ncnnn partitions of $\{1,\ldots,m\}$ on a circle} 
   \nonumber \\[-1mm]
     & & \quad
    \hbox{that are invariant under a reflection of type ${\mathcal R}_2$}  \;.
\label{def_Xm.new}
\end{eqnarray}
The values of $X(m)$ for $1\leq m \leq 16$
are displayed in Table~\ref{table_dimensions2}.
We furthermore denote by $\bm{X}(m)$ the set 
of all ${\mathcal R}_2$-invariant ncnnn partitions of $\{1,\ldots,m\}$ 
on a circle.
The first sets $\bm{X}(m)$ are easy to write down.
Using the delta-function shorthand \reff{def.delta.shorthand} for partitions,
we find that
$\bm{X}(m)= \{\bone\}$ for $m=1,2,3,4$,
$\bm{X}(5)= \{\bone,\delta_{24} \}$, 
$\bm{X}(6)= \{\bone,\delta_{25},\delta_{13}\delta_{46} \}$, 
$\bm{X}(7)= \{\bone,\delta_{26},\delta_{35},\delta_{26}\delta_{35},
\delta_{13}\delta_{57},\delta_{246} \}$, and  
$\bm{X}(8)= \{\bone,\delta_{27},\delta_{36},\delta_{27}\delta_{36},
\delta_{13}\delta_{68},\delta_{14}\delta_{58},\delta_{24}\delta_{57} \}$.

Because each equivalence class
modulo reflection contains either one or two elements,
according as those elements are or are not invariant under reflection,
we have
\be
   N_2(m) \;=\; {1\over 2}\left[ X(m)   + d_m \right]
   \;.
 \label{eq.N2mXm}
\ee

It turns out that the numbers $X(m)$ give us not only $N_2$ but also $N_3$:

\begin{lemma}\label{lemma1}
The number $N_3(m)$ of equivalence classes modulo a
reflection of type ${\mathcal R}_3$
of ncnnn partitions 
of the set $\{1,2,\ldots,m\}$ on a circle is given by  
\be
N_3(m) \;=\; \left\{ \begin{array}{ll}
    N_2(m) \;=\; {1\over 2}\left[ X(m)   + d_m \right] 
            & \quad \mbox{\rm  for $m$ odd} \\[2mm]
                 {1\over 2}\left[ X(m+2) + d_m \right] 
            & \quad \mbox{\rm for $m$ even}
                     \end{array}\right.
\label{def_N3_vs_X}
\ee
\end{lemma}

\proof
The equality for odd $m$ is trivial from \reff{property1}/\reff{eq.N2mXm}.
For even $m$ we start by defining the quantity 
\be
Y(m) \;=\; 2N_3(m) - d_m   \;,
\label{def_Ym}
\ee
which is equal to the number of ncnnn partitions
that are invariant under a reflection of type ${\mathcal R}_3$.
Let us now show that $Y(2\ell)=X(2\ell+2)$ for all $\ell\geq 1$. 
In this proof we shall denote by $\bm{Y}(m)$ the set of all 
${\mathcal R}_3$-invariant ncnnn partitions of $\{1,\ldots,m\}$ on a circle. 

The idea is to show that there is a bijection between the set 
$\bm{Y}(2\ell)$ of ${\mathcal R}_3$-invariant ncnnn partitions of 
$\{1,2,\ldots,\ell,\ell+1,\ldots,2\ell\}$ on a circle and the set
$\bm{X}(2\ell+2)$ of ${\mathcal R}_2$-invariant ncnnn partitions of 
$\{1,2,\ldots,\ell,\ell+1,(\ell+1)',\ell+2,\ldots,2\ell,1'\}$ on a circle.
The trick is to split the vertices $1$ and $\ell+1$ into two pairs 
of nearest-neighbor vertices $1,1'$ and $\ell+1, (\ell+1)'$, respectively.

This bijection is clear between the subset of $\bm{Y}(2\ell)$ with $1$ 
and $\ell+1$ singletons, and the subset of $\bm{X}(2\ell+2)$ for which
$1,1',(\ell+1),(\ell+1)'$ are all singletons. Using the delta-function 
shorthand \reff{def.delta.shorthand}, those partitions belonging to the above  
subsets have exactly the same expression (e.g., for $\ell=3$, 
the ncnnn partitions $\delta_{2,6}$ and $\delta_{3,5}$ belong to these
subsets). 

There is also a bijection between the subset of $\bm{Y}(2\ell)$ with 
$\ell+1$ a singleton and $1$ joined to some ${\mathcal R}_3$-symmetric block 
$B\cup B'$ with $B\subseteq \{3,4,\ldots,\ell\}$ and 
$B' = {\mathcal R}_3 \, B$, 
and the subset of $\bm{X}(2\ell+2)$ with both $(\ell+1)$ and $(\ell+1)'$ 
singletons, and the sites $1$ and $1'$ joined to the blocks $B$ and $B'$, 
respectively. This property is due to the facts that $1$ and $1'$ 
[resp.\ $\ell+1$ and $(\ell+1)'$] are nearest-neighbor vertices, and 
the partitions are non-crossing. 
For instance, if $\ell=3$, the partition $\delta_{1,3,5}$ in $\bm{Y}(6)$
corresponds uniquely to the partition $\delta_{1,3}\delta_{1',5}$ in 
$\bm{X}(8)$, and vice versa. 
A similar bijection can be shown when $1$ is a singleton, but $\ell+1$ is
joined to some ${\mathcal R}_3$-symmetric block. 

Finally, there is a bijection between the subset of $\bm{Y}(2\ell)$ with
both $1$ and $\ell+1$ joined to some ${\mathcal R}_3$-symmetric blocks
$B_1\cup B_1'$ and $B_2\cup B_2'$ (that might be the same), and
the subset of $\bm{X}(2\ell+2)$ characterized by the blocks 
$\{1\}\cup B_1$, $\{1'\}\cup B_1'$, $\{\ell+1\}\cup B_2$, and 
$\{(\ell+1)'\}\cup B_2'$. (If $1$ and $\ell+1$ belong to the same block,
the modifications are obvious.)

As the above subsets are non-intersecting and contain all partitions of
both $\bm{Y}(2\ell)$ and $\bm{X}(2\ell+2)$, we conclude that such 
bijection exists, and therefore, that $Y(2\ell)=X(2\ell+2)$, as claimed.
\qed

Henceforth we consider only reflections of type ${\mathcal R}_2$.
Our goal is to obtain a closed formula for the quantity $X(m)$.
Our results can be summarized in the following proposition:

\begin{proposition} \label{prop2}
The quantity $X(m)$ is given by
\be
X(m) \;=\; \left\{ \begin{array}{ll}
        1 & \qquad \text{if $m=1$} \\[2mm]
        \displaystyle
        \sum\limits_{k\geq 0} {\lfloor{m-1\over 2}\rfloor   \choose k}
                              {\lfloor{m-1\over 2}\rfloor -k\choose k + 
                                             I[\mbox{\rm $m$ is odd}]}
          & \qquad \text{if $m\geq 2$}
                   \end{array}\right.
\label{final_Xm}
\ee
The generating function for this sequence is 
\be
G(z) \;=\; \sum\limits_{m=1}^\infty X(m) z^m \;=\;
  {1\over 2z} \left[ {2z^3-z^2+1 \over \sqrt{(1+z^2)(1-3z^2)}} + 2z^2-1 \right]
   \;.
\label{final_GXm}
\ee
\end{proposition}

\begin{figure}
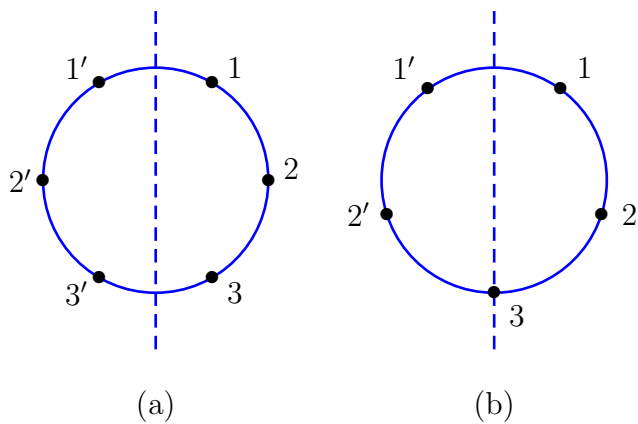

\centering
     \psset{unit=1.5cm}
     \pspicture(-1.0,-1.3)(6,2.5)
%%%%%\psframe  (-1.0,-1.3)(6,2.5)
     \rput{0}(-1,0){
        \pscircle[linecolor=blue,linewidth=1.0pt](2,1){1.01}
        \psline[linecolor=blue,linewidth=1.0pt,linestyle=dashed](2,-0.5)(2,2.5)
        \rput{0}(3.0,1.000000){$\bullet$}
        \rput{0}(2.5,1.866030){$\bullet$}
        \rput{0}(1.5,1.866030){$\bullet$}
        \rput{0}(1.0,1.000000){$\bullet$}
        \rput{0}(1.5,0.133975){$\bullet$}
        \rput{0}(2.5,0.133975){$\bullet$}

        \rput{0}(3.2, 1.1){$2$}
        \rput{0}(2.7, 2.0){$1$}
        \rput{0}(1.3, 2.0){$1'$}
        \rput{0}(0.8, 1.0){$2'$}
        \rput{0}(1.3, 0.0){$3'$}
        \rput{0}(2.7, 0.0){$3$}
        \rput{0}( 2.0,-1.0){(a)}
     }
     \rput{0}(3,0){
        \pscircle[linecolor=blue,linewidth=1.0pt](1,1){1.01}
        \psline[linecolor=blue,linewidth=1.0pt,linestyle=dashed](1,-0.5)(1,2.5)
        \rput{0}(1.0000000,0.000000){$\bullet$}
        \rput{0}(1.9510600,0.690983){$\bullet$}
        \rput{0}(1.5877900,1.809020){$\bullet$}
        \rput{0}(0.4122150,1.809020){$\bullet$}
        \rput{0}(0.0489435,0.690983){$\bullet$}
        \rput{0}( 1.2,-0.2){$3$}
        \rput{0}( 2.2, 0.7){$2$}
        \rput{0}( 1.8, 2.0){$1$}
        \rput{0}( 0.2, 2.0){$1'$}
        \rput{0}(-0.2, 0.7){$2'$}
        \rput{0}( 1.0,-1.0){(b)}
     }
     \endpspicture

\caption{\label{fig_labels}
   Vertex labels used in the proof of Proposition~\ref{prop2},
   for (a) $m$ even and (b) $m$ odd.
   Vertices with a prime are obtained from the corresponding unprimed
   vertices by reflection $\mathcal{R}_2$.
   When $m=2\ell+1$ is odd, the extra vertex (which is invariant 
   under reflection) is denoted $\ell+1$. 
}
\end{figure}

In what follows, it is convenient to change the notation for the vertices:
When $m=2\ell$ is even, we renumber the vertices
$1,\ldots,\ell,\ell+1,\ldots,2\ell$ as $1,\ldots,\ell,\ell',\ldots,1'$,
so that the vertices $j$ and $j'$ ($1\leq j \leq \ell$)
are transformed into each other under a reflection ${\mathcal R}_2$.
When $m=2\ell+1$ is odd, we renumber the vertices
$1,\ldots,\ell,\ell+1,\ell+2,\ldots,2\ell+1$
as $1,\ldots,\ell,\ell+1,\ell',\ldots,1'$,
so that the vertices $j$ and $j'$ ($1\leq j \leq \ell$)
are again related by reflection,
while the vertex $\ell+1$ is invariant under reflection.
See Figure~\ref{fig_labels} for a few examples.  

\proofof{Proposition~\ref{prop2}}
The idea is to obtain a recursion for the $X(m)$,
and then solve this recursion.
First, we want to obtain $X(2m+1)$ from $X(2m)$.
We use the notation $i \toto j$ to denote that $i$ and $j$ belong to
the same block of the partition being discussed.
We observe that for a partition in $\bm{X}(2m)$,
we have $1 \toto m$ if and only if $1' \toto m'$
(since the partition is ${\mathcal R}_2$-invariant).
We then split the set $\bm{X}(2m)$ into three subsets, as follows:
\begin{itemize}
\item[a)] $\bm{X}_a(2m)$ is the set of partitions in $\bm{X}(2m)$ such that
          there is no block containing both unprimed and primed vertices
          and such that $1 \not\toto m$;
\item[b)] $\bm{X}_b(2m)$ is the set of partitions in $\bm{X}(2m)$ such that
          there is no block containing both unprimed and primed vertices
          and such that $1 \toto m$;
\item[c)] $\bm{X}_c(2m)$ is the set of partitions in $\bm{X}(2m)$ such that
     there is at least one block containing both unprimed and primed vertices.
     [Note that any such block $B$ is necessarily ${\mathcal R}_2$-invariant,
      for otherwise $B$ would cross ${\mathcal R}_2 B$.]
\end{itemize}
We denote by $X_a(2m)$, $X_b(2m)$ and $X_c(2m)$ the number of elements
within the corresponding subset.
For instance, for $\bm{X}(8)$ we have the decomposition
$\bm{X}_a(8)=\{\bone,\delta_{13}\delta_{1'3'},\delta_{24}\delta_{2'4'}\}$,
$\bm{X}_b(8)=\{\delta_{14},\delta_{1'4'}\}$ and 
$\bm{X}_c(8)=\{\delta_{22'},\delta_{33'},\delta_{22'}\delta_{33'}\}$.

Let us suppose that $m\geq 3$. 
Going from $\bm{X}(2m)$ to $\bm{X}(2m+1)$ means that we have to add a new 
vertex labeled $m+1$ between $m$ and $m'$. 
Indeed, if $\mathcal{P}\in\bm{X}(2m)$, then it also belongs to $\bm{X}(2m+1)$
[with the new vertex $m+1$ becoming a singleton].
Furthermore, to any partition
$\mathcal{P}\in\bm{X}_a(2m) \cup \bm{X}_c(2m)$
we can adjoin a factor $\delta_{mm'}$
[i.e.\ contract the blocks containing $m$ and $m'$ into a single block]
and get a partition in $\bm{X}(2m+1)$
[with the new vertex $m+1$ again becoming a singleton];
please note that this is not possible if $\mathcal{P}\in\bm{X}_b(2m)$
since the new partition would include a nearest-neighbor connection
$1 \toto 1'$.  
Finally, if $\mathcal{P}\in\bm{X}_c(2m)$,
we can generate a new partition in $\bm{X}(2m+1)$
by joining the new vertex $m+1$ to the closest block containing
both primed and unprimed vertices.
It is not hard to see that in this way we generate {\em all}\/
the elements of $\bm{X}(2m+1)$, and that we generate each one
exactly once;  therefore,
\begin{eqnarray}
X(2m+1)  & = &  X(2m) + X_a(2m) + 2X_c(2m)  \;.
\end{eqnarray}

Now, $X_a(2m)+X_b(2m)$ is just the total number of partitions in 
$\bm{X}(2m)$ with no block containing both unprimed and primed vertices.
But this number is obviously equal to the number of ncnnn
partitions of the set $\{1,2,\ldots,m\}$ with free boundary conditions,
i.e.\ the Motzkin number $M_{m-1}$:
\be
X_a(2m)+X_b(2m) \;=\; M_{m-1}  \;.
\ee
On the other hand, $X_b(2m)$ can be interpreted as the number of
ncnnn partitions of $\{1,2,\ldots,m\}$ with free boundary 
conditions, such that $m$ and $1$ always belong to the same block. But this 
equivalent to the number of ncnnn partitions of 
$\{1,2,\ldots,m-1\}$ on a circle, so we have
\be
X_b(2m) \;=\; d_{m-1} \;=\; R_{m-1}
\ee
since $m\geq 3$.
Putting everything together we obtain the recursion
\be
X(2m+1) \;=\; 3X(2m) - \left( M_{m-1} + R_{m-1} \right)
    \qquad \hbox{for $m\geq 3$}
\ee
together with the initial conditions $X(1)=1$, $X(3)=1$, $X(5)=2$.
It is interesting to note that the same recursion holds also for $m=1,2$,
as can be checked from Tables~\ref{table_dimensions2}
and \ref{table_dimensions}.
Thus, our final result is
\be
X(2m+1) \;=\; 3 X(2m) - \left( M_{m-1} + R_{m-1} \right)
    \qquad \hbox{for $m\geq 1$}
\label{recurrence_X2m+1}
\ee 
with the initial condition $X(1)=1$.

The second half of the recursion can be obtained by considering how
to obtain the partitions in $\bm{X}(2m+2)$ by starting from those in 
$\bm{X}(2m+1)$.  In this case, the vertex $m+1$ is split into a pair of
nearest-neighbors $m+1$ and $(m+1)'$. Let us assume that $m\geq 2$. 

Consider a partition $\mathcal{P}\in\bm{X}(2m+1)$
in which the block containing the vertex $m+1$
is $\{m+1\} \cup B \cup B'$,
where $B$ contains only unprimed vertices and $B' = {\mathcal R}_2 B$
contains only primed vertices (note that $B$ could be the empty set).
We can then create a partition $\widetilde{\mathcal{P}} \in \bm{X}(2m+2)$
by creating blocks $\{m+1\} \cup B$ and $\{(m+1)'\} \cup B'$
and leaving all other blocks of $\mathcal{P}$ as is.
The partition $\widetilde{\mathcal{P}}$ has the property that,
if one contracts the blocks containing $m+1$ and $(m+1)'$
into a single block, one obtains $\mathcal{P}$;
moreover, $\widetilde{\mathcal{P}}$ is the {\em unique}\/ partition in
$\bm{X}(2m+2)$ with this property
(it is not hard to see that all other ${\mathcal R}_2$-invariant partitions
 with this property would contain either a nearest-neighbor connection
 or a crossing).
Moreover, every partition in $\bm{X}(2m+2)$ is obtained in this way
{\em except}\/ those in which $1 \toto m+1$
[and hence also $1' \toto (m+1)'$];
these latter partitions cannot be obtained
because contracting $m+1$ to $(m+1)'$ would create a
nearest-neighbor connection $1 \toto 1'$.
Now, the number of these extra partitions is equal to the number of ncnnn
partitions of $\{1,2,\ldots,m\}$ on a circle,
which is $d_m$;  and since $m\geq 2$, this also equals $R_m$
[cf.\ \reff{def_dm}].
We therefore conclude that
\be
X(2m+2) \;=\; X(2m+1) + R_m  \qquad\hbox{for $m\geq 2$}
\ee
together with the initial conditions $X(2)=X(4)=1$. It is
interesting to note that the same recursion holds also for $m=1$,
as can be checked from Table~\ref{table_dimensions2}. Thus, our 
final result is
\be
X(2m+2) \;=\; X(2m+1) + R_m  \qquad\hbox{for $m\geq 1$}
\label{recurrence_X2m+2}
\ee
with the initial condition $X(2)= 1$. 

By combining \reff{recurrence_X2m+1}/\reff{recurrence_X2m+2}
and \reff{recursion_Rm},
we can obtain the following recurrences for $X(m)$:
\begin{subeqnarray}
X(2m)   &=& 3 X(2m-2) - 2 R_{m-2}
    \qquad\qquad \hbox{for $m\geq 2$}
\slabel{recurrence_Xeven}\\
X(2m+1) &=& 3 X(2m-1) - R_{m} + R_{m-1} 
    \qquad \hbox{for $m\geq 2$}
\slabel{recurrence_Xodd}
\label{recurrence_Xm}
\end{subeqnarray}
with the initial conditions $X(m)=1$ for $m=1,2,3$.

A solution of \reff{recurrence_Xeven} can be obtained by defining 
$\bar{X}(m)=X(2m+2)$ for $m \ge 0$.
The recurrence \reff{recurrence_Xeven} can then be written
in terms of the $\bar{X}$ as
\be
\bar{X}(m) \;=\; 3 \bar{X}(m-1) - 2 R_{m-1}
   \qquad \hbox{for $m\geq 1$}
\ee
with the initial condition $\bar{X}(0)=1$.
{}From this equation we obtain the corresponding generating function 
\be
\bar{G}(z) \;=\; \sum\limits_{k=0}^\infty \bar{X}(m) z^m  
           \;=\; {1 - 2z R(z) \over 1- 3z } 
           \;=\; {1 \over \sqrt{(1+z)(1-3z)} }
   \;,
\label{def_barG}
\ee
where $R(z)$ is the generating function \reff{def_GRiordan}
for the Riordan numbers.

Let us now observe that \reff{def_barG} coincides with
the generating function of the central trinomial coefficients
\cite[p.~163]{Comtet_book}
\cite[sequence~{\tt A002426}]{Sloane_on-line},
so we have
\be
\bar{X}(m) \;=\; [t^m](1+ t + t^2)^m
   \qquad \hbox{for $m\geq 0$}  \;.
\ee
By a double application of the binomial theorem we arrive at the 
expressions
\be
\bar{X}(m) \;=\; \sum\limits_{k=0}^m {m\choose k} {m-k \choose k} 
   \qquad \hbox{for $m\geq 0$}  \;.
\label{final_barX}
\ee
Note that many terms on the right-hand side of \reff{final_barX} vanish, as 
${m-k\choose k}=0$ whenever $k>m/2$. In order to lighten the
notation, we will use the generic summation over nonnegative $k$,
and only at the end we will make the upper bound explicit. 

The formula for $X(2m)$ and its generating function $G_2(z)$ can be obtained 
from \reff{final_barX}/\reff{def_barG}: 
\begin{subeqnarray}
X(2m) &=& \sum\limits_{k\geq 0} {m-1 \choose k} {m-1-k \choose k}
       \slabel{final_Xeven_a} \\[2mm]
G_2(z)&=& \sum\limits_{m=1}^\infty X(2m)z^{2m} \;=\; 
          z^2 \bar{G}(z^2) \;=\; {z^2 \over \sqrt{(1+z^2)(1-3z^2)}}
       \slabel{final_Xeven_b}
\label{final_Xeven}
\end{subeqnarray}

A solution of \reff{recurrence_Xodd} can be obtained in a similar way
by defining $\widehat{X}(m) = X(2m+1)$ for $m \ge 1$
(we will treat $X(1) = 1$ separately).
The recursion \reff{recurrence_Xodd} can then be written as
\be
\widehat{X}(m) \;=\; 3\widehat{X}(m-1) - R_{m} + R_{m-1}
   \qquad \hbox{for $m\geq 2$}
\ee
with the initial condition $\widehat{X}(1)=1$.
{}From this equation we obtain the corresponding generating function 
\begin{eqnarray}
\widehat{G}(z) \;=\;  \sum\limits_{k=1}^\infty \widehat{X}(m) z^m  
           &=&  {R(z)(z-1)+1 \over 1- 3z } \nonumber \\ 
           &=& {2z \over (1+z)(1-3z) + (1-z)\sqrt{(1+z)(1-3z)} }
  \;\,.
\label{def_hatG}
\end{eqnarray}

We now observe that \reff{def_hatG} coincides with the
generating function of the next-to-central trinomial coefficients
\cite[sequence~{\tt A005717}]{Sloane_on-line},
so we have
\be
\widehat{X}(m) \;=\; [t^{m+1}](1+ t + t^2)^m \;=\; [t^{m-1}](1+ t + t^2)^m
   \qquad \hbox{for $m\geq 1$}
   \;.
\ee
By a double application of the binomial theorem we arrive at the 
expressions
\be
\widehat{X}(m) \;=\; \sum\limits_{k=0}^m {m\choose k} {m-k \choose k+1} 
           \;=\; \sum\limits_{k=0}^m {m\choose k} {m-k \choose k-1}
   \qquad \hbox{for $m\geq 1$}  \;.
\label{final_hatX}
\ee

The formula for $X(2m+1)$ and its generating function $G_1(z)$ can be obtained 
from \reff{final_hatX}/\reff{def_hatG}: 
\begin{subeqnarray}
X(2m+1)&=& \begin{cases}
              1  & \text{for $m=0$}  \\[2mm]
              \displaystyle
               { \sum\limits_{k\geq 0} {m \choose k} {m-k \choose k+1} }
                 & \text{for $m\geq 1$}
           \end{cases}
       \slabel{final_Xodd_a} \\[2mm]
G_1(z) &=& \sum\limits_{m=0}^\infty X(2m+1)z^{2m+1} \;=\; 
          z \,+\, z\widehat{G}(z^2)  \nonumber \\
       &=& z + {2z^3 \over (1+z^2)(1-3z^2) + (1-z^2)\sqrt{(1+z^2)(1-3z^2)}}
       \slabel{final_Xodd_b}
\label{final_Xodd}
\end{subeqnarray}

{}From \reff{final_Xeven_a}/\reff{final_Xodd_a} it is easy to obtain 
\reff{final_Xm}.
{}From \reff{final_Xeven_b}/\reff{final_Xodd_b} it is easy to obtain 
the formula \reff{final_GXm} for the
generating function $G(z)=G_1(z)+G_2(z)$.
\qed

By Proposition~\ref{prop2} together with \reff{eq.N2mXm}/\reff{def_N3_vs_X},
it is easy to compute closed expressions for $N_2(m)$ and $N_3(m)$
as well as for their generating functions. 
We have
\be
N_2(m) \;=\; \left\{ \begin{array}{ll}
        1 & \qquad \text{for $m=1$} \\[2mm]
        \displaystyle
        \smhalf R_m \,+\, \smhalf
        \sum\limits_{k\geq 0} {\lfloor{m-1\over 2}\rfloor   \choose k}
                              {\lfloor{m-1\over 2}\rfloor -k\choose k +
                                             I[\mbox{\rm $m$ is odd}]}
          & \qquad \text{for $m\geq 2$}
                   \end{array}\right.
\label{final_N2}
\ee
and
\be
N_3(m) \;=\; \left\{ \begin{array}{ll}
        1 & \qquad \text{for $m=1$} \\[2mm]
        \displaystyle
        \smhalf R_m \,+\, \smhalf
        \sum\limits_{k\geq 0} {\lfloor{m\over 2}\rfloor   \choose k}
                              {\lfloor{m\over 2}\rfloor -k\choose k +
                                             I[\mbox{\rm $m$ is odd}]}
          & \qquad \text{for $m \geq 2$}
                   \end{array}\right.
\label{final_N3}
\ee
The corresponding generating functions are
\begin{eqnarray}
F_2(z) \;=\; \sum\limits_{m=1}^\infty N_2(m)z^m
   & = & {1\over 2}\left[ G(z) + D(z) \right] \nonumber \\
   & = & {1\over 4z}\left[ {2z^3 -z^2 +1 \over \sqrt{(1+z^2)(1-3z^2)}} 
           - \sqrt{1-3z\over 1+z} + 2z(2z-1) \right] 
      \nonumber \\ \label{final_F2}\\
F_3(z) \;=\; \sum\limits_{m=1}^\infty N_3(m)z^m
   & = & {1\over 2}\left[ G_1(z) + {G_2(z) \over z^2} - 1 + D(z) \right]
      \nonumber \\
   & = & z -1 - {1\over 4z} \sqrt{1-3z\over 1+z} +
           {1+2z -z^2 \over 4z \sqrt{(1+z^2)(1-3z^2)}}
\label{final_F3}
\end{eqnarray}
Note that
\be
   F_3(z) - F_2(z)
   %%  \;=\; {1\over 2}\left[ (1-z^2) \bar{G}(z^2) - 1 \right]
   \;=\; z^2 \widehat{G}(z^2)  \;,
\ee
so that
\be
   N_3(2\ell+2) - N_2(2\ell+2)  \;=\;
   [t^{\ell-1}](1+ t + t^2)^\ell
   \qquad \hbox{for $\ell\geq 0$}
   \;.
\ee

\subsection[The dimensions $N_{\neq}(m)$ and $N_{=}(m)$]%
{The dimensions $\bm{N_{\neq}(m)}$ and $\bm{N_{=}(m)}$}

We can now prove the theorems stated in Section~\ref{sec.block2}. 

\proofof{Theorems~\protect\ref{theorem_Neq} and~\protect\ref{theorem_Nnoteq}}
As explained in Section~\ref{sec.block2}, we have
\begin{subeqnarray}
N_{=}(m)    &=& N_3(m+1) \\
N_{\neq}(m) &=& N_2(m+2)
 \label{eq.B35}
\end{subeqnarray}
The formulae \reff{def_Neq}/\reff{def_Nnoteq}
for $N_{=}(m)$ and $N_{\neq}(m)$
then follow trivially from the formulae \reff{final_N2}/\reff{final_N3}
for $N_2(m)$ and $N_3(m)$.
The results for the generating functions follow likewise from the identities
\begin{subeqnarray}
G_{=}   (z) &=& {1\over z}   \left[ F_3(z) - z \right]  \\
G_{\neq}(z) &=& {1\over z^2} \left[ F_2(z) - z -z^2 \right]
\end{subeqnarray}
\qed

{\bf Remark.}  We see from \reff{property1} and \reff{eq.B35} that
\begin{equation}
   N_{\neq}(m) \;=\; N_{=}(m+1) \quad\hbox{for $m$ odd}
\end{equation}
(cf.~Table~\ref{table_dimensions}).

\bigskip

\proofof{Corollary~\protect\ref{corollary_dimT}}
{}From Theorems~\ref{theorem_Neq} and~\ref{theorem_Nnoteq} we have
\begin{eqnarray}
\dim \T''(m) &=& N_{=}(m) + N_{\neq}(m) \nonumber \\
           &=& {1\over 2}(R_{m+1}+R_{m+2}) + {1\over 2} 
           \sum\limits_{k\geq 0} {m' \choose k} \left\{
           {m'-k \choose k+1} + {m'-k \choose k} \right\} \nonumber\\
           &=& {1\over 2}M_{m+1} + {1\over 2} 
           \sum\limits_{k\geq 0} {m' \choose k} 
           {m'-k +1\choose k+1} 
\end{eqnarray}
where we have used the relation \reff{recursion_Rm} and the
elementary recursion relation for the binomials;
this proves \reff{def_dimT_a}.

To prove \reff{def_dimT_b}, we start from \reff{final_hatX}
and add $\displaystyle{{m \choose k} {m-k \choose k}}$
to the summand on both sides;
using the elementary recursion for binomials,
we get
\be
\sum\limits_{k\geq 0} {m\choose k}{m-k+1 \choose k+1 } \;=\;
\sum\limits_{k\geq 0} {m\choose k}{m-k+1 \choose k } 
  \;.
\label{eq_binom}
\ee
Applying this with $m$ replaced by $m'$ shows the
the equivalence of \reff{def_dimT_b} with \reff{def_dimT_a}.

Finally, by adding \reff{def_dimT_a} and \reff{def_dimT_b},
dividing by 2, and using once again the
recursion relation for binomials, we prove \reff{def_dimT_c}.
\qed

\medskip

\proofof{\reff{def_SqFree}}
We replace $m+2$ by $m$ in \reff{def_dimT_b},
which entails replacing $m'$ by $m'-1$.
Expanding the binomials into factorials yields \reff{def_SqFree}.
\qed

%
% Appendix B
%
\section{Chromatic polynomials for the family $\bm{\widehat{S}_{m,n}}$}
\label{sec.transfer.bis}

We denote by $\widehat{S}_{m,n}$ the graph obtained from
the square-lattice strip of width $m$, length $n$
and free boundary conditions in both directions
by attaching one extra site to all the sites on the {\em top}\/ row
and another extra site to all the sites on the {\em bottom}\/ row
(see Figure~\ref{fig_hatSmn}).
This graph is obviously isomorphic to the graph $S_{n,m}$;
we have merely rotated the picture by 90 degrees.
In particular, the chromatic polynomials of $\widehat{S}_{m,n}$
and $S_{n,m}$ must coincide, i.e.
\begin{equation}
P_{\widehat{S}_{m,n}}(q)  \;=\; P_{S_{n,m}}(q)  \;,
\label{check_poly_bis}
\end{equation}
and we can use this as a check on the correctness of our computations.

Here we shall develop the transfer-matrix formalism
for the family $\widehat{S}_{m,n}$,
where $m$ is fixed (and small) and $n$ is arbitrary;
this is complementary to the formalism developed in
Section~\ref{sec.transfer}, which treats $S_{n,m}$
for $n$ fixed (and small) and $m$ arbitrary.

The graph $\widehat{S}_{1,n}$ is is simply the path $P_{n+2}$,
so its chromatic polynomial is
\be
P_{\widehat{S}_{1,n}}(q) \;=\; P_{P_{n+2}}(q) \;=\; q(q-1)^{n+1}  \;.
\ee

The chromatic polynomial of the graph $\widehat{S}_{2,n}$ can be
computed by noting that this graph is obtained by gluing together
$n-1$ squares $C_4$ and two triangles $C_3$ along edges.
Its chromatic polynomial is therefore
\be
P_{\widehat{S}_{2,n}}(q) \;=\;
   {P_{C_4}(q)^{n-1} \, P_{C_3}(q)^2   \over  [q(q-1)]^n}
               \;=\; q(q-1)(q-2)^2 (q^2 -3q+3)^{n-1}
   \;.
\ee

To handle the graphs $\widehat{S}_{m,n}$ with $m\geq 3$ we need a different 
technique. The idea is to use the standard transfer matrices 
$\T''(m_{\rm F})$ for square-lattice strips
with free boundary conditions\footnote{
   See Section~\ref{sec.setup.transfer} above,
   and see \cite{transfer1} for more details.
   Since $v=-1$, the horizontal transfer matrix $\H$ is a projection,
   so we can use the modified transfer matrix $\T' = \H\V\H$ and the
   modified basis vectors $\basisf_{\scrp} = \H\basise_{\scrp}$.
   Moreover, since the graphs $\widehat{S}_{m,n}$
   are invariant under reflection in the central vertical axis,
   we can use the symmetry-reduced transfer matrix $\T''$.
   These transfer matrices are precisely the ones that can be found
   in \cite[Section~5]{transfer1}
   (but denoted there $T$ rather than $\T''$).
}
but with different initial and final vectors
to take account of the extra sites attached to the top and bottom rows.
More specifically, we have
\be
P_{\widehat{S}_{m,n}} \;=\; \endv_{+}^{\rm T} \T''(m_{\rm F})^{n-1} \basisf_{+}
   \;,
\label{def_P_hatS}
\ee
where the initial vector $\basisf_{+}$ is given by  
\be
\basisf_{+} \;=\; \H \V \basise_{\rm contr}
\label{def_f_plus}
\ee
where $\basise_{\rm contr}$ is the basis vector for the partition
with all sites in the same block,
while the final vector $\endv_+$  
is defined by [compare to \reff{def_uT}] 
\be
   \endv_+^{\rm T} \basise_{\scrp}   \;=\;  q (q-1)^{|\scrp|}
   \;.
 \label{def_uT_plus}
\ee
The reason for this last formula is clear: the factor $q$ comes from the extra
vertex on top, and we obtain a factor of $q-1$ for each block of the
partition $\scrp$ (instead of a factor of~$q$) because each site on the top 
row is attached to the extra site, so that there are only $q-1$ allowed
colors for those sites (rather than $q$).

For instance,
for $m=3$ the chromatic polynomial of the graph $\widehat{S}_{3,n}$ is given
in the basis $\bm{B} = \left\{ \bone, \delta_{13}\right\}$ by\footnote{
   Please recall from Section~\ref{sec.numerical.bm}
   that we are using a shorthand notation for basis elements:
   we write the basis as a collection of vectors $\basise_{\mathcal P}$,
   but it should be understood that the basis vectors are actually
   $\basisf_{\mathcal P} =\H \basise_{\mathcal P}$.
   This convention applies throughout this Appendix.
}
\be
P_{\widehat{S}_{3,n}} \;=\; 
q(q-1)(q-2) \left( \begin{array}{c} q-2 \\ 1
                   \end{array} \right)^{\rm T} \cdot \T''(3_{\rm F})^{n-1} 
            \cdot 
            \left( \begin{array}{c} q-3 \\ q
                   \end{array} \right)
\ee
with
\be
\T''(3_{\rm F}) \;=\; \left( \begin{array}{cc}
                             q^3 - 5 q^2 + 10 q -8 & q^2 - 4 q + 5 \\
                             1                     & q -2
                          \end{array} \right)
  \;.
 \label{eq.T3F}
\ee
For $m=4$, the corresponding chromatic polynomial is given
in the basis $\bm{B} = \{ \bone, \delta_{13}+\delta_{24},\delta_{14} \}$ by
\be
P_{\widehat{S}_{4,n}} \;=\; 
q(q-1)(q-2) \left( \begin{array}{c} (q-2)^2 \\ 2(q-2) \\ q-3
                   \end{array} \right)^{\rm T} \cdot \T''(4_{\rm F})^{n-1}
            \cdot  
            \left( \begin{array}{c} q-4 \\ 1 \\ 1
                   \end{array} \right)
\ee
with
\be
  \T''(4_{\rm F}) =
  \left( \!\! \begin{array}{ccc}
q^4 - 7 q^3 + 21 q^2 - 32 q + 21  & 2 (q^3 - 6 q^2 + 14 q -12) &
                                                   q^3 - 7 q^2 + 19 q -20 \\
q -2                              & q^2 - 4 q +5               & 3 - q    \\
-1                                & -2(q-2)                    & q^2 -5 q + 7
                      \end{array} \!\! \right) .
 \label{eq.T4F}
\ee
We have performed the corresponding computation also for widths $m=5$ and 6,
and we have checked that the identity \reff{check_poly_bis}
holds for $1\leq m \leq 6$ and $1\leq n \leq 8$. 

We can alternatively obtain the chromatic polynomial of the strip graph 
$\widehat{S}_{m,n}$ by using a different technique, which is useful when
dealing with more complicated endgraphs.
Our starting graph consists of a strip together with attached endgraphs
at top and bottom;  each of these endgraphs consists of
one or more extra vertices and several extra edges. 
The idea is to use the deletion-contraction identity on the extra edges of the
endgraphs to get rid of the extra vertices.
At the end, the chromatic polynomial $P_{\widehat{S}_{m,n}}$ can be written 
as a linear combination of chromatic polynomials $P_{S_{m,n,j}}$ associated 
to different strip graphs $S_{m,n,j}$,
each of which is obtained from the $m_{\rm F} \times n_{\rm F}$ grid
by adding only extra {\em edges}\/ to the top and/or bottom rows.
(This approach is closely related to a method developed by
 Ro\v{c}ek, Shrock and Tsai \cite{Shrock_98}.)
These extra edges can be handled by inserting the appropriate operators
$\QQ_{ij} = I + v_{ij} \J_{ij}$ at left and/or right
in the standard expression \reff{Z_free_FK}
for the chromatic polynomial in terms of transfer matrices.
There is, however, one complication:
for $m \ge 4$, these extra edges can give rise to {\em crossing}\/ partitions
(for instance, if in a grid of width $m=4$ we add an extra edge 13
 on the bottom row {\em and}\/ an extra edge 24 on the top row).\footnote{
    Notice that this can happen even though the graph with the extra edges
    is planar (as it is in the just-mentioned example).
    To avoid crossing partitions we need to be able to draw the graph
    in a planar fashion {\em within the $m \times n$ box}\/.
}
We must therefore work in a basis that includes also
(non-nearest-neighbor) {\em crossing}\/ partitions.

For $m=3$ we have
\be
P_{\widehat{S}_{3,n}}(q) \;=\; (q-2)^2 \, P_{3_{\rm F}\times n_{\rm F}}(q)
- 2(q-2)\, P_{S_{3,n,a}}(q) + P_{S_{3,n,b}}(q)
\ee
where $3_{\rm F}\times n_{\rm F}$ is a square-lattice strip of width $m=3$, 
length $n$ and free boundary conditions; 
$S_{3,n,a}$ is the chromatic polynomial of the graph obtained by adding to
$3_{\rm F}\times n_{\rm F}$ an extra edge joining the sites $1$ and $3$ 
on the bottom row; and $S_{3,n,b}$ is the graph obtained by adding to
$S_{3,n,a}$ one further edge joining sites $1$ and $3$ on the top row.
The chromatic polynomials of these three graphs can then be obtained by 
a transfer-matrix formalism. For the graph $3_{\rm F}\times n_{\rm F}$, we
can use the standard method discussed in Section~\ref{sec.setup.transfer}.
For the other two graphs,
we obtain the corresponding chromatic polynomial by inserting
the operators $\QQ_{13}$ appropriately. Thus, the chromatic polynomial for
$\widehat{S}_{3,n}$ can be written as 
\begin{eqnarray}
P_{\widehat{S}_{3,n}}(q) &=& (q-2)^2 \, P_{3_{\rm F}\times n_{\rm F}}(q)
    -2(q-2) \, \endv^{\rm T}\cdot \T''(3_{\rm F})^{n-1}\cdot \QQ_{13} 
            \cdot \basisf_{\rm id}  \nonumber \\
  & & \quad 
+ \endv^{\rm T}\cdot \QQ_{13} \cdot \T''(3_{\rm F})^{n-1}\cdot \QQ_{13} 
      \cdot \basisf_{\rm id} 
\end{eqnarray} 
where $\T''(3_{\rm F})$ is given by \reff{eq.T3F}.
After some algebra we find
\begin{eqnarray}
P_{\widehat{S}_{3,n}}(q) &=& q(q-1)(q-2)\left\{ 
  (q-2)\left( \begin{array}{c} q-1 \\ 1 \end{array}\right)^{\rm T}  
  \cdot \T''(3_{\rm F})^{n-1} \cdot 
  \left( \begin{array}{c} 1 \\ 0 \end{array}\right) \right. \nonumber \\
& & \qquad +  \left.
 \left( \begin{array}{c} 3-2q \\ -2 \end{array}\right)^{\rm T} 
  \cdot \T''(3_{\rm F})^{n-1} \cdot 
  \left( \begin{array}{c} 1 \\ -1 \end{array}\right)
  \right\} 
   \;.
\end{eqnarray} 

For $m=4$ the situation is similar but more involved algebraically. 
We shall work in the basis
$\bm{B} =
 \{ \bone, \delta_{13}+\delta_{24},\delta_{14}, \delta_{13}\delta_{24} \}$,
where the last basis element corresponds to the crossing partition.
After application of the deletion-contraction identity, and some 
algebra, we find
\begin{eqnarray}
P_{\widehat{S}_{4,n}}(q) &=& (q-2)^2 \, P_{4_{\rm F}\times n_{\rm F}}(q)
-2(q-1)\,\endv^{\rm T}          \widetilde{\T}''(4_{\rm F})^{n-1} \QQ_{14} \basisf_{\rm id}
\nonumber \\
& & \quad
-2(q-1)\,\endv^{\rm T}           \widetilde{\T}''(4_{\rm F})^{n-1}
                   (\QQ_{13} + \QQ_{24}) \basisf_{\rm id}
        +\endv^{\rm T} \QQ_{14} \widetilde{\T}''(4_{\rm F})^{n-1} \QQ_{14} \basisf_{\rm id}
\nonumber\\
& & \quad
      +2 \endv^{\rm T} \QQ_{14} \widetilde{\T}''(4_{\rm F})^{n-1}
                   (\QQ_{13} + \QQ_{24})\basisf_{\rm id}
\nonumber\\
& & \quad
      + \endv^{\rm T}(\QQ_{13}+\QQ_{24}) \widetilde{\T}''(4_{\rm F})^{n-1} 
                   (\QQ_{13}+\QQ_{24}) \basisf_{\rm id}
\end{eqnarray}
with the transfer matrix
\be
  \widetilde{\T}''(4_{\rm F}) =
  \left( \!\! \begin{array}{ccc|c}
                   &   &   &  q^2 - 5q + 8 \\
                   & \T''(4_{\rm F})  &     &  q-3    \\
                   &   &   &  -1 \\
              \hline
               0   & 0   &   0  &  1  \vphantom{\Bigl[}
              \end{array} \!\! \right)
 \label{eq.T4Fwidetilde}
\ee
where the upper-left $3 \times 3$ submatrix is given by \reff{eq.T4F}.
Notice that we have symmetrized the operators $\QQ_{13}$ and $\QQ_{24}$
in order to use the symmetrized connectivity basis $\bm{B}$ as above
(i.e., classes of
non-nearest-neighbor connectivities that are invariant
under reflections with respect to the center of the strip).
The definition \reff{def_uT} of the left vector $\endv$
continues to be valid even in the presence of crossing partitions.
After some algebra we find
\begin{eqnarray}
P_{\widehat{S}_{4,n}}(q) &=& q(q-1)\left\{ 
(q-1)\left( 
\begin{array}{c} (q-1)^2 \\ 2(q-1) \\ q-2 \\ 1 \end{array}\right)^{\rm T}  
\cdot \widetilde{\T}''(4_{\rm F})^{n-1} \cdot 
\left( 
\begin{array}{c} q-7 \\ 2 \\ 2 \\ 0 \end{array}\right) \right. \nonumber \\
& & \qquad +  
 \left( 
\begin{array}{c} q^2-3q+3 \\ 2(q-1) \\ 0 \\ 1 \end{array}\right)^{\rm T} 
\cdot \widetilde{\T}''(4_{\rm F})^{n-1} \cdot 
\left( 
\begin{array}{c} 5 \\ -2 \\ -1 \\ 0 \end{array}\right) \nonumber \\
& & \qquad +  \left.
 2(q-2) \left( 
\begin{array}{c} q-1 \\ 1 \\ 1 \\ 0 \end{array}\right)^{\rm T} 
\cdot \widetilde{\T}''(4_{\rm F})^{n-1} \cdot 
\left( 
\begin{array}{c} 2 \\ -1 \\ 0 \\ 0 \end{array}\right) \
  \right\} 
   \;.
\end{eqnarray} 
Note that in this final expression the crossing partitions
play no role, because all three right vectors have zero
in the last entry, and the transfer matrix \reff{eq.T4Fwidetilde}
has zeros in the last row in all columns other than the last.
So we could have restricted attention to non-crossing partitions, after all!
We suspect that a cancellation of this type occurs whenever
the graph is planar, and that a transfer-matrix description
using only non-crossing partitions can be obtained by adding
extra layers at top and bottom
together with suitable $v=\infty$ vertical and/or horizontal edges.

With this method we have checked the identity \reff{check_poly_bis}
for $1\leq m \leq 4$ and $1\leq n \leq 8$. 

%%%%%%%%%%%%%%%%%%%%%%%%%%%%%%%
% BIBLIOGRAPHY
%%%%%%%%%%%%%%%%%%%%%%%%%%%%%%
%%\clearpage

\clearpage
%%%%%%%%%%%%%%%%%%%%%%%%%%%%%%%%%%%%%%%%%%%%%%%%%%%%%%%%%%%%%%%
%
% BEGIN TABLES
%
%%%%%%%%%%%%%%%%%%%%%%%%%%%%%%%%%%%%%%%%%%%%%%%%%%%%%%%%%%%%%%%
\addcontentsline{toc}{section}{Tables}

%%%%%%%%%%%%%%%%%%%%%%%%%%%%%%%%%%%%%%%%%%%%%%%%%%%%%%%%%%%%%%%
%
% TABLE 1: DIMENSIONS 
%
%%%%%%%%%%%%%%%%%%%%%%%%%%%%%%%%%%%%%%%%%%%%%%%%%%%%%%%%%%%%%%%
\begin{table}
\centering
\begin{tabular}{|r|r|r|r|r|r|r|}
\hline\hline
$m$& $B_m$      & $C_m$    &$M_{m-1}$ &SqFree$(m)$& $N_{\neq}(m-2)$
                                                  & $N_{=}(m-2)$\\
\hline\hline
1  &          1 &        1 &        1 &       1 &        &        \\
2  &          2 &        2 &        1 &       1 &        &        \\
3  &          5 &        5 &        2 &       2 &      1 &     1  \\
4  &         15 &       14 &        4 &       3 &      2 &     1  \\
5  &         52 &       42 &        9 &       7 &      4 &     3  \\
6  &        203 &      132 &       21 &      13 &      9 &     4  \\
7  &        877 &      429 &       51 &      32 &     21 &    11  \\
8  &       4140 &     1430 &      127 &      70 &     49 &    21  \\
9  &      21147 &     4862 &      323 &     179 &    124 &    55  \\
10 &     115975 &    16796 &      835 &     435 &    311 &   124  \\
11 &     678570 &    58786 &     2188 &    1142 &    815 &   327  \\
12 &    4213597 &   208012 &     5798 &    2947 &   2132 &   815  \\
13 &   27644437 &   742900 &    15511 &    7889 &   5712 &  2177  \\
14 &  190899322 &  2674440 &    41835 &   21051 &  15339 &  5712  \\
15 & 1382958545 &  9694845 &   113634 &   57192 &  41727 & 15465  \\
16 &10480142147 & 35357670 &   310572 &  155661 & 113934 & 41727  \\
\hline\hline
\end{tabular}
\vspace{1cm}
\caption{
   Dimension of the transfer matrix.
   For each strip width $m$ we give
   the number $B_m$ of all partitions,
   the number $C_m$ of non-crossing partitions,
   the number $M_{m-1}$ of non-crossing non-nearest-neighbor partitions
   with free boundary conditions,
   and the number $\hbox{\rm SqFree}(m)$ of equivalence classes
   of non-crossing non-nearest-neighbor partitions modulo reflection
   in the center of the strip. The dimension of the transfer matrix
   $\T''(m)$ associated to the strip $S_{m,n}$ is $\mbox{\rm SqFree}(m+2)$.
   We also show the dimensionalities $N_{\neq}(m)$ and $N_{=}(m)$
   of the transfer matrices $\T''_{\neq}(m)$ and $\T''_{=}(m)$, respectively. 
}
\label{table_dimensions}
\end{table}

\clearpage

%%%%%%%%%%%%%%%%%%%%%%%%%%%%%%%%%%%%%%%%%%%%%%%%%%%%%%%%%%%%%%%%%
%
% \neq BC's 
%
%%%%%%%%%%%%%%%%%%%%%%%%%%%%%%%%%%%%%%%%%%%%%%%%%%%%%%%%%%%%%%%%%
%
% TABLE 2: COEFFICIENTS b_k^\neq    
%
\begin{sidewaystable}
\centering
\begin{tabular}{r|rrrrrrrrrrrrrrrr}
\hline\hline\\[-4mm]
$m$& $b^{\neq}_0$   & $b^{\neq}_1$   & $b^{\neq}_2$   & $b^{\neq}_3$& 
     $b^{\neq}_4$   & $b^{\neq}_5$   & $b^{\neq}_6$   & $b^{\neq}_7$& 
     $b^{\neq}_8$   & $b^{\neq}_9$   & $b^{\neq}_{10}$& $b^{\neq}_{11}$& 
     $b^{\neq}_{12}$& $b^{\neq}_{13}$& $b^{\neq}_{14}$& $b^{\neq}_{15}$ 
\\[2mm]
\hline
% DO NOT EDIT *****************
$1$ & $1$ & $3$ & \multicolumn{1}{|r}{$0$} & $0$ & $0$ & $0$ & $0$ & $0$ & $0$ & $0$ & $0$ & $0$ & $0$ & $0$ & $0$ & $0$\\
\cline{4-4}
$2$ & $1$ & $5$ & $8$ & \multicolumn{1}{|r}{$2$} & $-4$ & $4$ & $4$ & $-24$ & $40$ & $0$ & $-128$ & $128$ & $752$ & $-3392$ & $6624$ & $-5184$\\
\cline{5-5}
$3$ & $1$ & $7$ & $19$ & $23$ & \multicolumn{1}{|r}{$7$} & $-11$ & $2$ & $21$ & $-19$ & $-59$ & $139$ & $16$ & $-347$ & $62$ & $335$ & $6886$\\
\cline{6-6}
$4$ & $1$ & $9$ & $34$ & $67$ & $67$ & \multicolumn{1}{|r}{$19$} & $-21$ & $-13$ & $40$ & $20$ & $-75$ & $-166$ & $654$ & $-754$ & $245$ & $1835$\\
\cline{7-7}
$5$ & $1$ & $11$ & $53$ & $143$ & $227$ & $195$ & \multicolumn{1}{|r}{$52$} & $-45$ & $-35$ & $20$ & $156$ & $-110$ & $-324$ & $139$ & $1833$ & $-5085$\\
\cline{8-8}
$6$ & $1$ & $13$ & $76$ & $259$ & $556$ & $749$ & $570$ & \multicolumn{1}{|r}{$138$} & $-101$ & $-78$ & $-20$ & $198$ & $381$ & $-975$ & $-646$ & $3534$\\
\cline{9-9}
$7$ & $1$ & $15$ & $103$ & $423$ & $1138$ & $2056$ & $2429$ & $1666$ & \multicolumn{1}{|r}{$367$} & $-256$ & $-127$ & $-139$ & $129$ & $1064$ & $-228$ & $-3872$\\
\cline{10-10}
$8$ & $1$ & $17$ & $134$ & $643$ & $2073$ & $4666$ & $7345$ & $7775$ & $4870$ & \multicolumn{1}{|r}{$965$} & $-666$ & $-223$ & $-221$ & $-485$ & $1841$ & $3088$\\
\cline{11-11}
$9$ & $1$ & $19$ & $169$ & $927$ & $3477$ & $9337$ & $18225$ & $25582$ & $24638$ & $14219$ & \multicolumn{1}{|r}{$2536$} & $-1811$ & $-350$ & $-188$ & $-1917$ & $1550$\\
\cline{12-12}
$10$ & $1$ & $21$ & $208$ & $1283$ & $5482$ & $17067$ & $39600$ & $68667$ & $87380$ & $77434$ & $41471$ & \multicolumn{1}{|r}{$6619$} & $-4973$ & $-559$ & $426$ & $-4676$\\
\cline{13-13}
$11$ & $1$ & $23$ & $251$ & $1719$ & $8236$ & $29126$ & $78121$ & $160546$ & $251621$ & $293864$ & $241705$ & $120781$ & \multicolumn{1}{|r}{$17181$} & $-13789$ & $-831$ & $2878$\\
% END OF TABLE DATA *****************
\hline\hline
\end{tabular}
\vspace{1cm}
\caption{
   Coefficients $b^{\neq}_k(m)$ of the large-$q$ expansion of the
   dominant eigenvalue $\lambda_{\star,\neq}$.
   For each $1\leq m \leq 11$, we include
   all coefficients $b^{\neq}_k(m)$ up to $k=15$.
   For the whole data set up to $k=40$,
   see the {\sc Mathematica} file {\tt data\_Neq.m}
   included in the on-line version of the paper at arXiv.org.
   Those data points below the staircase-like line satisfy  
   $m\geq m_{\rm min}^{\neq}(k)$ [cf.\ \protect\reff{def_m_min_neq}].
}
\label{table_coef_b_neq}
\end{sidewaystable}

%
% TABLE 3: COEFFICIENTS c_k^\neq   
%
\clearpage
\thispagestyle{empty}
\begin{sidewaystable}
\hspace*{-1cm}
%%%\centering
\begin{tabular}{r|rrrrrrrrrrrrrrr}
\hline\hline\\[-4mm]
$m$& $ c^{\neq}_1$     & $2c^{\neq}_2$     &  $3c^{\neq}_3$    & 
     $4c^{\neq}_4$     & $5c^{\neq}_5$     & $6c^{\neq}_6$     &  
     $7c^{\neq}_7$     & $8c^{\neq}_8$     & $9c^{\neq}_9$     & 
     $10c^{\neq}_{10}$ & $11c^{\neq}_{11}$ & $12c^{\neq}_{12}$ & 
     $13c^{\neq}_{13}$ & $14c^{\neq}_{14}$ & $15c^{\neq}_{15}$ \\[2mm]
\hline
% DO NOT EDIT *****************
$1$ & $-3$ &  \multicolumn{1}{|r}{$-9$} & $-27$ & $-81$ & $-243$ & $-729$ & $-2187$ & $-6561$ & $-19683$ & $-59049$ & $-177147$ & $-531441$ & $-1594323$ & $-4782969$ & $-14348907$\\
\cline{3-3}
$2$ & $-5$ & $-9$ &  \multicolumn{1}{|r}{$-11$} & $-9$ & $-15$ & $-57$ & $-75$ & $447$ & $3085$ & $10791$ & $28309$ & $71055$ & $212077$ & $720263$ & $2329749$\\
\cline{4-4}
$3$ & $-7$ & $-11$ & $-13$ &  \multicolumn{1}{|r}{$-15$} & $-7$ & $103$ & $588$ & $1969$ & $5387$ & $14529$ & $40286$ & $106983$ & $258381$ & $554704$ & $922742$\\
\cline{5-5}
$4$ & $-9$ & $-13$ & $-12$ & $-1$ &  \multicolumn{1}{|r}{$36$} & $128$ & $320$ & $943$ & $3849$ & $15852$ & $58115$ & $195236$ & $623874$ & $1913472$ & $5551353$\\
\cline{6-6}
$5$ & $-11$ & $-15$ & $-11$ & $9$ & $59$ &  \multicolumn{1}{|r}{$171$} & $500$ & $1681$ & $5488$ & $16305$ & $47487$ & $147099$ & $481457$ & $1575678$ & $4990814$\\
\cline{7-7}
$6$ & $-13$ & $-17$ & $-10$ & $19$ & $87$ & $232$ &  \multicolumn{1}{|r}{$596$} & $1683$ & $5273$ & $18063$ & $63270$ & $210208$ & $644943$ & $1849922$ & $5061480$\\
\cline{8-8}
$7$ & $-15$ & $-19$ & $-9$ & $29$ & $115$ & $287$ & $678$ &  \multicolumn{1}{|r}{$1845$} & $6012$ & $20881$ & $69692$ & $219419$ & $668588$ & $2007980$ & $5885716$\\
\cline{9-9}
$8$ & $-17$ & $-21$ & $-8$ & $39$ & $143$ & $342$ & $767$ & $2015$ &  \multicolumn{1}{|r}{$6508$} & $22629$ & $76763$ & $248394$ & $766372$ & $2251389$ & $6294362$\\
\cline{10-10}
$9$ & $-19$ & $-23$ & $-7$ & $49$ & $171$ & $397$ & $856$ & $2177$ & $7004$ &  \multicolumn{1}{|r}{$24707$} & $84890$ & $274117$ & $833229$ & $2418344$ & $6784758$\\
\cline{11-11}
$10$ & $-21$ & $-25$ & $-6$ & $59$ & $199$ & $452$ & $945$ & $2339$ & $7509$ & $26775$ &  \multicolumn{1}{|r}{$92599$} & $298952$ & $906755$ & $2626305$ & $7326784$\\
\cline{12-12}
$11$ & $-23$ & $-27$ & $-5$ & $69$ & $227$ & $507$ & $1034$ & $2501$ & $8014$ & $28833$ & $100330$ &  \multicolumn{1}{|r}{$324291$} & $980827$ & $2823500$ & $7820795$\\
% END OF TABLE DATA *****************
\hline\hline
\end{tabular}
\vspace{1cm}
\caption{
   Coefficients $k c^{\neq}_k(m)$ of the large-$q$ expansion of 
   $\log(q^{-m} \lambda_{\star}^{\neq})$ where $\lambda_{\star,\neq}$ is
   the dominant eigenvalue for $\neq$ boundary conditions. 
   For each $1\leq m \leq 11$, we include
   all coefficients $c^{\neq}_k(m)$ up to $k=15$.
   For the whole data set up to $k=40$,
   see the {\sc Mathematica} file {\tt data\_Neq.m}
   included in the on-line version of the paper at arXiv.org.
   Those data points below the staircase-like line satisfy
   $m\geq m_{\rm min}^{\neq}(k)$ [cf.\ \protect\reff{def_m_min_neq}].
}
\label{table_coef_c_neq}
\end{sidewaystable}

%
% TABLE 4: POLYNOMIALS C_k, E_k, F_k  
%
\clearpage
\thispagestyle{empty}
\begin{table}
\centering
\begin{tabular}{|r|rr|rr|rr|}
\hline\hline\\[-5mm]
$k$ & \multicolumn{2}{c|}{$C_k(m)=\alpha_k m + \beta_k$} 
  & \multicolumn{2}{|c|}{$E_k(m)=\gamma_k m + \delta_k$} 
  & \multicolumn{2}{|c|}{$F_k(m)=\epsilon_k m + \phi_k$} \\[2mm]
\cline{2-7}\\[-5mm]
  & $k \alpha_k $ & $k \beta_k $ & $k \gamma_k $ & $k \delta_k $ & 
    $k \epsilon_k $ & $k \phi_k $ \\[2mm]
\hline
% DO NOT EDIT *****************
$1$ &$-2$ &$-1$ &$-3$ &$4$ &$1$ &$-5$\\
$2$ &$-2$ &$-5$ &$-13$ &$12$ &$11$ &$-17$\\
$3$ &$1$ &$-16$ &$-48$ &$61$ &$49$ &$-77$\\
$4$ &$10$ &$-41$ &$-181$ &$264$ &$191$ &$-305$\\
$5$ &$28$ &$-81$ &$-658$ &$934$ &$686$ &$-1015$\\
$6$ &$55$ &$-98$ &$-2164$ &$2643$ &$2219$ &$-2741$\\
$7$ &$89$ &$55$ &$-6142$ &$5401$ &$6231$ &$-5346$\\
$8$ &$162$ &$719$ &$-13989$ &$4128$ &$14151$ &$-3409$\\
$9$ &$505$ &$2459$ &$-19281$ &$-26021$ &$19786$ &$28480$\\
$10$ &$2058$ &$6195$ &$31592$ &$-165918$ &$-29534$ &$172113$\\
$11$ &$7742$ &$15168$ &$414158$ &$-652120$ &$-406416$ &$667288$\\
$12$ &$25291$ &$46102$ &$2389460$ &$-2362893$ &$-2364169$ &$2408995$\\
$13$ &$73552$ &$171677$ &$11542242$ &$-9775073$ &$-11468690$ &$9946750$\\
$14$ &$197755$ &$647719$& & & & \\
$15$ &$508036$ &$2234229$& & & & \\
$16$ &$1264258$ &$6898415$& & & & \\
$17$ &$2984620$ &$19118828$& & & & \\
% END OF TABLE DATA *****************
\hline\hline
\end{tabular}
\vspace{1cm}
\caption{
   Polynomials $C_k(m)$, $E_k(m)$ and $F_k(m)$ representing for $m\ge k+1$
   the coefficients
   $c_k^\sharp(m)$ [cf.\ \protect\reff{def_ckm}],
   $e_k(m)$ [cf.~\protect\reff{def_series_el}] and 
   $f_k(m)$ [cf.~\protect\reff{def_series_fl}], respectively. 
   All these polynomials are of degree $1$ in $m$, and their coefficients 
   multiplied by $k$ are always integers.
}
\label{table_new_Ck}
\end{table}

%
% TABLE 5: COEFFICIENTS c_k^\neq - MyF[k,L]  
%
\clearpage
\begin{sidewaystable}
\centering
\begin{tabular}{r|rrrrrrrrrrrrrrr}
\hline\hline\\[-4mm]
$m$& $\Delta^{\neq}_1$ & $\Delta^{\neq}_2$& $\Delta^{\neq}_3$& 
     $\Delta^{\neq}_4$ & $\Delta^{\neq}_5$& 
     $\Delta^{\neq}_6$ & $\Delta^{\neq}_7$& $\Delta^{\neq}_8$& 
     $\Delta^{\neq}_9$ & $\Delta^{\neq}_{10}$ &
     $\Delta^{\neq}_{11}$ & $\Delta^{\neq}_{12}$ & 
     $\Delta^{\neq}_{13}$ & 
     $\Delta^{\neq}_{14}$ & $\Delta^{\neq}_{15}$ 
\\[2mm]
\hline\\[-4mm]
% DO NOT EDIT *****************
$1$ & $0$ &  \multicolumn{1}{|r}{$-1$} & \multicolumn{1}{r|}{$-4$} & $-\frac{25}{2}$ & $-38$ & $-\frac{343}{3}$ & $-333$ & $-\frac{3721}{4}$ & $-\frac{7549}{3}$ & $-\frac{33651}{5}$ & $-18187$ & $-\frac{301417}{6}$ & $-141504$ & $-\frac{5628443}{14}$ & $-\frac{17091172}{15}$\\[2mm]
\cline{3-3}\cline{5-6}
$2$ & $0$ & $0$ &  \multicolumn{1}{|r}{$1$} & $3$ & \multicolumn{1}{r|}{$2$} & $-\frac{23}{2}$ & $-44$ & $-\frac{149}{2}$ & $-\frac{128}{3}$ & $48$ & $-213$ & $-\frac{8543}{4}$ & $-8208$ & $-23069$ & $-\frac{920552}{15}$\\[2mm]
\cline{4-4}\cline{7-8}
$3$ & $0$ & $0$ & $0$ &  \multicolumn{1}{|r}{$-1$} & $-2$ & $6$ & \multicolumn{1}{r|}{$38$} & $\frac{191}{2}$ & $157$ & $216$ & $172$ & $-\frac{3748}{3}$ & $-10304$ & $-49020$ & $-\frac{567119}{3}$\\[2mm]
\cline{5-5}\cline{9-10}
$4$ & $0$ & $0$ & $0$ & $0$ &  \multicolumn{1}{|r}{$1$} & $1$ & $-13$ & $-53$ & \multicolumn{1}{r|}{$-70$} & $\frac{285}{2}$ & $1089$ & $\frac{7995}{2}$ & $12153$ & $\frac{67819}{2}$ & $\frac{256996}{3}$\\[2mm]
\cline{6-6}\cline{11-12}
$5$ & $0$ & $0$ & $0$ & $0$ & $0$ &  \multicolumn{1}{|r}{$-1$} & $0$ & $19$ & $56$ & $-18$ & \multicolumn{1}{r|}{$-581$} & $-\frac{4243}{2}$ & $-4460$ & $-4344$ & $14427$\\[2mm]
\cline{7-7}\cline{13-13}
$6$ & $0$ & $0$ & $0$ & $0$ & $0$ & $0$ &  \multicolumn{1}{|r}{$1$} & $-1$ & $-24$ & $-48$ & $150$ & \multicolumn{1}{r|}{$1030$} & $2458$ & $\frac{2239}{2}$ & $-14731$\\[2mm]
\cline{8-8}\cline{14-14}
$7$ & $0$ & $0$ & $0$ & $0$ & $0$ & $0$ & $0$ &  \multicolumn{1}{|r}{$-1$} & $2$ & $28$ & $30$ & $-310$ & \multicolumn{1}{r|}{$-1381$} & $-1716$ & $6349$\\[2mm]
\cline{9-9}\cline{15-15}
$8$ & $0$ & $0$ & $0$ & $0$ & $0$ & $0$ & $0$ & $0$ &  \multicolumn{1}{|r}{$1$} & $-3$ & $-31$ & $-3$ & $483$ & \multicolumn{1}{r|}{$1545$} & $-277$\\[2mm]
\cline{10-10}\cline{16-16}
$9$ & $0$ & $0$ & $0$ & $0$ & $0$ & $0$ & $0$ & $0$ & $0$ &  \multicolumn{1}{|r}{$-1$} & $4$ & $33$ & $-32$ & $-655$ & \multicolumn{1}{r|}{$-1453$}\\[2mm]
\cline{11-11}
$10$ & $0$ & $0$ & $0$ & $0$ & $0$ & $0$ & $0$ & $0$ & $0$ & $0$ &  \multicolumn{1}{|r}{$1$} & $-5$ & $-34$ & $74$ & $813$\\[2mm]
\cline{12-12}
$11$ & $0$ & $0$ & $0$ & $0$ & $0$ & $0$ & $0$ & $0$ & $0$ & $0$ & $0$ &  \multicolumn{1}{|r}{$-1$} & $6$ & $34$ & $-122$\\[2mm]
% END OF TABLE DATA *****************
\hline\hline
\end{tabular}
\vspace{1cm}
\caption{
   Coefficients $\Delta^{\neq}_k(m)$ 
   for $\neq$ boundary conditions [cf.\ \protect\reff{def_Delta}]
   for $1\leq k \leq 15$ and $1\leq m\leq 11$. 
   Those data points below the lower staircase-like line satisfy
   $m\geq m_{\rm min}^{\neq}(k)$ [cf.\ \protect\reff{def_m_min_neq}]
   and are therefore zero.
   Those data points between the two staircase-like lines can be
   fitted to a polynomial Ansatz
   [cf.\ \protect\reff{eq.Deltaneq.1}--\protect\reff{eq.Deltaneq.6}].   
}
\label{table_diff_coef_c_neq}
\end{sidewaystable}

\clearpage
%%%%%%%%%%%%%%%%%%%%%%%%%%%%%%%%%%%%%%%%%%%%%%%%%%%%%%%%%%%%%%%%%
%
% = BC's 
%
%%%%%%%%%%%%%%%%%%%%%%%%%%%%%%%%%%%%%%%%%%%%%%%%%%%%%%%%%%%%%%%%%
%
% TABLE 6: COEFFICIENTS b_k^=    
%
\thispagestyle{empty}
\begin{sidewaystable}
\centering
\begin{tabular}{r|rrrrrrrrrrrrrrrr}
\hline\hline\\[-4mm]
$m$& $b^{=}_0$   & $b^{=}_1$   & $b^{=}_2$   & $b^{=}_3$& 
     $b^{=}_4$   & $b^{=}_5$   & $b^{=}_6$   & $b^{=}_7$& 
     $b^{=}_8$   & $b^{=}_9$   & $b^{=}_{10}$& $b^{=}_{11}$& 
     $b^{=}_{12}$& $b^{=}_{13}$& $b^{=}_{14}$& $b^{=}_{15}$ 
\\[2mm]
\hline
% DO NOT EDIT *****************
$1$ & $1$ &  \multicolumn{1}{|r}{$2$} & $0$ & $0$ & $0$ & $0$ & $0$ & $0$ & $0$ & $0$ & $0$ & $0$ & $0$ & $0$ & $0$ & $0$\\
\cline{3-3}
$2$ & $1$ & $5$ &  \multicolumn{1}{|r}{$7$} & $0$ & $0$ & $0$ & $0$ & $0$ & $0$ & $0$ & $0$ & $0$ & $0$ & $0$ & $0$ & $0$\\
\cline{4-4}
$3$ & $1$ & $7$ & $19$ &  \multicolumn{1}{|r}{$22$} & $2$ & $-10$ & $20$ & $-8$ & $-59$ & $111$ & $226$ & $-1478$ & $3029$ & $-1922$ & $-359$ & $-26461$\\
\cline{5-5}
$4$ & $1$ & $9$ & $34$ & $67$ &  \multicolumn{1}{|r}{$66$} & $11$ & $-33$ & $18$ & $74$ & $-109$ & $-175$ & $410$ & $884$ & $-3801$ & $4508$ & $-7375$\\
\cline{6-6}
$5$ & $1$ & $11$ & $53$ & $143$ & $227$ & \multicolumn{1}{|r}{$194$} & $41$ & $-79$ & $-25$ & $166$ & $92$ & $-567$ & $-49$ & $1555$ & $-957$ & $-366$\\
\cline{7-7}
$6$ & $1$ & $13$ & $76$ & $259$ & $556$ & $749$ &  \multicolumn{1}{|r}{$569$} & $124$ & $-166$ & $-149$ & $215$ & $541$ & $-495$ & $-1832$ & $2683$ & $2725$\\
\cline{8-8}
$7$ & $1$ & $15$ & $103$ & $423$ & $1138$ & $2056$ & $2429$ &  \multicolumn{1}{|r}{$1665$} & $350$ & $-361$ & $-366$ & $1$ & $1282$ & $766$ & $-4301$ & $-1274$\\
\cline{9-9}
$8$ & $1$ & $17$ & $134$ & $643$ & $2073$ & $4666$ & $7345$ & $7775$ &  \multicolumn{1}{|r}{$4869$} & $945$ & $-820$ & $-744$ & $-603$ & $1495$ & $4652$ & $-4003$\\
\cline{10-10}
$9$ & $1$ & $19$ & $169$ & $927$ & $3477$ & $9337$ & $18225$ & $25582$ & $24638$ &  \multicolumn{1}{|r}{$14218$} & $2513$ & $-2023$ & $-1294$ & $-1843$ & $-155$ & $10811$\\
\cline{11-11}
$10$ & $1$ & $21$ & $208$ & $1283$ & $5482$ & $17067$ & $39600$ & $68667$ & $87380$ & $77434$ &  \multicolumn{1}{|r}{$41470$} & $6593$ & $-5252$ & $-2094$ & $-3658$ & $-6157$\\
\cline{12-12}
$11$ & $1$ & $23$ & $251$ & $1719$ & $8236$ & $29126$ & $78121$ & $160546$ & $251621$ & $293864$ & $241705$ &  \multicolumn{1}{|r}{$120780$} & $17152$ & $-14144$ & $-3152$ & $-5277$\\
\cline{13-13}
$12$ & $1$ & $25$ & $298$ & $2243$ & $11903$ & $47088$ & $143119$ & $339167$ & $628439$ & $901807$ & $975816$ & $750058$ &  \multicolumn{1}{|r}{$351253$} & $44217$ & $-38862$ & $-4178$\\
% END OF TABLE DATA *****************
\hline\hline
\end{tabular}
\vspace{1cm}
\caption{
   Coefficients $b^{=}_k(m)$ of the large-$q$ expansion of the
   dominant eigenvalue $\lambda_{\star, =}$.
   For each $1\leq m \leq 12$, we include
   all coefficients $b^{=}_k(m)$ up to $k=15$.
   For the whole data set up to $k=40$,
   see the {\sc Mathematica} file {\tt data\_Eq.m}
   included in the on-line version of the paper at arXiv.org.
   Those data points below the staircase-like line satisfy  
   $m\geq m_{\rm min}^{=}(k)$ [cf.\ \protect\reff{def_m_min_eq}].
}
\label{table_coef_b_eq}
\end{sidewaystable}

%
% TABLE 7: COEFFICIENTS c_k^=   
%
\clearpage
\begin{sidewaystable}
\centering
\begin{tabular}{r|rrrrrrrrrrrrrrr}
\hline\hline\\[-4mm]
$m$& $ c^{=}_1$     & $2c^{=}_2$     &  $3c^{=}_3$    & 
     $4c^{=}_4$     & $5c^{=}_5$     & $6c^{=}_6$     &  
     $7c^{=}_7$     & $8c^{=}_8$     & $9c^{=}_9$     & 
     $10c^{=}_{10}$ & $11c^{=}_{11}$ & $12c^{=}_{12}$ & 
     $13c^{=}_{13}$ & $14c^{=}_{14}$ & $15c^{=}_{15}$ \\[2mm]
\hline
% DO NOT EDIT *****************
$1$ &  \multicolumn{1}{|r}{$-2$} & $-4$ & $-8$ & $-16$ & $-32$ & $-64$ & $-128$ & $-256$ & $-512$ & $-1024$ & $-2048$ & $-4096$ & $-8192$ & $-16384$ & $-32768$\\
\cline{2-2}
$2$ & $-5$ &  \multicolumn{1}{|r}{$-11$} & $-20$ & $-23$ & $25$ & $286$ & $1255$ & $4273$ & $12580$ & $32989$ & $76885$ & $153502$ & $229315$ & $72061$ & $-1244900$\\
\cline{3-3}
$3$ & $-7$ & $-11$ &  \multicolumn{1}{|r}{$-10$} & $-7$ & $-37$ & $-134$ & $-63$ & $1209$ & $5732$ & $17459$ & $58194$ & $250586$ & $1073039$ & $3960589$ & $12879935$\\
\cline{4-4}
$4$ & $-9$ & $-13$ & $-12$ &  \multicolumn{1}{|r}{$-5$} & $31$ & $206$ & $691$ & $1491$ & $2643$ & $5947$ & $17371$ & $54094$ & $211345$ & $1009373$ & $4696918$\\
\cline{5-5}
$5$ & $-11$ & $-15$ & $-11$ & $9$ &  \multicolumn{1}{|r}{$64$} & $171$ & $367$ & $1233$ & $5650$ & $22100$ & $70983$ & $202131$ & $541595$ & $1384095$ & $3268114$\\
\cline{6-6}
$6$ & $-13$ & $-17$ & $-10$ & $19$ & $87$ &  \multicolumn{1}{|r}{$226$} & $603$ & $1875$ & $5705$ & $16563$ & $51940$ & $180694$ & $630656$ & $2058179$ & $6217815$\\
\cline{7-7}
$7$ & $-15$ & $-19$ & $-9$ & $29$ & $115$ & $287$ &  \multicolumn{1}{|r}{$685$} & $1829$ & $5760$ & $20581$ & $73102$ & $235991$ & $690896$ & $1919073$ & $5277046$\\
\cline{8-8}
$8$ & $-17$ & $-21$ & $-8$ & $39$ & $143$ & $342$ & $767$ &  \multicolumn{1}{|r}{$2007$} & $6535$ & $22939$ & $76796$ & $242598$ & $746287$ & $2255267$ & $6547532$\\
\cline{9-9}
$9$ & $-19$ & $-23$ & $-7$ & $49$ & $171$ & $397$ & $856$ & $2177$ &  \multicolumn{1}{|r}{$7013$} & $24667$ & $84527$ & $274501$ & $841744$ & $2438686$ & $6732003$\\
\cline{10-10}
$10$ & $-21$ & $-25$ & $-6$ & $59$ & $199$ & $452$ & $945$ & $2339$ & $7509$ &  \multicolumn{1}{|r}{$26765$} & $92654$ & $299360$ & $905793$ & $2614923$ & $7310959$\\
\cline{11-11}
$11$ & $-23$ & $-27$ & $-5$ & $69$ & $227$ & $507$ & $1034$ & $2501$ & $8014$ & $28833$ &  \multicolumn{1}{|r}{$100341$} & $324219$ & $980385$ & $2825208$ & $7834970$\\
\cline{12-12}
$12$ & $-25$ & $-29$ & $-4$ & $79$ & $255$ & $562$ & $1123$ & $2663$ & $8519$ & $30891$ & $108072$ &  \multicolumn{1}{|r}{$349582$} & $1054405$ & $3021143$ & $8327541$\\
% END OF TABLE DATA *****************
\hline\hline
\end{tabular}
\vspace{1cm}
\caption{
   Coefficients $k c^{=}_k(m)$ of the large-$q$ expansion of 
   $\log(q^{-m} \lambda_{\star}^{=})$ where $\lambda_{\star, =}$ is
   the dominant eigenvalue for $=$ boundary conditions. 
   For each $1\leq m \leq 12$, we include
   all coefficients $c^{=}_k(m)$ up to $k=15$.
   For the whole data set up to $k=40$,
   see the {\sc Mathematica} file {\tt data\_Eq.m}
   included in the on-line version of the paper at arXiv.org.
   Those data points below the staircase-like line satisfy
   $m\geq m_{\rm min}^{=}(k)$ [cf.\ \protect\reff{def_m_min_eq}].
}
\label{table_coef_c_eq}
\end{sidewaystable}

%
% TABLE 8: COEFFICIENTS c_k^= - MyF[k,L]  
%
\clearpage
\begin{sidewaystable}
\centering
\begin{tabular}{r|rrrrrrrrrrrrrrr}
\hline\hline\\[-4mm]
$m$& $\Delta^{=}_1$ & $\Delta^{=}_2$& $\Delta^{=}_3$& 
     $\Delta^{=}_4$ & $\Delta^{=}_5$& 
     $\Delta^{=}_6$ & $\Delta^{=}_7$& $\Delta^{=}_8$& 
     $\Delta^{=}_9$ & $\Delta^{=}_{10}$ &
     $\Delta^{=}_{11}$ & $\Delta^{=}_{12}$ & 
     $\Delta^{=}_{13}$ & 
     $\Delta^{=}_{14}$ & $\Delta^{=}_{15}$ 
\\[2mm]
\hline\\[-4mm]
% DO NOT EDIT *****************
$1$ &  \multicolumn{1}{|r|}{$1$} & $\frac{3}{2}$ & $\frac{7}{3}$ & $\frac{15}{4}$ & $\frac{21}{5}$ & $-\frac{7}{2}$ & $-\frac{272}{7}$ & $-\frac{1137}{8}$ & $-\frac{3476}{9}$ & $-\frac{9277}{10}$ & $-\frac{24958}{11}$ & $-\frac{25163}{4}$ & $-\frac{253421}{13}$ & $-\frac{430929}{7}$ & $-\frac{925011}{5}$\\[2mm]
\cline{2-4}
$2$ & $0$ &  \multicolumn{1}{|r}{$-1$} & \multicolumn{1}{r|}{$-2$} & $-\frac{1}{2}$ & $10$ & $\frac{137}{3}$ & $146$ & $\frac{1615}{4}$ & $\frac{3037}{3}$ & $\frac{11339}{5}$ & $4203$ & $\frac{28409}{6}$ & $-6882$ & $-\frac{485584}{7}$ & $-\frac{4495201}{15}$\\[2mm]
\cline{3-3}\cline{5-6}
$3$ & $0$ & $0$ &  \multicolumn{1}{|r}{$1$} & $1$ & \multicolumn{1}{r|}{$-8$} & $-\frac{67}{2}$ & $-55$ & $\frac{1}{2}$ & $\frac{586}{3}$ & $509$ & $1800$ & $\frac{128611}{12}$ & $52362$ & $\frac{388515}{2}$ & $\frac{9121598}{15}$\\[2mm]
\cline{4-4}\cline{7-8}
$4$ & $0$ & $0$ & $0$ &  \multicolumn{1}{|r}{$-1$} & $0$ & $14$ & \multicolumn{1}{r|}{$40$} & $\frac{31}{2}$ & $-204$ & $-848$ & $-2615$ & $-\frac{23293}{3}$ & $-19580$ & $-30669$ & $28703$\\[2mm]
\cline{5-5}\cline{9-10}
$5$ & $0$ & $0$ & $0$ & $0$ &  \multicolumn{1}{|r}{$1$} & $-1$ & $-19$ & $-37$ & \multicolumn{1}{r|}{$74$} & $\frac{1123}{2}$ & $1555$ & $\frac{4929}{2}$ & $166$ & $-\frac{36057}{2}$ & $-\frac{301259}{3}$\\[2mm]
\cline{6-6}\cline{11-12}
$6$ & $0$ & $0$ & $0$ & $0$ & $0$ &  \multicolumn{1}{|r}{$-1$} & $2$ & $23$ & $24$ & $-198$ & \multicolumn{1}{r|}{$-880$} & $-\frac{2859}{2}$ & $1359$ & $15995$ & $62358$\\[2mm]
\cline{7-7}\cline{13-13}
$7$ & $0$ & $0$ & $0$ & $0$ & $0$ & $0$ &  \multicolumn{1}{|r}{$1$} & $-3$ & $-26$ & $-2$ & $340$ & \multicolumn{1}{r|}{$1071$} & $335$ & $-\frac{16133}{2}$ & $-34229$\\[2mm]
\cline{8-8}\cline{14-14}
$8$ & $0$ & $0$ & $0$ & $0$ & $0$ & $0$ & $0$ &  \multicolumn{1}{|r}{$-1$} & $4$ & $28$ & $-28$ & $-486$ & \multicolumn{1}{r|}{$-1062$} & $1822$ & $16601$\\[2mm]
\cline{9-9}\cline{15-15}
$9$ & $0$ & $0$ & $0$ & $0$ & $0$ & $0$ & $0$ & $0$ &  \multicolumn{1}{|r}{$1$} & $-5$ & $-29$ & $65$ & $623$ & \multicolumn{1}{r|}{$798$} & $-4970$\\[2mm]
\cline{10-10}\cline{16-16}
$10$ & $0$ & $0$ & $0$ & $0$ & $0$ & $0$ & $0$ & $0$ & $0$ &  \multicolumn{1}{|r}{$-1$} & $6$ & $29$ & $-108$ & $-739$ & \multicolumn{1}{r|}{$-242$}\\[2mm]
\cline{11-11}
$11$ & $0$ & $0$ & $0$ & $0$ & $0$ & $0$ & $0$ & $0$ & $0$ & $0$ &  \multicolumn{1}{|r}{$1$} & $-7$ & $-28$ & $156$ & $823$\\[2mm]
\cline{12-12}
$12$ & $0$ & $0$ & $0$ & $0$ & $0$ & $0$ & $0$ & $0$ & $0$ & $0$ & $0$ &  \multicolumn{1}{|r}{$-1$} & $8$ & $26$ & $-208$\\[2mm]
% END OF TABLE DATA *****************
\hline\hline
\end{tabular}
\vspace{1cm}
\caption{
   Coefficients $\Delta^{=}_k(m)$ 
   for $=$ boundary conditions [cf.\ \protect\reff{def_Delta}]
   for $1\leq k \leq 15$ and $1\leq m\leq 12$. 
   Those data points below the lower staircase-like line satisfy
   $m\geq m_{\rm min}^{=}(k)$ [cf.\ \protect\reff{def_m_min_eq}]
   and are therefore zero.
   Those data points between the two staircase-like lines can be
   fitted to a polynomial Ansatz
   [cf.\ \protect\reff{eq.Deltaeq.1}--\protect\reff{eq.Deltaeq.3}].
}
\label{table_diff_coef_c_eq}
\end{sidewaystable}

%%%%%%%%%%%%%%%%%%%%%%%%%%%%%%%%%%%%%%%%%%%%%%%%%%%%%%%%%%%%%%%%%
%
% = - \neq BC'S 
%
%%%%%%%%%%%%%%%%%%%%%%%%%%%%%%%%%%%%%%%%%%%%%%%%%%%%%%%%%%%%%%%%%
%
% TABLE 9: COEFFICIENTS d_k 
%
\clearpage
\thispagestyle{empty}
\begin{sidewaystable}
\centering
\begin{tabular}{r|rrrrrrrrrrrrr}
\hline\hline\\[-4mm]
$m$& $d_1$    & $d_2$    & $d_3$    &  $d_4$ & $d_5$    & 
     $d_6$    & $d_7$    & $d_8$    &  $d_9$ & $d_{10}$ &
     $d_{11}$ & $d_{12}$ & $d_{13}$  
\\[2mm]
\hline\\[-4mm]
% DO NOT EDIT *****************
$1$ & $0$ & $0$ & $0$ & $0$ & $0$ & $0$ & $0$ & $0$ & $0$ & $0$ & $0$ & $0$ & $0$\\
\cline{2-2}
$2$ & \multicolumn{1}{r|}{$2$} & $-4$ & $4$ & $4$ & $-24$ & $40$ & $0$ & $-128$ & $128$ & $752$ & $-3392$ & $6624$ & $-5184$\\
\cline{3-3}
$3$ & $5$ & \multicolumn{1}{r|}{$-1$} & $-18$ & $29$ & $40$ & $-170$ & $-87$ & $1494$ & $-3376$ & $1984$ & $694$ & $33347$ & $-197615$\\
\cline{4-4}
$4$ & $8$ & $12$ & \multicolumn{1}{r|}{$-31$} & $-34$ & $129$ & $100$ & $-576$ & $-230$ & $3047$ & $-4263$ & $9210$ & $-64867$ & $238639$\\
\cline{5-5}
$5$ & $11$ & $34$ & $-10$ & \multicolumn{1}{r|}{$-146$} & $64$ & $457$ & $-275$ & $-1416$ & $2790$ & $-4719$ & $5981$ & $47388$ & $-300709$\\
\cline{6-6}
$6$ & $14$ & $65$ & $71$ & $-235$ & \multicolumn{1}{r|}{$-343$} & $876$ & $857$ & $-3329$ & $809$ & $13186$ & $-49587$ & $89187$ & $-19$\\
\cline{7-7}
$7$ & $17$ & $105$ & $239$ & $-140$ & $-1153$ & \multicolumn{1}{r|}{$298$} & $4073$ & $-2598$ & $-12636$ & $26869$ & $5107$ & $-135403$ & $331386$\\
\cline{8-8}
$8$ & $20$ & $154$ & $521$ & $382$ & $-1980$ & $-2811$ & \multicolumn{1}{r|}{$7091$} & $9054$ & $-29248$ & $-9219$ & $120387$ & $-160314$ & $-41468$\\
\cline{9-9}
$9$ & $23$ & $212$ & $944$ & $1655$ & $-1762$ & $-9261$ & $2518$ & \multicolumn{1}{r|}{$35377$} & $-15992$ & $-117784$ & $152402$ & $218878$ & $-752711$\\
\cline{10-10}
$10$ & $26$ & $279$ & $1535$ & $4084$ & $1481$ & $-17296$ & $-21291$ & $60152$ & \multicolumn{1}{r|}{$83981$} & $-236658$ & $-170823$ & $927069$ & $-457202$\\
\cline{11-11}
$11$ & $29$ & $355$ & $2321$ & $8155$ & $10890$ & $-20003$ & $-74321$ & $29922$ & $301365$ & \multicolumn{1}{r|}{$-105940$} & $-1047643$ & $906441$ & $2674932$\\
% END OF TABLE DATA *****************
\hline\hline
\end{tabular}
\vspace{1cm}
\caption{
   Coefficients $d_\ell(m)$ [cf.\ \protect\reff{def_series_dl}]
   for $1\leq \ell \leq 13$ and $1\leq m\leq 11$. 
   For the whole data set up to $\ell=20$,
   see the {\sc Mathematica} file {\tt data\_Diff.m}
   included in the on-line version of the paper at arXiv.org.
   Those data points below the lower staircase-like line satisfy
   $m\geq \ell+1$. 
}
\label{table_diff_coef_d}
\end{sidewaystable}

%
% TABLE 10: COEFFICIENTS ell * e_ell(k)   
%
\clearpage
\thispagestyle{empty}
\begin{sidewaystable}
\centering
\begin{tabular}{r|rrrrrrrrrrrrr}
\hline\hline\\[-4mm]
$m$& $ e_1$     & $2e_2$     & $3e_3$     & 
     $4e_4$     & $5e_5$     & $6e_6$     &  
     $7e_7$     & $8e_8$     & $9e_9$     & 
     $10e_{10}$ & $11e_{11}$ & $12e_{12}$ & 
     $13e_{13}$  \\[2mm]
\hline
% DO NOT EDIT *****************
$1$ & $0$ & $0$ & $0$ & $0$ & $0$ & $0$ & $0$ & $0$ & $0$ & $0$ & $0$ & $0$ & $0$\\
\cline{2-2}
$2$ & \multicolumn{1}{r|}{$-2$} & $-12$ & $-44$ & $-128$ & $-352$ & $-1056$ & $-3488$ & $-11584$ & $-36512$ & $-109792$ & $-326944$ & $-993536$ & $-3077440$\\
\cline{3-3}
$3$ & $-5$ & \multicolumn{1}{r|}{$-27$} & $-86$ & $-251$ & $-910$ & $-3690$ & $-13669$ & $-44019$ & $-125411$ & $-312552$ & $-557969$ & $452254$ & $12297228$\\
\cline{4-4}
$4$ & $-8$ & $-40$ & \multicolumn{1}{r|}{$-131$} & $-456$ & $-1753$ & $-6283$ & $-19874$ & $-54888$ & $-117392$ & $-59295$ & $1300181$ & $10433589$ & $57898495$\\
\cline{5-5}
$5$ & $-11$ & $-53$ & $-179$ & \multicolumn{1}{r|}{$-641$} & $-2361$ & $-8087$ & $-24847$ & $-63873$ & $-106973$ & $100877$ & $1976447$ & $11655889$ & $52017110$\\
\cline{6-6}
$6$ & $-14$ & $-66$ & $-227$ & $-822$ & \multicolumn{1}{r|}{$-3014$} & $-10335$ & $-31584$ & $-80518$ & $-146018$ & $-21616$ & $1451227$ & $9451641$ & $45388641$\\
\cline{7-7}
$7$ & $-17$ & $-79$ & $-275$ & $-1003$ & $-3672$ & \multicolumn{1}{r|}{$-12505$} & $-37600$ & $-93611$ & $-159962$ & $63596$ & $2351805$ & $15369983$ & $78870073$\\
\cline{8-8}
$8$ & $-20$ & $-92$ & $-323$ & $-1184$ & $-4330$ & $-14669$ & \multicolumn{1}{r|}{$-43735$} & $-107776$ & $-180512$ & $85408$ & $2646349$ & $16541923$ & $80306297$\\
\cline{9-9}
$9$ & $-23$ & $-105$ & $-371$ & $-1365$ & $-4988$ & $-16833$ & $-49877$ & \multicolumn{1}{r|}{$-121773$} & $-199559$ & $118720$ & $3077172$ & $19166619$ & $94490152$\\
\cline{10-10}
$10$ & $-26$ & $-118$ & $-419$ & $-1546$ & $-5646$ & $-18997$ & $-56019$ & $-135762$ & \multicolumn{1}{r|}{$-218831$} & $150012$ & $3489075$ & $21529295$ & $105609348$\\
\cline{11-11}
$11$ & $-29$ & $-131$ & $-467$ & $-1727$ & $-6304$ & $-21161$ & $-62161$ & $-149751$ & $-238112$ & \multicolumn{1}{r|}{$181594$} & $3903607$ & $23921635$ & $117192631$\\
% END OF TABLE DATA *****************
\hline\hline
\end{tabular}
\vspace{1cm}
\caption{
   Coefficients $\ell e_\ell(m)$ [cf.\ \protect\reff{def_series_el}]
   for $1\leq \ell \leq 13$ and $1\leq m\leq 11$. 
   For the whole data set up to $\ell=20$,
   see the {\sc Mathematica} file {\tt data\_Diff.m}
   included in the on-line version of the paper at arXiv.org.
   Those data points below the lower staircase-like line satisfy
   $m\geq \ell+1$. 
}
\label{table_diff_coef_e}
\end{sidewaystable}

%
% TABLE 11: COEFFICIENTS Delta^(E)_\ell(k) 
%
\clearpage
\begin{sidewaystable}
\centering
\begin{tabular}{r|rrrrrrrrrrrrr}
\hline\hline\\[-4mm]
$m$& $\Delta^{(e)}_1$    & $\Delta^{(e)}_2$& $\Delta^{(e)}_3$&
     $\Delta^{(e)}_4$    & $\Delta^{(e)}_5$&
     $\Delta^{(e)}_6$    & $\Delta^{(e)}_7$& $\Delta^{(e)}_8$&
     $\Delta^{(e)}_9$    & $\Delta^{(e)}_{10}$ &
     $\Delta^{(e)}_{11}$ & $\Delta^{(e)}_{12}$ &
     $\Delta^{(e)}_{13}$ \\[2mm]
\hline\\[-3mm]
% DO NOT EDIT *****************
$1$ & \multicolumn{1}{r|}{$-1$} & $\frac{1}{2}$ & $\frac{-13}{3}$ & $\frac{-83}{4}$ & $\frac{-276}{5}$ & $\frac{-479}{6}$ & $\frac{741}{7}$ & $\frac{9861}{8}$ & $\frac{45302}{9}$ & $\frac{67163}{5}$ & $\frac{237962}{11}$ & $\frac{-26567}{12}$ & $\frac{-1767169}{13}$\\[2mm]
\cline{2-2}\cline{3-4}
$2$ & \multicolumn{1}{r|}{$0$} & $1$ & \multicolumn{1}{r|}{$-3$} & $\frac{-15}{2}$ & $6$ & $\frac{629}{6}$ & $485$ & $\frac{6133}{4}$ & $3119$ & $\frac{-3529}{5}$ & $-45740$ & $\frac{-1136521}{4}$ & $-1260527$\\[2mm]
\cline{3-3}\cline{5-6}
$3$ & $0$ & \multicolumn{1}{r|}{$0$} & $-1$& $7$ & \multicolumn{1}{r|}{$26$} & $\frac{53}{2}$ & $-92$ & $\frac{-1545}{2}$ & $\frac{-13849}{3}$ & $-24141$ & $-104393$ & $\frac{-4353233}{12}$ & $-965725$\\[2mm]
\cline{4-4}\cline{7-8}
$4$ & $0$ & $0$ & \multicolumn{1}{r|}{$0$} & $1$ & $-11$ & $-45$ & \multicolumn{1}{r|}{$-101$} & $\frac{-765}{2}$ & $-1583$ & $\frac{-3949}{2}$ & $26879$ & $\frac{1619321}{6}$ & $1654200$\\[2mm]
\cline{5-5}\cline{9-9}
$5$ & $0$ & $0$ & $0$ & \multicolumn{1}{r|}{$0$} & $-1$ & $15$ & $66$ & \multicolumn{1}{r|}{$243$} & $1717$ & $\frac{21767}{2}$ & $50707$ & $\frac{345247}{2}$ & $313921$\\[2mm]
\cline{6-6}\cline{10-10}
$6$ & $0$ & $0$ & $0$ & $0$ & \multicolumn{1}{r|}{$0$} & $1$ & $-19$ & $-89$ & \multicolumn{1}{r|}{$-479$} & $-4525$ & $-34691$ & $\frac{-420371}{2}$ & $-1083826$\\[2mm]
\cline{7-7}\cline{11-11}
$7$ & $0$ & $0$ & $0$ & $0$ & $0$ & \multicolumn{1}{r|}{$0$} & $-1$ & $23$ & $114$ & \multicolumn{1}{r|}{$837$} & $9529$ & $83888$ & $603804$\\[2mm]
\cline{8-8}\cline{12-12}
$8$ & $0$ & $0$ & $0$ & $0$ & $0$ & $0$ & \multicolumn{1}{r|}{$0$} & $1$ & $-27$ & $-141$ & \multicolumn{1}{r|}{$-1345$} & $-17572$ & $-173582$\\[2mm]
\cline{9-9}\cline{13-13}
$9$ & $0$ & $0$ & $0$ & $0$ & $0$ & $0$ & $0$ & \multicolumn{1}{r|}{$0$} & $-1$ & $31$ & $170$ & \multicolumn{1}{r|}{$2031$} & $29619$\\[2mm]
\cline{10-10}\cline{14-14}
$10$ & $0$ & $0$ & $0$ & $0$ & $0$ & $0$ & $0$ & $0$ & \multicolumn{1}{r|}{$0$} & $1$ & $-35$ & $-201$ & \multicolumn{1}{r|}{$-2923$}\\[2mm]
\cline{11-11}
$11$ & $0$ & $0$ & $0$ & $0$ & $0$ & $0$ & $0$ & $0$ & $0$ & \multicolumn{1}{r|}{$0$} & $-1$ & $39$ & $234$\\[2mm]
% END OF TABLE DATA *****************
\hline\hline
\end{tabular}
\vspace{1cm}
\caption{
   Coefficients $\Delta^{(e)}_\ell(m)$ [cf.\ \protect\reff{def_deltaE}]
   for $1\leq \ell \leq 13$ and $1\leq m\leq 11$. 
   For the whole data set up to $\ell=20$,
   see the {\sc Mathematica} file {\tt data\_Diff.m}
   included in the on-line version of the paper at arXiv.org.
   Those data points below the lower staircase-like line satisfy
   $m\geq \ell+1$ and are therefore zero.
   Those data points between the two staircase-like lines can be
   fitted to a polynomial Ansatz [cf.\ \protect\reff{eq.DeltaE}].
}
\label{table_diff_coef_deltae}
\end{sidewaystable}

%
% TABLE 12: COEFFICIENTS f_ell(k)   
%
\clearpage
\thispagestyle{empty}
\begin{sidewaystable}
\centering
{\small 
\begin{tabular}{r|rrrrrrrrrrrr}
\hline\hline\\[-4mm]
$m$& $ f_1$     & $2f_2$     & $3f_3$     & 
     $4f_4$     & $5f_5$     & $6f_6$     &  
     $7f_7$     & $8f_8$     & $9f_9$     & 
     $10f_{10}$ & $11f_{11}$ & $12f_{12}$  \\[2mm]
\hline\\[-2mm]
% DO NOT EDIT *****************
$1$ & $-\frac{5}{2}$ & $-\frac{77}{12}$ & $-\frac{135}{8}$ & $-\frac{32651}{720}$ & $-\frac{35755}{288}$ & $-\frac{20877127}{60480}$ & $-\frac{3359377}{3456}$ & $-\frac{10035074977}{3628800}$ & $-\frac{426428659}{53760}$ & -$\frac{2193911030309}{95800320}$ & $-\frac{231533054159}{3483648}$ & $-\frac{506578309836952357}{2615348736000}$ \\[2mm]
\cline{2-2}
$2$ & \multicolumn{1}{r|}{$-3$} & $2$ & $\frac{57}{2}$ & $\frac{673}{6}$ & $\frac{1433}{4}$ & $\frac{2349}{2}$ & $\frac{97907}{24}$ & $\frac{5001829}{360}$ & $\frac{3524187}{80}$ & $\frac{9408107}{72}$ & $\frac{542977339}{1440}$ & $\frac{6664350307}{6048}$ \\[2mm]
\cline{3-3}
$3$ & $-2$ & \multicolumn{1}{r|}{$16$} & $\frac{149}{2}$ & $240$ & $888$ & $\frac{14699}{4}$ & $\frac{41798}{3}$ & $\frac{136816}{3}$ & $\frac{1047455}{8}$ & $\frac{985253}{3}$ & $\frac{1821506}{3}$ & $-\frac{65347931}{240}$ \\[2mm]
\cline{4-4}
$4$ & $-1$ & $27$ & \multicolumn{1}{r|}{$119$} & $453$ & $\frac{3573}{2}$ & $6450$ & $\frac{40759}{2}$ & $\frac{168316}{3}$ & $\frac{482555}{4}$ & $\frac{842209}{12}$ & $-\frac{2524997}{2}$ & $-10308961$ \\[2mm]
\cline{5-5}
$5$ & $0$ & $38$ & $168$ & \multicolumn{1}{r|}{$650$} & $\frac{4845}{2}$ & $8258$ & $\frac{50561}{2}$ & $65330$ & $112542$ & $-\frac{980089}{12}$ & $-1917212$ & $-11481293$ \\[2mm]
\cline{6-6}
$6$ & $1$ & $49$ & $217$ & $841$ & \multicolumn{1}{r|}{$3101$} & $10564$ & $\frac{64367}{2}$ & $82297$ & $151507$ & $38929$ & $-1393622$ & $-\frac{18512379}{2}$ \\[2mm]
\cline{7-7}
$7$ & $2$ & $60$ & $266$ & $1032$ & $3787$ & \multicolumn{1}{r|}{$12792$} & $\frac{76563}{2}$ & $95448$ & $165848$ & $-42865$ & $-2280408$ & $-15142278$ \\[2mm]
\cline{8-8}
$8$ & $3$ & $71$ & $315$ & $1223$ & $4473$ & $15011$ & \multicolumn{1}{r|}{$44502$} & $109787$ & $\frac{374067}{2}$ & $-62624$ & $-\frac{5139139}{2}$ & $-16296427$ \\[2mm]
\cline{9-9}
$9$ & $4$ & $82$ & $364$ & $1414$ & $5159$ & $17230$ & $50733$ & \multicolumn{1}{r|}{$123950$} & $\frac{413135}{2}$ & $-94033$ & $-\frac{5984927}{2}$ & $-18892310$ \\[2mm]
\cline{10-10}
$10$ & $5$ & $93$ & $413$ & $1605$ & $5845$ & $19449$ & $56964$ & $138101$ & \multicolumn{1}{r|}{$226340$} & $-123242$ & $-\frac{6792897}{2}$ & $-21230139$ \\[2mm]
\cline{11-11}
$11$ & $6$ & $104$ & $462$ & $1796$ & $6531$ & $21668$ & $63195$ & $152252$ & $246126$ & \multicolumn{1}{r|}{$-152761$} & $-\frac{7606543}{2}$ & $-23597380$ \\[2mm]
% END OF TABLE DATA *****************
\hline\hline
\end{tabular}
}
\vspace{1cm}
\caption{
   Coefficients $k f_\ell(m)$ [cf.\ \protect\reff{def_series_fl}]
   for $1\leq \ell \leq 12$ and $1\leq m\leq 11$. 
   For the whole data set up to $\ell=20$,
   see the {\sc Mathematica} file {\tt data\_Diff.m}
   included in the on-line version of the paper at arXiv.org.
   Those data points below the lower staircase-like line satisfy
   $m\geq \ell+1$. 
}
\label{table_diff_coef_f}
\end{sidewaystable}

%
% TABLE 13: COEFFICIENTS Delta^(f)_\ell(k) 
%
\clearpage
\thispagestyle{empty}
\begin{sidewaystable}
\centering
{\small
\begin{tabular}{r|rrrrrrrrrrrr}
\hline\hline\\[-4mm]
$m$& $\Delta^{(f)}_1$    & $\Delta^{(f)}_2$& $\Delta^{(f)}_3$&
     $\Delta^{(f)}_4$    & $\Delta^{(f)}_5$&
     $\Delta^{(f)}_6$    & $\Delta^{(f)}_7$& $\Delta^{(f)}_8$&
     $\Delta^{(f)}_9$    & $\Delta^{(f)}_{10}$ &
     $\Delta^{(f)}_{11}$ & $\Delta^{(f)}_{12}$ \\[2mm]
\hline\\[-3mm]
% DO NOT EDIT *****************
$1$ & \multicolumn{1}{r|}{$\frac{3}{2}$} & $-\frac{5}{24}$ & $\frac{89}{24}$ & $\frac{49429}{2880}$ & $\frac{58997}{1440}$ & $\frac{10693433}{362880}$ & $-\frac{6417937}{24192}$ & $-\frac{49015644577}{29030400}$ & $-\frac{3021208819}{483840}$ & $-\frac{15853024855589}{958003200}$ & $-\frac{1140319275215}{38320128}$ & $-\frac{623813932276888357}{31384184832000}$\\[2mm]
\cline{2-2}\cline{3-4}
$2$ & \multicolumn{1}{r|}{$0$} & $-\frac{3}{2}$ & \multicolumn{1}{r|}{$\frac{5}{2}$} & $\frac{211}{24}$ & $\frac{1}{4}$ & $-\frac{1045}{12}$ & $-\frac{10411}{24}$ & $-\frac{3959651}{2880}$ & $-\frac{639991}{240}$ & $\frac{1268867}{720}$ & $\frac{68414609}{1440}$ & $\frac{20691736771}{72576}$\\[2mm]
\cline{3-3}\cline{5-6}
$3$ & $0$ & \multicolumn{1}{r|}{$0$} & $\frac{3}{2}$ & $-7$ & \multicolumn{1}{r|}{$-31$} & $-\frac{965}{24}$ & $\frac{251}{3}$ & $\frac{4921}{6}$ & $\frac{114917}{24}$ & $\frac{73472}{3}$ & $\frac{316126}{3}$ & $\frac{1058694949}{2880}$\\[2mm]
\cline{4-4}\cline{7-8}
$4$ & $0$ & $0$ & \multicolumn{1}{r|}{$0$} & $-\frac{3}{2}$ & $\frac{23}{2}$ & $\frac{105}{2}$ & \multicolumn{1}{r|}{$\frac{229}{2}$} & $\frac{8731}{24}$ & $\frac{17353}{12}$ & $\frac{38897}{24}$ & $-\frac{55295}{2}$ & $-\frac{815320}{3}$\\[2mm]
\cline{5-5}\cline{9-9}
$5$ & $0$ & $0$ & $0$ & \multicolumn{1}{r|}{$0$} & $\frac{3}{2}$ & $-16$ & $-\frac{151}{2}$ & \multicolumn{1}{r|}{$-252$} & $-1652$ & $-\frac{254681}{24}$ & $-50220$ & $-\frac{2069443}{12}$\\[2mm]
\cline{6-6}\cline{10-10}
$6$ & $0$ & $0$ & $0$ & $0$ & \multicolumn{1}{r|}{$0$} & $-\frac{3}{2}$ & $\frac{41}{2}$ & $100$ & \multicolumn{1}{r|}{$479$} & $4402$ & $34326$ & $\frac{5039659}{24}$\\[2mm]
\cline{7-7}\cline{11-11}
$7$ & $0$ & $0$ & $0$ & $0$ & $0$ & \multicolumn{1}{r|}{$0$} & $\frac{3}{2}$ & $-25$ & $-126$ & \multicolumn{1}{r|}{$-824$} & $-9344$ & $-\frac{167015}{2}$\\[2mm]
\cline{8-8}\cline{12-12}
$8$ & $0$ & $0$ & $0$ & $0$ & $0$ & $0$ & \multicolumn{1}{r|}{$0$} & $-\frac{3}{2}$ & $\frac{59}{2}$ & $\frac{307}{2}$ & \multicolumn{1}{r|}{$\frac{2631}{2}$} & $\frac{34655}{2}$\\[2mm]
\cline{9-9}\cline{13-13}
$9$ & $0$ & $0$ & $0$ & $0$ & $0$ & $0$ & $0$ & \multicolumn{1}{r|}{$0$} & $\frac{3}{2}$ & $-34$ & $-\frac{365}{2}$ & \multicolumn{1}{r|}{$-1982$}\\[2mm]
\cline{10-10}
$10$ & $0$ & $0$ & $0$ & $0$ & $0$ & $0$ & $0$ & $0$ & \multicolumn{1}{r|}{$0$} & $-\frac{3}{2}$ & $\frac{77}{2}$ & $213$\\[2mm]
\cline{11-11}
$11$ & $0$ & $0$ & $0$ & $0$ & $0$ & $0$ & $0$ & $0$ & $0$ & \multicolumn{1}{r|}{$0$} & $\frac{3}{2}$ & $-43$\\[2mm]
% END OF TABLE DATA *****************
\hline\hline
\end{tabular}
}
\vspace{1cm}
\caption{
   Coefficients $\Delta^{(f)}_\ell(m)$ [cf.\ \protect\reff{def_deltaF}]
   for $1\leq \ell \leq 12$ and $1\leq m\leq 11$. 
   For the whole data set up to $\ell=20$,
   see the {\sc Mathematica} file {\tt data\_Diff.m}
   included in the on-line version of the paper at arXiv.org.
   Those data points below the lower staircase-like line satisfy
   $m\geq \ell+1$ and are therefore zero.
   Those data points between the two staircase-like lines can be
   fitted to a polynomial Ansatz [cf.\ \protect\reff{eq.DeltaF}].
}
\label{table_diff_coef_deltaf}
\end{sidewaystable}

%%%%%%%%%%%%%%%%%%%%%%%%%%%%%%%%%%%%%%%%%%%%%%%%%%%%%%%%%%%%%%%%%
% 
% TABLE 14: NUMBER OF CLASSES OF PARTITIONS 
%
%%%%%%%%%%%%%%%%%%%%%%%%%%%%%%%%%%%%%%%%%%%%%%%%%%%%%%%%%%%%%%%%%
\clearpage
\begin{table}
\centering
\begin{tabular}{|r|r|r|r|r|}
\hline\hline
$m$& $d_m$  & $X(m)$  &  $N_2(m)$ &  $N_3(m)$  \\
\hline\hline
1  &      1 &    1 &      1 &      1  \\
2  &      1 &    1 &      1 &      1  \\
3  &      1 &    1 &      1 &      1  \\
4  &      3 &    1 &      2 &      3  \\
5  &      6 &    2 &      4 &      4  \\
6  &     15 &    3 &      9 &     11  \\
7  &     36 &    6 &     21 &     21  \\
8  &     91 &    7 &     49 &     55  \\
9  &    232 &   16 &    124 &    124  \\
10 &    603 &   19 &    311 &    327  \\
11 &   1585 &   45 &    815 &    815  \\
12 &   4213 &   51 &   2132 &   2177  \\
13 &  11298 &  126 &   5712 &   5712  \\
14 &  30537 &  141 &  15339 &  15465  \\
15 &  83097 &  357 &  41727 &  41727  \\
16 & 227475 &  393 & 113934 & 114291  \\
\hline\hline
\end{tabular}
\vspace{1cm}
\caption{
   Counts of certain classes of
   non-crossing non-nearest-neighbor (ncnnn) partitions
   of the set $\{1,2,\ldots,m\}$ on a circle.
   For each $m$, we show the total number $d_m$ of ncnnn partitions
   (which equals the Riordan number $R_m$ when $m \ge 2$);
   the number $X(m)$ of ncnnn partitions
   that are invariant under reflection with respect 
   an axis going between vertices $1$ and $m$;
   the number $N_2(m)$ of equivalence classes of ncnnn partitions
   modulo reflection with respect an axis going between vertices $1$ and $m$;
   and the number $N_3(m)$ of equivalence classes of ncnnn partitions
   modulo reflection with respect an axis going through vertex $1$.
   Note that $R_0 = 1$ and $R_1 = 0$.
}
\label{table_dimensions2}
\end{table}
\clearpage

%%%%%%%%%%%%%%%%%%%%%%%%%%%%%%%%%%%%%%%%%%%%%%%%%%%%%%%%%%%%%%%
%
% FIGURES
%
%%%%%%%%%%%%%%%%%%%%%%%%%%%%%%%%%%%%%%%%%%%%%%%%%%%%%%%%%%%%%%%
\addcontentsline{toc}{section}{Figures}

%
% FIGURE 1: Family S_{m,n} 
%

\clearpage
%
% FIGURE 2: Theta graphs 
%

%
% FIGURE 3: bi-fans and bi-pyramids (physics) 
%
\begin{figure}
\centering
     \psset{xunit=1.5cm}
     \psset{yunit=1.5cm}
     \pspicture(0,-1.0)(7,3)
%%%%%\psframe  (0,-1.0)(7,3)
     \rput{0}(0,0){
        \psline[linecolor=blue,linewidth=1.0pt](1,0)(1,3)
        \psline[linecolor=blue,linewidth=1.0pt](0,1.5)(1,0)(2,1.5)
        \psline[linecolor=blue,linewidth=1.0pt](0,1.5)(1,1)(2,1.5)
        \psline[linecolor=blue,linewidth=1.0pt](0,1.5)(1,2)(2,1.5)
        \psline[linecolor=blue,linewidth=1.0pt](0,1.5)(1,3)(2,1.5)
        \multirput{0}(1,0)(0,1){4}{$\bullet$}
        \multirput{0}(0,1.5)(2,0){2}{$\bullet$}
        \rput{0}(1.0,-0.8){(a)}
     }
     \rput{0}(4,0){
        \psline[linecolor=blue,linewidth=1.0pt](1,0)(1,3)
        \psline[linecolor=blue,linewidth=1.0pt](0,1.5)(1,0)(2,1.5)
        \psline[linecolor=blue,linewidth=1.0pt](0,1.5)(1,1)(2,1.5)
        \psline[linecolor=blue,linewidth=1.0pt](0,1.5)(1,2)(2,1.5)
        \psline[linecolor=blue,linewidth=1.0pt](0,1.5)(1,3)(2,1.5)
        \psarc[linecolor=blue,linewidth=1.0pt] (1,1.5){2.25}{270}{90}
        \multirput{0}(1,0)(0,1){4}{$\bullet$}
        \multirput{0}(0,1.5)(2,0){2}{$\bullet$}
        \rput{0}(1.0,-0.8){(b)}
     }
     \endpspicture
\caption{\label{fig_bipyr}
(a) Bi-fan $P_4 + \bar{K}_2$. (b) Bipyramid $C_4 + \bar{K}_2$.
}
\end{figure}

\clearpage
%
% FIGURE 4: Sq_1Fx100F
%

\clearpage
%
% FIGURE 5: ZEROS theta graphs  
%

\clearpage
%
% FIGURE 6: Family S_{m,n}^{=} 
%

%%%%%%%%%%%\clearpage
%
% FIGURE 7: Family S_{m,n}^{\neq} 
%
\begin{figure}
\centering
     \psset{unit=1.5cm}
     \pspicture(0,0)(10,4)
     \psline[linecolor=blue,linewidth=1.0pt](1,0)(1,2)
     \psline[linecolor=blue,linewidth=1.0pt](2,0)(2,2)
     \psline[linecolor=blue,linewidth=1.0pt](3,0)(3,2)
     \psline[linecolor=blue,linewidth=1.0pt](4,0)(4,2)

     \psline[linecolor=blue,linewidth=1.0pt](1,0)(4,0)
     \psline[linecolor=blue,linewidth=1.0pt](1,1)(4,1)
     \psline[linecolor=blue,linewidth=1.0pt](1,2)(4,2)

     \psline[linecolor=blue,linewidth=1.0pt](0,1)(1,0)
     \psline[linecolor=blue,linewidth=1.0pt](0,1)(1,1)
     \psline[linecolor=blue,linewidth=1.0pt](0,1)(1,2)

     \psline[linecolor=blue,linewidth=1.0pt](5,1)(4,0)
     \psline[linecolor=blue,linewidth=1.0pt](5,1)(4,1)
     \psline[linecolor=blue,linewidth=1.0pt](5,1)(4,2)

     \psarc[linecolor=blue,linewidth=1.0pt](2.5,1){2.5}%
          {0}{180}

     \pscircle[linecolor=blue,linewidth=1.0pt](8,1.5){0.5}
     \pscircle[linecolor=blue,linewidth=1.0pt](8,1.5){1}
     \pscircle[linecolor=blue,linewidth=1.0pt](8,1.5){1.5}

     \psline[linecolor=blue,linewidth=1.0pt](8.5,1.5)(9.5,1.5)
     \psline[linecolor=blue,linewidth=1.0pt](7.5,1.5)(6.5,1.5)
     \psline[linecolor=blue,linewidth=1.0pt,linestyle=dashed]%
          (8.25,1.93301)(8.75,2.79904)
     \psline[linecolor=blue,linewidth=1.0pt,linestyle=dashed]%
          (7.75,1.93301)(7.25,2.79904)
     \psline[linecolor=blue,linewidth=1.0pt](7.75,1.06699)(7.25,0.200962)
     \psline[linecolor=blue,linewidth=1.0pt](8.25,1.06699)(8.75,0.200962)

     \multirput{0}(1,0)(0,1){3}{$\bullet$}
     \multirput{0}(2,0)(0,1){3}{$\bullet$}
     \multirput{0}(3,0)(0,1){3}{$\bullet$}
     \multirput{0}(4,0)(0,1){3}{$\bullet$}
     \multirput{0}(0,1)(5,0){2}{$\bullet$}

     \multirput[c]{0}(8.5,1.5)(0.5,0){3}{$\bullet$}
     \multirput[c]{0}(7.5,1.5)(-0.5,0){3}{$\bullet$}
     \multirput[c]{0}(8.25,1.93301)(0.25,0.433013){3}{$\bullet$}
     \multirput[c]{0}(7.75,1.93301)(-0.25,0.433013){3}{$\bullet$}
     \multirput[c]{0}(7.75,1.06699)(-0.25,-0.433013){3}{$\bullet$}
     \multirput[c]{0}(8.25,1.06699)(0.25,-0.433013){3}{$\bullet$}

     \rput{0}( 1  ,-0.4){$1$}
     \rput{0}( 2  ,-0.4){$2$}
     \rput{0}( 3  ,-0.4){$3$}
     \rput{0}( 4  ,-0.4){$4$}
     \rput{0}( 5.2, 1){$5$}
     \rput{0}(-0.2, 1){$0$}

     \rput{0}(8.80,3.00){$0$}
     \rput{0}(9.70,1.50){$1$}
     \rput{0}(8.80,0.00){$2$}
     \rput{0}(7.20,0.00){$3$}
     \rput{0}(6.30,1.50){$4$}
     \rput{0}(7.20,3.00){$5$}

     \rput{0}( 8  ,-1.0){(b)}
     \rput{0}( 2.5,-1.0){(a)}
     \endpspicture
     \vspace*{2cm}
\caption{
   (a) Square-lattice strip $S_{4,3}^{\neq}$ of width $m=4$ and length $n=3$
      with the two extra sites (0 and 5) joined by a $v=-1$ edge.
   (b) Alternate representation of $S_{4,3}^{\neq}$ in terms of a
      square-lattice strip of width $m+2=6$, length $n=3$
      and cylindrical boundary conditions, with edge weights $v \to +\infty$
      on the dashed edges.
  \label{fig_SmnNotEq}
}
\end{figure}

\clearpage
%
% FIGURE 8: Family \widehat{S}_{m,n} 
%
\begin{figure}
\centering
     \psset{xunit=1.5cm}
     \psset{yunit=1.5cm}
     \pspicture(0,0)(4,5)
     \psline[linecolor=blue,linewidth=1.0pt](0,1)(4,1)
     \psline[linecolor=blue,linewidth=1.0pt](0,2)(4,2)
     \psline[linecolor=blue,linewidth=1.0pt](0,3)(4,3)
     \psline[linecolor=blue,linewidth=1.0pt](0,4)(4,4)

     \psline[linecolor=blue,linewidth=1.0pt](0,1)(0,4)
     \psline[linecolor=blue,linewidth=1.0pt](1,1)(1,4)
     \psline[linecolor=blue,linewidth=1.0pt](2,1)(2,4)
     \psline[linecolor=blue,linewidth=1.0pt](3,1)(3,4)
     \psline[linecolor=blue,linewidth=1.0pt](4,1)(4,4)

     \psline[linecolor=blue,linewidth=1.0pt](0,1)(2,0)(4,1)
     \psline[linecolor=blue,linewidth=1.0pt](1,1)(2,0)(3,1)
     \psline[linecolor=blue,linewidth=1.0pt](2,0)(2,1)

     \psline[linecolor=blue,linewidth=1.0pt](0,4)(2,5)(4,4)
     \psline[linecolor=blue,linewidth=1.0pt](1,4)(2,5)(3,4)
     \psline[linecolor=blue,linewidth=1.0pt](2,5)(2,4)

     \multirput{0}(0,1)(0,1){4}{$\bullet$}
     \multirput{0}(1,1)(0,1){4}{$\bullet$}
     \multirput{0}(2,1)(0,1){4}{$\bullet$}
     \multirput{0}(3,1)(0,1){4}{$\bullet$}
     \multirput{0}(4,1)(0,1){4}{$\bullet$}

     \multirput{0}(2,0)(0,5){2}{$\bullet$}
     \endpspicture
     \vspace*{2cm}
\caption{\label{fig_hatSmn}
   Square-lattice strip $\widehat{S}_{5,4}$ 
   of width $m=5$ and length $n=4$ with extra sites at top and bottom.
}
\end{figure}

\clearpage
%
% Sq_1F 
%

\clearpage
%
% Sq_2F 
%

\clearpage
%
% Sq_3F 
%

\clearpage
%
% Sq_4F 
%

\clearpage
%
% Sq_5F 
%

\clearpage
%
% Sq_6F 
%

\clearpage
%
% Sq ALL
%

\clearpage

%
% FREE energies 
%

%
% FREE energies 
%

\clearpage

\addcontentsline{toc}{section}{References}
\begin{thebibliography}{99}

\bibitem{Bakaev_94}  A.V. Bakaev and V.I. Kabanovich,
   Series expansions for the $q$-colour problem on the
   square and cubic lattices,
   {\em J. Phys. A: Math. Gen.}\/ {\bf 27}, 6731--6739 (1994).

\bibitem{Baxter_86}  R.J. Baxter,
   $q$ colourings of the triangular lattice,
   {\em J. Phys. A: Math. Gen.}\/ {\bf 19}, 2821--2839 (1986).

\bibitem{Baxter_87}  R.J. Baxter,
  Chromatic polynomials of large triangular lattices,
  {\em J. Phys. A: Math. Gen.}\/ {\bf 20}, 5241--5261 (1987).

\bibitem{Baxter_76}  R.J. Baxter, S.B. Kelland and F.Y. Wu,
   Equivalence of the Potts model or Whitney polynomial with an ice-type model,
   {\em J. Phys. A: Math. Gen.}\/ {\bf 9}, 397--406 (1976).

\bibitem{Beraha_79}  S. Beraha and J. Kahane,
   Is the four-color conjecture almost false?,
   {\em J. Combin. Theory B}\/ {\bf 27}, 1--12 (1979).

\bibitem{BKW_75}  S. Beraha, J. Kahane and N.J. Weiss,
   Limits of zeroes of recursively defined polynomials,
   {\em Proc. Nat. Acad. Sci. USA}\/ {\bf 72}, 4209 (1975).

\bibitem{BKW_78}  S. Beraha, J. Kahane and N.J. Weiss,
   Limits of zeroes of recursively defined families of polynomials,
   in {\em Studies in Foundations and Combinatorics}\/
   (Advances in Mathematics Supplementary Studies, Vol.~1),
   ed.~G.-C. Rota, pp.~213--232
   (Academic Press, New York, 1978).

\bibitem{Beraha_80}  S. Beraha, J. Kahane and N.J. Weiss,
   Limits of chromatic zeros of some families of maps,
   {\em J. Combin. Theory B}\/ {\bf 28}, 52--65 (1980).

\bibitem{Bernhart_99} F.R. Bernhart,
  Catalan, Motzkin, and Riordan numbers,
  {\em Discrete Math.}\/ {\bf 204}, 73--112 (1999).

\bibitem{Birkhoff_12}  G.D. Birkhoff,
   A determinant formula for the number of ways of coloring a map,
   {\em Ann. Math.}\/ {\bf 14}, 42--46 (1912).

\bibitem{Blote_82}  H.W.J. Bl\"ote and M.P. Nightingale,
   Critical behaviour of the two-dimensional Potts model with a
   continuous number of states: A finite size scaling analysis,
   {\em Physica A}\/ {\bf 112}, 405--465 (1982).

\bibitem{gen_theta}  J.I. Brown, C.A. Hickman, A.D. Sokal and D.G. Wagner,
   On the chromatic roots of generalized theta graphs,
   {\em J. Combin. Theory B}\/ {\bf 83}, 272--297 (2001),
   math.CO/0012033 at arXiv.org.

\bibitem{Tutte_sq_02} S.-C. Chang, J. Salas  and R. Shrock, 
   Exact Potts model partition functions for strips of the square lattice,
   {\em J. Stat. Phys.}\/ {\bf 107}, 1207--1253 (2002), 
   cond-mat/0108144 at arXiv.org.

\bibitem{Chang-Shrock_00_TRI_TEMP}  S.-C. Chang and R. Shrock,
   Exact Potts model partition function on strips
   of the triangular lattice,
   {\em Physica A}\/ {\bf 286}, 189--238 (2000),
   cond-mat/0004181 at arXiv.org.

\bibitem{Chang-Shrock_01_TEMP}  S.-C. Chang and R. Shrock,
   Exact Potts model partition functions on wider arbitrary-length strips
   of the square lattice,
   {\em Physica A}\/ {\bf 296}, 234--288 (2001),
   cond-mat/0011503 at arXiv.org.

\bibitem{Comtet_book} L. Comtet, {\em Advanced Combinatorics}\/
  (D. Reidel, Dordrecht, 1974).

\bibitem{Corless_96}  R.M. Corless, G.H. Gonnet, D.E.G. Hare,
   D.J. Jeffrey and D.E. Knuth,
   On the Lambert $W$ function,
   {\em Adv. Comput. Math.}\/ {\bf 5}, 329--359 (1996).

\bibitem{deBruijn_61}  N.G. De Bruijn, {\em Asymptotic Methods in Analysis}\/,
   $2^{nd}$ ed. (North-Holland, Amsterdam, 1961).

\bibitem{Donaghey_77}  R. Donaghey and L.W. Shapiro,
   Motzkin numbers,
   {\em J. Combin. Theory A}\/ {\bf 23}, 291--301 (1977).

\bibitem{Dubail_Jacobsen_Saleur_09}
   J. Dubail, J.L. Jacobsen, and H. Saleur,
   Conformal two-boundary loop model on the annulus,
   {\em Nucl. Phys. B}\/ {\bf 813}, 430--459 (2009),
   arXiv:0812:2746 at arXiv.org.

\bibitem{Edwards-Sokal}  R.G. Edwards and A.D. Sokal,
   Generalization of the Fortuin-Kasteleyn-Swendsen-Wang representation
   and Monte Carlo algorithm,
   {\em Phys. Rev. D}\/ {\bf 38}, 2009--2012 (1988).

\bibitem{Enting_78}  I.G. Enting,
   Generalised M\"obius functions for rectangles on the square lattice,
   {\em J. Phys. A: Math. Gen.}\/ {\bf 11}, 563--568 (1978).

\bibitem{Enting_96}  I.G. Enting,
   Series expansions from the finite lattice method,
   {\em Nucl. Phys. B -- Proc. Suppl.}\/ {\bf 47}, 180--187 (1996).

\bibitem{Enting_05}  I.G. Enting,
   Inclusion-exclusion relations for series expansions of boundary effects
    using the finite lattice method,
   {\em J. Stat. Mech.}\/ (2005), P12007.

\bibitem{Enting_06}  I.G. Enting,
   Surface magnetisation of the 3-state Potts model with free and fixed
    boundaries on the square lattice,
   {\em J. Phys.: Conf. Ser.}\/ {\bf 42}, 83--94 (2006).

\bibitem{FP_paper2} R. Fern\'andez and A. Procacci,
   Regions without complex zeros for chromatic polynomials
   on graphs with bounded degree,
   {\em Combin. Probab. Comput.}\/ {\bf 17}, 225--238 (2008),
   arXiv:0704.2617 at arXiv.org.

\bibitem{Fortuin_72}  C.M. Fortuin and P.W. Kasteleyn,
   On the random-cluster model. I. Introduction and relation to other models,
   {\em Physica}\/ {\bf 57}, 536--564 (1972).

\bibitem{Golub_book} G.H. Golub and C.F. Van Loan, {\em Matrix
   Computations}, 3rd edition (The Johns Hopkins University Press,
   Baltimore, 1996).

\bibitem{Gouyou_88}  D. Gouyou-Beauchamps and G. Viennot,
   Equivalence of the two-dimensional directed animal problem
   to a one-dimensional path problem,
   {\em Adv. Appl. Math.}\/ {\bf 9}, 334--357 (1988).

\bibitem{Graham_94}  R.L. Graham, D.E. Knuth and O. Patashnik,
  {\em Concrete Mathematics: A Foundation for Computer Science}\/,
  2nd ed.~(Addison-Wesley, Reading, Mass., 1994).

\bibitem{JPS_08}  B. Jackson, A. Procacci and A.D. Sokal,
   Complex zero-free regions at large $|q|$ for multivariate Tutte polynomials
   (alias Potts-model partition functions) with general complex edge weights,
   preprint (October 2008), arXiv:0810.4703 at arXiv.org.

\bibitem{Jacobsen_10} J.L. Jacobsen,
   Bulk, surface and corner free energy series for the chromatic polynomial 
   on the square and triangular lattices
   {\em J. Phys. A: Math. Theor.}\/ {\bf 43}, 315002 (2010),
   arXiv:1005.3609 at arXiv.org.

\bibitem{Jacobsen_RSOS}  J.L. Jacobsen, J.-F. Richard and J. Salas,
   Complex-temperature phase diagram of Potts and RSOS models,
   {\em Nucl. Phys. B}\/ {\bf 743}, 153--206 (2006),
   cond-mat/0511059 at arXiv.org.

\bibitem{transfer2} J.L. Jacobsen and J. Salas,
   Transfer matrices and partition-function zeros for antiferromagnetic
   Potts models.  II.~Extended results for square-lattice chromatic polynomial,
   {\em J. Stat. Phys.}\/ {\bf 104}, 701--723 (2001),
   cond-mat/0011456 at arXiv.org.

\bibitem{transfer4} J.L. Jacobsen and J. Salas,
   Transfer matrices and partition-function zeros for antiferromagnetic
   Potts models.  IV.~Chromatic polynomial with cyclic boundary conditions,
   {\em J. Stat. Phys.}\/ {\bf 122}, 705-760 (2006),  
   cond-mat/0407444 at arXiv.org.

\bibitem{Jacobsen-Salas_toroidal}  J.L. Jacobsen and J. Salas,
   Phase diagram of the chromatic polynomial on a torus,
   {\em Nucl. Phys. B}\/ {\bf 783}, 238--296 (2007), 
   cond-mat/0703228 at arXiv.org.

\bibitem{Jacobsen-Salas_flow}  J.L. Jacobsen and J. Salas,
   Is the five-flow conjecture almost false?,
   preprint (September 2010),
   arXiv:1009.4062 at arXiv.org.

\bibitem{transfer3}
   J.L. Jacobsen, J. Salas and A.D. Sokal,
   Transfer matrices and partition-function zeros for antiferromagnetic
   Potts models.  III.~Triangular-lattice chromatic polynomial,
   {\em J. Stat. Phys.}\/ {\bf 112}, 921--1017 (2003),
   cond-mat/0204587 at arXiv.org.

\bibitem{Jacobsen_Saleur_08a}
   J.L. Jacobsen and H. Saleur, 
   Conformal boundary loop models, 
   {\em Nucl. Phys. B}\/ {\bf 788}, 137--166 (2008),
   math-ph/0611078 at arXiv.org. 

\bibitem{Jacobsen_Saleur_08b}
   J.L. Jacobsen and H. Saleur, 
   Combinatorial aspects of boundary loop models, 
   {\em J. Stat. Mech.}\/ (2008), P01021,
   arXiv:0709.0812 at arXiv.org.

\bibitem{Jacobsen_Saleur_08c}
   J.L. Jacobsen and H. Saleur, 
   Boundary chromatic polynomial,  
   {\em J. Stat. Phys.}\/ {\bf 132}, 707--719 (2008),
   arXiv:0803.2665 at arXiv.org.

\bibitem{Kasteleyn_69}  P.W. Kasteleyn and C.M. Fortuin,
   Phase transitions in lattice systems with random local properties,
   {\em J. Phys. Soc. Japan}\/ {\bf 26} (Suppl.), 11--14 (1969).

\bibitem{Kim-Enting_79}  D. Kim and I.G. Enting, 
   The limit of chromatic polynomials, 
   {\em J. Combin. Theory B}\/ {\bf 26}, 327--336 (1979). 

\bibitem{Kato_80}  T. Kato, {\em Perturbation Theory for Linear Operators}\/,
   $2^{nd}$ ed., corrected printing (Springer-Verlag, Berlin--New York, 1980).

\bibitem{Klazar_98}  M. Klazar,
   On trees and noncrossing partitions,
   {\em Discrete Appl. Math.}\/ {\bf 82}, 263--269 (1998).

\bibitem{Neef-Enting_77} T. de Neef and I.G. Enting, 
   Series expansions from the finite--lattice method,
   {\em J. Phys. A: Math. Gen.}\/ {\bf 10}, 801--805 (1977).

\bibitem{Potts_52}  R.B. Potts,
   Some generalized order-disorder transformations,
   {\em Proc. Cambridge Philos. Soc.}\/ {\bf 48}, 106--109 (1952).

\bibitem{Privman_90} V. Privman, Finite-size scaling theory,
 in {\em Finite Size Scaling and Numerical Simulation of Statistical Physics}\/,
 V. Privman (editor), pp.~1--98 (World Scientific, Singapore, 1990).

\bibitem{Procacci_03} A. Proccaci, B. Scoppola, and V. Gerasimov,
   Potts model on infinite graphs and the limit of chromatic polynomials,
   Commun. Math. Phys. {\bf 235}, 215--231 (2003),
   cond-mat/0201183 at arXiv.org.

\bibitem{Read_Royle} R.C. Read and G.F. Royle, Chromatic roots of
   families of graphs, in {\em Graph Theory, Combinatorics and Applications}\/,
   vol.~2, Y.~Alavi, G.~Chartrand, O.R.~Oellermann and A.J.~Schwenk (editors),
   pp.~1009--1029 (Wiley, New York, 1991).

\bibitem{Riordan_75}  J. Riordan,
    Enumeration of plane trees by branches and endpoints,
    {\em J. Combin. Theory A}\/ {\bf 19}, 214--222 (1975).

\bibitem{Shrock_98}  M. Ro\v{c}ek, R. Shrock and S.--H. Tsai, 
   Chromatic polynomials for $J(\prod H)I$ strip graphs and their
   asymptotic limits,  
   {\em Physica A}\/ {\bf 259}, 367--387 (1998),
   cond-mat/9807106 at arXiv.org.

\bibitem{transfer1}  J. Salas and A.D. Sokal,
   Transfer matrices and partition-function zeros for antiferromagnetic
   Potts models.  I.~General theory and square-lattice chromatic polynomial,
   {\em J. Stat. Phys.}\/ {\bf 104}, 609--699 (2001),
   cond-mat/0004330 at arXiv.org.

\bibitem{transfer5}  J. Salas and A.D. Sokal,
   Transfer matrices and partition-function zeros for antiferromagnetic
   Potts models.  V.~Further results for the square-lattice chromatic 
   polynomial, {\em J. Stat. Phys.}\/ {\bf 135}, 279--373 (2009),
   arXiv:0711.1738 at arXiv.org.

\bibitem{Shrock_00_TEMP}  R. Shrock, Exact Potts model partition functions
   on ladder graphs, {\em Physica A}\/ {\bf 283}, 388--446 (2000),
   cond-mat/0001389  at arXiv.org.

\bibitem{Shrock_BCC99}  R. Shrock,
   Chromatic polynomials and their zeros and asymptotic limits for
   families of graphs,
   %% 17th British Combinatorial Conference (Canterbury, 1999).
   {\em Discrete Math.}\/ {\bf 231}, 421--446 (2001),
   cond-mat/9908387 at arXiv.org.

\bibitem{Shrock_97a}  R. Shrock and S.-H. Tsai,
   Asymptotic limits and zeros of chromatic polynomials and
   ground state entropy of Potts antiferromagnets,
   {\em Phys. Rev. E}\/ {\bf 55}, 5165--5178 (1997),
   cond-mat/9612249 at arXiv.org.

\bibitem{Shrock_97c}  R. Shrock and S.-H. Tsai,
   Families of graphs with $W_r(\{G\},q)$ functions that are nonanalytic
   at $1/q=0$,
   {\em Phys. Rev. E}\/ {\bf 56}, 3935--3943 (1997),
   cond-mat/9707096 at arXiv.org.

\bibitem{Shrock_98b}  R. Shrock and S.-H. Tsai,
   Ground state degeneracy of Potts antiferromagnets: cases with
   non-compact $W$ boundaries having multiple points at $1/q=0$, 
   {\em J. Phys. A: Math. Gen.}\/ {\bf 31}, 9641--9665 (1998),
   cond-mat/9810057 at arXiv.org.

\bibitem{Simion_91}  R. Simion and D. Ullman,
    On the structure of the lattice of noncrossing partitions,
    {\em Discrete Math.}\/ {\bf 98}, 193--206 (1991).

\bibitem{Sloane_on-line}
   N.J.A. Sloane, The On-Line Encyclopedia of Integer Sequences,
   \url{http://www.research.att.com/~njas/sequences/index.html}

\bibitem{Sokal_chromatic_bounds}  A.D. Sokal,
   Bounds on the complex zeros of (di)chromatic polynomials and
   Potts-model partition functions,
   {\em Combin. Probab. Comput.}\/ {\bf 10}, 41--77 (2001),
   cond-mat/9904146 at arXiv.org.

\bibitem{Sokal_chromatic_roots}  A.D. Sokal,
   Chromatic roots are dense in the whole complex plane,
   {\em Combin. Probab. Comput.}\/ {\bf 13}, 221--261 (2004),
   cond-mat/0012369 at arXiv.org.

\bibitem{Sokal_bcc2005}  A.D. Sokal, The multivariate Tutte polynomial
   (alias Potts model) for graphs and matroids,
   in Bridget S.~Webb (editor),
   {\em Surveys in Combinatorics, 2005}\/, pp.~173--226
   (Cambridge University Press, Cambridge--New York, 2005),
   math.CO/0503607 at arXiv.org.

\bibitem{Stanley_86}  R.P. Stanley, {\em Enumerative Combinatorics}\/,
   vol. 1 (Wadsworth \& Brooks/Cole, Monterey, CA, 1986).
   Reprinted by Cambridge University Press, 1999.

\bibitem{Stanley_99}  R.P. Stanley, {\em Enumerative Combinatorics}\/,
   vol. 2 (Cambridge University Press, Cambridge--New York, 1999).

\bibitem{Tutte_47}  W.T. Tutte,
   A ring in graph theory,
   {\em Proc. Cambridge Philos. Soc.}\/ {\bf 43}, 26--40 (1947).

\bibitem{Tutte_54}  W.T. Tutte,
   A contribution to the theory of chromatic polynomials,
   {\em Canad. J. Math.}\/ {\bf 6}, 80--91 (1954).

\bibitem{Whitney_32a}  H. Whitney,
   A logical expansion in mathematics,
   {\em Bull. Amer. Math. Soc.}\/ {\bf 38}, 572--579 (1932).

\bibitem{Wu_82}  F.Y. Wu, The Potts model,
   {\em Rev. Mod. Phys.}\/ {\bf 54}, 235--268 (1982);
   erratum {\bf 55}, 315 (1983).

\bibitem{Wu_84}  F.Y. Wu, Potts model of magnetism (invited),
   {\em J. Appl. Phys.}\/ {\bf 55}, 2421--2425 (1984).

\bibitem{Yang-Lee_52}  C.N. Yang and T.D. Lee,
   Statistical theory of equations of state and phase transitions.\  
   I.~Theory of condensation,
   {\em Phys. Rev.}\/ {\bf 87}, 404--409 (1952)

\end{thebibliography}
\end{document}